\UseRawInputEncoding
\documentclass[twocolumn]{aastex701} %  % manuscript for double-space % 
\usepackage{CJKutf8,CJK}    % or simply CJK for legacy encodings

\usepackage[printonlyused]{acronym}
\usepackage{calc}
\newlength{\okinalen}
\newcommand{\oumuamua}{1I/\hbox to.666\okinalen{\hss`\hss}Oumuamua}
\newcommand{\borisov}{2I/Borisov}

\newcommand{\objname}{3I/ATLAS} % 5/18/2024 COC

\newcommand{\objnameFull}{3I/ATLAS, also designated C/2025 N1 (ATLAS)} % 7/17/2024 COC/MKF
\newcommand{\rubinobs}{NSF-DOE Vera C. Rubin Observatory}

\usepackage{booktabs}
\usepackage{graphicx}
\usepackage[outline]{contour} % 8/7/2022 COC -- outline text
\usepackage[percent]{overpic} % 8/14/2021 COC superimpose text on pictures

\newcommand{\labelcolor}{yellow}

\newcommand{\labelpicA}[5]{
\begin{overpic}[width=#4\linewidth]{#1}
	\put (5,7) {\huge\color{\labelcolor} \textbf{\contour{black}{#2}}}
	\put (55,8) {\large\color{\labelcolor} \textbf{\contour{black}{#3}}}
	\put (2,55) {\includegraphics[width=0.13\linewidth]{#5}}
    \put (0,0) {\includegraphics[width=0.31\linewidth]{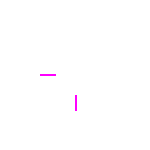}}
\end{overpic}
}

\newcommand{\labelpicB}[5]{
\begin{overpic}[width=#4\linewidth]{#1}
	\put (5,7) {\huge\color{\labelcolor} \textbf{\contour{black}{#2}}}
	\put (40,8) {\large\color{\labelcolor} \textbf{\contour{black}{#3}}}
	\put (2,38) {\includegraphics[width=0.11\linewidth]{#5}}
    \put (0,0) {\includegraphics[width=0.18\linewidth]{reticle_150X150.png}}
\end{overpic}
}

\newcommand{\labelpicBb}[5]{
\begin{overpic}[width=#4\linewidth]{#1}
	\put (5,7) {\huge\color{\labelcolor} \textbf{\contour{black}{#2}}}
	\put (25,8) {\large\color{\labelcolor} \textbf{\contour{black}{#3}}}
	\put (2,38) {\includegraphics[width=0.11\linewidth]{#5}}
    \put (0,0) {\includegraphics[width=0.18\linewidth]{reticle_150X150.png}}
\end{overpic}
}

\newcommand{\labelpicD}[5]{
\begin{overpic}[width=#4\linewidth]{#1}
	\put (5,7) {\huge\color{\labelcolor} \textbf{\contour{black}{#2}}}
	\put (40,8) {\large\color{\labelcolor} \textbf{\contour{black}{#3}}}
	\put (2,77) {\includegraphics[width=0.2\linewidth]{#5}}
\end{overpic}
}

\revised{March 23, 2026}
\submitjournal{ApJL}

\begin{document}

\title{
NSF-DOE Vera C. Rubin Observatory Observations of Interstellar Comet 3I/ATLAS (C/2025 N1)
}

\newcommand{\linccfw}{LSST Interdisciplinary Network for Collaboration and Computing Frameworks, 933 North Cherry Avenue, Tucson, AZ 85721, USA}
\newcommand{\diracuw}{Dept. of Astronomy \& the DiRAC Institute, University of Washington, Box 351580, Seattle, WA 98195, USA}
\newcommand{\nau}{Department of Astronomy and Planetary Science, Northern Arizona University, Flagstaff, USA}
\newcommand{\harvard}{Center for Astrophysics $\vert$ Harvard \& Smithsonian, 60 Garden St, Cambridge, MA 02138, USA}
\newcommand{\aerospaceillinois}{The Grainger College of Engineering, Department of Aerospace Engineering, University of Illinois Urbana-Champaign, Urbana, USA}
\newcommand{\astroillinois}{Department of Astronomy, University of Illinois at Urbana-Champaign, Urbana, IL 61801, USA}
\newcommand{\rubinobschile}{Vera C. Rubin Observatory, Avenida Juan Cisternas \#1500, La Serena, Chile}
\newcommand{\rubinprojectoffice}{Rubin Observatory Project Office, 950 N Cherry Ave, Tucson, AZ 85719, USA}
\newcommand{\planetaryscienceinst}{Planetary Science Institute, 1700 East Fort Lowell Rd., Suite 106, Tucson, AZ 85719, USA}
\newcommand{\slac}{SLAC National Accelerator Laboratory, 2575 Sand Hill Rd., Menlo Park, CA 94025, USA}
\newcommand{\kavli}{Kavli Institute for Particle Astrophysics and Cosmology, Stanford University, 2575 Sand Hill Road, M/S 29, Menlo Park, CA 94025, USA}
\newcommand{\kavlistanford}{Kavli Institute for Particle Astrophysics and Cosmology, Physics and Astrophysics Building, 452 Lomita Mall, Stanford, CA 94305-4085, USA} % pjm said this 3/21/2026 COC
\newcommand{\STScI}{Space Telescope Science Institute, 3700 San Martin Drive, Baltimore, MD 21218, USA}
\newcommand{\SWRI}{Southwest Research Institute, 1301 Walnut Street, Suite 400, Boulder, CO 80302, USA}
\newcommand{\linea}{Laborat\'{o}rio Interinstitucional de e-Astronomia - LIneA, Av. Pastor Martin Luther King Jr, 126 Del Castilho, Nova Am\'{e}rica Offices, Torre 3000/sala 817 CEP: 20765-000, Brazil}
\newcommand{\NOIRlab}{NSF National Optical-Infrared Astronomy Research Laboratory, 950 North Cherry Avenue, Tucson, AZ 85719, USA}
\newcommand{\qubuk}{Astrophysics Research Centre, School of Mathematics and Physics, Queen's University Belfast, BT7 1NN, UK}
\newcommand{\uofcanterbury}{School of Physical and Chemical Sciences --- Te Kura Mat\={u}, University of Canterbury, Private Bag 4800, Christchurch 8140, New Zealand}
\newcommand{\uofhelsinki}{Department of Physics, P.O.\ Box 64, 00014 University of Helsinki, Finland}
\newcommand{\physoxford}{Department of Physics, University of Oxford, Denys Wilkinson Building, Keble Road, Oxford, OX1 3RH, UK}
\newcommand{\physharvard}{Department of Physics, Harvard University, 17 Oxford St., Cambridge MA 02138, USA}
\newcommand{\ucfphys}{Department of Physics, University of Central Florida, 4111 Libra Drive, Physical Sciences Bldg. 430,
Orlando, FL 32816-2385, USA}
\newcommand{\princeton}{Department of Astrophysical Sciences, Princeton University, Princeton, NJ 08544, USA}
\newcommand{\marylandastro}{Department of Astronomy, University of Maryland
College Park, MD 20742, USA}
\newcommand{\dukephys}{Department of Physics, Duke University, Durham, NC 27708, USA144}
\newcommand{\dukeece}{Department of Electrical and Computer Engineering, Duke University, Durham, NC 27708, USA}
\newcommand{\ncsa}{National Center for Supercomputing Applications, University of Illinois Urbana-Champaign, Urbana, IL 61801, USA} % looked up to triple-check 3/24/2026 COC
\newcommand{\edinburgh}{Institute for Astronomy, University of Edinburgh, Royal Observatory Edinburgh, Blackford Hill, Edinburgh, EH9 3HJ, UK}
\newcommand{\seti}{SETI Institute, 339 Bernardo Ave, Suite 200, Mountain View, CA 94043, USA}
\newcommand{\lpl}{Lunar and Planetary Laboratory, University of Arizona, 1629 E University Blvd, Tucson, AZ 85721}
\newcommand{\lco}{Las Cumbres Observatory, 6740 Cortona Drive Suite 102, Goleta, CA 93117, USA}
\newcommand{\mcti}{Observatório Nacional/MCTI, R. General José Cristino 77, CEP 20921-400 Rio de Janeiro - RJ, Brazil}
\newcommand{\saclay}{Universit\'e Paris-Saclay, CNRS/IN2P3, IJCLab, 91405 Orsay, France}
\newcommand{\aoipoland}{Astronomical Observatory Institute, Adam Mickiewicz University, S{\l}oneczna 36, 60-286 Pozna\'n, Poland}
\newcommand{\brookhaven}{Brookhaven National Laboratory, Upton, NY 11973, USA}
\newcommand{\camino}{Departamento de Astronom\'ia, Universidad de Chile, Camino del Observatorio 1515, Las Condes, Santiago, Chile}
\newcommand{\dstparthenope}{Department of Science and Technology, Parthenope University of Naples, Centro Direzionale C4 Island, 80143, Naples, Italy}
\newcommand{\davis}{Department of Physics and Astronomy, University of California, Davis, One Shields Avenue, Davis, CA 95616, USA}
\newcommand{\aura}{Association of Universities for Research in Astronomy, 1331 Pennsylvania Ave NW, Washington, DC 20004}
\newcommand{\mpc}{Minor Planet Center -- Center for Astrophysics $\vert$ Harvard \& Smithsonian, 60 Garden St., MS 15, Cambridge (MA), USA}
\newcommand{\sorbonne}{Sorbonne Université, CNRS/IN2P3, Laboratoire de Physique
Nucléaire et de Hautes Energies (LPNHE), FR-75005 Paris, France}
\newcommand{\rubinops}{Vera C.\ Rubin Observatory/NSF NOIRLab, 950 N.\ Cherry Ave., Tucson, AZ  85719, USA}
\newcommand{\nrc}{National Research Council of Canada, Herzberg Astronomy and Astrophysics Research Centre, 5071 West Saanich Road, Victoria, BC V8T 1E7, Canada}
\newcommand{\ukansas}{Department of Physics and Astronomy, University of Kansas, Malott Hall Room 1082, 1251 Wescoe Hall Drive, Lawrence, KS, 66045, USA}
\newcommand{\uwisconsin}{Department of Physics, University of Wisconsin, Madison, 1150 University Avenue, Madison, WI 53706, USA}
% Repeated affiliations via ChatGPT 3/23/2026 COC
\newcommand{\uvicphys}{Department of Physics and Astronomy, University of Victoria, Elliott Building, 3800 Finnerty Road, Victoria, BC V8P 5C2, Canada}

\newcommand{\amuastro}{Astronomical Observatory Institute, Faculty of Physics and Astronomy Adam Mickiewicz University, ul. Słoneczna 36, 60-286 Poznań, Poland}

\newcommand{\byuphys}{Brigham Young University, Department of Physics and Astronomy, N283 ESC, Provo, UT 84602, USA}

\newcommand{\cnrsccinp}{CNRS, CC-IN2P3, 21 Avenue Pierre de Coubertin, CS70202, F-69627 Villeurbanne Cedex, France}

\newcommand{\capodimonte}{INAF, Osservatorio Astronomico di Capodimonte, Salita Moiariello, 16, Naples, I-80131, Italy}

\newcommand{\jplcaltech}{Jet Propulsion Laboratory, California Institute of Technology, 4800 Oak Grove Dr., Pasadena, CA 91109, USA}

% Paper lead; cutouts; morphology, thermo, ...; processing; ToO coordination; ...
\correspondingauthor{Colin Orion Chandler}
\email{coc123@uw.edu}
\author[0000-0001-7335-1715]{Colin Orion Chandler}
\email{coc123@uw.edu}
\affiliation{\linccfw}
\affiliation{\diracuw}
\affiliation{\nau}

% NOTE: MUST DISABLE THE FOLLOWING LINE FOR LaTeX SUBMISSION TO JOURNAL (corresponding author ONLY); PDF version MUST include this line!
% \input{authors}

% linking, fitting, science, Rubin calibration, ...
\author[0000-0003-0743-9422]{Pedro H. Bernardinelli} % linking, science, ...
\email{phbern@uw.edu}
\altaffiliation{DiRAC Post-doctoral Fellow}
\affiliation{\diracuw}

% everything
\author[0000-0003-1996-9252]{Mario Juri\'c}
\email{mjuric@uw.edu}
\affiliation{\diracuw}

% Introduction, citations, author management, early lightcurve and other colin assigned sidequests :), colin's 3I observing team etc.+++
\author[0009-0004-0059-3229]{Devanshi Singh}
\email{ds2004@uw.edu}
\affiliation{\diracuw}

% Activity consulation, dust modeling, Colin reality checks, ...
\author[0000-0001-7225-9271]{Henry H. Hsieh}
\affiliation{\planetaryscienceinst}
 \email{hhsieh@psi.edu}  

% Much, much reprocessing and reprocessing support
\author[0000-0001-8708-251X]{Ian Sullivan}
\email{sullii@uw.edu}
\affiliation{\diracuw}

\author[0000-0001-5916-0031]{R. Lynne Jones}
\email{ljones.uw@gmail.com}
\affiliation{\rubinobschile}
\affiliation{Aston Carter, Victoria BC, Canada} % from Aerotek, per Google Form 3/19/2026 COC

% Butler catalog help
\author[0009-0005-5452-0671]{Jacob A. Kurlander}
\email{jkurla@uw.edu}
\affiliation{\diracuw}

% lightcurve+
\author[0009-0007-1972-5975]{Dmitrii Vavilov}
\email{vavilov@uw.edu}
\affiliation{\diracuw}

% orbit fitting
\author[0000-0002-1398-6302]{Siegfried Eggl}
\email{eggl@illinois.edu}
\affiliation{\aerospaceillinois}
\affiliation{\astroillinois}

% orbit fitting
% ...

% Punch List
% Henry: new H_V and uncertainty throughout because we have new H_r
% Henry: decide if we want subsubsec:coma refs to point somewhere new?
% corresponding new diameter estimates throughout
% Astrometric residual plots: Rahil to replace - done
% Rahil: Orbit Determination - done
% Pedro: replace the two horizontal panel plot that has the Rubin astrometry histograms under them
% Joe Murtagh: fold in new Discoverability section text, figure?
% Pedro: colors: add column to table with measures; re-derive S'.
% Colors to abstract, summary

\newcommand{\colorGminR}{$0.657\pm0.013$}
\newcommand{\colorRminI}{$0.235\pm0.018$}
\newcommand{\colorIminZ}{$0.147\pm0.042$}
\newcommand{\colorZminY}{$0.047\pm0.052$}
\newcommand{\Hvmag}{$13.32\pm 0.16$ mag}
\newcommand{\Himag}{$12.93 \pm 0.16$ mag}
\newcommand{\Hrmag}{$13.16 \pm 0.16$ mag}
\newcommand{\radNucl}{($6.6 \pm 0.5$)~km}
% \newcommand{\hR}{$13.32\pm 0.16$}

%%%%%%%%%%%%%%%%%%%%%%%%%%%%%%%%%%%%%
\begin{abstract}
We report on the observation and measurement of astrometry, photometry, morphology, and activity of the interstellar object \objnameFull{} with the NSF-DOE Vera C.\ Rubin Observatory. Comet \objname{}, the third known interstellar object, was discovered on UT 2025 July 1. Rubin Observatory had coincidentally collected images of the object's region of the sky during routine commissioning. Facilitated by Rubin's high resolution and large aperture, we successfully recovered object detections from Rubin observations spanning UT 2025 June 21 (10 days before discovery, when \objname{} was 4.5~au from the Sun) through the date of discovery, and we acquired additional images through UT 2025 July 20 as part of commissioning. %Facilitated by Rubin's high resolution and large aperture, we report on the detection of cometary activity as early as June 21st, and observe it throughout. 
We measure on-sky locations of \objname{} in Rubin $ugrizy$ bands, with a typical precision of $\sim 70$~mas, and briefly describe the reason this is coarser than our measured static source astrometric precision of $\sim3$~mas in Rubin images. We measure $grizy$ magnitudes of \objname{} photometry at $\sim 0.01$~mag precision, detecting no short-term photometric variability above 0.01 mag. We derive an estimated near-nucleus dust-to-nucleus scattering cross-section ratio of $\eta\gtrsim13$ on UT 2025 July 2 based on Rubin photometry and an upper limit nucleus size computed from Hubble Space Telescope observations. We find Rubin colors of $g-r=$ (\colorGminR{}) mag, $r-i=$ (\colorRminI{}) mag, $i-z=$ (\colorIminZ{}) mag, $z-y=$ (\colorZminY{}) mag. % Pedro colors here
%We derive a lower-limit $V$-band absolute magnitude of $H_V$ = (13.7 $\pm$ 0.2)~mag, and an equivalent upper-limit effective nucleus radius of $\sim$ (6.6 $\pm$ 0.5)~km. 
These data represent the earliest observations of this object by a large ($\gtrsim$8-meter class) telescope and illustrate the type of measurements (and discoveries) Rubin's \ac{LSST} will begin to provide after it begins in early 2026.
\end{abstract}

\keywords{
Comae (271), 
Comet tails (274),
Interstellar objects (52) 
}

\acresetall % our new friend; resets memory of used acronyms for TeX package 11/1/2023 COC

%%%%%%%%%%%%%%%%%%%%%%%%%%%%%%%%%%%%%%%%
\section{Introduction}\label{sec:intro}

\setcounter{footnote}{0}

\subsection{Background}\label{subsec:background}

\begin{figure}
    \centering    
    \labelpicD{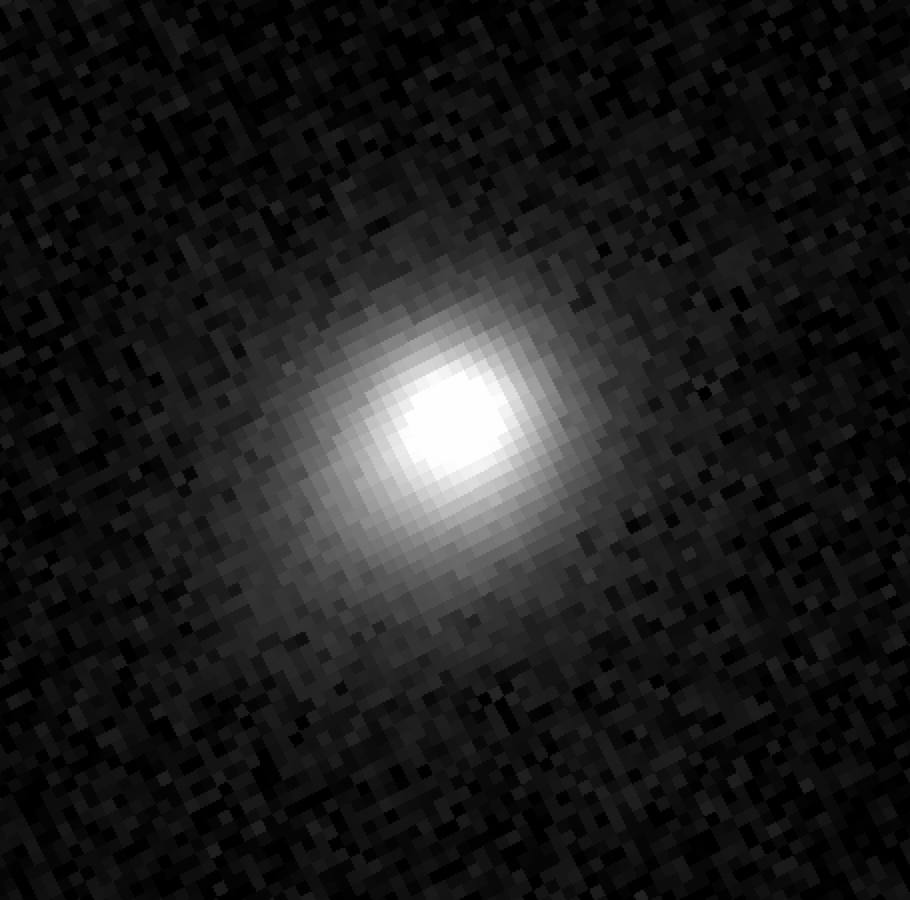}{}{}{1.0}{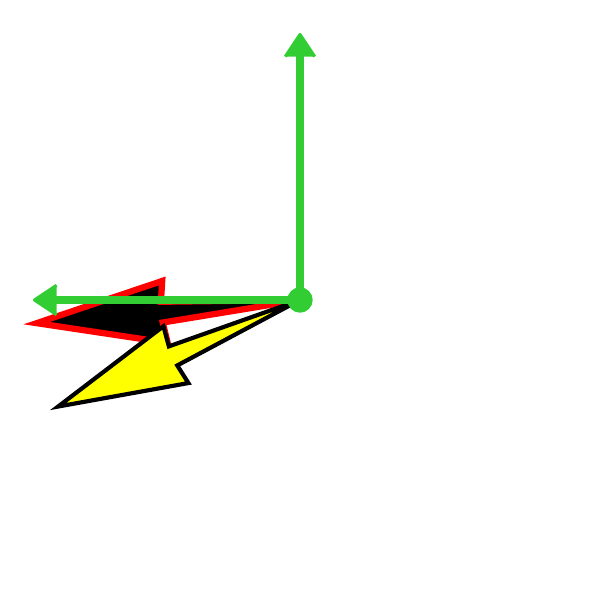
}
    \caption{\objname{} minimum co-addition from UT 2025 July 3, 21$\times$30~s $r$-band images acquired by \rubinobs{}. The FOV is $\sim15\arcsec \times 15\arcsec$, with north up and east left. 
    The anti-solar (yellow arrow) and anti-motion (red-outlined black arrow) directions are shown.
    }
    \label{fig:prettypicture}
\end{figure}

\acp{ISO} are small bodies detected in the solar system but not gravitationally bound to the Sun. 
They are presumed to be passing through our planetary system but to have originated in exoplanetary systems in our Galaxy. 
Studies of these objects can reveal the bulk compositions and surface processing of extrasolar planetesimals, as well as the dynamical footprints of the planetary systems that ejected them \citep{fitzsimmonsreviewcomets3, seligmanmororeview, jewitteseligman, mororeview}. % TODO add trilling 2017?
However, our knowledge of this population remains sparse, as only three have been identified: \oumuamua{} \citep{Williams2017}, \borisov{} \citep{Borisov2019}, and now, \objname{} (Figure \ref{fig:prettypicture}), discovered on UT 2025 July 1 \citep{Denneau20253I}. %\citep{deen2025cbet5578}.

The first interstellar object, \oumuamua{}, discovered in October 2017, did not show any visible coma or outgassing \citep{meech2017_oumuamua, ye2017_oumuamua, jewitt2017_oumuamua, trilling2018_oumuamua}, yet the object exhibited significant non-gravitational acceleration \citep{micheli2018_oumuamua, oumuamuaissi2019_oumuamua} and an unusually large light curve amplitude indicating an elongated or flattened, tumbling shape \citep{Bannister:2017, drahus2017_oumuamua, knight2017_oumuamua, belton2018_oumuamua, bolin2018_1I, fraser2018_oumuamua, mcneill2018_oumuamua, maschenko2019_oumuamua, vavilov2019_oumuamua}. 
\borisov{}, discovered in August 2019, became the second interstellar object identified in our Solar System. In contrast to \oumuamua{}'s visibly inert nature, \borisov{} displayed a pronounced dust and gas coma, making it the first interstellar comet \citep{jewitt2019_borisov, fitzsimmons2019_borisov, bolin2020_borisov, ye2020_borisov, mckay, guzik2020_borisov, hui, kimanisotropy, cremonese, Yang_2021, Deam:2025, leon2019, leon2020}. 
Unlike other inner Solar System comets, \borisov{} had a much higher \( \mathrm{CO}/\mathrm{H}_2\mathrm{O} \) abundance ratio \citep{bodewits2020_borisov, cordiner}. 

%Pan-STARRS should be an acronym 

\subsection{\objname{}}\label{subsec:3I_intro}

The third interstellar object, \objnameFull{}, was first identified on UT 2025 July 1 \citep{Denneau20253I} by the \ac{ATLAS} \citep{ATLASmainpaperperchance}. It quickly became clear that the object was of interstellar origin, with an eccentricity exceeding 6\footnote{\url{https://ssd.jpl.nasa.gov/tools/sbdb\_lookup.html\#/?sstr=C\%2F2025\%20N1}}.
Numerous follow-up observers quickly identified the presence of a faint cometary tail, establishing the object as a weakly active cometary body \citep[e.g.,][]{deen2025cbet5578,Jewitt2025,Alarcon2025}. % ,bolin2025_3I,seligman2025_3I,opitom_3I}. % added Deen here 12/13/2025 COC; COC adjusted to just have the "quick reports" (around object discovery time) 2/17/2026 COC
Multi-filter imaging and spectroscopy by multiple observers \citep{alvarezcandal2025_3I,bolin2025_3I,champagne2025_3I,deLaFuenteMarcos2025,opitom_3I,seligman2025_3I} indicated that the object had a moderately red spectral gradient of $S'\sim18$\%/100~nm, at least some of which was likely due to a dust coma given the object's visible activity at the time of those observations.

Meanwhile, no gas emission was detected in \ac{VLT}/\ac{MUSE} observations on UT 2025 July 3 \citep{opitom_3I}, when the object was at a heliocentric distance of $r_h=4.47$~au. \citet{bolin2025_3I} computed \objname{}, then at $r_H=4.43$~AU, had a non-nucleus-subtracted $A(0^\circ)f\rho$ of $280.8\pm3.2\ \mathrm{cm}$ and estimated a dust mass-loss rate of $0.1$--$1.0\ \mathrm{kg\,s}^{-1}$, noting that this level of activity was comparable to that of \borisov{} \citep{prodan2024} % TODO what is the value?
but far exceeding that of \oumuamua{}. % TODO what is the value?

Nucleus measurements have also been attempted, but should be regarded with caution, given the object's visible activity.
%The strongest constraint on the nucleus size to date is likely provided by precovery observations on UT 2025 May 22 obtained by the \ac{ZTF}, from which 
\citet{seligman2025_3I} used precovery observations obtained by the \ac{ZTF} on UT 2025 May 22 to derive a lower limit $V$-band absolute magnitude of $H_V=12.4$~mag, corresponding to an upper limit effective radius of $r_n\sim10$~km, assuming a comet-like albedo ($p_V = 0.05$).
However, the strongest constraint on the nucleus size to date is likely provided by 
\citet{jewitt2025_hst3I} who constrained the nucleus size to $r\leq2.8$~km based on a fit to the surface brightness distribution of the inner coma as observed by the \ac{HST} on UT 2025 July 21.
The discrepancy between these size estimates indicates that there was already substantial coma present at the time of those early \ac{ZTF} precovery observations.

\citet{seligman2025_3I} reported measured $g'-r'$ colors from  \ac{ZTF} precovery data that were nearly solar, suggesting that the object's nucleus may be less red than the ejected dust, given later observations (which include increased flux contributions from a by-then detectable coma) of redder spectral slopes for the object. However, as the \ac{ZTF} multi-filter observations were not simultaneous (and therefore subject to potential light curve variability), and substantial unresolved coma is present even in those earlier data (as discussed above), these results should be regarded as uncertain. 

Meanwhile, time-series observations have revealed a largely flat light curve \citep[amplitude $<0.2$~mag; e.g.,][]{seligman2025_3I,champagne2025_3I,deLaFuenteMarcos2025}, suggesting that the object could be nearly spherical, the rotational pole could be close to the line of sight with respect to the Earth, nucleus rotational light curve variations might be damped by a nearly steady-state coma, or a combination of all of these factors. 
Given that substantial coma was already present by the time 3I was discovered, however, the most likely explanation is that reported time-series photometry were dominated by relatively steady-state coma flux with minimal to no signatures of any rotational variation from the nucleus.
%As the coma is already resolved, it is likely that the latter explanation is dominant, and the nucleus is considerably smaller than 10~km in radius. 

Based on the published data from the \ac{ATLAS} survey’s discovery of \objname{} at the time,  \citet{loeb2025commentdiscoverypreliminarycharacterization} notes that \objname{}'s earlier estimated nucleus size of $r_n \sim 10\ \mathrm{km}$ implies a local number density $n_{0} \sim 3\times10^{-4}\,\mathrm{au}^{-3}$, which would overproduce interstellar mass. (\citealt{jewitt2025_hst3I}, with \ac{HST}, later found $0.22 < r_n < 2.8$~km.) 
To resolve this mass-budget conflict, they concluded that \objname{} must either be a comet with a small nucleus ($r_n \lesssim 0.6\ \mathrm{km}$) or a rare $\sim 10\ \mathrm{km}$ body with $n_{0} \lesssim 5\times10^{-8}\,\mathrm{au}^{-3}$. \citet{seligman2025_3I} concluded that \objname{} has a number density \(n_0 \sim 10^{-3}\,\mathrm{au}^{-3}\). % TODO so what about this Seligman statement? Hanging

\citet{hopkins2025differentstar3iatlascontext} made use of \textit{Gaia} DR3 and \objname{}'s orbital arc (fit as of 2025 July 3) to show \objname{}'s velocity is outside of the main moving-group structures in the \={O}tautahi\textendash Oxford \ac{ISO} model, unlike those of \oumuamua{} and \borisov{}.
\objname{} has a high velocity perpendicular to the Galactic plane, creating its unusually Southern direction of approach for an ISO. They rule out any common origin with \oumuamua{} or \borisov{} based on kinematic exclusion, and derive a broad age range for \objname{} as most likely older than 7 Gyr, underscoring the large uncertainty in ISO age estimates \citep{hopkins1, marcetaseligman}; this age is supported by the subsequent $\sim$ 3 - 11 Gyr estimate of \citet{taylor2025kinematicage3iatlasimplications}.
 % or maybe its the USS Enterprise (NCC-1701) ^^ yes please

%%%%%%%%%%%%%%%%%%%%%%%%%%%%%%%%%%%%%%%%%%%%%%%%%%%%%%%%%%%%%%%%%%
\section{The Vera C. Rubin Observatory}\label{subsec:rubinobs}

The \rubinobs{} \citep{ivezic2019}\footnote{\url{https://rubinobservatory.org}} resides atop Cerro Pach\'on in Chile, hosting the 8.4~m Simonyi Survey Telescope \citep{2024SPIE13094E..09S}. It is equipped with the 3.2~gigapixel \ac{LSST} Camera \citep[LSSTCam; ][]{LSSTCam, Roodman2024}, which covers a $9.6~\mathrm{deg}^2$ field of view sampled at $0.2\arcsec$ per pixel to a depth of \(m_r \sim 24\) with a 30~s exposure. Rubin Observatory (site code X05) will begin the \ac{LSST}, an ambitious ten-year imaging survey of the southern sky, in early 2026. \citet{ivezic2019} provides an overview of Rubin and the \ac{LSST}, including the science drivers that led to its specific configuration. During its main survey, the \ac{LSST} will obtain observations from near-UV to near-IR ($ugrizy$ bandpasses) with approximately 30-second exposures, imaging the entire visible southern sky repeatedly every few nights. \citet{Bianco2022}providese a more detailed overview of the survey strategy. 

Rubin Observatory, which began acquiring images with the full LSSTCam in 2025 April, was in its commissioning period, carrying out \ac{SV} observations, until the end of 2025 \citep[see Section 6 of][]{SITCOMTN-005}. During this period, data quality and acquisition rate were variable. Summit systems, including the \ac{AOS}, were still being refined, and observing procedures, such as interrupts for \ac{ToO} (Figure \ref{fig:gallery4}), were still being tested. Not all data products that will be available during \ac{LSST} are currently available. In particular, the \ac{SV} survey has not yet progressed far enough to have sufficient images available to create templates for difference imaging across its footprint.

The \ac{SV} survey began limited operations on 2025 June 20, running for approximately 50\% of the available nightly telescope time %5 hours out of the 10-hour nights over 
and covering approximately 1,447 $\mathrm{deg}^2$. This first night of the \ac{SV} survey coincidentally covered the position of \objname{} -- the first detection of \objname{} (and any \ac{ISO}) with Rubin (Figure \ref{fig:gallery}a). This observation, along with the following nights (UT 2025 June 21-24), provide the earliest high-resolution, deep observations of \objname{} and a unique opportunity to study its brightness and morphology at larger distances before its perihelion passage (UT 2025 October 29).

%%%%%%%%%%%%%%%%%%%%%%%%%%%%%%%%%%%%%%%%%%%%%%%%%%%%%%%

\newcommand{\figsize}{0.31}
\begin{figure*}
    \centering
    % NOTE: \INPUT FILES ARE AUTOMATICALLY REWRITTEN
        % \input{gallery}
        %% Generated on 2025-10-06T22:08:47.790978+00:00 UTC
\begin{tabular}{ccc}
\labelpicA{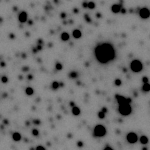}{a}{2025-06-21}{\figsize}{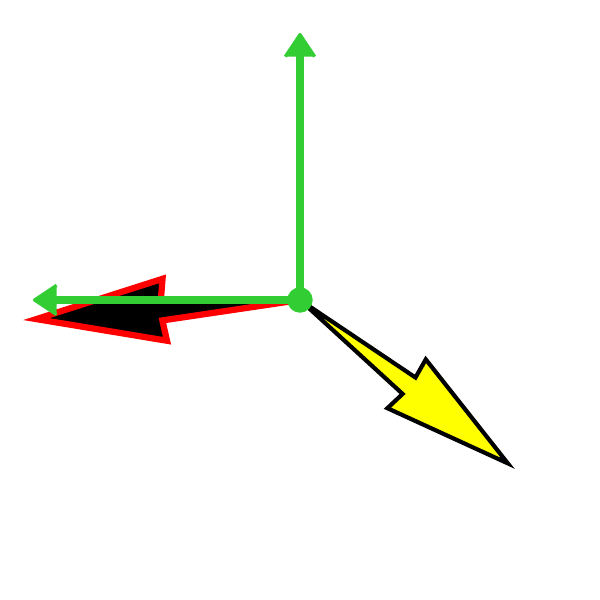} & \labelpicA{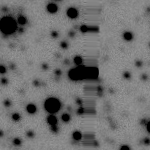}{b}{2025-06-22}{\figsize}{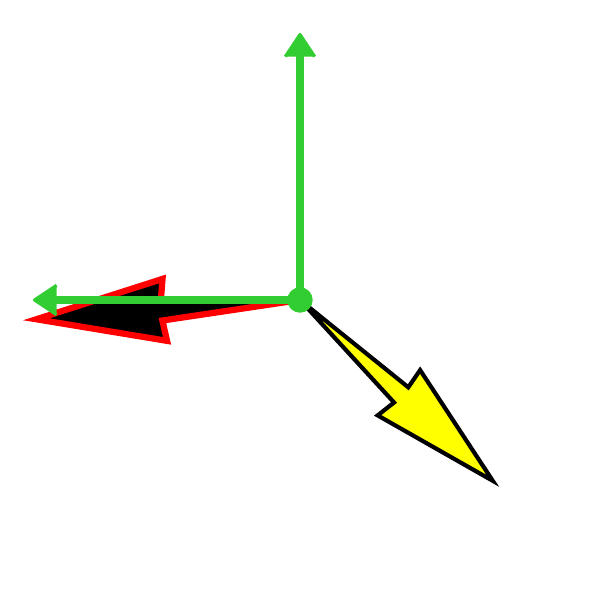} & \labelpicA{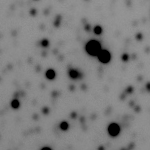}{c}{2025-06-22}{\figsize}{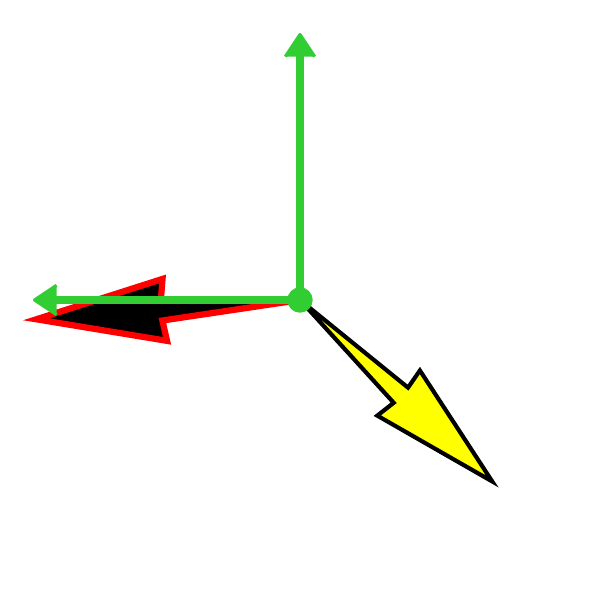} \\
\labelpicA{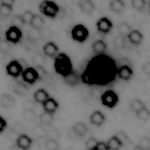}{d}{2025-06-24}{\figsize}{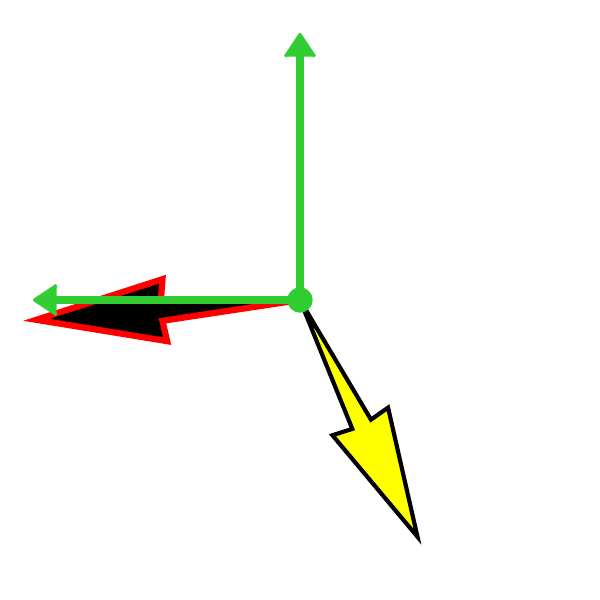} & \labelpicA{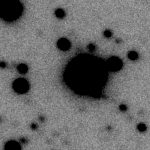}{e}{2025-06-30}{\figsize}{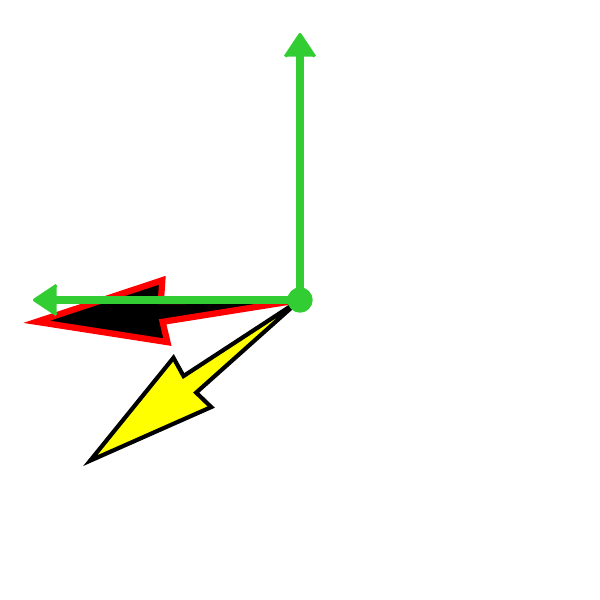} & \labelpicA{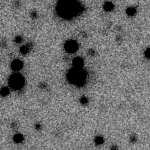}{f}{2025-07-02}{\figsize}{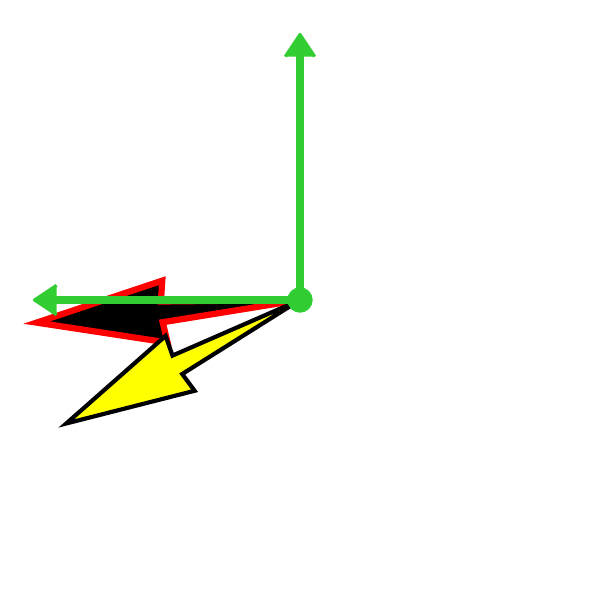} \\
\labelpicA{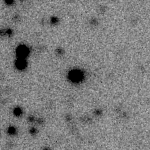}{g}{2025-07-02}{\figsize}{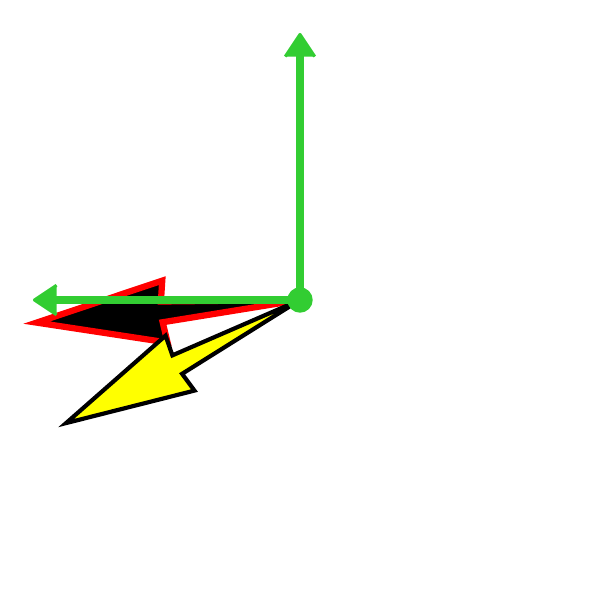} & \labelpicA{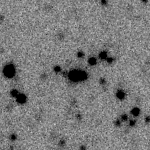}{h}{2025-07-02}{\figsize}{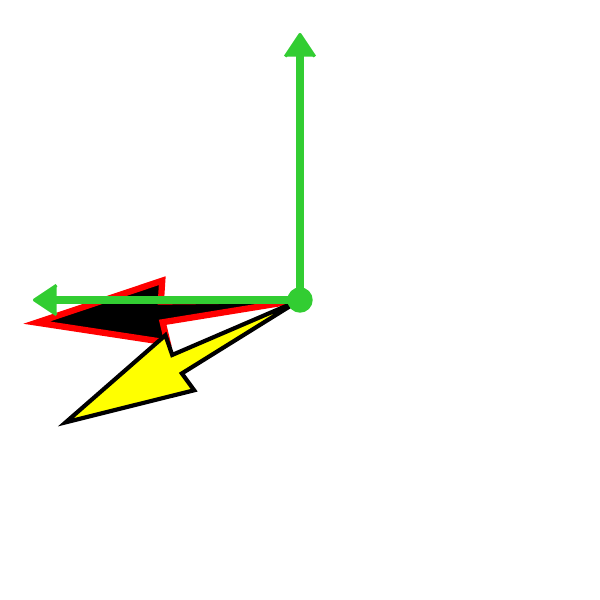} & \labelpicA{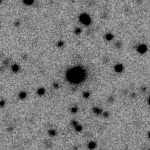}{i}{2025-07-02}{\figsize}{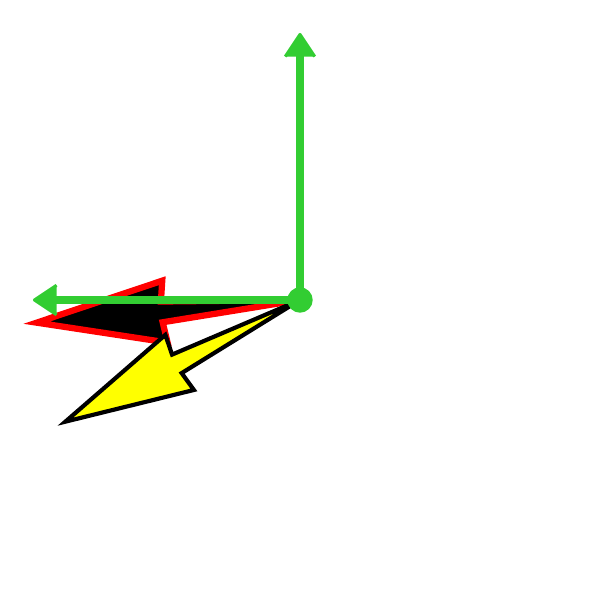} \\
\end{tabular}

%% Label-to-date mapping:
%% \textbf{(a)} 2025-06-21 08:11:32
%% \textbf{(b)} 2025-06-22 02:32:47
%% \textbf{(c)} 2025-06-22 03:07:49
%% \textbf{(d)} 2025-06-24 03:07:46
%% \textbf{(e)} 2025-06-30 02:25:46
%% \textbf{(f)} 2025-07-02 00:44:25
%% \textbf{(g)} 2025-07-02 01:20:33
%% \textbf{(h)} 2025-07-02 02:31:16
%% \textbf{(i)} 2025-07-02 03:33:02

    \caption{Gallery of serendipitous observations of \objname{} from the \rubinobs{}.  All images are $30 \arcsec\times 30\arcsec$ and have been reprojected so that north appears up, and east to the left (green arrows). The anti-solar (yellow, black-outlined arrow) and anti-motion (black, red-outlined arrow) directions are indicated.  All dates and times are \ac{TAI}.
    \textbf{(a)} 2025 June 21 08:11:32. 
    \textbf{(b)} 2025 June 22 02:32:47. An area of roughly vertical saturation masking can be seen near the center of the frame; \objname{} is not within the masking, but the nearby blended star is.  
    \textbf{(c)} 2025 June 22 03:07:49.
    % \textbf{(d)} 2025 June 30 02:25:46. \objname{} is in front of a saturated star.
    \textbf{(d)} 2025 June 24 03:07:46. The image was unintentionally out of focus.  
    \textbf{(e)} 2025 June 30 02:26:26. \objname~is adjacent to the saturated star at the center. 
    \textbf{(f)} 2025 July 02 00:44:25.
    \textbf{(g)} 2025 July 02 01:20:33.
    \textbf{(h)} 2025 July 02 02:31:16.
    \textbf{(i)} 2025 July 02 03:33:02.
        }
    \label{fig:gallery}
\end{figure*}

%%%%%%%%%%%%%%%%%%%%%%%%%%%%%%%%%%%%%%%%%%%%%
\section{Observations}\label{sec:observations}

In order to search for \objname{} in Rubin data, we derived interpolated and extrapolated \objname{} positions from the ephemeris provided by the JPL Scout service \citep{maybeSCOUT} -- while \objname{} was on the \ac{MPC} \ac{NEOCP} with the designation A11pl3Z -- once for the date/time of every LSSTCam and LSST Commissioning Camera \citep[LSSTComCam;][]{LSSTComCam} visit from the Rubin \texttt{exposure} table, which includes science and non-science (e.g., engineering) images. For each visit, we checked to see if \objname{} was within $\sim9^{\circ}$ of any of the pointing (boresight) center on-sky coordinates. In addition, we used the available astrometric data from Scout before UT July 2 to re-fit the orbit using the orbit fitting software \texttt{Layup}. These solutions were used to propagate \objname{}'s ephemerides to all LSSTCam visits using the survey simulation software \texttt{Sorcha} \citep{Merritt2025, Holman2025} and the publicly available \texttt{lsstcam\_20250930} cadence simulation \citep{SITCOMTN-005, RTN-011}.

\begin{figure*}
\begin{interactive}{animation}{3I_SV_update.mp4}
\includegraphics[width=\linewidth]{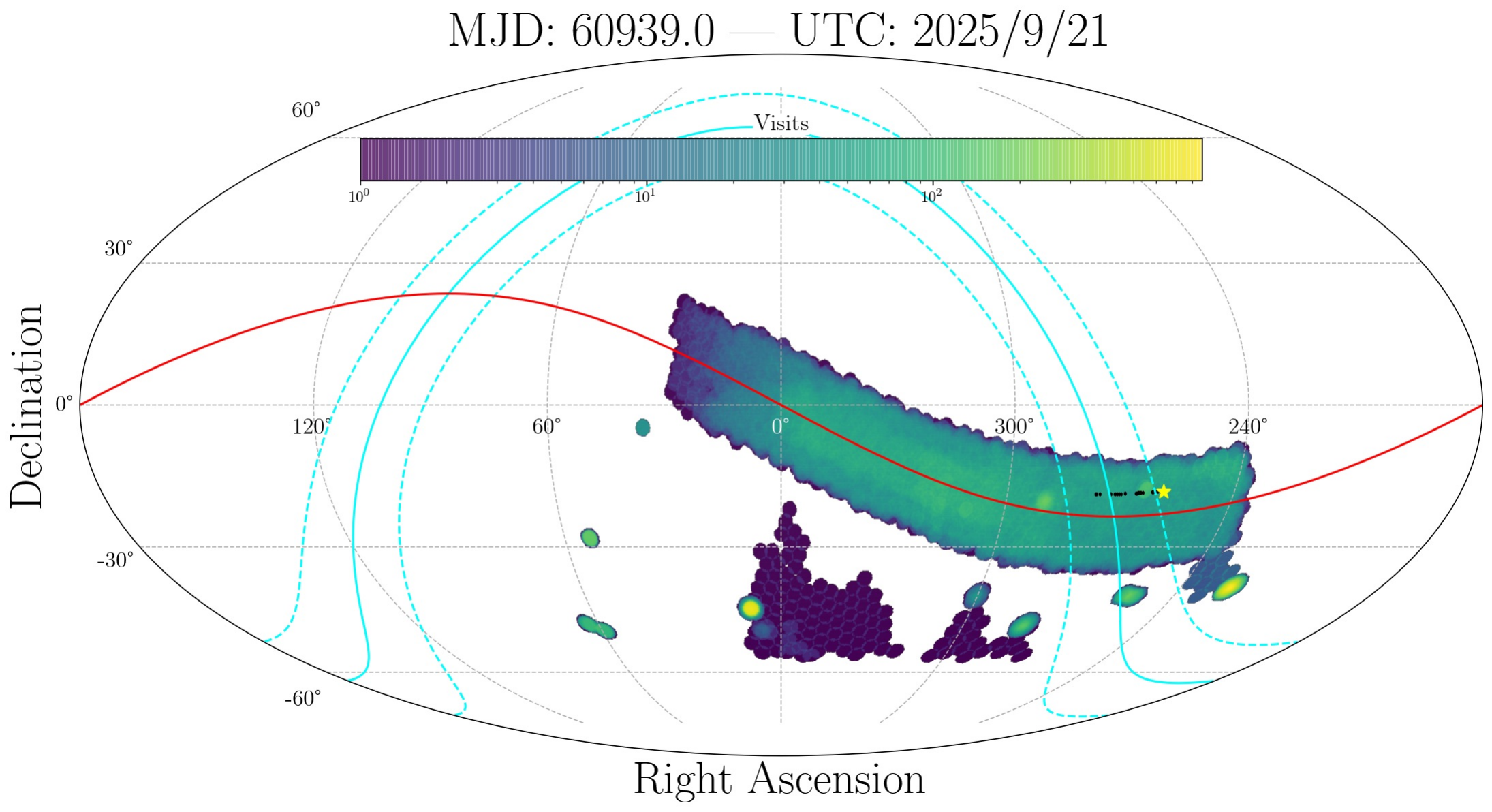}
\end{interactive}
\caption{Sky-map of the final \ac{SV} observations distributions of \objname{}. The \ac{SV} here is simulated from recorded pointings already enacted at the time of writing and nominal operations thereafter. The black points represent individual observations of \objname{}, with the yellow star highlighting its final observation in the \ac{SV} in UT 2025 July 20. The red solid line marks the ecliptic plane, and the solid and dashed blue lines mark the galactic plane and $\pm10^{\circ}$ from the plane, respectively.
(This static figure represents the final frame of an animation available in the online HTML version of the final article, showing the progression of \objname{} detections during the nominal \ac{SV}. Each frame of the animation represents one additional night of observations, with the \texttt{HEALPy} sky-map updated accordingly. The plot title updates to reflect the night of each observation, and new black points appear as previous observations and are added to the map. The yellow star moves with each frame, representing the most up-to-date observation of \objname{} within the \ac{SV}. The total duration of the animation is $\sim$28~s.)
\label{fig:sorcha}}
\end{figure*}

For the \texttt{Sorcha} simulations, we generated a synthetic \objname{}, adopting this best-fit set of orbital elements together with the slope-inferred colors (measured colors were not yet available) in Section \ref{subsec:colors}, an absolute magnitude $H_r = 12.5$, and a phase curve parameter $G = 0.15$. Given the uncertainty in how \objname{} sublimation and dust production vary with changing heliocentric distances at the time of writing, we made no attempt to model the flux enhancement from reflected sunlight by \objname{}'s coma/tail and model \objname{}'s apparent brightness as an airless non-active body. This configuration successfully reproduces all existing \ac{SV} detections, and the cumulative sky-map of these detections is shown in Figure \ref{fig:sorcha} (with an animation available in the online version). When the simulation was extended forward using the actual \ac{SV} cadence, incorporating weather losses, maintenance downtime, and the evolving on-sky geometry of \objname{} relative to Rubin, \texttt{Sorcha} shows that \objname{} does not enter any further accessible \ac{SV} pointings after UT 2025 July 20. Both methodologies used here to identify \objname{} therefore confirm that all recoverable appearances of \objname{} in the \ac{SV} data have already been obtained.
% Colin double-checked this on the RSP 12/13/2025 COC

%Using the methods described above, 
In total, we identified nine \ac{SV} observations that serendipitously contained \objname{} within detector bounds between UT 2025 June 21 and UT 2025 July 2 (the serendipitous observations timespan), although several of these detections are significantly blended with background sources (Figure~\ref{fig:gallery}). An additional 32 images (Figure \ref{fig:gallery2}) were acquired on UT 2025 July 3 during camera calibration procedures, where a composite image constructed from these data is shown in Figure \ref{fig:prettypicture}.
That sequence was a planned part of test operations for that night, but --- noticing the opportunity --- the Observing Specialist shifted the telescope pointing to coincide with \objname{}’s position. Additional images were acquired over the next week, some as part of \ac{ToO} testing (Figure \ref{fig:gallery3}). A concerted \ac{ToO} test with \objname{} was carried out on UT 2025 July 13 that resulted in over 35 images spanning numerous filters (Figure \ref{fig:gallery4}).  Several additional nights of observations serendipitously captured \objname{} between the following night (UT 2025 July 14), with the final images acquired UT 2025 July 20 (Figure \ref{fig:gallery5}).

Details of all observations, including information about observing geometry, are summarized in Table~\ref{tab:observations}, where we note that, unless stated otherwise, all date/times are given in UTC. Rubin Observatory records observations in \ac{TAI}, which does not account for leap seconds. Throughout this manuscript, \ac{TAI} is UTC + 37~s for all stamps; additional Rubin time information is available in \citet{hoblittITTN009SummitTime2022}. 
In all cases, we used comparison observations of the areas of the sky where \objname{} was detected at times when the object was absent to verify the correct identification of each candidate detection.

% Generated: 2025-11-14 03:19:21Z
\setlength{\tabcolsep}{3pt}
\begin{longtable}{rcccccccccccc}
\caption{Table of Observations} \label{tab:observations} \\
\toprule\toprule
Row & Date Midpoint & Visit ID & Chip(s) & RA$^a$ & Dec$^b$ & eRA$^c$ & eDec$^d$ & Band & Mag$^e$ & eMag$^f$ & Phase$^g$ & r$^h$ \\
 & (TAI) &  &  & ($^\circ$) & ($^\circ$) & ($\arcsec$) & ($\arcsec$) &  &  &  & ($^\circ$) & (au) \\
\midrule
\endfirsthead
\caption[]{Table of Observations} \\
\toprule
Row & Date Midpoint & Visit ID & Chip(s) & RA$^a$ & Dec$^b$ & eRA$^c$ & eDec$^d$ & Band & Mag$^e$ & eMag$^f$ & Phase$^g$ & r$^h$ \\
 & (TAI) &  &  & ($^\circ$) & ($^\circ$) & ($\arcsec$) & ($\arcsec$) &  &  &  & ($^\circ$) & (au) \\
\midrule
\endhead
\midrule
\multicolumn{13}{r}{Continued on next page} \\
\midrule
\endfoot
\bottomrule
\multicolumn{13}{l}{$^a$Measured Right ascension; $^b$Measured declination; $^c$ Measured Right ascension uncertainty; $^d$Measured Declination} \\
\multicolumn{13}{l}{ uncertainty; $^e$Measured apparent aperture magnitude in specified band; $^f$Measured apparent aperture magnitude uncertainty} \\
\multicolumn{13}{l}{ in specified band; $^g$Solar phase angle (Sun-target-observer); $^h$Heliocentric distance}
\endlastfoot
1 & 2025-06-21 08:11:32 & 2025062000620 & 104 & 276.513300 & -18.755701 & 0.043 & 0.050 & \textit{z} & $\cdots$ & $\cdots$ & 1.630 & 4.837 \\
2 & 2025-06-22 02:32:47 & 2025062100383 & 38 & $\cdots$ & $\cdots$ & $\cdots$ & $\cdots$ & \textit{g} & $\cdots$ & $\cdots$ & 1.470 & 4.812 \\
3 & 2025-06-22 03:07:49 & 2025062100431 & 38 & $\cdots$ & $\cdots$ & $\cdots$ & $\cdots$ & \textit{r} & $\cdots$ & $\cdots$ & 1.465 & 4.811 \\
4 & 2025-06-24 03:07:46 & 2025062300417 & 77 & $\cdots$ & $\cdots$ & $\cdots$ & $\cdots$ & \textit{i} & $\cdots$ & $\cdots$ & 1.120 & 4.744 \\
5 & 2025-06-30 02:26:26 & 2025062900393 & 9 & $\cdots$ & $\cdots$ & $\cdots$ & $\cdots$ & \textit{i} & $\cdots$ & $\cdots$ & 1.684 & 4.545 \\ % manually done 12/13/2025 COC
6 & 2025-07-02 00:44:25 & 2025070100200 & 119 & 271.447823 & -18.684875 & 0.084 & 0.077 & \textit{i} & 17.722 & 0.05 & 2.233 & 4.480 \\
7 & 2025-07-02 01:20:33 & 2025070100248 & 119 & 271.434746 & -18.684645 & 0.063 & 0.067 & \textit{z} & 17.610 & 0.10 & 2.240 & 4.479 \\
8 & 2025-07-02 02:31:16 & 2025070100350 & 119 & 271.409146 & -18.684044 & 0.050 & 0.060 & \textit{z} & 17.542 & 0.10 & 2.255 & 4.478 \\
9 & 2025-07-02 03:33:02 & 2025070100399 & 119 & 271.386660 & -18.683526 & 0.049 & 0.046 & \textit{i} & 17.709 & 0.05 & 2.268 & 4.476 \\
10 & 2025-07-03 01:53:12 & 2025070200188 & 6 & $\cdots$ & $\cdots$ & $\cdots$ & $\cdots$ & \textit{i} & $\cdots$ & $\cdots$ & 2.557 & 4.445 \\
11 & 2025-07-03 03:03:56 & 2025070200274 & 6 & 270.873594 & -18.670730 & 0.039 & 0.041 & \textit{z} & $\cdots$ & $\cdots$ & 2.573 & 4.444 \\
12 & 2025-07-03 06:25:00 & 2025070200479 & 94 & $\cdots$ & $\cdots$ & $\cdots$ & $\cdots$ & \textit{r} & $\cdots$ & $\cdots$ & 2.618 & 4.439 \\
13 & 2025-07-03 06:27:17 & 2025070200480 & 94 & 270.798445 & -18.668739 & 0.086 & 0.100 & \textit{r} & $\cdots$ & $\cdots$ & 2.618 & 4.439 \\
14 & 2025-07-03 06:32:03 & 2025070200481 & 94 & 270.796728 & -18.668700 & 0.080 & 0.106 & \textit{r} & 17.934 & 0.03 & 2.619 & 4.439 \\
15 & 2025-07-03 06:36:22 & 2025070200482 & 94 & 270.795081 & -18.668671 & 0.073 & 0.076 & \textit{r} & 17.942 & 0.03 & 2.620 & 4.439 \\
16 & 2025-07-03 06:48:53 & 2025070200483 & 94 & 270.790488 & -18.668570 & 0.089 & 0.096 & \textit{r} & 17.941 & 0.03 & 2.623 & 4.438 \\
17 & 2025-07-03 06:49:27 & 2025070200484 & 94 & 270.790243 & -18.668547 & 0.084 & 0.104 & \textit{r} & 17.946 & 0.03 & 2.623 & 4.438 \\
18 & 2025-07-03 06:50:17 & 2025070200485 & 94 & 270.789975 & -18.668545 & 0.075 & 0.060 & \textit{r} & 17.947 & 0.03 & 2.623 & 4.438 \\
19 & 2025-07-03 06:50:52 & 2025070200486 & 94 & 270.789724 & -18.668519 & 0.064 & 0.077 & \textit{r} & 17.950 & 0.03 & 2.623 & 4.438 \\
20 & 2025-07-03 06:51:47 & 2025070200487 & 94 & 270.789388 & -18.668523 & 0.069 & 0.070 & \textit{r} & 17.956 & 0.03 & 2.624 & 4.438 \\
21 & 2025-07-03 06:52:21 & 2025070200488 & 94 & 270.789171 & -18.668512 & 0.050 & 0.073 & \textit{r} & 17.954 & 0.03 & 2.624 & 4.438 \\
22 & 2025-07-03 06:53:16 & 2025070200489 & 94 & 270.788852 & -18.668532 & 0.058 & 0.074 & \textit{r} & 17.953 & 0.03 & 2.624 & 4.438 \\
23 & 2025-07-03 06:53:51 & 2025070200490 & 94 & 270.788645 & -18.668511 & 0.051 & 0.064 & \textit{r} & 17.949 & 0.03 & 2.624 & 4.438 \\
24 & 2025-07-03 06:54:40 & 2025070200491 & 94 & 270.788325 & -18.668504 & 0.051 & 0.063 & \textit{r} & 17.948 & 0.03 & 2.624 & 4.438 \\
25 & 2025-07-03 06:55:15 & 2025070200492 & 94 & 270.788133 & -18.668489 & 0.050 & 0.060 & \textit{r} & 17.952 & 0.03 & 2.624 & 4.438 \\
26 & 2025-07-03 06:58:56 & 2025070200493 & 94 & 270.786767 & -18.668463 & 0.062 & 0.065 & \textit{r} & 17.951 & 0.03 & 2.625 & 4.438 \\
27 & 2025-07-03 06:59:30 & 2025070200494 & 94 & 270.786568 & -18.668443 & 0.063 & 0.070 & \textit{r} & 17.955 & 0.03 & 2.625 & 4.438 \\
28 & 2025-07-03 07:00:20 & 2025070200495 & 94 & 270.786210 & -18.668402 & 0.067 & 0.074 & \textit{r} & 17.953 & 0.03 & 2.626 & 4.438 \\
29 & 2025-07-03 07:00:54 & 2025070200496 & 94 & 270.786006 & -18.668399 & 0.051 & 0.073 & \textit{r} & 17.948 & 0.03 & 2.626 & 4.438 \\
30 & 2025-07-03 07:01:46 & 2025070200497 & 94 & 270.785735 & -18.668441 & 0.058 & 0.074 & \textit{r} & 17.951 & 0.03 & 2.626 & 4.438 \\
31 & 2025-07-03 07:02:20 & 2025070200498 & 94 & 270.785494 & -18.668414 & 0.053 & 0.072 & \textit{r} & 17.948 & 0.03 & 2.626 & 4.438 \\
32 & 2025-07-03 07:07:48 & 2025070200499 & 94 & 270.783508 & -18.668359 & 0.051 & 0.073 & \textit{r} & 17.947 & 0.03 & 2.627 & 4.438 \\
33 & 2025-07-03 07:08:22 & 2025070200500 & 94 & $\cdots$ & $\cdots$ & $\cdots$ & $\cdots$ & \textit{r} & 17.962 & 0.03 & 2.627 & 4.438 \\
34 & 2025-07-03 07:09:07 & 2025070200501 & 94 & $\cdots$ & $\cdots$ & $\cdots$ & $\cdots$ & \textit{r} & 17.954 & 0.03 & 2.627 & 4.438 \\
35 & 2025-07-03 07:09:41 & 2025070200502 & 94 & $\cdots$ & $\cdots$ & $\cdots$ & $\cdots$ & \textit{r} & 17.951 & 0.03 & 2.628 & 4.438 \\
36 & 2025-07-03 07:10:31 & 2025070200503 & 94 & $\cdots$ & $\cdots$ & $\cdots$ & $\cdots$ & \textit{r} & 17.952 & 0.03 & 2.628 & 4.438 \\
37 & 2025-07-03 07:11:05 & 2025070200504 & 94 & $\cdots$ & $\cdots$ & $\cdots$ & $\cdots$ & \textit{r} & 17.956 & 0.03 & 2.628 & 4.438 \\
38 & 2025-07-03 07:11:55 & 2025070200505 & 94 & $\cdots$ & $\cdots$ & $\cdots$ & $\cdots$ & \textit{r} & 17.967 & 0.03 & 2.628 & 4.438 \\
39 & 2025-07-03 07:12:30 & 2025070200506 & 94 & $\cdots$ & $\cdots$ & $\cdots$ & $\cdots$ & \textit{r} & 17.955 & 0.03 & 2.628 & 4.438 \\
40 & 2025-07-03 07:13:19 & 2025070200507 & 94 & $\cdots$ & $\cdots$ & $\cdots$ & $\cdots$ & \textit{r} & 17.952 & 0.03 & 2.628 & 4.438 \\
41 & 2025-07-03 07:13:53 & 2025070200508 & 94 & $\cdots$ & $\cdots$ & $\cdots$ & $\cdots$ & \textit{r} & 17.952 & 0.03 & 2.628 & 4.438 \\
42 & 2025-07-04 06:36:15 & 2025070300497 & 98 & 270.262204 & -18.654113 & 0.055 & 0.059 & \textit{r} & $\cdots$ & $\cdots$ & 2.944 & 4.405 \\
43 & 2025-07-05 01:23:40 & 2025070400162 & 137,144 & $\cdots$ & $\cdots$ & $\cdots$ & $\cdots$ & \textit{y} & 17.461 & 0.10 & 3.204 & 4.379 \\
44 & 2025-07-05 01:56:38 & 2025070400210 & 137,144 & 269.827654 & -18.641391 & 0.057 & 0.064 & \textit{y} & 17.476 & 0.10 & 3.212 & 4.379 \\
45 & 2025-07-06 23:53:08 & 2025070600062 & 41 & $\cdots$ & $\cdots$ & $\cdots$ & $\cdots$ & \textit{y} & $\cdots$ & $\cdots$ & 3.872 & 4.315 \\
46 & 2025-07-07 00:44:03 & 2025070600140 & 41 & $\cdots$ & $\cdots$ & $\cdots$ & $\cdots$ & \textit{y} & $\cdots$ & $\cdots$ & 3.884 & 4.314 \\
47 & 2025-07-12 01:12:32 & 2025071100318 & 42 & 265.854190 & -18.488537 & 0.042 & 0.043 & \textit{i} & $\cdots$ & $\cdots$ & 5.729 & 4.147 \\
48 & 2025-07-12 01:46:44 & 2025071100364 & 42 & $\cdots$ & $\cdots$ & $\cdots$ & $\cdots$ & \textit{z} & $\cdots$ & $\cdots$ & 5.738 & 4.146 \\
49 & 2025-07-12 04:39:06 & 2025071100563 & 59 & 265.768030 & -18.484531 & 0.060 & 0.072 & \textit{i} & $\cdots$ & $\cdots$ & 5.784 & 4.142 \\
50 & 2025-07-12 05:13:58 & 2025071100611 & 59 & $\cdots$ & $\cdots$ & $\cdots$ & $\cdots$ & \textit{z} & $\cdots$ & $\cdots$ & 5.794 & 4.142 \\
51 & 2025-07-13 00:35:43 & 2025071200343 & 93 & $\cdots$ & $\cdots$ & $\cdots$ & $\cdots$ & \textit{r} & $\cdots$ & $\cdots$ & 6.102 & 4.115 \\
52 & 2025-07-13 03:13:21 & 2025071200518 & 30 & 265.203945 & -18.457186 & 0.085 & 0.075 & \textit{g} & 18.262 & 0.03 & 6.145 & 4.111 \\
53 & 2025-07-13 03:14:43 & 2025071200520 & 3,6 & $\cdots$ & $\cdots$ & $\cdots$ & $\cdots$ & \textit{g} & $\cdots$ & $\cdots$ & 6.145 & 4.111 \\
54 & 2025-07-13 03:18:35 & 2025071200522 & 30 & 265.201767 & -18.457068 & 0.056 & 0.060 & \textit{i} & 17.389 & 0.05 & 6.146 & 4.111 \\
55 & 2025-07-13 03:19:57 & 2025071200524 & 3,4,6,7 & $\cdots$ & $\cdots$ & $\cdots$ & $\cdots$ & \textit{i} & 17.385 & 0.05 & 6.147 & 4.111 \\
56 & 2025-07-13 03:23:44 & 2025071200526 & 30 & $\cdots$ & $\cdots$ & $\cdots$ & $\cdots$ & \textit{r} & 17.675 & 0.03 & 6.148 & 4.111 \\
57 & 2025-07-13 03:25:05 & 2025071200528 & 3,4,6,7 & $\cdots$ & $\cdots$ & $\cdots$ & $\cdots$ & \textit{r} & $\cdots$ & $\cdots$ & 6.148 & 4.111 \\
58 & 2025-07-13 03:28:52 & 2025071200530 & 30 & $\cdots$ & $\cdots$ & $\cdots$ & $\cdots$ & \textit{y} & 17.204 & 0.10 & 6.149 & 4.111 \\
59 & 2025-07-13 03:30:13 & 2025071200532 & 3,4,6,7 & $\cdots$ & $\cdots$ & $\cdots$ & $\cdots$ & \textit{y} & 17.232 & 0.10 & 6.150 & 4.111 \\
60 & 2025-07-13 03:33:59 & 2025071200534 & 30 & 265.195258 & -18.456752 & 0.059 & 0.054 & \textit{z} & 17.265 & 0.10 & 6.151 & 4.111 \\
61 & 2025-07-13 03:38:59 & 2025071200538 & 30 & $\cdots$ & $\cdots$ & $\cdots$ & $\cdots$ & \textit{g} & 18.279 & 0.03 & 6.152 & 4.111 \\
62 & 2025-07-13 03:39:38 & 2025071200539 & 30,31 & $\cdots$ & $\cdots$ & $\cdots$ & $\cdots$ & \textit{g} & 18.290 & 0.03 & 6.152 & 4.111 \\
63 & 2025-07-13 03:44:01 & 2025071200542 & 30 & $\cdots$ & $\cdots$ & $\cdots$ & $\cdots$ & \textit{i} & 17.402 & 0.05 & 6.153 & 4.111 \\
64 & 2025-07-13 03:44:39 & 2025071200543 & 31 & $\cdots$ & $\cdots$ & $\cdots$ & $\cdots$ & \textit{i} & 17.378 & 0.05 & 6.154 & 4.111 \\
65 & 2025-07-13 03:48:58 & 2025071200546 & 30 & $\cdots$ & $\cdots$ & $\cdots$ & $\cdots$ & \textit{r} & 17.637 & 0.03 & 6.155 & 4.110 \\
66 & 2025-07-13 03:49:36 & 2025071200547 & 31 & $\cdots$ & $\cdots$ & $\cdots$ & $\cdots$ & \textit{r} & 17.664 & 0.03 & 6.155 & 4.110 \\
67 & 2025-07-13 03:53:56 & 2025071200550 & 30 & $\cdots$ & $\cdots$ & $\cdots$ & $\cdots$ & \textit{y} & 17.176 & 0.10 & 6.156 & 4.110 \\
68 & 2025-07-13 03:54:33 & 2025071200551 & 31 & $\cdots$ & $\cdots$ & $\cdots$ & $\cdots$ & \textit{y} & 17.231 & 0.10 & 6.156 & 4.110 \\
69 & 2025-07-13 03:58:53 & 2025071200554 & 30 & 265.184742 & -18.456200 & 0.064 & 0.061 & \textit{z} & 17.274 & 0.10 & 6.157 & 4.110 \\
70 & 2025-07-13 03:59:31 & 2025071200555 & 31 & $\cdots$ & $\cdots$ & $\cdots$ & $\cdots$ & \textit{z} & 17.258 & 0.10 & 6.158 & 4.110 \\
71 & 2025-07-13 04:03:46 & 2025071200558 & 30 & 265.182679 & -18.456142 & 0.075 & 0.074 & \textit{g} & 18.268 & 0.03 & 6.159 & 4.110 \\
72 & 2025-07-13 04:04:24 & 2025071200559 & 31 & $\cdots$ & $\cdots$ & $\cdots$ & $\cdots$ & \textit{g} & 18.282 & 0.03 & 6.159 & 4.110 \\
73 & 2025-07-13 04:08:42 & 2025071200562 & 30 & $\cdots$ & $\cdots$ & $\cdots$ & $\cdots$ & \textit{i} & 17.416 & 0.05 & 6.160 & 4.110 \\
74 & 2025-07-13 04:09:20 & 2025071200563 & 31 & $\cdots$ & $\cdots$ & $\cdots$ & $\cdots$ & \textit{i} & 17.371 & 0.05 & 6.160 & 4.110 \\
75 & 2025-07-13 04:13:34 & 2025071200566 & 30 & 265.178568 & -18.455915 & 0.060 & 0.058 & \textit{r} & 17.611 & 0.03 & 6.161 & 4.110 \\
76 & 2025-07-13 04:14:11 & 2025071200567 & 31 & 265.178324 & -18.455923 & 0.063 & 0.061 & \textit{r} & 17.656 & 0.03 & 6.162 & 4.110 \\
77 & 2025-07-13 04:18:26 & 2025071200570 & 30 & $\cdots$ & $\cdots$ & $\cdots$ & $\cdots$ & \textit{y} & 17.132 & 0.10 & 6.163 & 4.110 \\
78 & 2025-07-13 04:19:04 & 2025071200571 & 31 & $\cdots$ & $\cdots$ & $\cdots$ & $\cdots$ & \textit{y} & 17.234 & 0.10 & 6.163 & 4.110 \\
79 & 2025-07-13 04:23:19 & 2025071200574 & 30 & $\cdots$ & $\cdots$ & $\cdots$ & $\cdots$ & \textit{z} & 17.264 & 0.10 & 6.164 & 4.110 \\
80 & 2025-07-13 04:23:57 & 2025071200575 & 31 & 265.174176 & -18.455696 & 0.061 & 0.063 & \textit{z} & 17.257 & 0.10 & 6.164 & 4.110 \\
81 & 2025-07-13 04:28:07 & 2025071200578 & 30 & $\cdots$ & $\cdots$ & $\cdots$ & $\cdots$ & \textit{g} & 18.239 & 0.03 & 6.165 & 4.110 \\
82 & 2025-07-13 04:28:45 & 2025071200579 & 31 & $\cdots$ & $\cdots$ & $\cdots$ & $\cdots$ & \textit{g} & 18.285 & 0.03 & 6.165 & 4.110 \\
83 & 2025-07-13 04:33:06 & 2025071200582 & 30 & 265.170335 & -18.455511 & 0.054 & 0.058 & \textit{i} & 17.411 & 0.05 & 6.167 & 4.109 \\
84 & 2025-07-13 04:33:44 & 2025071200583 & 31 & 265.170039 & -18.455497 & 0.048 & 0.055 & \textit{i} & 17.380 & 0.05 & 6.167 & 4.109 \\
85 & 2025-07-13 04:37:54 & 2025071200586 & 30 & 265.168291 & -18.455395 & 0.055 & 0.062 & \textit{r} & 17.582 & 0.03 & 6.168 & 4.109 \\
86 & 2025-07-13 04:38:32 & 2025071200587 & 31 & 265.168023 & -18.455378 & 0.048 & 0.045 & \textit{r} & 17.650 & 0.03 & 6.168 & 4.109 \\
87 & 2025-07-13 04:42:43 & 2025071200590 & 30 & $\cdots$ & $\cdots$ & $\cdots$ & $\cdots$ & \textit{y} & $\cdots$ & $\cdots$ & 6.169 & 4.109 \\
88 & 2025-07-13 04:43:20 & 2025071200591 & 31 & $\cdots$ & $\cdots$ & $\cdots$ & $\cdots$ & \textit{y} & $\cdots$ & $\cdots$ & 6.169 & 4.109 \\
89 & 2025-07-14 04:22:11 & 2025071300432 & 102 & $\cdots$ & $\cdots$ & $\cdots$ & $\cdots$ & \textit{i} & $\cdots$ & $\cdots$ & 6.552 & 4.077 \\
90 & 2025-07-14 04:56:49 & 2025071300480 & 102 & $\cdots$ & $\cdots$ & $\cdots$ & $\cdots$ & \textit{z} & $\cdots$ & $\cdots$ & 6.561 & 4.076 \\
91 & 2025-07-19 00:57:02 & 2025071800083 & 22 & 261.530749 & -18.245460 & 0.053 & 0.057 & \textit{g} & $\cdots$ & $\cdots$ & 8.488 & 3.916 \\
92 & 2025-07-19 00:57:45 & 2025071800084 & 177 & 261.530413 & -18.245412 & 0.082 & 0.078 & \textit{g} & $\cdots$ & $\cdots$ & 8.488 & 3.916 \\
93 & 2025-07-19 23:25:43 & 2025071900042 & 66 & 260.929564 & -18.205121 & 0.042 & 0.044 & \textit{r} & $\cdots$ & $\cdots$ & 8.868 & 3.885 \\
94 & 2025-07-19 23:42:18 & 2025071900063 & 66 & 260.922118 & -18.204647 & 0.038 & 0.039 & \textit{i} & $\cdots$ & $\cdots$ & 8.873 & 3.885 \\
95 & 2025-07-20 00:05:16 & 2025071900088 & 123 & 260.911789 & -18.203959 & 0.086 & 0.075 & \textit{u} & $\cdots$ & $\cdots$ & 8.880 & 3.884 \\
96 & 2025-07-20 03:33:14 & 2025071900336 & 173 & $\cdots$ & $\cdots$ & $\cdots$ & $\cdots$ & \textit{g} & $\cdots$ & $\cdots$ & 8.939 & 3.880 \\
97 & 2025-07-20 04:15:08 & 2025071900384 & 80 & 260.799042 & -18.196257 & 0.087 & 0.092 & \textit{u} & $\cdots$ & $\cdots$ & 8.951 & 3.879 \\
\end{longtable}

 % 11/13/2025 COC new table

%%%%%%%%%%%%%%%%%%%%%%%%%%%%%%%%%%%%%%%%%%%%%
\section{Data Processing}\label{sec:reduction}

Rubin's observing facilities are designed to be supported by a largely automated data management system \citep{Juric2017}. Once the \ac{LSST} survey begins, products enabling Solar System science, including moving object detection, image differencing, astrometry, photometry, and morphological classification (e.g., ``extendedness''), will be automatically available on minute-to-daily timescales. Further details on the \ac{LSST} algorithms and pipelines are provided in \cite{PSTN-019} and \cite{Bosch_HSC}. Their present performance on LSSTComCam data, which is similar to LSSTCam but smaller in scale, is summarized in \cite{RTN-095}. Detailed information on expected data products can be found in the \ac{LSST} Data Products Definition Document \citep[DPDD;][]{Juric2023} or a summary by \cite{Graham2022}.

At the time of \objname{}'s discovery, only a portion of this system had been deployed and verified. This includes the data acquisition and storage system (allowing Rubin to take and store the images and image metadata) and an automated ``Nightly Validation'' pipeline supporting daily data reduction for quality assurance purposes that includes \ac{ISR}, \ac{WCS} determination, and detection of isolated bright ($>20\sigma$) sources. Calibration production pipelines (e.g., for generating and applying biases, darks, flats, etc.) have also been commissioned. While Rubin's solar system pipelines have recently been demonstrated with Rubin First Look\footnote{\url{https://rubinobservatory.org/news/rubin-first-look}} showcase observations, they were not yet active when \objname{} was discovered, as it was too early into the \ac{SV} survey for sufficient data to be available to enable image differencing and, consequently, moving object detection.

For this reason, much of our data reduction in this work has been carried out with custom pipeline runs, custom configurations, with occasional custom-written software to backfill for features not yet commissioned or unavailable for other reasons at the time of \objname{}'s appearance. In particular, much of our analysis was conducted on science images (rather than difference images) and relied on the Rubin deblender \citep{melchior2018_scarlet} to mitigate confusion in crowded regions. As the location of \objname{} was sufficiently well constrained, we only calibrated images for the 4k $\times$ 4k CCD detector where the object was expected to be, rather than the entire 188 array of detectors that is active in a typical commissioning visit.

\begin{figure*}
    \centering
    \includegraphics[width=0.45\textwidth, trim=48 48 48 48, clip]{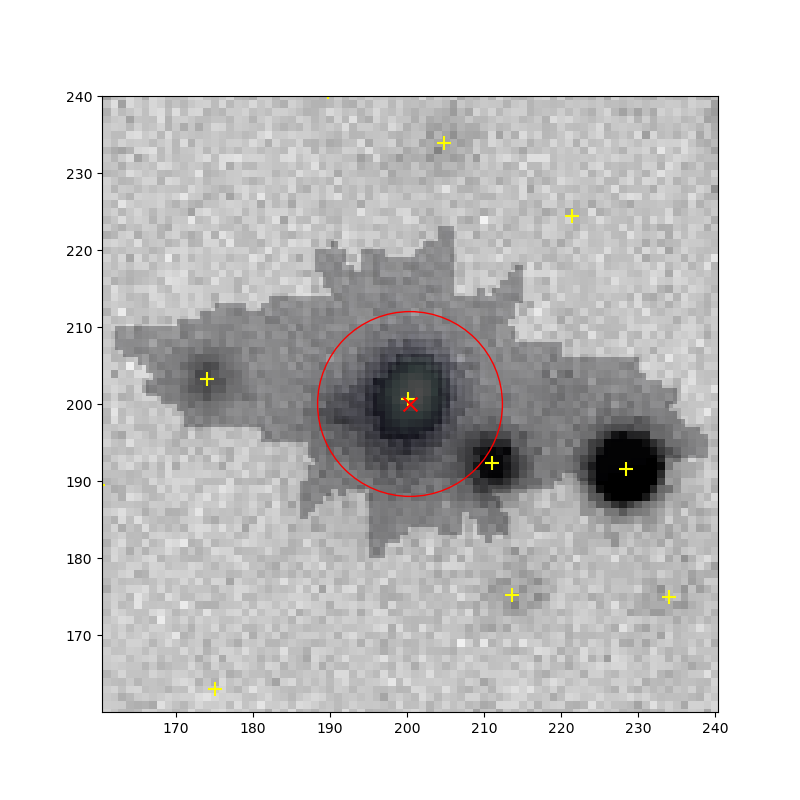}
    \includegraphics[width=0.45\textwidth, trim=48 48 48 48, clip]{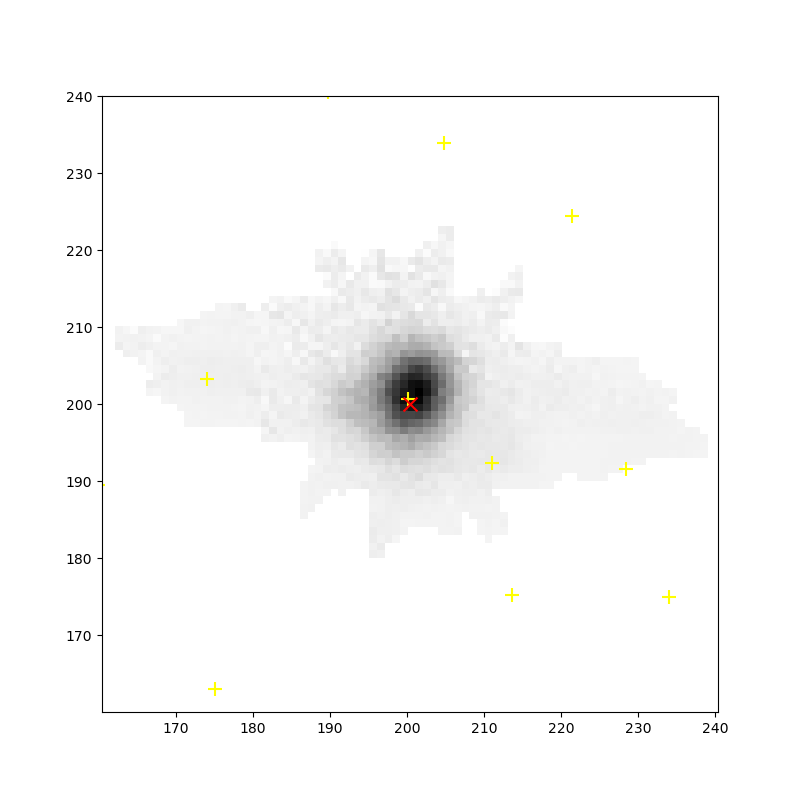}
    \caption{Deblending \objname{} and background sources in a Rubin calibrated exposure (visit 2025070100200, detector 119) imaged UT 2025 July 2 00:44:25 (TAI). On the left is a cutout from a crowded field
    %({\tt visit\_id=2025070100200}) 
    centered on \objname{}. The red ring represents the \ac{LSST} Science Pipeline's aperture photometry radius (see Section~\ref{subsec:lightcurve}). In the image, \objname{} is seen nestled between four sources. %  (two to the left, two to the right) removed 11/14/2025 COC per Reviewer
    The brighter three % (one to the left, and two to the right) removed per Reviewer 11/14/2025
    have been detected as separate and deblended (marked with yellow plus symbols). The fourth source (just to the left of \objname{}, and inside the red ring) has not. The resulting {\em footprint} (the dark gray region in the left panel) -- a set of pixels deemed by the deblender to contain \objname{}'s flux -- is shown on the right. The deblender successfully removed the flux belonging to the three detected stars, preventing major biases in photometry and astrometry. However, the flux due to the fainter undetected star is still present. This illustrates both the power and the caveats of deblender applications: while the worst effects of crowding are mitigated, some low-level residual flux from faint blended sources likely remains.}
    \label{fig:deblender}
\end{figure*}

For the data presented here, we specifically removed instrument signatures (flats, darks, biases) from the raw images, fit and subtracted the sky background, computed the \ac{WCS} solution using the Gaia DR3 catalog, and photometrically calibrated the images. We also ran a dedicated {\tt SingleFrameDetectAndMeasure} \ac{LSST} pipeline task to detect, deblend, and measure all sources with significance $> 5\sigma$. As the fields with \objname{} were fairly crowded, this task frequently failed to converge due to overly large blends resulting from numerous overlapping sources. This was a software element that had not yet been fully tuned at this stage of Rubin commissioning, and has since been greatly improved as a result of this work. When successful, the measurements of detected sources include both astrometry (centroids) and photometry (both \ac{PSF} and aperture fluxes). 
Because of the special challenges presented by the data analyzed here, and also to better understand what specialized automated data handling may be required in similar situations in the future, we also performed a more customized photometric analysis (described in detail in Section~\ref{subsec:photmeasurements}) to maximize the quality of our \objname{}-specific results. 
An example of a successful processing run, including the appearance of a deblended source footprint, is shown in Figure~\ref{fig:deblender}.

%%%%%%%%%%%%%%%%%%%%%%%%%%%%%%%%%%%%%%%%%%%%%%%%%%
\section{Analysis and Results} \label{sec:results}

%%%%%%%%%%%%%%%%%%%%%%%
% 12/13/2025 COC: skipping this for now and will return post-Rahil. Noting there are conflicting offsets reported as-is.
\subsection{Astrometry}
\label{subsec:astrometricCode}

The Rubin science pipeline stack has not yet been optimized for analyses of crowded fields. As a result, we developed a small custom centroid and magnitude extractor to handle the case of \objname{}. 
%We extracted the astrometry of \objname{} in two different ways. First,
Our code derives positions by extracting cutouts of $30\times30$ pixels centered at the nominal position of an object and obtaining windowed centroid measurements (\texttt{XWIN} and \texttt{YWIN}) following a procedure analogous to that of the \textsc{Source Extractor} \citep{Bertin1996} package. These measurements are then translated to celestial coordinates using the full image \ac{WCS} fitted by the \ac{LSST} pipelines.

To validate our measurement code, in addition to using it to analyze \objname{}, we also apply it to stars in images for which the Rubin detection and measurement task successfully obtained a solution, as well as to known asteroids in all images. 
% The results are presented in Figure \ref{fig:centroiding}. 
The star centroids are in excellent agreement with the LSST stack measurements, with maximum offsets of $\sim$5~mas in both R.A. and Dec. 
%showing differences less than 5~mas per coordinate. 
For asteroids, the median difference between our measured position and the JPL Horizons-computed ephemerides is on the order of $1 \sigma$, with an overall systematic offset of $-20$~mas in R.A. and $-34$~mas in Dec.
%\subsection{Astrometry of \objname{}}
% Astrometric results for \objname{} for images where the object was sufficiently deblended using the method described above are included in Table~\ref{tab:observations}.

% Colin commenting this out 12/13/2025 
% We compare our measured positions to JPL Horizons ephemerides (obtained on UT 2025 July 14), noting a +113~mas residual in R.A. and $-$80~mas in Dec. 
While typically R.A. offsets can be attributed to errors in the timing for moving objects (the timing for Rubin is well-established; \citealt{hoblittITTN009SummitTime2022}), and effects such as Differential Chromatic Refraction (DCR) can cause shifts in both R.A. and Dec, our analysis does not show similarly prominent biases when comparing the offset between \objname{} and the other asteroids.
%; in fact, the bias in R.A. changes sign (in the same visits), thus requiring a different explanation.
The systematic shift is also present on the subset of images successfully measured with the \ac{LSST} Science Pipelines, whose performance has been extensively tested as part of the \ac{LSST} Data Preview 1 (DP1) release \citep{RTN-095}. In particular, DP1 demonstrated that the \ac{LSST} Commissioning Camera achieved residuals of order $5$~mas for high-\ac{SNR} sources, consistent with the limits of atmospheric turbulence; a similar performance is expected here for LSSTCam. 

\subsubsection{Rubin Astrometric Calibration}
% Finished rewriting through here 12/13/2025 COC

\begin{figure*}
    \centering
         \includegraphics[width=0.49\linewidth]{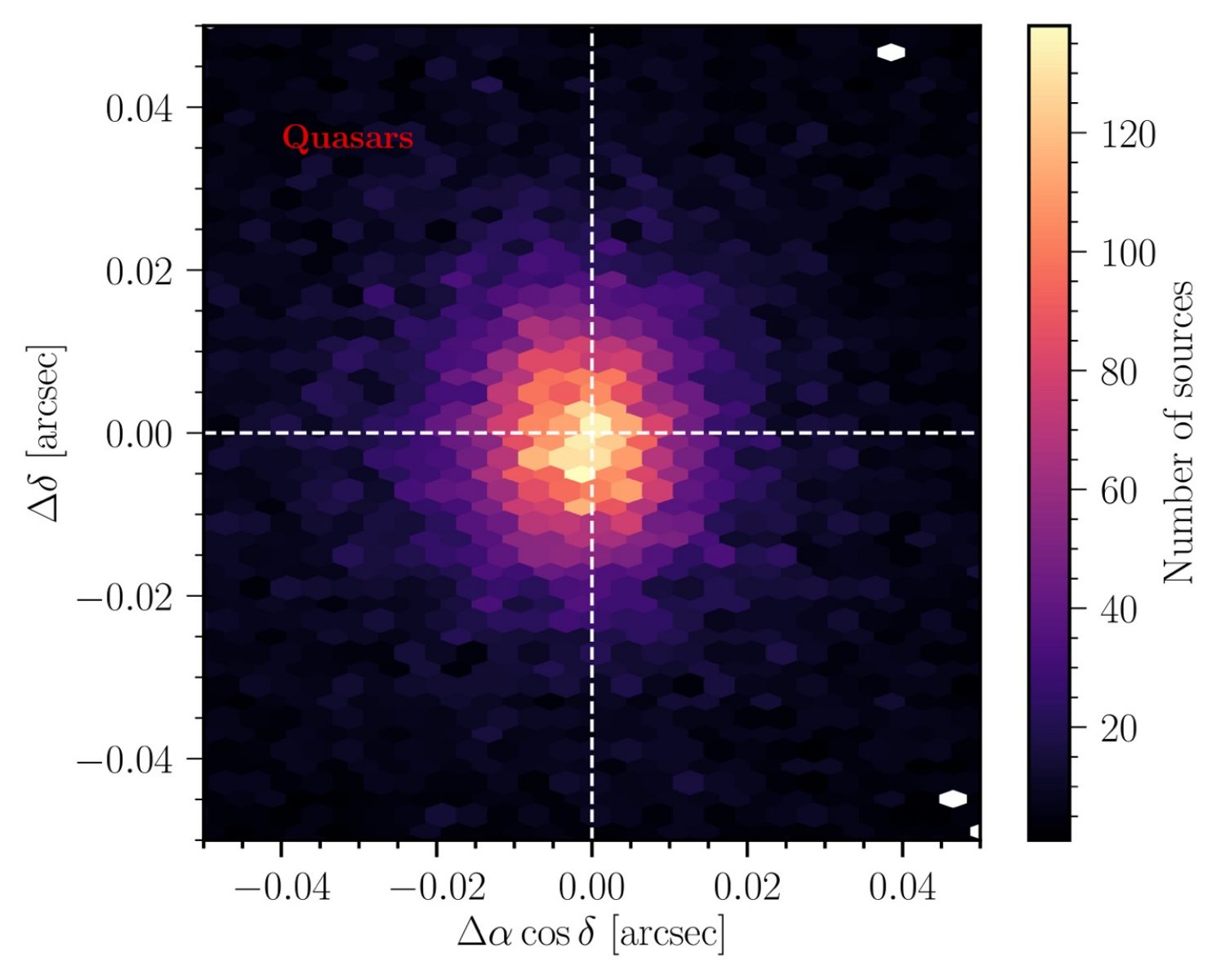} \includegraphics[width=0.49\linewidth]{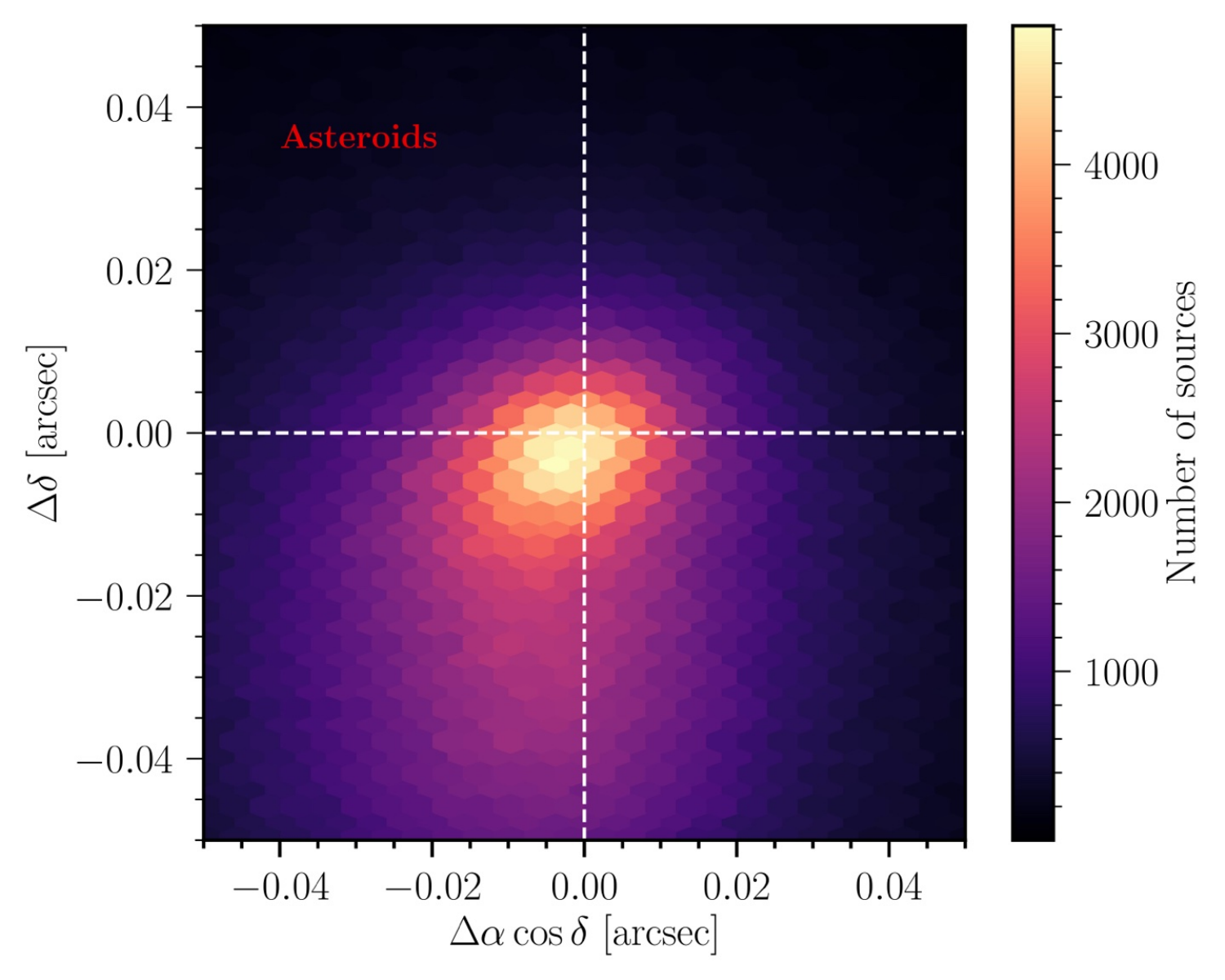} \\
    \caption{Two-dimensional histograms of the astrometric residuals between Rubin observations of objects matched to the Quaia quasar catalog (left) and Rubin observations of known asteroids with positions derived by the JPL Horizons ephemeris service (right). In both cases, the color scale indicates the number of sources per bin, and the dashed lines intersect at the origin.}
    \label{fig:newast}
\end{figure*}

To evaluate the apparent bias we observed in LSSTCam's astrometry, we verified the Rubin pipeline-derived source positions in two complementary ways: static-source position validation and measurement of moving objects.

First, we assessed whether we could accurately determine the positions of static sources. Using catalogs derived from difference imaging produced as part of the \ac{SV} commissioning efforts, we compared the positions of known quasars in Rubin data with the well-established Quaia quasar catalog \citep{quaia}. Since quasars are photometrically variable, they appear in difference images, thus proving to be ideal sources for this experiment. 

We applied a simple positional crossmatch between the sources detected in each difference imaging catalog to the quasars, and considered only sources whose on-sky separation was $d \leq 0.5\arcsec$, as any larger positional difference indicates a potential mismatch. The result was a catalog of approximately 65,000 identified sources (Figure \ref{fig:newast}, left panel). We find that $\langle \Delta \alpha \cos \delta \rangle = -2.6 \, \mathrm{mas}$ and $\langle \Delta \delta \rangle = -0.6 \, \mathrm{mas}$, indicating an excellent agreement between these two independent catalogs. 

Next, we assessed whether our asteroid measurements were equally well calibrated. We used data from 2.5 million observations of roughly 200 thousand unique, previously known asteroids that were measured by the LSST Solar System association pipeline during \ac{SV} efforts. We queried their positions as determined by the JPL Horizons ephemeris service at the midpoint time of each Rubin image taken of each asteroid (Figure \ref{fig:newast}, right panel). As with the quasars, we only considered sources with $d \leq 0.5\arcsec$, where we have $\langle \Delta \alpha \cos\delta \rangle = -8.5 \, \mathrm{mas}$ and $\langle \Delta \delta \rangle = -21.7 \, \mathrm{mas}$. This analysis reveals a small but measurable bias in these asteroid measurements. The Rubin offsets reported here are below the noise floor of Section \ref{sec:manual_ast}; further characterization of these offsets will be left to a future publication and is beyond the scope of this work.

\subsubsection{Curated Astrometric Measurements}\label{sec:manual_ast}

We manually produced our own bespoke, curated astrometric measurements (rows of Table \ref{tab:observations} with reported positions) from a subset of the available images, following a traditional, non-pipeline astrometry approach. 
%Optimal
We first identified viable images by visually inspecting each exposure and rejecting those where \objname{} was (1) notably involved (blended) with nearby background sources, (2) very close to the edge of the array (i.e., chip boundary), or (3) contaminated by obvious artifacts (e.g., scattered light).

We produced $600\times600$ pixel ($\sim 126\arcsec \times 126\arcsec$) image cutouts centered on the object's ephemeris position. We astrometrically solved these images using all Gaia DR3 sources down to magnitude $G=21.0$ via a quadratic polynomial model fit. We rejected Gaia sources showing residuals $>0.2\arcsec$ from the solution since they were likely contaminated by other nearby blended sources, especially at times when \objname{} was crossing crowded fields near the galactic center.

We measured the position of \objname{} on each cut-out using a Gaussian PSF fit following the ``zero-aperture astrometry correction'' technique \citep{2004DPS....36.3416T,2021Icar..35814276F}. We repeated the fit with five different circular apertures, with radii ranging from two to six pixels, and linearly extrapolated to the theoretical location of a zero pixel aperture, in both R.A.\ and Dec.

Following the methodology outlined in \citet{2022PSJ.....3..156F}, we assigned an astrometric uncertainty to each measured position. For stellar sources, the \citet{2022PSJ.....3..156F} approach is designed to provide conservative uncertainty estimates. However, for cometary objects, the non-Gaussian nature of the target's PSF can cause a slight underestimation of the object centroid component of the error budget. The reduced $\chi^2$ from the orbit fitting in Section \ref{subsec:orbitfitting} indicates that the assigned errors are, nevertheless, realistic and properly capture the intrinsic astrometric noise.

% Rahil
% Skipping rewrite here per Rahil 12/13/2025 COC
\subsection{Orbit Determination} \label{subsec:orbitfitting}
% %%%%%%%%%% THIS ONE IN CHI
% \begin{figure}
%     \centering
%         \includegraphics[width=1.0\linewidth]{3I_postfit_LSST.png} \\ \includegraphics[width=1.0\linewidth]{3I_postfit_LSST_zoom.png} \\
%     \caption{Astrometric residuals of \objname{} based on data available at the \acf{MPC} up until 2025 October 29.
%     \textbf{Top: }Residuals in Right Ascension scaled by cos(Dec) vs Declination plotted in units of the goodness of fit parameter $\chi$. Regions of constant $\chi=1,3$ are also shown. Points corresponding to residuals from LSST measurements are marked with rings around them.
%     \textbf{Bottom: } Zoomed in view of top panel.
%     }
%     \label{fig:astrometry_residuals}
% \end{figure}

%%%%%%%%%% THIS ONE IN ARCSEC
\begin{figure}
    \centering
        \includegraphics[width=1.0\linewidth]{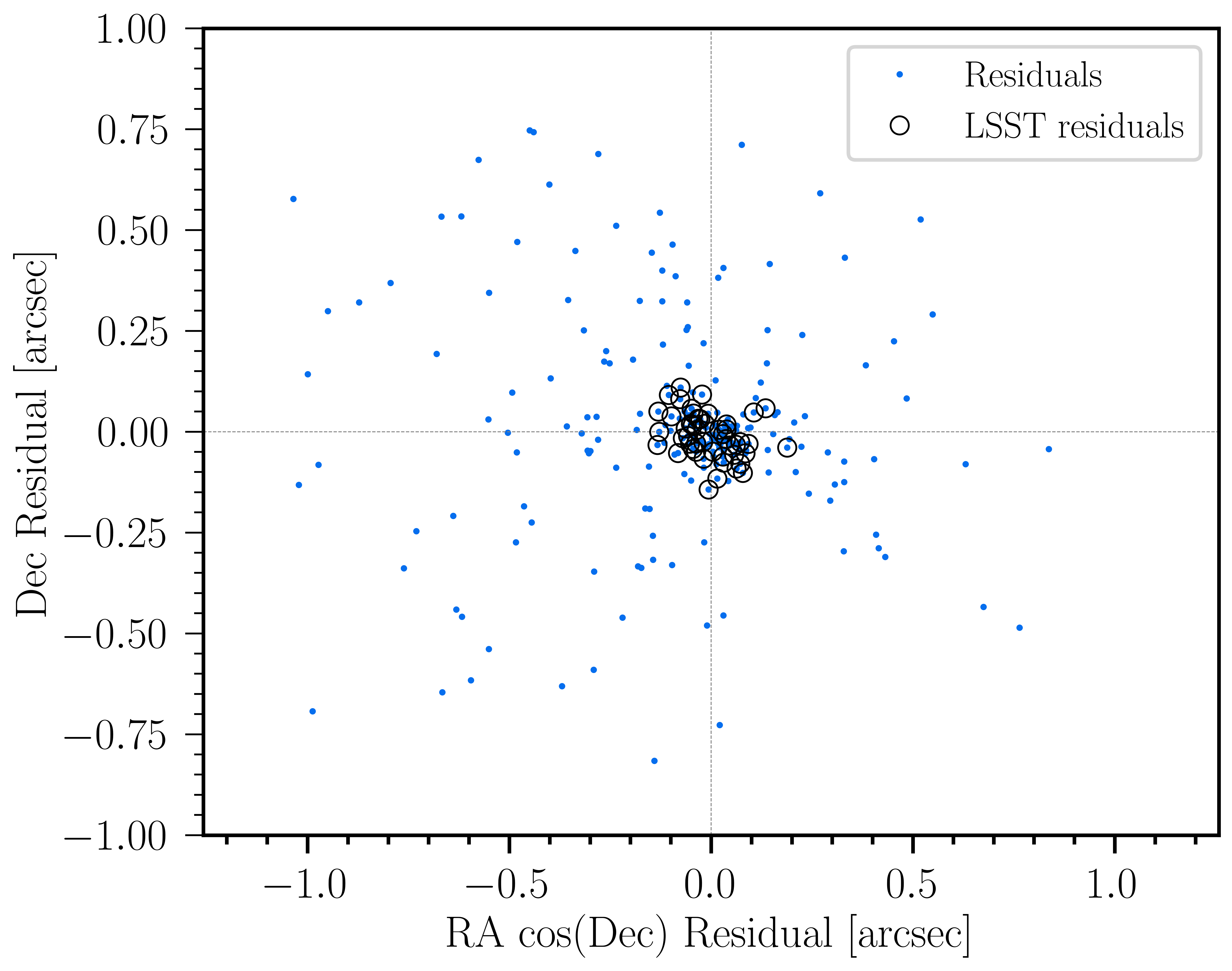} \\ \includegraphics[width=1.0\linewidth]{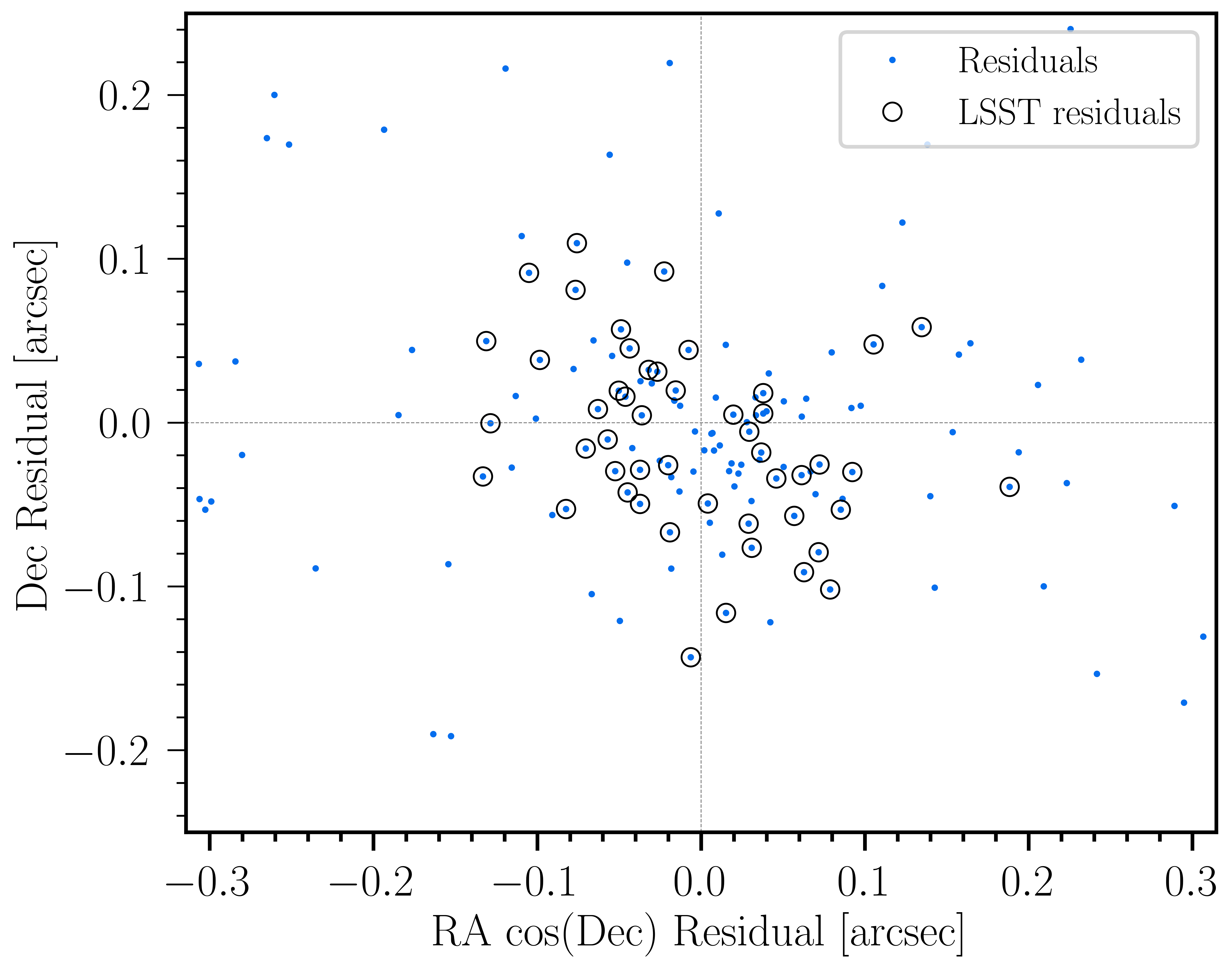} \\
    \caption{Astrometric residuals of \objname{} based on data available at the \acf{MPC} up until 2025 October 29.
    \textbf{Top: }Residuals in Right Ascension scaled by $\cos\left(\mathrm{Dec}\right)$ vs residuals in Declination. Points corresponding to residuals from LSST measurements are marked with rings around them.
    \textbf{Bottom: } Zoomed in view of top panel.
    }
    \label{fig:astrometry_residuals}
\end{figure}

To assess the impact of our curated astrometric measurements of \objname{}, we compute a new orbit solution using all available astrometry up to perihelion (2025 October 29) and perform a residual analysis using the \ac{GRSS} \citep{makadia2025gauss}. \ac{GRSS} is a user-friendly, open-source Python package with a C++ binding for high precision propagation and orbit fitting of small bodies in the Solar System. Astrometric measurements of asteroids by other observatories are commonly weighted according to the \citet{verevs2017statistical} scheme, and astrometry catalogs are debiased as suggested in \citet{eggl2020star}.
% However, since \objname{} shows signs of cometary activity, an alternative weighting scheme has been applied, where all astrometry, including Rubin astrometry, was weighted at 1\arcsec{} (likely higher than the actual astrometric performance of the telescope).
However, since \objname{} shows clear signs of cometary activity, a conservative astrometric rejection scheme must be used to reduce the effects of any centroiding and timing errors in the astrometry. The results of this orbit fit are presented in Figure \ref{fig:astrometry_residuals}. The LSST astrometry is conservatively weighed at 0.25~arcsec for orbit determination. The estimation leads to a reduced chi-squared, $\chi^2_\nu  = 0.1$ (see Figure \ref{fig:astrometry_residuals}), indicating a successful (albeit conservative) fit.

For the LSST observations, the postfit residuals in right ascension are $-5\pm70$~mas and those in declination are $-10\pm55$~mas, indicating that the curated LSST astrometry fits very well with respect to their uncertainties. The somewhat linear trend in the residuals is aligned with the direction of the tail and is a manifestation of the extended nature of \objname{}. It is important to note that the spread in the residuals is only a fraction of the measurement uncertainty and, therefore, is not statistically significant. Conversely, if all available astrometry at the MPC is accepted at face value with a weight of 1~arcsec, the LSST residuals in right ascension are 211$\pm$176~mas and those in declination are $-197\pm91$~mas. This more than one order of magnitude increase in the residuals highlights the dangers of ill-weighted astrometry in the orbit determination process, especially for extended objects such as \objname{}.

No evidence of significant non-gravitational acceleration can be found with a pre-perihelion dataset alone, but we note that an orbit solution containing post-perihelion data will need non-gravitational accelerations as part of the estimate.
% We have also verified these results using \textsc{Layup} and the \ac{MPC} orbit fitting software.

%%%%%%%%%%%%%
\subsection{Photometry}\label{subsec:photometry}

\subsubsection{Photometric Measurements}\label{subsec:photmeasurements}

As described in Section~\ref{sec:reduction}, the LSST pipeline performs automated photometric analyses, but, due to the special challenges presented by the \objname{} data reported in this work, we also conducted a more customized photometric analysis. 
We report photometry in an aperture of exactly $3.0\arcsec$ radius ($\sim$15 pixels), measured using our custom difference imaging and subtracting a sky background based on the clipped mean in an annulus from $6.0\arcsec$ to $8.0\arcsec$. This non-pipeline approach is necessary for several reasons, including the pipeline's limited performance on crowded fields (which is actively being improved) and the need to avoid self-subtraction of real flux from \objname{}. For each science image, we check for images suitable for template creation by requiring that the exposures (1) be in the same band as the science image, (2) show a similar PSF (and on-detector position, when possible), and (3) have acquisition times sufficiently separated in time (e.g., several minutes) such that \objname{} would be at least two aperture radii ($6 \arcsec$) from its location on the science image.
% \begin{itemize}
%     \item In the same band as the science image
%     \item Showing a similar PSF (and on-detector position when possible)
%     \item Sufficiently separated in time (e.g. several minutes) that 3I is at least two aperture radii (6 arcsec) from its location on the science image
% \end{itemize}

Prior to image subtraction, we used bilinear interpolation to resample the science and template-creation images onto a consistent astrometric grid using stereographic projection with celestial north up and a pixel size of exactly $0\farcs2$. For each science image, we constructed a clipped median stack of all template-creation images that met the above criteria. We matched the template image PSF to the science image by convolving with an optimized kernel \citep{alardlupton} composed of basis functions that involved Gaussians with $\sigma$ equal to 4.0, 2.0, and 1.0 pixels ($0\farcs8$, $0\farcs4$, and $0\farcs2$) multiplied by polynomials up to fourth, fourth, and fifth order, respectively. We did not attempt photometry for any science image that had no corresponding template-creation images, or when \objname{} fell on a background source too bright for adequate subtraction.

We photometrically calibrated our astrometrically resampled images using Pan-STARRS DR2 data obtained from the VizieR catalog service \citep{ochsenbein2000_vizier}. Calibrating large-aperture photometry is challenging in dense fields with no isolated stars. We avoided calibration stars brighter than magnitude 15.5 (15.0 in the $g$-band) due to concerns about saturation and nonlinearity. Subject to these maximum brightness constraints, we experimented using calibration sets in different magnitude ranges, from broad and faint (e.g., mag 18.0 to 15.5) to narrow and bright (mag 16.0 to 15.5). The calibration changed systematically as the mean magnitude of the calibration stars increased, indicating that contamination within our $3.0\arcsec$ photometric aperture may have persisted even in the brightest magnitude range. 

The problem was most pronounced in the reddest bands ($z$ and $y$), as the fields are most crowded in these bands. To address this problem, we performed an alternative calibration without clipping applied to the background calculation -- i.e., we deliberately included stellar flux in the ``sky'' measurement to statistically remove a mean density of background stars from the photometric aperture. We calculated the median photometric calibration in flux space rather than magnitude space to avoid problems from occasional negative fluxes that may result from this approach. Using this method, the calibration still changed systematically with the magnitude of the calibration stars, but now with the opposite sign --- an expected result that may be caused by clustering of the brighter stars. Both methods appeared to converge (from opposite directions) toward a similar calibration as we increased the mean brightness of the stars being used, with $\sim$0.05 mag agreement for bluer bands. For the $z$ and $y$ bands in the densest fields, the disagreement was still $\sim 0.20$ mag for the brightest cohort of stars. Hence, we take the average of the two calibration methods as our final calibration, and report approximately half the difference as our systematic error: 0.03 mag in $g$- and $r$-band, 0.05 mag in $i$, and 0.10 mag in $z$ and $y$. The internal consistency of our photometry makes it clear that systematic errors dominate the random error in every case, so we have not attempted to rigorously quantify the latter.

%We measure and report both aperture (fixed at 12 pixels or $2\farcs4$) and PSF magnitudes measured by the \ac{LSST} pipelines. As part of the image processing, the \ac{LSST} stack attempts to deblend sources using the Scarlet Lite algorithm, based on Scarlet \citep{melchior2018_scarlet}, thus mitigating to an extent the effects of blending in crowded fields (we verify this assumption further below). To the computed errors, we also add in quadrature, a 0.01\, mag noise floor for the photometric measurements, the design requirement for Rubin \citep{ivezic2019}. We note that this is a conservative estimate, as photometric performance of 0.005 mag has already been demonstrated with the \ac{LSST} ComCam \citep{RTN-095}.

% done through here 12/13/2025 COC

\subsubsection{Photometric variability}\label{subsec:lightcurve}

\begin{figure*}
    \centering
         \includegraphics[width=\linewidth]{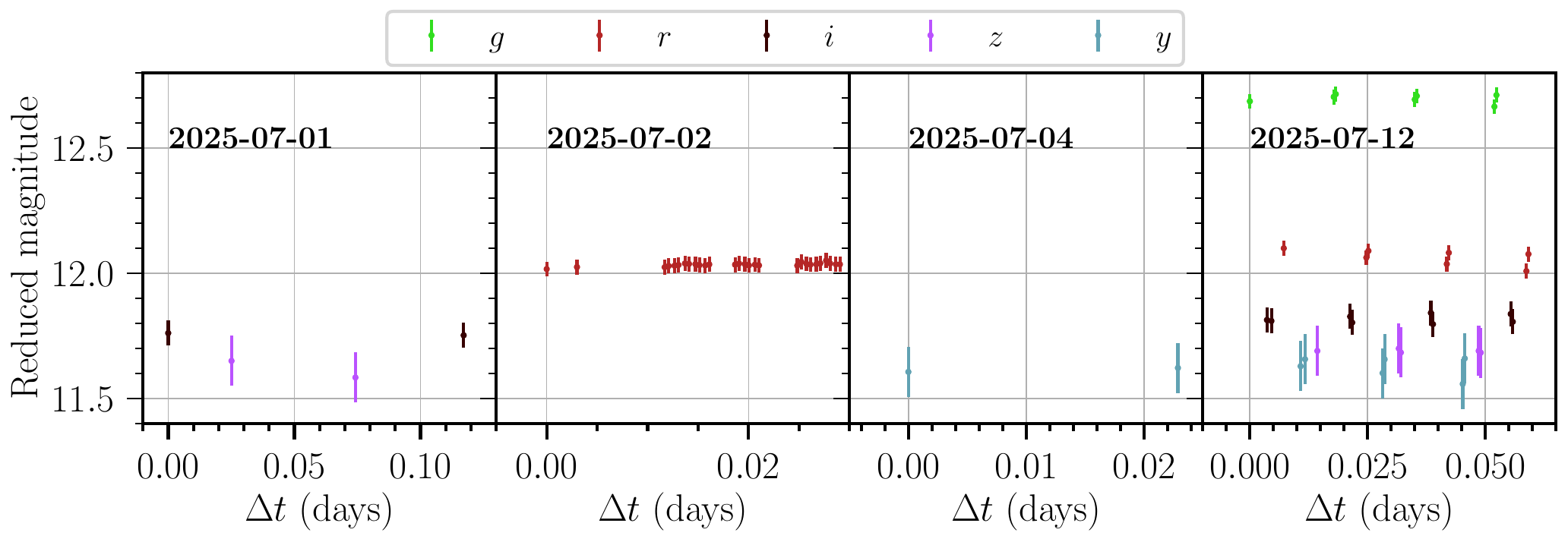}\\
         \includegraphics[width=\linewidth]{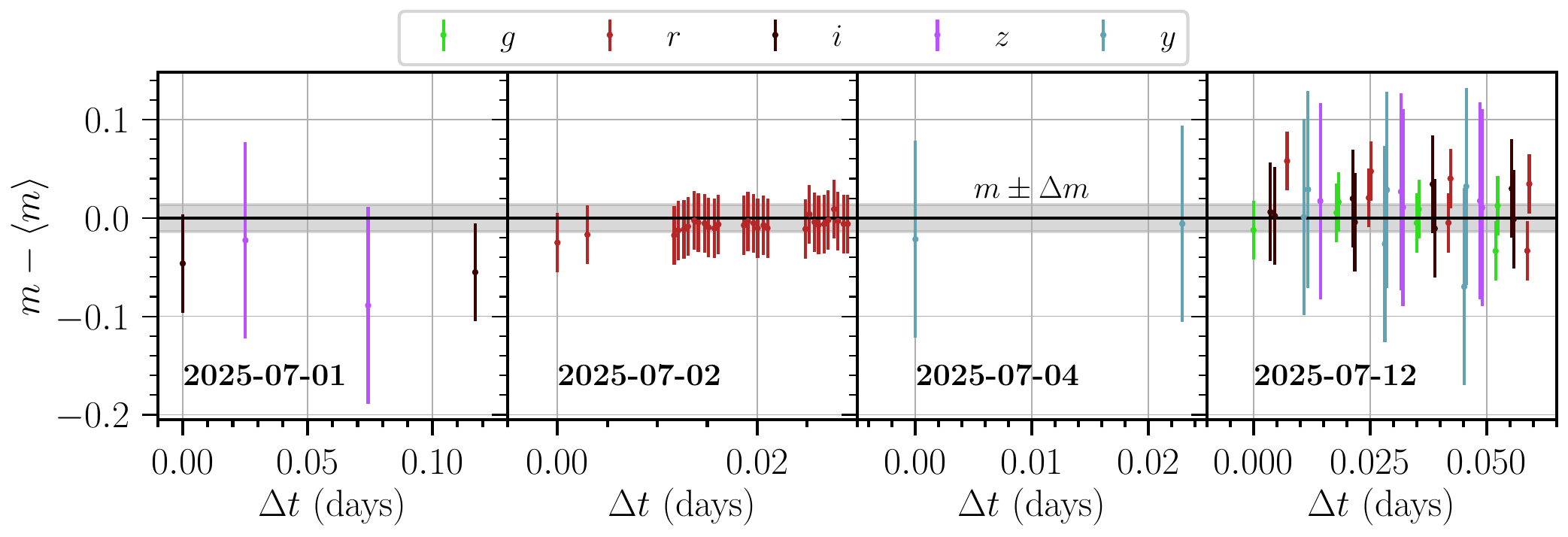}\\
        \vspace{10pt}
          \includegraphics[width=0.32\linewidth]{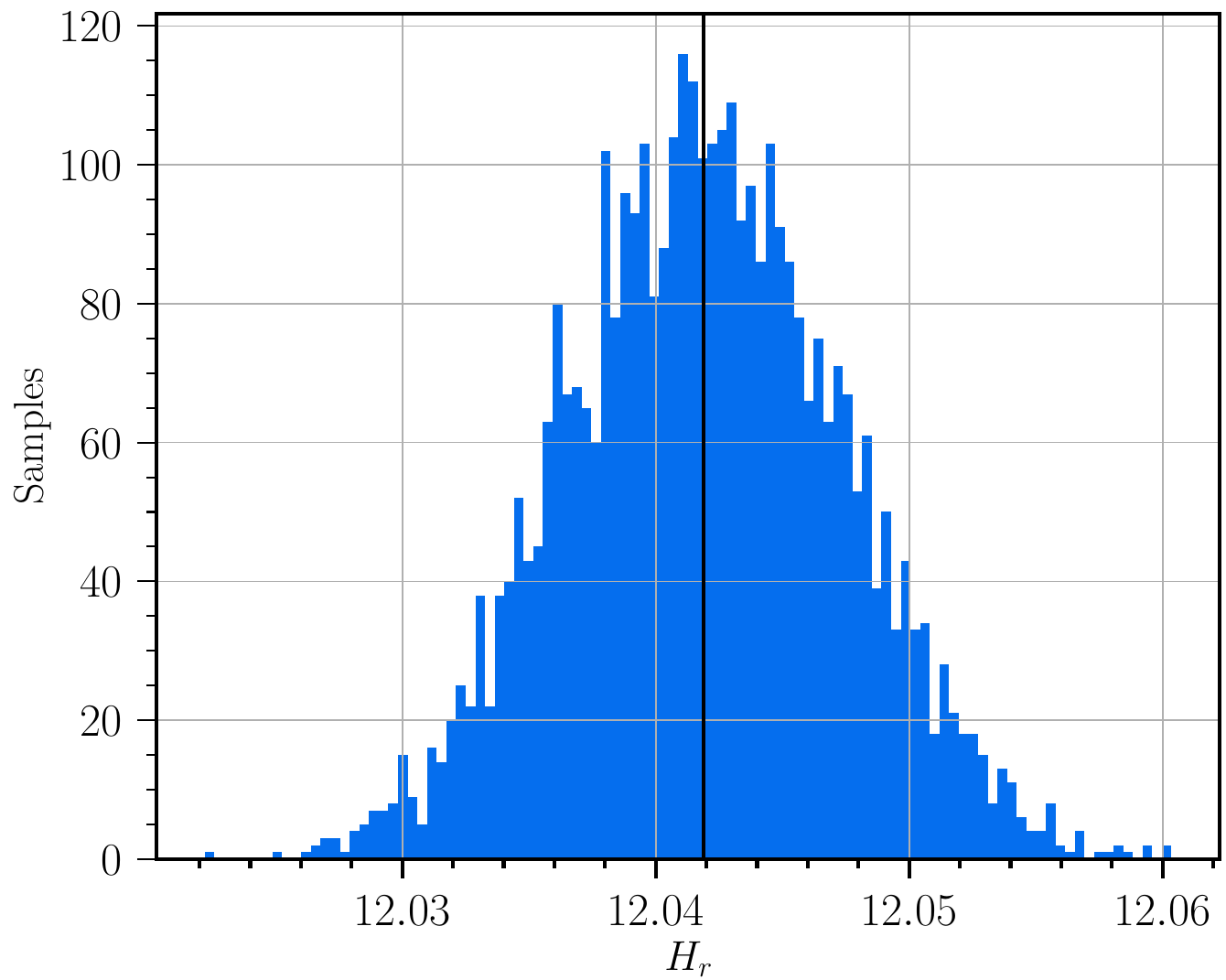} 
        \includegraphics[width=0.32\linewidth]{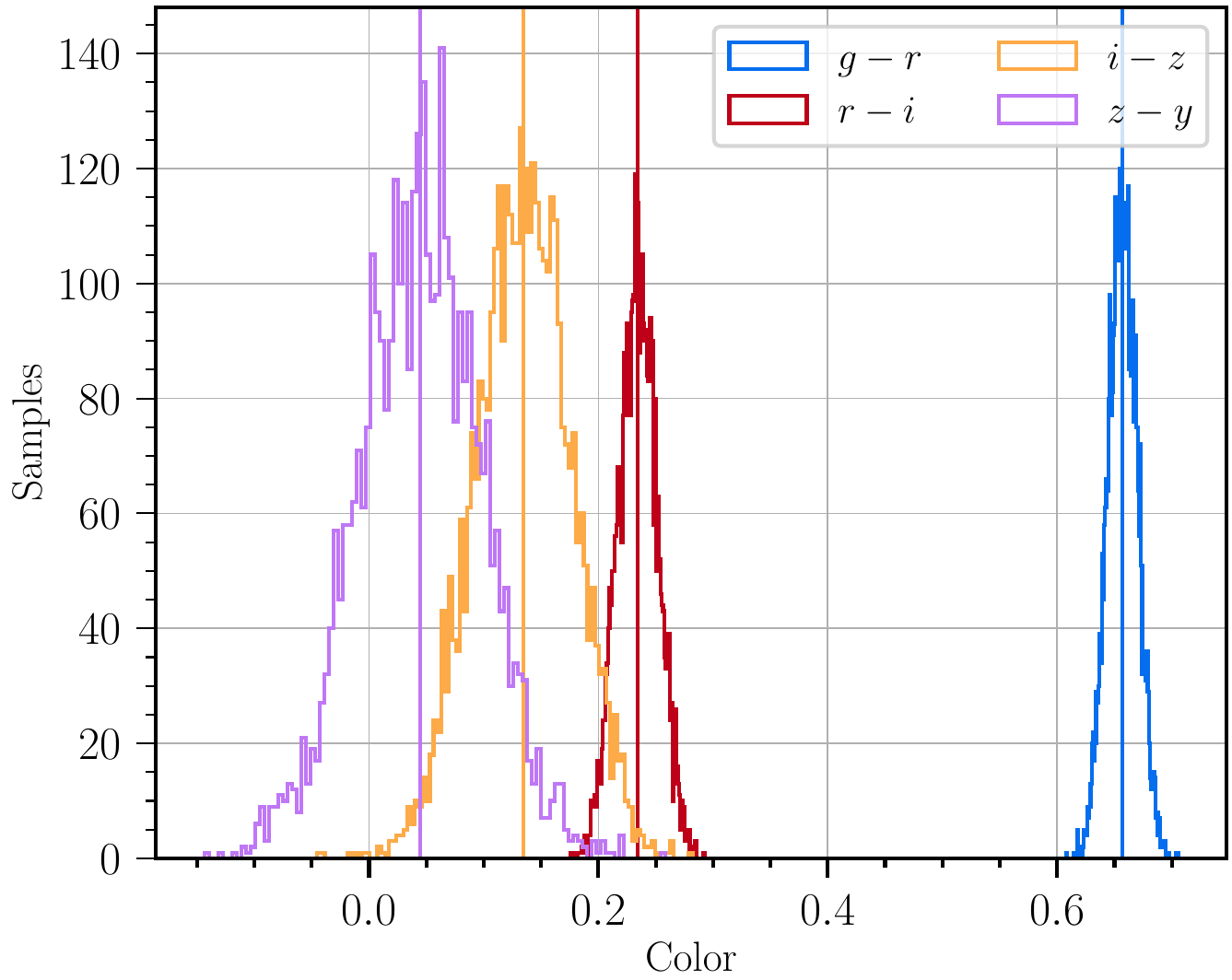} 
        \includegraphics[width=0.32\linewidth]{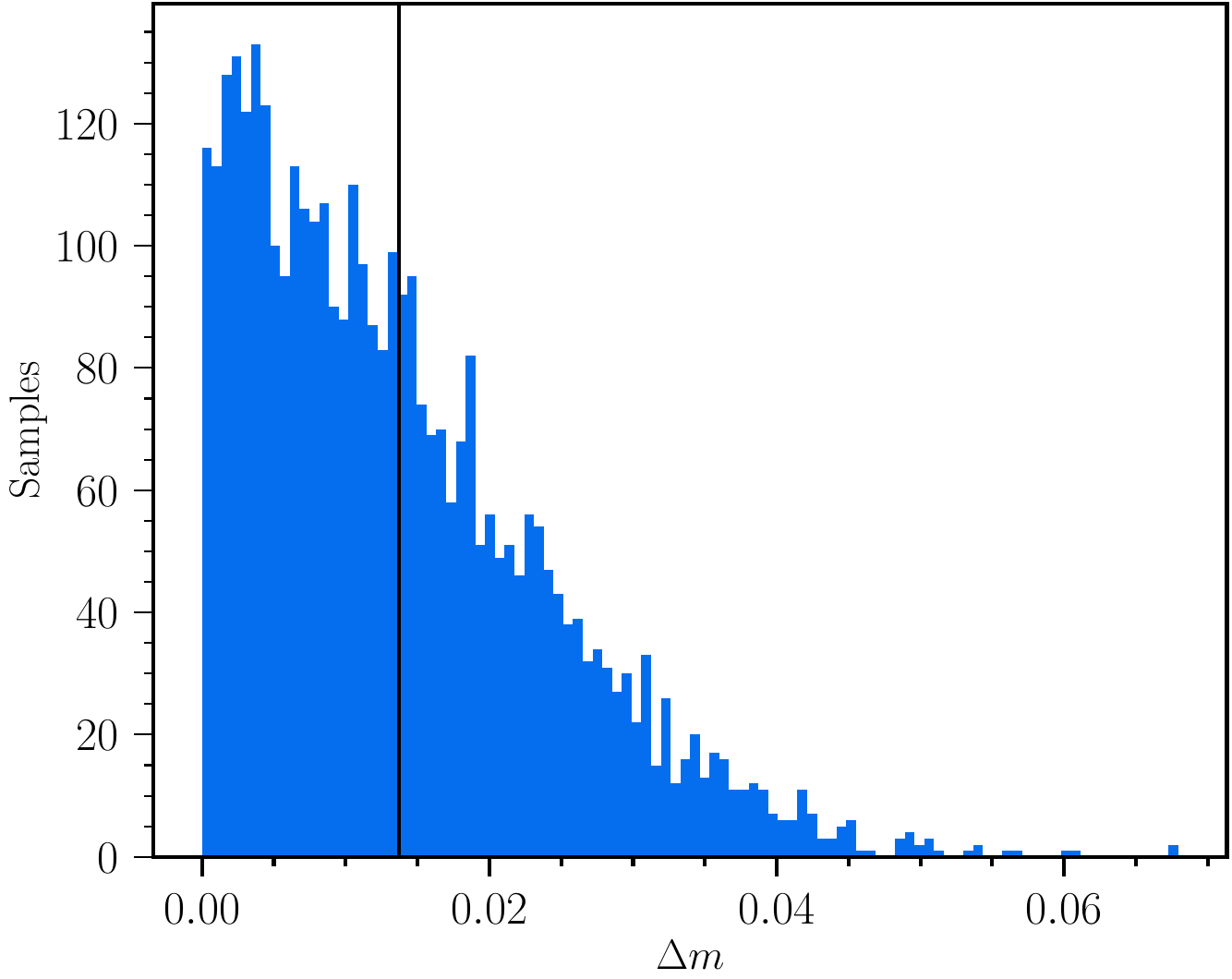} 
    \caption{\textbf{Top:} The two horizontal panels show time series photometry of \objname{} taken on four different nights and in the LSST $grizy$ bands. The top panel shows the reduced magnitude (where distance to the observer and the Sun is subtracted), while the bottom panel shows the same value, but where the mean magnitude per band is also subtracted, i.e., any excess scatter is due to lightcurve variations. The gray shaded area in the bottom panel represents the 68\% region of the estimated variability $\langle \Delta m \langle$. \textbf{Bottom left:} Probability distribution of the mean magnitude yielded by this procedure, with the vertical line indicating the mean measurement $\langle{H_r}\rangle = 12.042 \pm 0.005\, \mathrm{mag}$. \textbf{Bottom center:} Probability distribution of the color pairs $g-r$, $r-i$, $i-z$ and $z-y$. The mean value (vertical line) is indicated; all values are summarized in Table \ref{tab:3I_colors}. \textbf{Bottom right:} The probability distribution of the peak-to-peak light curve amplitude (in magnitudes) $\Delta m \equiv 2.5 \log_{10} [(1+A)/(1-A)]$ for the flux measurements over this timespan, with $\langle \Delta m \rangle  = 0.014 \pm 0.010 $.}
    \label{fig:lightcurve_random}
\end{figure*}

The sequence of observations taken on UT 2025 July 3 gave us a unique opportunity to measure the variability of \objname{} on approximately minute timescales. We applied the Bayesian methodology of \cite{bernardinelli2023photometry} to the fluxes from Section \ref{subsec:photometry}. We assumed every measurement at time $\mu$ and band $b$ had a flux measurement $f_{\mu, b} = \langle{f_b}\rangle [1 + A h(\phi_\mu)]$, where $\langle{f_b}\rangle$ is the object's mean flux at band $b$, $A$ is its light curve semi-amplitude, and $h (\phi)$ is a time-varying function with phase $0 \leq \phi \leq 1$. We marginalized over $\phi$ for each visit, thus assuming that the object was seen at a random point of its light curve, which, for simplicity, we used $h(\phi) = \sin (2\pi \phi)$. In practice, this procedure yields a statistically rigorous Monte Carlo sampling of the probability distribution of the object's mean flux and its variability without detailed modeling of its light curve (\emph{e.g.}, phase folding or shape modeling). We present the results of this analysis in the bottom panel of Figure \ref{fig:lightcurve_random}. Our photometry is consistent with $A$ on the order of 6\%. This means that we can exclude variations bigger than $\approx 0.1 \, \mathrm{mag}$ in the timescale of these observations. Incidentally, this also demonstrates good photometric stability of the Rubin system even in such a crowded field and for a moving source.
% 0.14 to 0.1 after discussing with Pedro 6/24/2026 COC

%Given the multi-band nature of the data, this analysis also yields robust colors for \objname{}, which are presented in Table \ref{tab:3I_colors}.

% We end with a word of caution regarding deriving even lower-level variability or a light curve period from the photometry presented here. Despite our efforts at deblending, these measurements are still likely to be contaminated (at some unquantified level) by the presence of background stars, and signals from \objname{} passing on top of stars may still be present in its light curve.

While techniques such as difference imaging \citep{alardlupton,zogy,sedaghat} would be ideal to extract the photometry of this object, at the time these images were acquired, there were few visits in each of these fields (including those with the object). Because of this, high-quality templates are more challenging to construct and may not yield image differences that improve upon the measurements presented above. The UT 2025 July 3 data include 28 $r$-band observations of \objname{} within an hour (see Figure \ref{fig:lightcurve_random}), which can also be challenging due to the potential for self-subtraction, as the object moves $\lesssim 1 \, \mathrm{PSF}$ across two pairs of visits. Another option, which we leave for future work, would be to explore techniques such as scene modeling photometry \citep[\emph{e.g.,}][]{bernardinelli2023photometry}, which can measure the target at multiple images while simultaneously accounting for contamination of background sources.  

%%%%%%%%%%%%%%%%%%%%%%%%%%%%%%%%%%%%%%%%%%%%%%%%%%%%
\subsubsection{Colors} \label{subsec:colors}

% The vast majority of the observations reported in this work were not obtained in the nominal \ac{LSST} two-filter pairs, as the observatory was still in the commissioning phase. Thus, there were very few observations taken in different filters within a short period of time (cf Table~\ref{tab:observations}). This makes determining the colors from Rubin data alone insufficiently robust, given the evolution of activity and colors with changing distance and corresponding activity. This is coupled with additional uncertainty on potential rotational variation and potential contamination by background sources (as discussed in Section~\ref{subsec:lightcurve}. 
Here, we derive \objname{} colors using a model that leverages the previously measured object slope and directly from our own measurements. 
First, recalling that \citet{seligman2025_3I} and \citet{opitom_3I} independently reported spectral gradients of $S'\sim18$\%/100~nm for \objname{} (Section~\ref{subsec:3I_intro}), we use this spectral gradient to compute equivalent colors for the object in the LSST $ugrizy$ filter set (which differs from the SDSS filter system).  
Using the {\tt spectroscopy} module\footnote{\url{https://sbpy.readthedocs.io/en/stable/sbpy/spectroscopy/index.html}} in {\tt sbpy} and effective central wavelengths for LSST filters from the LSST website\footnote{\url{https://rubinobservatory.org/for-scientists/rubin-101/instruments}}, we compute colors for \objname{} for each filter pair given a spectral gradient of $S'=18\%$/100~nm (normalized at the central wavelength of the $g'$-band filter, 480.69~nm) and report these in Table~\ref{tab:3I_colors}. 
For reference, in the same table, we also show solar colors in LSST filters that were computed from a \citet{kurucz1993_modelatmospheres1} model spectrum of the Sun for use in the computation of the object's colors from its spectral gradient.

\setlength{\tabcolsep}{5pt}
\setlength{\extrarowheight}{0em}
\begin{table}[htb!]
\caption{Solar, Predicted, and Measured Colors}
\centering
\smallskip
%\footnotesize
\begin{tabular}{cccc}
\hline\hline
\multicolumn{1}{c}{Color}
 % &~~~~~~~~~~~
 & \multicolumn{1}{c}{Solar}
 % &~~~~~~~~~
 & \multicolumn{1}{c}{\objname{}$^a$}
 % &~~~~~~~~~
 & \multicolumn{1}{c}{\objname{}$^b$}
 \\[2pt]
\hline
$(u-g)_{\rm LSST}$ & 1.154 & 1.384 & $\cdots$ \\ % 
$(g-r)_{\rm LSST}$ & 0.436 & 0.677 & \colorGminR{} \\ % 0.657 $\pm$ 0.013  \\ %  
$(r-i)_{\rm LSST}$ & 0.112 & 0.299 & \colorRminI{} \\ % 0.235 $\pm$ 0.018  \\ %  
$(i-z)_{\rm LSST}$ & 0.011 & 0.147 & \colorIminZ{} \\ % 0.133 $\pm$ 0.042  \\ %  
$(z-y)_{\rm LSST}$ & 0.009 & 0.125 & \colorZminY{} \\ % 0.047 $\pm$ 0.052  \\ %  
\hline
\hline
\multicolumn{4}{l}{$^a$ Expected colors in LSST filters for \objname{} } \\
\multicolumn{4}{l}{~~~assuming a spectral gradient of $S'=18$\%.} \\
\multicolumn{4}{l}{$^b$ Our measured colors of \objname{}.} \\
% \multicolumn{5}{l}{~~~a spectral gradient of $S'=18$\%} 
\\
\end{tabular}
\label{tab:3I_colors}
\end{table}

% PEDRO COLOR SECTION HERE 12/13/2025 COC
%%%%%%%%%%%%%%%%%%%%%%%%%%%%%%%%%%%%%

%%%%%%%%%%%%%%%%%%%%%%%%%%%%%%%%%%%%%%%%%%%%%%%%%%%%%%

% NOTE: \INPUT FILES ARE AUTOMATICALLY REWRITTEN
% TODO: add RA, Dec, uncertainties on both, magnitudes, mag uncertainty, Pedro/Colin classification flag (read: probable blending), ...

% NOTE: 7/2 was when we started our search!

\subsection{Morphology} \label{subsec:morphology}
% done through here 12/13/2025 COC

%%%%%%
\subsubsection{Overview}\label{subsubsec:morph_overview}

\renewcommand{\figsize}{0.24}
\begin{figure*}
    \centering
    \begin{tabular}{c}
        \includegraphics[width=\figsize\linewidth]{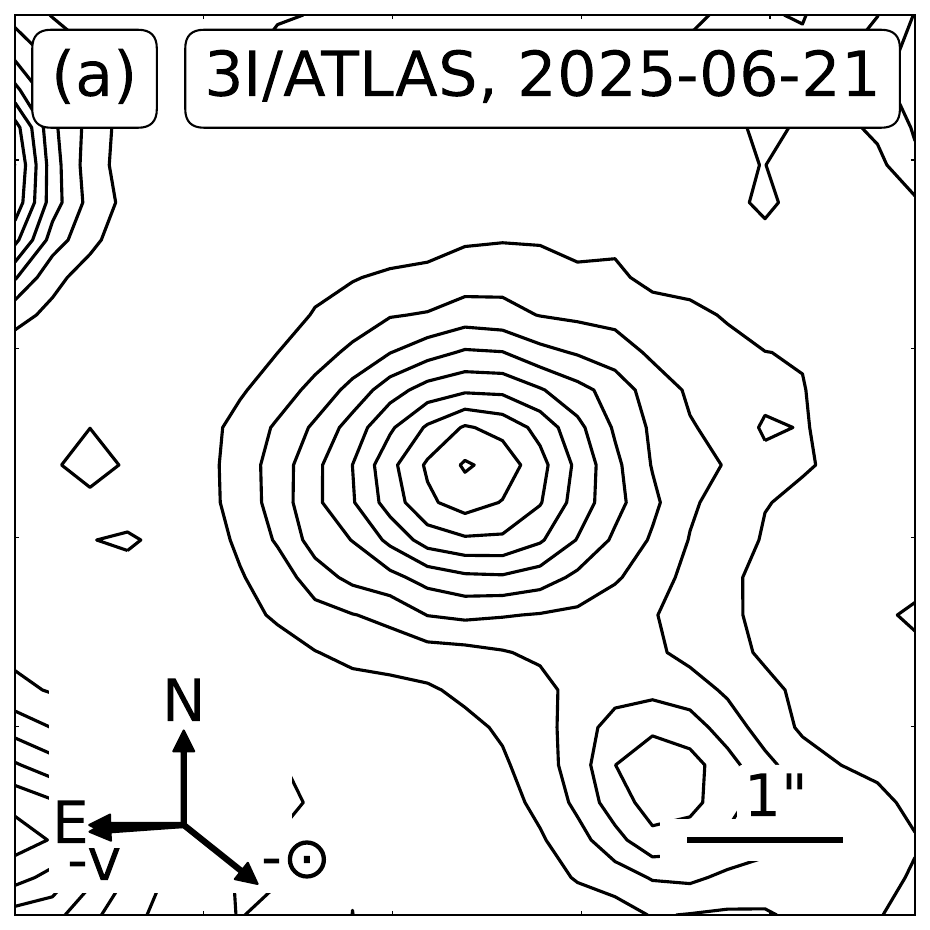} 
        \includegraphics[width=\figsize\linewidth]{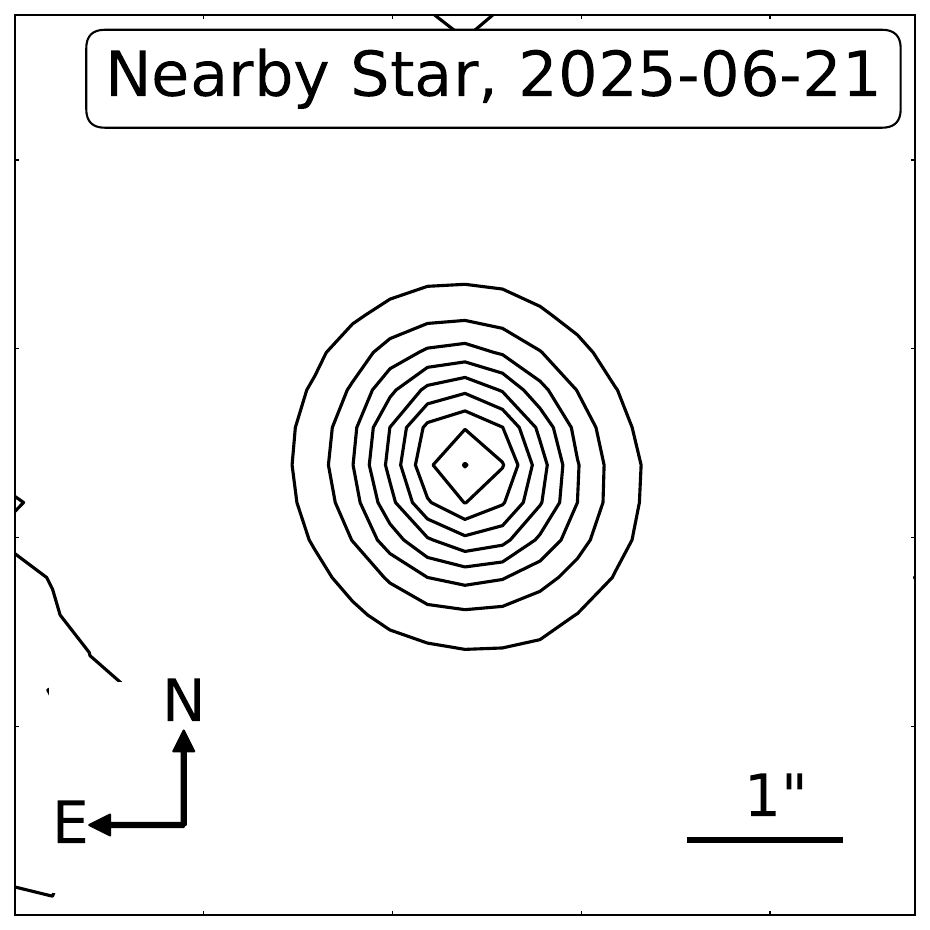}$~~$
        \includegraphics[width=\figsize\linewidth]{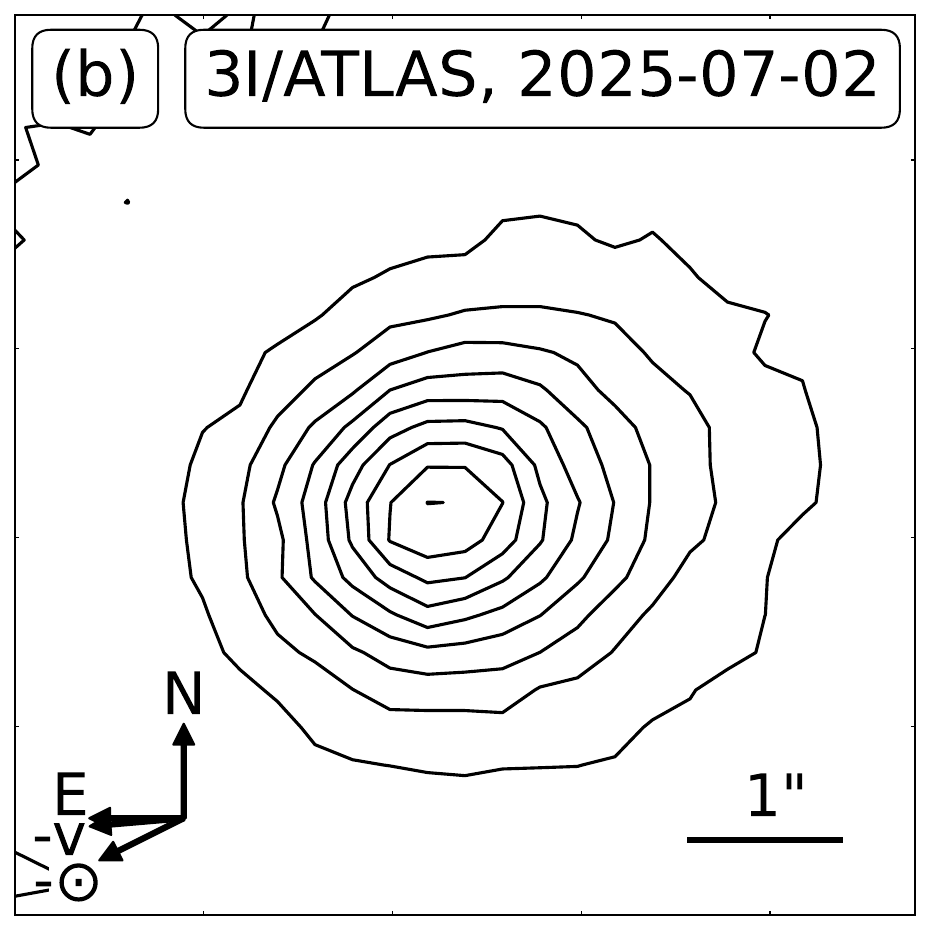}
        \includegraphics[width=\figsize\linewidth]{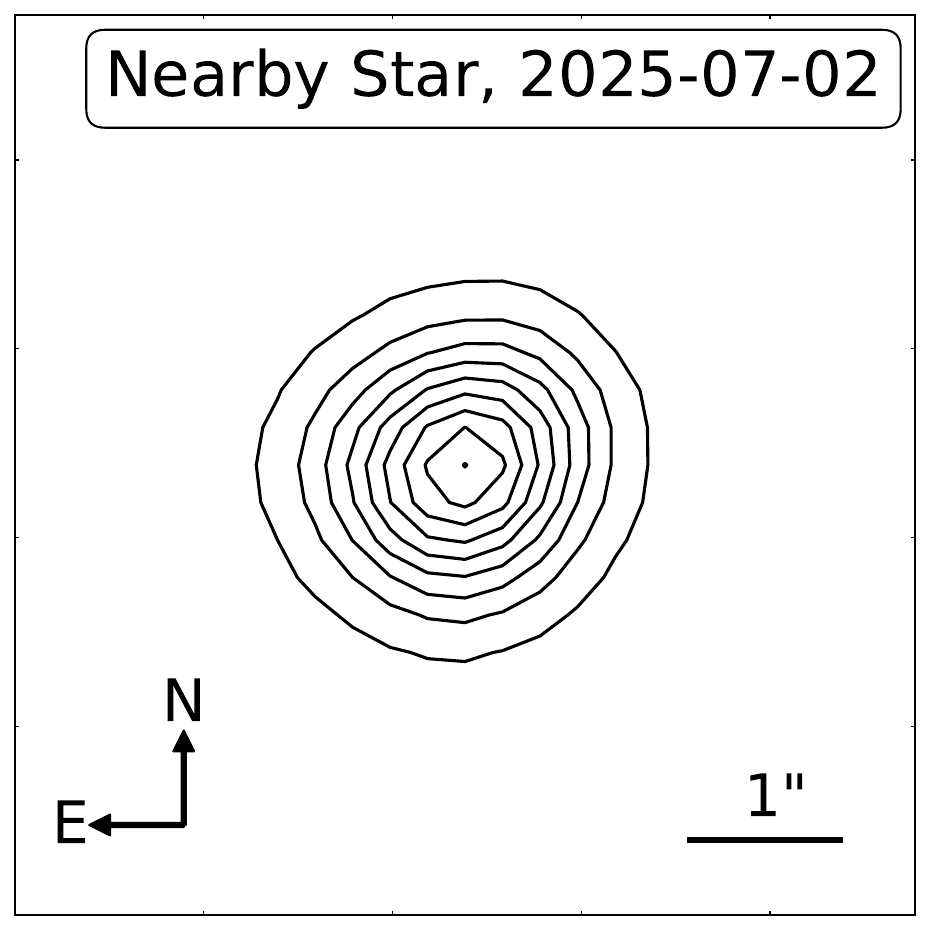}
        \\
        %\includegraphics[width=\figsize\linewidth]{n20250621.3I.contour.img.png} 
        %\includegraphics[width=\figsize\linewidth]{n20250621.star01.contour.img.png}$~~$
        %\includegraphics[width=\figsize\linewidth]{n20250702.3I.contour.img.png}
        %\includegraphics[width=\figsize\linewidth]{n20250702.star05.contour.img.png}
        %\\
        \includegraphics[width=\figsize\linewidth]{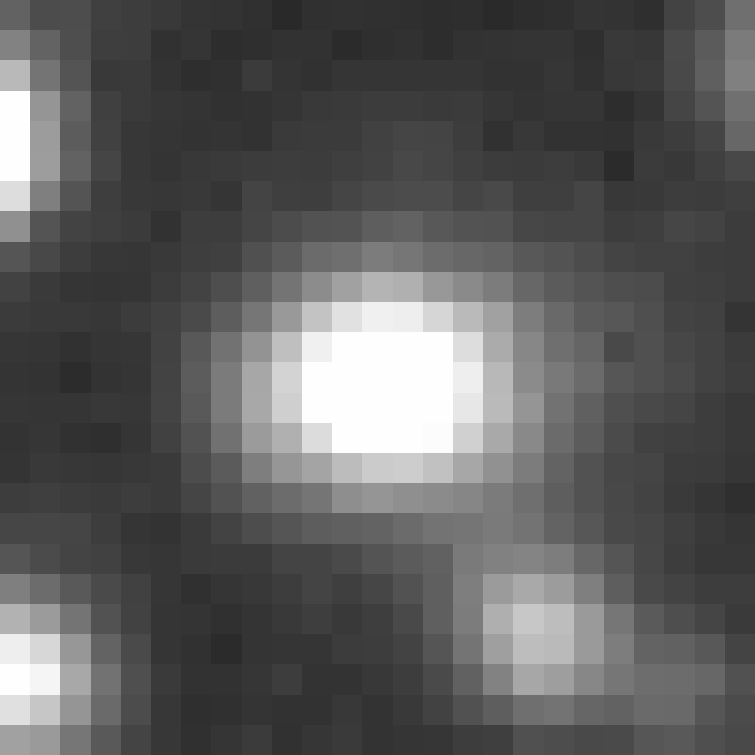} 
        \includegraphics[width=\figsize\linewidth]{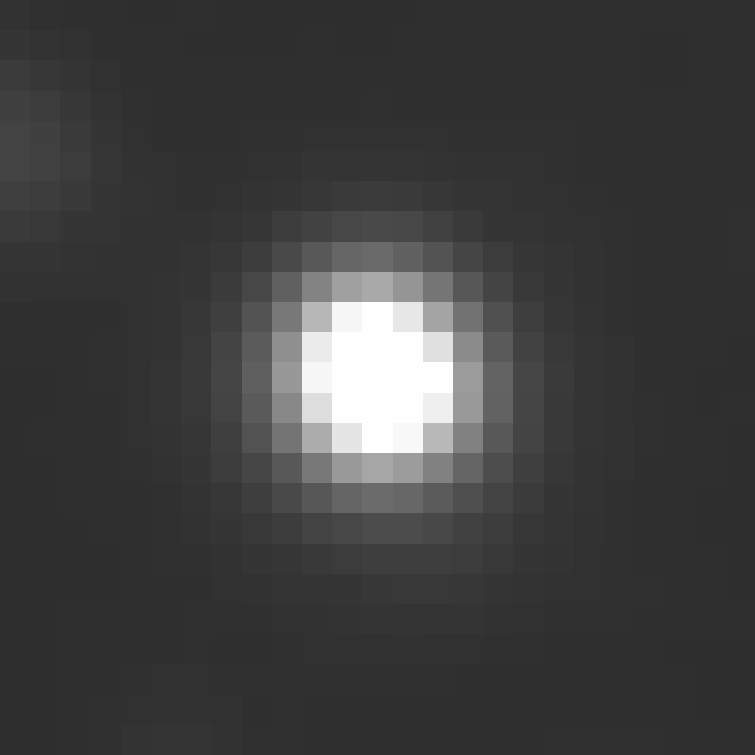}$~~$
        \includegraphics[width=\figsize\linewidth]{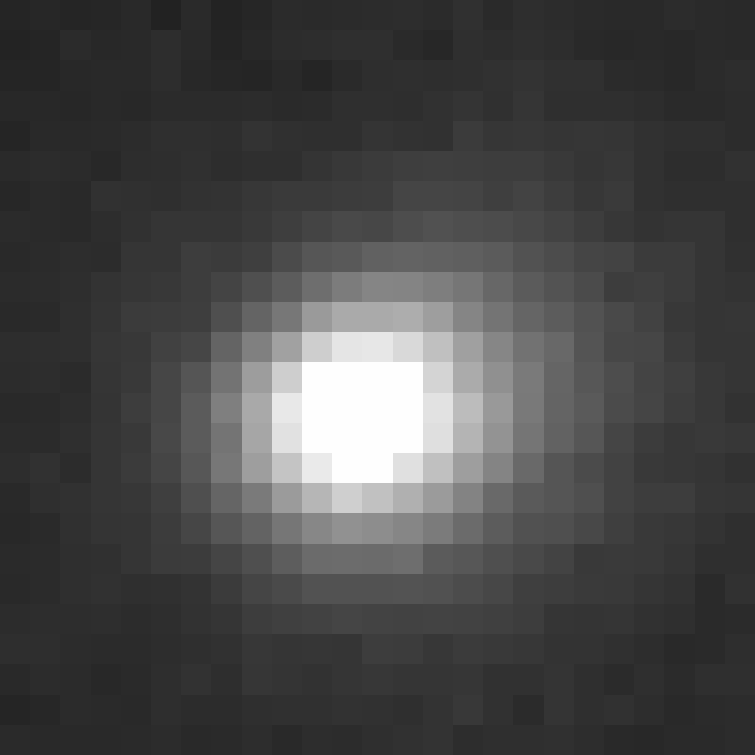}
        \includegraphics[width=\figsize\linewidth]{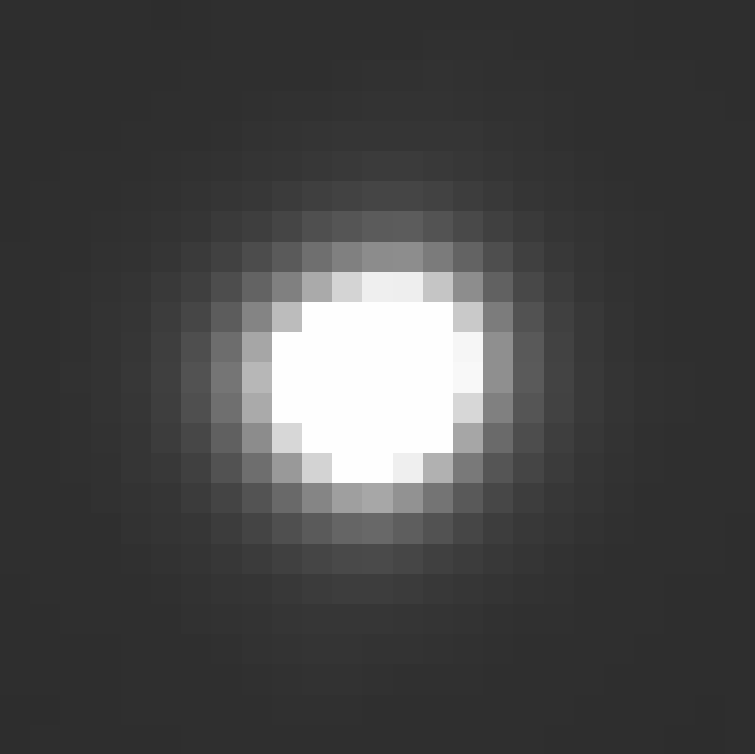}
        \\
    \end{tabular}
    \caption{Contour plots (top row; using 10 logarithmically-spaced contour levels ranging from the peak value of each image to the average background level in each image) of LSST images (bottom row) of \objname{} (left panel) and a nearby (within 1~arcmin of the target) reference field star from the same image (right) from (a) UT 2025 June 21 and (b) UT 2025 July 2, centered on the target. The FOV of each panel is $5\arcsec \times 5\arcsec$. North (N), east (E), and the position angles of the anti-solar vector ($-\odot$) and negative heliocentric velocity vector ($-v$) as projected on the sky are indicated as labeled. Corresponding grayscale images on which each contour plot is based are shown below each contour plot.
    }
    \label{fig:contours}
\end{figure*}

To study \objname{}'s morphology, we first construct contour plots of detections from UT 2025 June 21 at 08:11:33 and UT 2025 July 2 at 03:33:02, where the object was largely isolated from background sources (Figure~\ref{fig:contours}). On June 21, the target appeared somewhat extended in the East-West direction, the direction of its apparent non-sidereal motion. Therefore, this morphology may have been due to trailing (of \objname{}) in these sidereally tracked images. On July 2, the object appeared to have a notably extended and asymmetric coma, with a \ac{PA} at $\sim290^{\circ}$ east of north, very nearly opposite the direction of the projected anti-solar vector \ac{PA}. In other words, rather than pointing away from the Sun as is most commonly the case, the tail appeared to be pointing toward the Sun. 
To more quantitatively analyze the morphology of \objname{}, we first perform a radial surface brightness profile analysis (Section~\ref{subsec:radial_profiles}), followed by an analysis of one-dimensional surface brightness profiles measured perpendicular to the direction of the object's motion (Section~\ref{subsec:oned_profiles}).
Finally, we discuss \objname{}'s apparent sunward tail in Section~\ref{subsubsec:tail}.

\subsubsection{Radial Surface Brightness Profile Analysis}\label{subsec:radial_profiles}

\begin{figure}
    \centering
    \begin{tabular}{c}
            \includegraphics[width=1.0\linewidth]{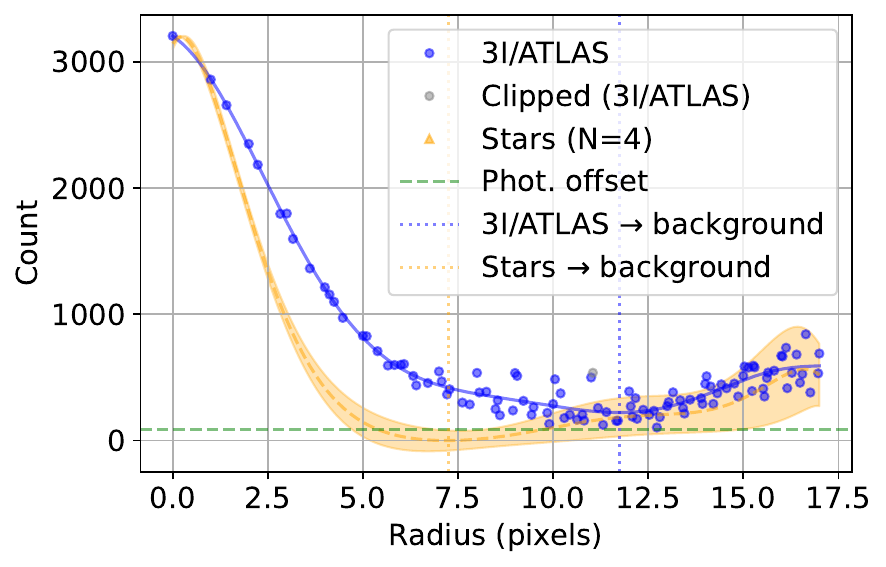} \\ \includegraphics[width=1.0\linewidth]{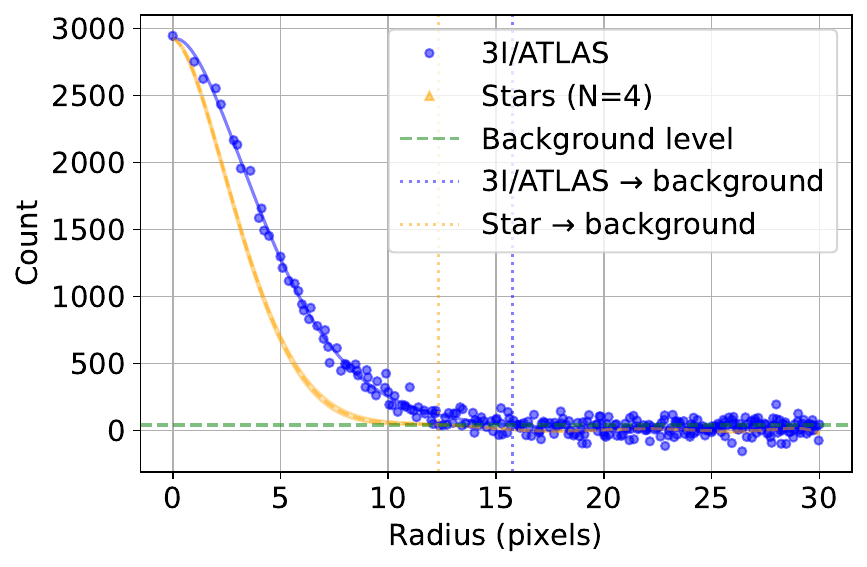} \\
    \end{tabular}
    \caption{Radial profile measurements of \objname{} (blue circles) and a set of four vetted comparison stars (yellow shaded region), along with polynomial fits (solid blue line and dashed yellow line, respectively). Data points beyond $3\sigma$ from the polynomial fit were clipped (grey markers, when present), and the function was re-fit. The signal floor is indicated by a horizontal green dashed line. Vertical dashed lines indicate the radial extent of \objname{} (blue) and the mean of the comparison stars fit (yellow). 
    \textbf{Top: } 2025-06-21. Coma extent at 11.75 px: $\sim$6,520~km. $r_H=$ 4.834~au. The signal floor is below zero flux, so a photometric offset has been applied to the fit. 
    \textbf{Bottom: } 2025-07-02. Coma extent at 18.64 px: $\sim$9,380~km. $r_H=$ 4.476~au. Residual background signal was measurable, so additional background subtraction was incorporated into the measurements.     
    } % 6,520.84~km, 9,383.96
    \label{fig:radialprofiles}
\end{figure}

We perform a radial profile analysis of \objname{} (Figure \ref{fig:radialprofiles}) when it was relatively isolated from background sources, despite the object transiting a crowded field. We carried out our analyses with $600\times600$ pixel ($120\arcsec \times 120\arcsec$) cutouts extracted via the Rubin Science Platform and \texttt{astropy}'s \texttt{Cutout2D} function, preserving an adequate \ac{WCS} for the purpose (e.g., matching to Gaia stars). The two \objname{}-isolated images were from UT 2025 June 21 and UT 2025 July 2. 

For each image, we prescribe an aperture sufficient to capture all of the flux of \objname{} discernible from the background (or noise floor; see below); these were 20 and 30 pixels ($4\arcsec$ and $6\arcsec$), respectively. We also randomly selected several background stars throughout the field, also in relative isolation, though the field was sufficiently crowded that only a handful of isolated stars were usable. Larger cutouts might have provided additional stars; however, proximate stars are more likely to conform to the same general shape, given that some data were acquired during engineering time. We summed the flux of \objname{} and the field stars in concentric radial annuli.

In Figure \ref{fig:radialprofiles} we show the azimuthally averaged radial profile of the reflected light from \objname{} to show the point spread function compared to that of nearby field stars. The excess of the radial profile compared to that of the nearby stars is indicative of cometary activity. However, the Rubin \ac{LSST} images are not tracking the motion of \objname{} and the exposure time of each image is 30~s. Therefore, there will be both trailing loss and smearing of the radial profile with respect to that of the comparison star. To quantify the magnitude of the effect, we first consider the object's apparent sky rates of motion of $\sim 60\arcsec/\mathrm{hr}$ (UT 2025 June 21, Table \ref{tab:observations}) and $\sim75\arcsec/\mathrm{hr}$ (UT 2025 July 2). Thus, in each exposure, the object had moved $\sim0.5\arcsec$ (2.5 pixels) and $\sim0.625\arcsec$ (3.125 pixels), respectively. At first glance, this appears to be comparable in magnitude to the offset between the comparison star and the \objname{} profile (see the region between 0.5 - 7.5 pixels in Figure \ref{fig:radialprofiles}, top panel). However, we note that the smearing of the profile should \textit{only} operate along the direction of motion. Should there be just a trailing starlike nucleus with no coma, we should see at least a few data points representing pixel intensities of 3I/ATLAS that lie along the stellar PSF curve (belonging to pixels lying perpendicular to the motion). We do not see such points, which strongly support the presence of a coma. %Therefore, the azimuthal averaging should reduce this effect, and we conclude that there is still evidence for cometary activity.

These images were fully processed by the LSST Science pipelines \citep{PSTN-019} and, consequently, should have already had their background values subtracted. However, (a) commissioning is ongoing, and (b) crowded fields have proven challenging due to the absence of discernible background (i.e., little to no area on the sky without one or more sources in the field). In the case of our two isolated images, the June 21 image underwent background subtraction that resulted in appreciable ``negative flux,'' and the July 2 image still had some background flux measurable in our aperture.

To address the discrepant background levels, we fit a polynomial function to each set of measurements and derive a limit determined by the inflection point (i.e., where the curve reaches its floor). For UT 2025 June 21, \objname{} had an apparent radial extent of 11.75 pixels (2.35$\arcsec$, or $\sim6,520$~km). On UT 2025 July 2, the extent was 18.64~pixels $\approxeq$ 3.73$\arcsec$ $\approxeq$ $9,384$~km. These should be considered very conservative lower limits as the object appeared nearly head-on, so the tail could have extended far along the $z$-axis (as projected on the sky).

%%%%
\subsubsection{One-Dimensional Surface Brightness Profile Analysis}\label{subsec:oned_profiles}

In an effort to specifically quantitatively characterize the amount of coma present on these different dates, following \citet{luu1992_neoprofiles} and \citet{hsieh2005_phaethon}, we performed an analysis in which we compare one-dimensional surface brightness profiles of the comet, as measured perpendicularly to the direction of apparent non-sidereal motion, to those of template field stars with varying amounts of synthetically added spherically symmetric and steady-state coma following a $r^{-1}$ radial surface brightness profile, where $r$ is the angular distance from the photocenter as projected on the sky.  In this modeling analysis, coma levels were parameterized by $\eta=C_c/C_n$, where $C_c$ and $C_n$ were the total scattering cross-sections from the coma and nucleus (assuming that both have the same effective albedos), respectively, where we used a reference photometry aperture of $r_{\rm phot}=5\farcs5$ for measuring these fluxes.

For measuring one-dimensional surface brightness profiles, we rotated images of \objname{} and a selected well-isolated field star in each image to align the direction of motion horizontally in the image frame, where in this case, the non-sidereal motion of \objname{} was very nearly exactly West to East on June 21 and July 2, and so the orientations of the rotated images are virtually identical to those of the images in Figure~\ref{fig:gallery} and contour plots in Figure~\ref{fig:contours}.  We then averaged pixel values over horizontal rows over the entire widths of the object and reference stars, and subtract sky background sampled from nearby areas of blank sky.  Object and model profiles were then normalized to unity at their peaks and plotted together (Figure~\ref{fig:sbps}) to search for the closest matching model profile to each object profile.

\renewcommand{\figsize}{1.0}
\begin{figure}
    \centering
    \begin{tabular}{c}
        \includegraphics[width=\figsize\linewidth]{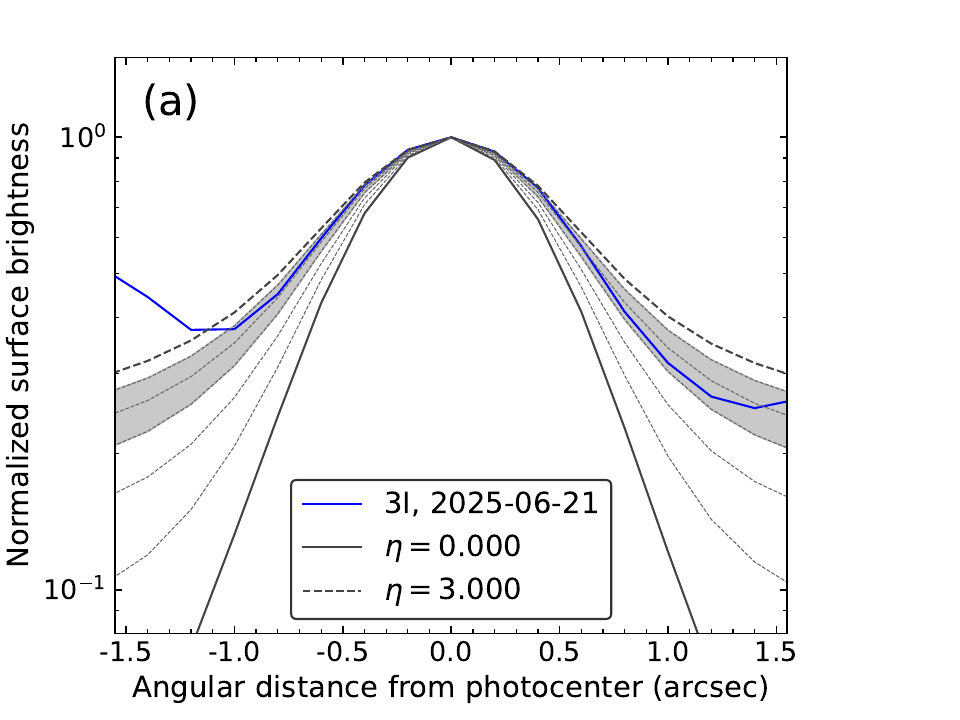} \\
        \includegraphics[width=\figsize\linewidth]{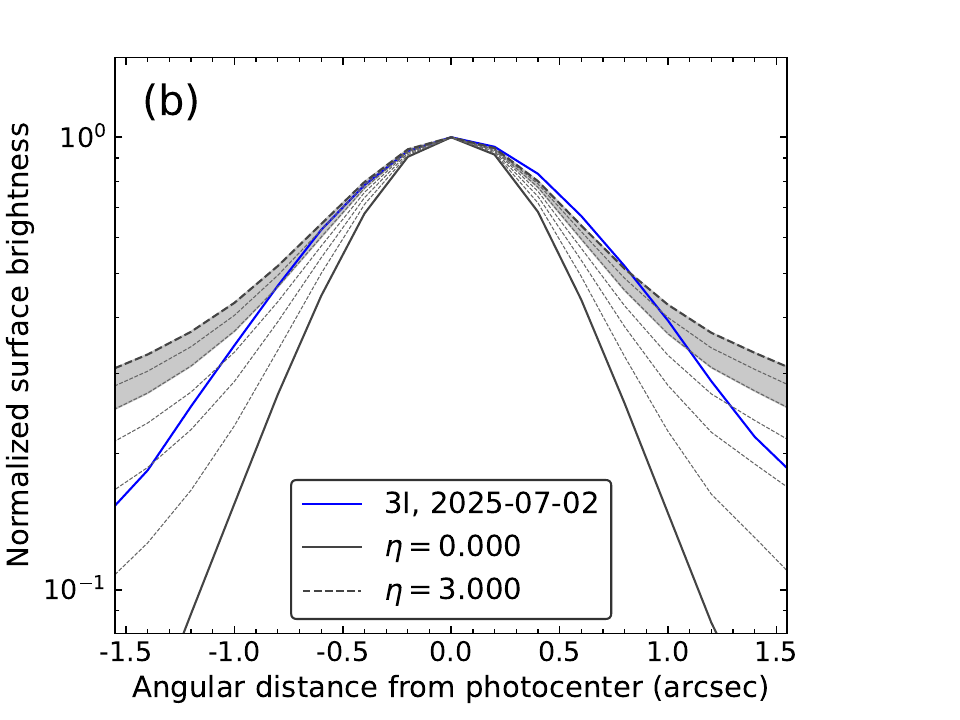} \\
    \end{tabular}
    \caption{One-dimensional surface brightness profiles (measured perpendicular to the direction of apparent non-sidereal motion) of \objname{} (solid blue lines) and seeing-convolved model nuclei with linearly spaced coma levels (i.e., $\eta_i=0.5(i-1)$) ranging from $\eta=0$ (solid black lines) to $\eta=3$ (dashed black line, with profiles corresponding to intermediate $\eta$ values shown as dotted black lines) for data from (a) UT 2025 June 21, and (b) UT 2025 July 2.  Estimated uncertainty regions, i.e., the regions bound by the model profiles adjacent to the best-matching model profile in each panel, are shaded in gray.
    }
    \label{fig:sbps}
\end{figure}

Adding synthetic coma using the method described above has less effect on surface brightness profile cores then it does on profile wings, but of course the profiles are more impacted by noise farther from a particular source's photocenter.  As such, we focus on examining the deviation of the comet profile close enough to the center where the signal is relatively strong, profiles vary relatively smoothly, and the $r^{-1}$ coma profile assumption used in our model is more likely to be true, but far enough away that the model profiles using different coma levels show visible separation from one another.  For profile plots on both dates, this range of interest is approximately $0\farcs4$ to $0\farcs7$ from the photocenter.

We find that coma is clearly present for \objname{} in the June 21 08:11:33 image (Figure~\ref{fig:sbps}a) at a level of $\eta\sim2.0\pm0.5$, where the uncertainty is estimated based on visually identifying the synthetic coma profiles that bound the comet profile.  Meanwhile, we find an approximate coma level of $\eta\sim2.5\pm0.5$ for the July 2 03:33:02 image (Figure~\ref{fig:sbps}b), indicating that the coma may have increased slightly over the 11 days between the two observations (although we also note that the coma level estimates are also consistent within their estimated uncertainties between the two dates).  That said, we also see a larger extension of the surface brightness profile on the north side of the comet, which is corroborated by the visible asymmetry of the comet in its contour plot from that date (Figure~\ref{fig:contours}b). As such, we consider the estimated increase in coma level of $\Delta\eta\sim0.5\pm0.5$ between June 21 and July 2 to be a lower limit. %We also hypothesize that the lack of visible asymmetry in \objname{}'s coma on June 21 is because activity had started recently enough at that time that the asymmetry had not yet had enough time to develop.

\subsubsection{Tail Analysis}\label{subsubsec:tail}

As noted by several observers \citep{Alarcon2025,bolin2025_3I,Jewitt2025,opitom_3I,seligman2025_3I}, \objname{} is observed to exhibit a short sunward-pointing tail in imaging data obtained since its discovery. 
Cometary dust tails typically extend away from the Sun in three-dimensional space due to radiation pressure acting on dust grains, and thus, this morphology for 3I is not the nominal expectation.  However, an antisolar dust tail can appear on the sunward side of a comet nucleus as projected on the sky given certain viewing geometries, while actual sunward dust tails can be produced by non-isotropic dust emission. 
%While such a morphology is certainly unusual—given that dust tails are typically directed antisolar due to radiation pressure acting on dust grains—it is 
Notably, the latter is not without precedent among distant active bodies, where \citet{farnham2021_C2014UN271} reported a similar sunward enhancement in comet C/2014 UN$_{271}$ (Bernardinelli–Bernstein), which they interpreted as the result of the slow ejection of relatively large dust particles predominantly from the sunlit hemisphere. 
%This expected behavior is confirmed by 

\begin{figure}
    \centering
    \begin{tabular}{c}
        \includegraphics[width=\figsize\linewidth]{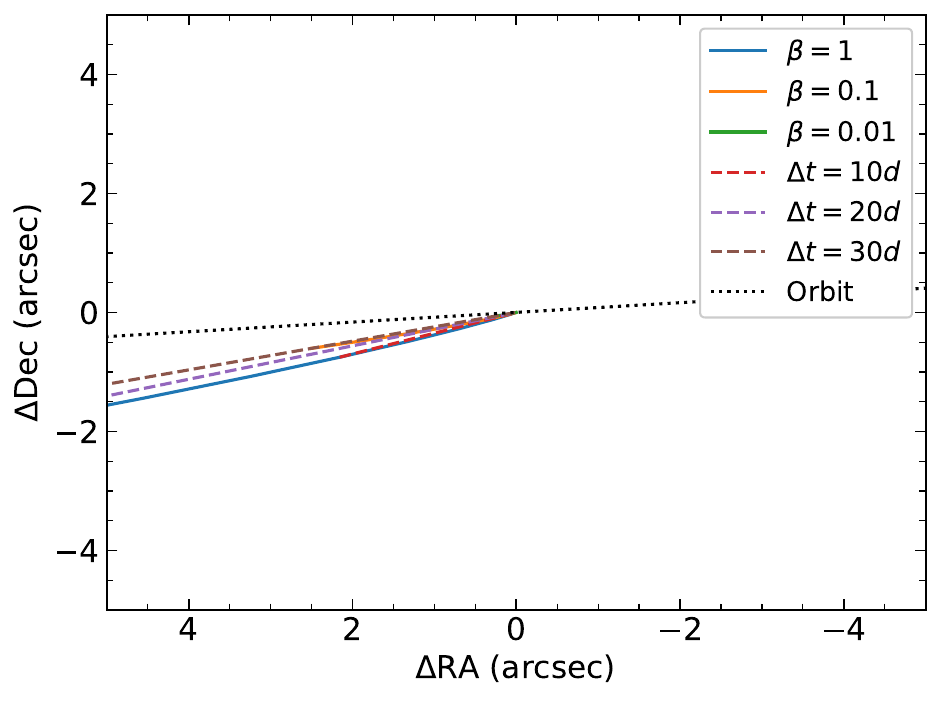} \\
    \end{tabular}
    \caption{Synchrone-syndyne grid for \objname{} on UT 2025 July 2 for isotropic dust emission. The position of the object is at the center of the plot (at $\Delta$RA = $\Delta$Dec = 0~arcsec), where (essentially overlapping) solid colored lines correspond to syndynes, along which particles of specified sizes should be found (where particle sizes are parameterized by $\beta$; particle radii, $a$, in $\mu$m, are approximately given by $a=\beta^{-1}$), dashed colored lines correspond to synchrones, along which particles ejected at specified times should be found (synchrones shown for 10, 20, and 30 days prior to the current epoch), and the dotted black line shows the projection of the orbital plane on the sky.}
    \label{fig:dustmodels}
\end{figure}

Simple Finson-Probstein-style dust modeling \citep{finson1968_cometdustmodeling1,finson1968_cometdustmodeling2} carried out using the {\tt SynGenerator} class\footnote{\url{https://sbpy.readthedocs.io/en/latest/sbpy/dynamics.html\#dust-syndynes-and-synchrones}} in {\tt sbpy} \citep{mommert2019_sbpy} to generate and plot syndynes and synchrones for \objname{} shows that, for isotropic dust emission (which is typically assumed), any dust tail should have a PA of $\sim$100$^{\circ}$ east of north (Figure~\ref{fig:dustmodels}), nominally ruling out the tail's sunward direction (PA~$\,\sim290^{\circ}$) as a viewing geometry effect, and implying instead that it is the result of
%\objname{}'s approximately sunward tail (PA~$\,\sim290^{\circ}$) strongly implies the presence of 
anisotropic dust emission \citep[e.g.,][]{hsieh2011_176p,farnham2021_C2014UN271,hsieh2025_358p}, specifically with an average direction in the orbit plane (which is close to the ecliptic plane), given the sunward tail's approximate orientation along the direction of heliocentric motion.
Although the sunward feature observed in \objname{} is broadly consistent with anisotropic dust emission, such a morphology can also arise from gas jets, as observed in comets like C/1996 B2 (Hyakutake), where strong sunward CN and C$_2$ emissions were detected in narrowband images \citep{Rodionov1998_B2}. CN detection at \objname{}'s distances during our observations ($\sim5$ -- $\sim4$~au would be very exceptional \citep{ahearn1995}, however, and optical spectroscopy of \objname{} as of 2025 July 30 has not detected any gas \citep{opitom_3I,alvarezcandal2025_3I}.

%If preferential dust ejection along the ecliptic plane can be confirmed, one means that such behavior might come about is if the object has an in-plane, sunward-facing rotational pole

We note that the implication of anistropic dust emission close to the ecliptic plane raises the possibility of a nearly in-plane rotation pole, since the time-averaged direction of a non-equatorial jet is equivalent to the direction of the nearest rotation pole \citep[e.g., see][]{hsieh2011_176p}. Such geometry could, in principle, allow pole-on viewing orientations with respect to the Earth \citep[e.g.,][]{buie2018_lucytrojanLCs}, which would in turn be consistent with only minor light curve variations being reported for \objname{} thus far in this work (Section~\ref{subsec:lightcurve}) and by others \citep[e.g.,][]{seligman2025_3I}.  That said, the small light curve ranges observed thus far could also be partially or even entirely due to the damping effects of a steady-state coma on the observable light curve variations of an underlying rotating nucleus \citep[e.g.,][]{hsieh2011_176p}.  
%Further investigation of these issues would be extremely useful for disentangling these different effects on \objname{}'s observed lightcurve.
Finally, it is important to keep in mind that the detailed geometry of dust emission depends critically on nucleus shape. Small cometary nuclei are frequently irregular or contact binary objects, often characterized by deep concavities capable of producing focused sunward jets. For example, in comet 67P/Churyumov–Gerasimenko, such topography-induced sunward jets were well documented by Rosetta \citep[e.g.,][]{shi2018_67Pconcavities}. Without knowledge of \objname{}’s nucleus shape, no firm conclusion on the physical cause of its coma morphology can be drawn.

\subsection{Nucleus Analysis}\label{subsec:nucleus}

%The comet had a measured $z'$-band magnitude of $m_z=18.488\pm0.006$ for the June 21 08:11:33 detection and a measured $i'$-band magnitude of $m_i=18.343\pm0.004$ for the July 2 03:33:02 detection.
The well-isolated July 2 03:33:02 detection of \objname{} (see Figure~\ref{fig:contours}b) had a measured $i$-band aperture magnitude of $m_i=17.633\pm0.006$ as measured in a circular aperture with a radius of $3\farcs4$ (17 pixels, or 8,600~km at the geocentric distance of the comet).  We note that this value differs slightly from the aperture magnitude reported for that detection in Table~\ref{tab:observations} given the slightly different aperture sizes used for the photometry listed in the table ($3\farcs0$ for all measurements) and for the customized measurement discussed here.

%We also derived interpolated/extrapolated reflectance from a 2nd order spline fitting with natural extrapolation at both ends at LSST wavelengths from the MuSCAT 4-color photometry provided by \citet{seligman2025_3I}. These are within 0.025 mag to the colors from linear approximation of the spectral slope, while the difference is around 0.1 mag for $u'-g'=1.475$ and $g'-r'=0.748$. The latter differences are consequences of  the spectral reflectance decreasing slightly more than a linear trend in the u–r wavelength range toward bluer wavelengths. 

%\red{need to figure out issue with normalization wavelength for spectral gradient}

%\red{...} we computed that a spectral slope of 18\%/100~nm corresponds to $r'-i'=0.30$ and $r'-z'=0.45$, and as such, we found estimated equivalent $r'$-band magnitudes of $r'=19.01$~mag and $r'=18.65$~mag on June 21 and July 2, respectively.  

On July 2, \objname{} had a heliocentric distance of $r_h=4.48$~au, a geocentric distance of $\Delta=3.47$~au, and a solar phase angle of $\alpha=2.3^{\circ}$.
In order to estimate the object's nucleus size, we first compute a reduced magnitude, $m(1,1,\alpha)$, i.e., normalizing the measured apparent aperture magnitude to $r_h=\Delta=1$~au using
\begin{equation}
    m_i(1,1,\alpha) = m_i(r_h,\Delta,\alpha) - 5\log(r_h\Delta)
\end{equation}
Given that the nucleus and dust are expected to have different phase darkening behavior, we then partition this reduced magnitude into its nucleus and dust components using the coma level estimated from our surface brightness profile analysis for July 2 of $\eta=2.5\pm0.5$, %(Section~\ref{subsubsec:coma}), 
where the nucleus is assumed to account for 1/(3.5$\pm$0.5) of the observed flux and the dust coma is assumed to account for 2.5/(3.5$\pm$0.5) of the observed flux, obtaining a nuclear magnitude of $m_{i,n}(1,1,\alpha)=13.04\pm0.16$ and a dust coma magnitude of $m_{i,d}(1,1,\alpha)=12.04\pm0.16$.

Adopting Jupiter-family comet (JFC) nuclei as a potentially reasonable proxy for the nucleus of \objname{}, we use the median linear phase coefficient and standard deviation of $\beta=0.046\pm0.017$~mag~deg$^{-1}$ found from measurements of a large sample of JFC nuclei by \citet{kokotanekova2017_jfcrotation}.  Correcting for $\alpha=2.3^{\circ}$ therefore results in an absolute magnitude of $H_{i,n} = m_{i,n}(1,1,0)=12.93\pm0.16$.

Meanwhile, we use the Schleicher-Marcus phase function\footnote{\url{https://asteroid.lowell.edu/comet/dustphase.html}} \citep[sometimes also referred to as the Halley-Marcus phase function;][]{schleicher2011_sw3,schleicher1998_halley,marcus2007_cometphasefunction} for determining the phase darkening correction of the dust component of this observation, finding that the expected observed flux at $\alpha=2.3^{\circ}$ is 0.91 of the expected observed flux at $\alpha=0^{\circ}$.  We therefore obtain $H_{i,d} = m_{i,d}(1,1,0)=11.94\pm0.16$.

%H_r = 12.042+/-0.005 (new Hr from Pedro, 12/10/25)

Lastly, we combine the distance- and phase-corrected magnitudes of \objname{}'s nucleus and dust components using
\begin{equation}
    m_{i,{\rm tot}} = -2.5 \log \left(10^{-0.4m_{i,n}(1,1,0)} + 10^{-0.4m_{i,d}(1,1,0)}\right)
\end{equation}
obtaining a total absolute magnitude of $m_{i,{\rm tot}}(1,1,0)=11.57\pm0.12$.  We note that the above calculations assume that the dust coma is optically thin, and utilize the {\tt uncertainties} python package for the calculation and propagation of uncertainties\footnote{\url{https://pythonhosted.org/uncertainties/}}.

%we calculate an equivalent absolute $i'$-band magnitude of $H_i=11.416\pm0.006$ (assuming $G=0.15$).  
%Using the coma level estimated from our surface brightness profile analysis for July 2 of $\eta=2.5\pm0.5$, we estimate a corresponding $i'$-band absolute magnitude for \objname{}'s nucleus of $H_{i,n}=(12.4\pm0.2)$~mag.

In order to estimate \objname{}'s nucleus size from the absolute magnitude found above, we first convert the $i$-band absolute magnitude to $r$-band using the object's measured $(r-i)_{\rm LSST}$ color (Table~\ref{tab:3I_colors}), obtaining
%\red{$H_{r,n} = H_{i,n}+0.299=13.23\pm0.16$}.  
$H_{r,n} = H_{i,n}+(0.235\pm0.018)=13.16\pm0.16$.  
We then compute an equivalent $V$-band absolute magnitude, $H_{V,n}$, using
\begin{equation}
    H_{V,n} = H_{r,n} + 0.73(V-R) - 0.088
\end{equation}
%r-i = 0.235+/-0.018
from \citet{jordi2006_filtertransformations}, where we use 
%\red{$V-R=0.716$}, as computed using \objname{}'s reported spectral slope of $S'=18$\%/100~nm.  
$V-R=(0.333\pm0.008)$ based on measured colors reported in Section~\ref{subsec:colors}, and assume that the difference between $r'$ and $r_{\rm LSST}$ is negligible for the purposes of this calculation.
We obtain $H_{V,n}=(13.32\pm0.16)$~mag, or $\sim1$~mag fainter than the $H_V$ value ($H_V>12.4$) estimated by \citet{seligman2025_3I} based on \ac{ZTF} precovery data.  Finally, we can estimate the effective nucleus radius using
\begin{equation}
    r_n = \left( {2.24\times10^{22}\over p_V} \times 10^{0.4(m_{\odot,V} - H_V)} \right)^{1/2}
\end{equation}
where we use $m_{\odot,V}=-26.71\pm0.03$ for the apparent $V$-band magnitude of the Sun \citep{hardorp1980_sun3} and assume a $V$-band geometric albedo of $p_V=0.05$ \citep[typical of reddened comet nuclei;][]{knight2024_cometnuclei}, finding $r_n=(6.6\pm0.5)$~km, about 2/3 of the previous nucleus size upper limit estimate by \citet{seligman2025_3I}.  
Given that this nucleus size is based on a coma-to-nucleus ratio that we expect to be underestimated %(see Section~\ref{subsubsec:coma}), 
we regard this result as an upper limit to \objname{}'s true nucleus size, where this has since been confirmed by \citet{jewitt2025_hst3I} who constrained the nucleus size to $r\leq2.8$~km based on a fit to the surface brightness distribution of the inner coma as observed by the Hubble Space Telescope on UT 2025 July 21. Later (post-perihelion), the nucleus was measured to be $1.3 \pm 0.2$~km \citep{manToHui2026}. 
%We further note that it is based on assumptions of spherically symmetric, steady-state coma, which we know is not the case (discussed below in Section~\ref{subsubsec:tail}), as well as identical geometric albedos for coma grains and the nucleus, for which we have no direct constraints.  As such, while we consider this result to be an advance over previous work, we note that further work to directly observe \objname{}'s nucleus with high-resolution imaging by space telescopes like the \textit{Hubble Space Telescope} and \textit{JWST} will be extremely useful for confirming or refining its size.

%\red{This is an upper limit, see C/2017 K2, where JWST finds radius $< 4.2$ km (Woodward et al. 2025) vs. 9 km by this method (Jewitt et al. 2017). JWST is surely observing 3I.}

Adopting the nucleus size upper limit found by \citet{jewitt2025_hst3I} as the true nucleus size for the purposes of the following analysis, we can estimate the potential mass loss rate, $\dot M$, from $\eta$ using
\begin{equation}
    {\dot M} = {(1.1\times10^{-3})\pi\rho_d{\bar a}\eta r_{\rm obj}^2\over pr_h^{0.5}\Delta}
\end{equation}
\citep{luu1992_neoprofiles}, where $\rho_d$ is the average bulk dust grain density, $\bar a$ is the weighted mean grain radius, $r_{\rm obj}$ is the object's effective radius, $p=5\farcs5$ is the angular photometry radius in arcseconds, and $r_h=4.48$~au and $\Delta=3.47$~au are the heliocentric and geocentric distances in au on July 2, keeping in mind that if we use $r_{\rm obj}\sim2.8$~km instead of our originally calculated $r_{\rm obj}\sim6.6$~km, we obtain $\eta\sim13$ instead of our originally calculated $\eta\sim2.5$ (on July 2).  In addition to the uncertainty on our nucleus size estimate, there are currently no useful empirical constraints on $\rho_d$ and $\bar a$.  While $\rho_d$ may vary by a factor of a few among different materials, $\bar a$ dominates the uncertainty as it can vary by several orders of magnitude (e.g., from $\mu$m to mm scales or larger).  For illustration, for arbitrary assumptions of $\rho=1000$~kg~m$^{-3}$ and $\bar a=1$~$\mu$m, we obtain ${\dot M}\sim10$~kg~s$^{-1}$, but $\bar a=10$~$\mu$m would result in ${\dot M}\sim100$~kg~s$^{-1}$.  Constraints on particle sizes (which may vary with heliocentric distance), e.g., via detailed dust modeling, will be critical for deriving more realistic mass-loss rates for \objname{} in the months following these reported observations, as it becomes more active.

For reference, we also calculate 
the $A(\alpha=0^{\circ})f\rho$ parameter \citep[hereafter, $Af\rho$;][]{ahearn1984_bowell}, which is frequently used to parameterize the dust content of cometary comae, and is given by
\begin{equation}
A{(\alpha=0^{\circ})}f\rho = {(2r_h\Delta)^2\over \rho_{ap}}10^{0.4[m_\odot-m_{d}(r_h,\Delta,0)]}
\label{equation:afrho}
\end{equation}
where $r_h$ is in au, $\Delta$ is in cm, $\rho_{ap}$ is the physical radius in cm of the photometry aperture used to measure the magnitude of the comet at the distance of the comet, $m_{\odot}$ is the apparent magnitude of the Sun at $\Delta=1$~au in the same filter used to observe the comet, and $m_d(r_h,\Delta,0)$ is the phase-angle-corrected (to $\alpha=0^{\circ}$) apparent magnitude of the dust with the flux contribution of the nucleus \citep[assuming the nucleus size upper limit from][]{jewitt2025_hst3I} subtracted from the measured total magnitude. For this calculation, the nucleus contribution to the flux is calculated assuming a linear phase function and median linear phase coefficient from a sample of JFC nuclei from \citet{kokotanekova2017_jfcrotation}, and the nucleus-subtracted magnitude of the dust is corrected to $\alpha=0^{\circ}$ using the Schleicher-Marcus phase function as discussed above.  Using parameter values specified above, we compute $Af\rho=(334\pm13)$~cm based on our $i_{\rm LSST}$-band detection on UT 2025 July 2, which is slightly larger than the $i'$-band $Af\rho=(288\pm5)$~cm value (using a photometry aperture with a radius of $4''$, or about 10,000~km at the distance of the comet) reported by \citet{bolin2025_3I}.

% \subsubsection{Probabilistic Tail Model} \label{subsec:probabilistictail}
% tex cut from here 7/17/2025 COC

%%%%%%%%%

\section{Discussion}\label{sec:discussion}

\subsection{Discoverability of \objname{}}\label{sec:discoverability}

The  \ac{SV} survey is intended to behave much like the operations LSST survey strategy. In the  \ac{SV} survey, observations are typically obtained in pairs of exposures separated by approximately 30 minutes, as is the case in the operations survey. However, the  \ac{SV} survey is attempting to gather on the order of 1-2 years worth of operations LSST images (about 150 exposures) within 2-3 months of commissioning time. To achieve this goal, the pointings within the main  \ac{SV} footprint are generally covered every night, rather than every few nights as in the operations survey. Additionally, the primary \ac{SV} survey footprint is much smaller than the full LSST footprint. The primary \ac{SV} survey covers a region along the ecliptic plane, stretching across a wide variety of galactic latitudes, including the Galactic plane itself. This provides tests of pipelines in both crowded and less dense fields, as well as the maximum opportunity to commission the Solar System linking pipelines.

The \ac{SV} survey is both similar and dissimilar to the nominal upcoming \ac{LSST} \citep[baseline\_v4.0;][]{peter_yoachim_2025_14920193, jones_2025_15128504} during comparable periods (i.e. between May and August, centered on 3I’s discovery). 
The \ac{SV} footprint is much more compact, predominantly following the ecliptic plane, with the exception of the \acp{DDF}.
While the exact frequency and quality of images may differ both \ac{SV} and \ac{LSST} provide relatively uniform coverage over the regions of sky included.

An early simulation of the entire \ac{SV} survey period was released alongside \citet{SITCOMTN-005}. This simulation was created by the Feature Based Scheduler \citep{peter_yoachim_2025_15742506, 2019AJ....157..151N}, much as the baseline operations survey is simulated. As the \ac{SV} survey progresses, these simulations are updated by including information about exposures acquired to-date, obtained from the \ac{LSST} Consolidated Database \citep{DMTN-227}, alongside attempts to improve the fidelity of the observatory model. The already-acquired exposure information is fed into the scheduler, then the simulation is continued for the remainder of the \ac{SV} period. One such simulation was created containing the pointing history of the \ac{SV} survey as of the start of observing on UT 2025 July 14 (`sv\_20250714'), which was used to evaluate both the likelihood of acquiring \objname{} in already acquired visits and the possibility of future detections.

%\subsubsection{Discoverability of \objname{}}
We used the ISO survey simulation framework outlined in \citet{dorsey2025_isomodel} to simulate the probabilistic observability of \objname{} in the \ac{LSST} \ac{SV} survey.
We simulate 10,000 clones, each sampled from the covariance matrix of \objname{}'s orbit (retrieved from JPL Horizons on 2025 July 15), and assess which \ac{SV} pointings would have observed an object of similar absolute magnitude and color (Table \ref{tab:3I_colors}), applying the \ac{LSST} \ac{SSP} criteria for moving object discovery.
For this analysis, we use the preliminary value of $H_r\sim12.5$ for \objname{} that was available at the time of this analysis.
% Colours used: H_r,u-r,g-r,i-r,z-r,y-r,GS12.5,1.59,0.436,-0.112,-0.123,-0.132,0.15
% Result is not particularly sensitive to colour, btw
As anticipated, all clones were detected and subsequently discovered within the \ac{SV} survey: the apparent magnitude of \objname{} ($m_r\sim18$ mag) is 5 magnitudes brighter than the median limiting magnitude of \ac{LSST} in $r$-band (S\ref{sec:intro}). 
For 95\% of the clones, the first successful observation occurred on UT 2025 June 21 (MJD 60847.155) and discovery occurred $\sim$13 days later, on UT 2025 July 4 (MJD 60860.998).
None of the clones were discovered earlier than UT 2025 July 2 (the date of discovery by \ac{ATLAS}).
This simulation demonstrates that had the \ac{SV} begun two weeks earlier and pipelines run as in \ac{LSST}, \ac{LSST} may have indeed discovered \objname{} prior to \ac{ATLAS}.

%\subsubsection{Future Prospects in the SV Survey}
%\label{sec:sorcha_future}

\subsection{Thermophysical Modeling}\label{subsec:thermo}

To get a general sense of the rate of temperature change for an airless body at a given distance, we make use of a simple thermophysical model, following \citet{hsiehMainbeltCometsPanSTARRS12015} and \citet{chandlerCometaryActivityDiscovered2020b}, whereby a body's heliocentric distance as a function of temperature is given by

\begin{equation}
    R(T) = \sqrt{\frac{F_\odot (1-A)}{\chi\left[\epsilon\sigma T^4 + L f_\mathrm{D}\dot{m}_\mathrm{S}(T)\right]}} \, \mathrm{au.},
\end{equation}

\noindent Here, $F_\odot=1360\ \mathrm{W\ m^{-2}}$ is the solar flux at 1 au, $A=0.05$ the assumed Bond albedo, $\epsilon=0.9$ the assumed infrared emissivity of the ice, $T_\mathrm{eq}$ the equilibrium temperature, $\sigma=5.670\times10^{-8}\ \mathrm{W\ m^{-2}\ K^{-4}}$ the Stefan–Boltzmann constant, $L$ the latent heat of sublimation,  $f_\mathrm{D}$ is the ``diffusion barrier factor'' accounting for emission blocked by overlaying material like regolith, $\dot{m}_\mathrm{S}(T)$ is the mass-loss rate as illustrated in \citep{chandlerCometaryActivityDiscovered2020b}. The parameter $\chi$ captures rotational and axial‐tilt effects on insolation: $\chi=1$ corresponds to a ``slab'' perpetually facing the Sun (resulting in maximum heating), $\chi=\pi$ to a rapid rotator with zero tilt, and $\chi=4$ is the value we adopt for an isothermal, fast‐rotating body in thermodynamic equilibrium.

\begin{figure*}
    \centering
    \includegraphics[width=0.6\linewidth]{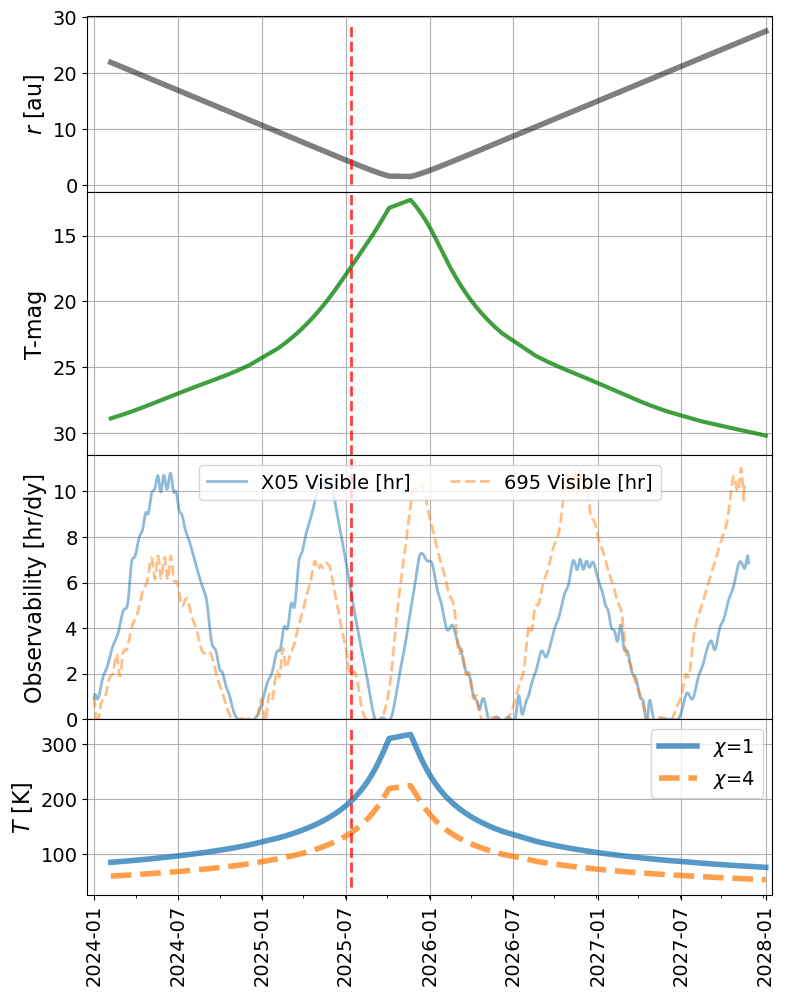} % renamed for arXiv 7/16/2025 COC
    \caption{\objname{} metrics plotted versus time from UT 2024 January 1 to UT 2028 January 1. 
    The vertical dashed red line indicates UT 2025 July 12; this is the date elements were retrieved from JPL Horizons \citep{giorginiJPLsOnLineSolar1996}.  
    The $r$ [au] panel indicates the heliocentric distance of \objname{}. 
    The T-mag panel indicates the JPL Horizons-computed total magnitude of \objname{}; T-mag should be considered a \textit{very} rough estimate, and is included here to provide a general sense of expected brightness behavior rather than a quantitative or predictive tool. % (in $V$-band). 
    The Observability [hr/dy] panel describes the number of nighttime hours \objname{} spends above the horizon for a given UT night at the specified observatory. Kitt Peak National Observatory (695) and Rubin Observatory (X05) are chosen as representative Northern and Southern Hemisphere sites, respectively. 
    The T [K] panel provides a simple temperature estimate for a representative airless body, where $\chi=4$ represents the isothermal case (e.g., fast-rotator) and $\chi=1$ corresponds to the ``flat slab'' scenario, which is the thermophysical extreme. These lines are to provide context only and do not represent actual temperatures on \objname{}.
    }
    \label{fig:observability}
\end{figure*}

Figure \ref{fig:observability} shows the results of this modeling, along with several other metrics useful for understanding the discovery and future observations of \objname{}. \objname{} was at its peak observability (number of hours at night an object is above the horizon) during 2025 June, and it will be less observable from \rubinobs{} during perihelion passage as compared to Northern Hemisphere sites. We also note that \objname{} becomes too bright (i.e., saturates LSSTCam) for Rubin to observe in any filter except possibly for \textit{u} and \textit{y} (solar system objects are typically much fainter in these filters) until roughly 2026 July.

\subsection{Implications for the ISO Luminosity Function}
\label{sec:sfd}

The estimated nuclear size of \objname{} (\Hvmag{}, \Hrmag{}, \Himag{}) %($H_{V,n}= 13.7 \pm 0.2$ mag or $H_{r,n} = 13.2 \pm 0.2$ mag) 
from \ac{LSST} photometry has important implications for the size-frequency distribution of the galactic population of \acp{ISO}.
The detections of \oumuamua{} and 2I/Borisov as significantly smaller interstellar objects ($r\sim100$~m and $r\sim400$~m respectively) implied an intrinsically steep size-frequency distribution \citep[$q_s\sim3$--4;][]{jewitt2020_borisovnucleus}, which would agree with historical non-detections of \acp{ISO} in Solar System surveys.
This constraint is consistent with other small body populations within the Solar System, from the  ``primordial'' \citep[e.g., large hot classical \acp{KBO};][]{petit2023_hotkbos} to the more collisionally evolved \citep[e.g., main-belt asteroids;][]{alvarezcandal2006}.
However, as \objname{} has an absolute magnitude much lower than either of its predecessors, its discovery strongly disfavors a steep slope, instead favoring slopes $q_s\lesssim3$ \citep{dorsey2025_isomodel}.
This apparent contradiction can be resolved by instead considering that the intrinsic \ac{ISO} \ac{SFD}, which is cumulative across all donor planetary systems, is likely not parameterized by a single slope power law.
It is rather a piecewise function of several such power laws with ``breaks'' --- forming a ``wavy'' \ac{SFD} --- similar to that demonstrated in our own Solar System \citep{bottke2023_kbevolution}. 
If the latter, \objname{} provides the first evidence to begin constraining the shape of the \acp{ISO} population \ac{SFD}.

%%%%%%%%%%%%%%%%%%%%%%%%%%%%%%%r
\section{Summary and Conclusions} \label{sec:summary}

We report the analysis of the third interstellar object, \objname{}, obtained from images taken during commissioning of the Rubin Observatory, from UT 2025 June 20 to UT 2025 July 20, coinciding with the start of the Observatory’s \ac{SV} survey. 
At the time of these observations, although only a portion of the data analysis and management system was deployed and verified, we obtained science-quality photometric and astrometric measurements via a combination of deploying Rubin data management software and bespoke and/or manual techniques. Even in a state of early commissioning, our analysis shows that Rubin Observatory is already capable of delivering high-grade science photometric and astrometric observations of Solar System small bodies.  
\\

Below, we summarize our key findings:
\begin{itemize}

\item Rubin images clearly show the interstellar object \objname{} with a dust coma, indicative of activity. These observations provide the earliest high-resolution evidence of detected cometary activity.

\item We detect no short-term photometric variability, %. A sequence of measurements taken on UT 2025 July 2 
constraining apparent brightness variations of \objname{} to less than 0.1 mag on timescales of a week.

\item Rubin multi-filter photometry is consistent with previous observations in the literature, with significantly smaller error bars. We report measured colors of $g-r=$ (\colorGminR{}) mag, $r-i=$ (\colorRminI{}) mag, $i-z=$ (\colorIminZ{}) mag, $z-y=$ (\colorZminY{}) mag.

\item In Rubin's observations leading up to (and including) the discovery of \objname{}, the coma’s radius appeared to increase slightly over the 11 days from $\sim$ 6,520~km (UT 2025 June 21) to $\sim$9,380~km (UT 2025 July 02) as measured from azimuthally averaged radial profiles. We estimate an increase in coma level of $\Delta \eta$ $\sim$0.5 between 2025 June 21 and 2025 July 2, which we assume to be a lower limit for several reasons, including that \objname{} was observed nearly head-on (very low phase angle) and the tail could have extended far along the \textit{z}-axis, as projected on the sky.

\item We obtain a lower-limit $V$-band absolute magnitude of $H_V=$
%(13.7\pm0.2) $
\Hvmag{} and an equivalent upper-limit effective radius of % $r_n\sim(6.6\pm0.5)$ km 
\radNucl{} for the nucleus, assuming a spherically symmetric steady-state coma, which we note is substantially larger than a later \ac{HST}-derived nucleus size upper limit of $r\leq2.8$~km.  Using the \ac{HST}-derived nucleus size, we estimate a near-nucleus dust-to-nucleus scattering cross-section ratio of $\eta\gtrsim13$ for \objname{} on UT 2025 July 2.  We estimate a mass loss rate ranging from 10 to 100 kg/s, depending on the grain sizes assumed to dominate, and we compute Af$\rho$ = (315 $\pm$ 15) cm for data obtained on UT 2025 July 2.

\item If the Rubin \ac{SSP} pipelines had been processing the commissioning data in real time, our modeling shows that there were sufficient \ac{SV} observations to identify \objname{} as a moving object.

% \item If the nominal \ac{SV} survey strategy continues as planned,  \objname{} should be observed in ten or more observations in each filter through mid-August 2025. Future \ac{SV} observations will likely be able to monitor the coma color as \objname{} moves towards perihelion. 

\item Analysis of the derived astrometry suggests that for bright high-\ac{SNR} and extended active small bodies, the combination of Rubin's large aperture and LSSTCam's pixel scale is less impacted by asymmetrical coma, providing precise positions for deriving accurate orbital parameters. 

\item Positional measurements with Rubin are highly precise (order 3~mas) when evaluating Gaia quasars, but a notable ($\sim$20~mas) declination offset caused by external factors comprises roughly 25\% of the positional uncertainty for \objname{}, with the balance due to the cometary nature of the object.
\end{itemize}

The spatial number density of \acp{ISO} remains poorly constrained, with forthcoming surveys expected to significantly advance our understanding of this population. Given its large étendue, the \ac{LSST} is expected to play a major role in this endeavor with \citet{dorsey2025_isomodel} estimating that it will identify $\sim$5-50 interstellar objects over its nominal 10-year lifetime. The discovery of a third macroscopic body originating from outside of our Solar System by \ac{ATLAS} lends further credence to this view, providing tantalizing evidence that a large reservoir of similar bodies may exist within reach of the discovery capabilities of Rubin \citep{cooklsstiso,engelhardtlsstiso,marcetaseligman,dorsey2025_isomodel}. If this is correct, we are about to enter a decade of large-scale discoveries, follow-up, and characterization of this exciting new population.

%%%%%%%%%%%%%%%%%%%%%%%%%%%%%%%%%%
% \section*{Acknowledgements}
\begin{acknowledgments}

% IMPORTANT MUST HAVE BLANK LINE ABOVE THIS ONE!

C.O.C. thanks Arthur and Jeanie Chandler for their ongoing support. 

This material is based upon work supported in part by the National Science Foundation through Cooperative Agreements AST-1258333 and AST-2241526 and Cooperative Support Agreements AST-1202910 and 2211468 managed by the Association of Universities for Research in Astronomy (AURA), and the Department of Energy under Contract No. DE-AC02-76SF00515 with the SLAC National Accelerator Laboratory managed by Stanford University. Additional Rubin Observatory funding comes from private donations, grants to universities, and in-kind support from LSST-DA Institutional Members.

C.O.C., W.B., M.W., M.T., and D.O. acknowledge support through Schmidt Sciences LLC. 
C.O.C, P.H.B, M.J., D.S., D.V., D.W., and J.R.A.D. acknowledge support from the DiRAC Institute in the Department of Astronomy at the University of Washington. The DiRAC Institute is supported through generous gifts from the Charles and Lisa Simonyi Fund for Arts and Sciences, and the Washington Research Foundation. M.J. acknowledges the tremendous patience and unwavering support from Dea, Lyra, and Mila throughout the two-decade journey that led to this moment.

C.O.C. and A.J.C. gratefully acknowledge support from the NSF (grant No. AST-2107800).  C.O.C. gratefully acknowledges support from the NASA CSSFP (grant No. 80NSSC26K0380). % 3/17/2026 COC

A.J.C gratefully acknowledges support from the Department of Energy under award DE-SC0011665.

M.J.H. gratefully acknowledges support from the NSF (grant No. AST-2206194), the NASA YORPD Program (grant No. 80NSSC22K0239), and the Smithsonian Scholarly Studies Program (2022, 2023).

S.E. acknowledges support by the National Science Foundation under Grant AST-2307570. R.M. acknowledges funding from a NASA Space Technology Graduate Research Opportunities (NSTGRO) award, NASA contract No. 80NSSC22K1173.

{\tt sbpy} is supported by NASA Planetary Data Archiving, Restoration, and Tools (PDART) Grant Numbers 80NSSC18K0987 and 80NSSC22K0143.

D.E. is supported by the National Science Foundation under Grant No. AST-2307569.

%%%%%%%%%%%%% BEGIN SSSC ACKS %%%%%%%%%%
Gy.M.Sz acknowledges support from SNN-147362 and the ADVANCED-153410 of the National Research, Development and Innovation Office (NKFIH, Hungary), and the ESA PRODEX Experiment Agreements No. 4000137122 No. 4000149203.

M.E.S. acknowledges support in part from UK Science and Technology Facilities Council (STFC) grants ST/V000691/1 and ST/X001253/1.

This work was supported by an LSST Discovery Alliance LINCC Frameworks Incubator grant [2023-SFF-LFI-01-Schwamb]. Support was provided by Schmidt Sciences.

This work was supported by an LSST Discovery Alliance LINCC Frameworks Incubator grants [2025-SFF-LFI-01-Holman/2025-SFF-LFI-11-Schwamb]. Support was provided by Schmidt Sciences.

M.T.B. appreciates support by the Rutherford Discovery Fellowships from New Zealand Government funding, administered by the Royal Society Te Ap\={a}rangi. 

B. Carry is supported by INSU PNP.

T.D. acknowledges support from the McDonnell Center for the Space Sciences at Washington University in St. Louis.

R.C.D. is supported by grant \#361233 awarded by the Research Council of Finland to M. Granvik.

D.M. is supported by The Science Fund of the Republic of Serbia through Project No. 7453 \emph{Demystifying enigmatic visitors of the near-Earth region (ENIGMA)}.

The views expressed in this article are those of the authors and do not reflect the official policy or position of the U.S. Naval Academy, Department of the Navy, the Department of Defense, or the U.S. Government.

This material is based in part on work done by M. Womack while serving at the National Science Foundation.

E.W. is supported by the National Science Foundation Graduate Research Fellowship Program under Grant No. 1000383199. Any opinions, findings, and conclusions or recommendations expressed in this material are those of the author(s) and do not necessarily reflect the views of the National Science Foundation.

F.B.R. acknowledges CNPq grant 316604/2023-2.

R. Malhotra acknowledges research funding from NASA grant 80NSSC21K0593.

C.L.P. thanks the FAPERJ/PDR-10 E-26/200.107/2025 and FAPERJ 200.108/2025. 

The authors acknowledge support from the DiRAC Institute in the Department of Astronomy at the University of Washington. The DiRAC Institute is supported through generous gifts from the Charles and Lisa Simonyi Fund for Arts and Sciences, Janet and Lloyd Frink, and the Washington Research Foundation.

The Catalina Sky Survey is funded by NASA's Planetary Defense Coordination Office, under grant 80NSSC24K1187.

CEH is supported by an LSST-DA Catalyst Fellowship, made possible by Grant 62192 from the John Templeton Foundation.

M.J.P. acknowledges support from the NASA Minor Planet Center Award 80NSSC22M0024.

A.R.G.J. thanks FAPEMIG for the financial support APQ-02987-24.

D.Z.S. is supported by an NSF Astronomy and Astrophysics Postdoctoral Fellowship under award AST-2303553. This research award is partially funded by a generous gift of Charles Simonyi to the NSF Division of Astronomical Sciences. The award is made in recognition of significant contributions to Rubin Observatory’s Legacy Survey of Space and Time.

This material is based upon work supported by the National Science Foundation under grant No.\ AST-2406527.

J. M. acknowledges support from the Department for the Economy (DfE) Northern Ireland postgraduate studentship scheme and travel support from the STFC for UK participation in LSST through grant ST/S006206/1. J. M. thanks the LSST-DA Data Science Fellowship Program, which is funded by LSST-DA, the Brinson Foundation, the WoodNext Foundation, and the Research Corporation for Science Advancement Foundation; his participation in the program has benefited this work.

J.L. acknowledges support from the Agencia Estatal de Investigacion del Ministerio de Ciencia e Innovaci\'on (AEI-MCINN) under grant "Hydrated Minerals and Organic Compounds in Primitive Asteroids" with reference PID2020-120464GB-100.

F.S. acknowledges support from the NASA Minor Planet Center Award 80NSSC22M0024.

Work supported by Department of Energy contract DE-AC02-76SF00515.

S.L. is supported by the U.S. Department of Energy under grant number DE-1161130-116-SDDTA and under Contract No. DE-AC02-76SF00515 with the SLAC National Accelerator Laboratory.

The work of A.A.P.M. was supported by the U.S. Department of Energy under contract number DE-AC02-76SF00515. AAPM  thanks the Department of Physics and the Laboratory of Particle Astrophysics and Cosmology of Harvard University, the Cosmology Group at Boston University, and the Department of Physics at Washington University in St. Louis for their hospitality during the preparation of this paper.

M.J.J. acknowledges support for the current research from the National Research Foundation (NRF) of Korea under the programs 2022R1A2C1003130 and RS-2023-00219959.

C.W.W. was supported by Department of Energy, grant DE-SC0010007.

D.F. conducted this research at the Jet Propulsion Laboratory, California Institute of Technology, under a contract with the National Aeronautics and Space Administration (80NM0018D0004).

T.S. is supported by the National Science Foundation Graduate Research Fellowship under Grant No. DGE-2146755 and at SLAC National Accelerator Laboratory under Department of Energy Contract No. DE-AC02-76SF00515.

R.M. acknowledges support through Schmidt Sciences LLC.

M.G. acknowledges support by the Department of Energy award number DE-SC0023528. Any opinions, findings, conclusions or recommendations expressed in this material are those of the authors and do not necessarily reflect the views of the Department of Energy. 

P.G. is supported by a Royal Society Leverhulme Trust Senior Fellowship and the UKRI Science and Technology Facilities Council.

H.A.R.D. is supported by the Australian Government through an Australian Research Council Linkage Infrastructure, Equipment and Facilities grant LE220100007 as well as the National Collaborative Research Infrastructure Strategy (NCRIS).

% SSSC LSST Solar System Readiness Sprint Funding Acknowledgements as we wouldn't have been ready for this without those meetings

This work was supported in part by the LSST Discovery Alliance Enabling Science grants program, the B612 Foundation, the University of Washington's DiRAC Institute, the Planetary Society, Karman+, and Breakthrough Listen through generous support of the LSST Solar System Readiness Sprints. Breakthrough Listen is managed by the Breakthrough Initiatives, sponsored by the Breakthrough Prize Foundation (\url{http://www.breakthroughinitiatives.org}). 

% 3/18/2026 COC
C.F. acknowledges support from the BASAL Centro de Astrofísica y Tecnologías Afines (CATA) ANID BASAL project FB210003.

The work of S.G. is supported by NOIRLab, which is managed by the Association of Universities for Research in Astronomy (AURA) under a cooperative agreement with the US National Science Foundation. S.G. acknowledges support from the National Science Foundation under grant No. AST-2307569, which is partially funded by a generous gift of Charles Simonyi to the NSF Division of Astronomical Sciences. The award is made in recognition of significant contributions to Rubin Observatory’s Legacy Survey of Space and Time. Any opinions, findings, and conclusions or recommendations expressed in this material are those of the author(s) and do not necessarily reflect the views of the National Science Foundation. S.G. also acknowledges support from NASA Solar System Workings grant 80NSSC22K0978.

%%%%%%%%%%%%% END SSSC ACKS %%%%%%%%%%%%

T.T. acknowledges support from DOE grant DE-SC0009999 and NSF grant AST-2205095.

L.T.S.C. is partially supported by Spanish Ministerio de Ciencia, Innovacion y Universidades under grant PID2021-123012 and the funding from the MAD4SPACE-CM TEC-2024/TEC-182 project funded by Comunidad de Madrid.

J.I.B.C. acknowledges grants 305917/2019-6, 306691/2022-1(CNPq) and 201.681/2019 (Rio de Janeiro State Research Support Foundation, FAPERJ).

This manuscript has been authored in part by Fermi Forward Discovery Group, LLC under Contract No. 89243024CSC000002 with the U.S. Department of Energy, Office of Science, Office of High Energy Physics.

C.L. acknowledges support from DOE grant DE-SC0009999.

K.M. is funded by the European Union - NextGenerationEU - ASTROVAR - uniri-iz-25-107.

S.A. is supported by an LSST-DA Catalyst Fellowship (Grant 62192 from the John Templeton Foundation to LSST-DA). S.A. also gratefully acknowledges support from Stanford University, the United States Department of Energy, and a generous grant from Fred Kavli and The Kavli Foundation.

This work has been supported by the French National Institute of Nuclear and Particle Physics (IN2P3) through dedicated funding provided by the National Center for Scientific Research (CNRS).

E.K.U. acknowledges Cosmic Frontier support from DOE grant DE-SC0007881.

\end{acknowledgments}

\vspace{5mm}
\facilities{
% HST(STIS), 
% Swift(XRT and UVOT), 
% AAVSO, 
% CTIO:1.3m, 
% CTIO:1.5m, 
% CTIO:4m (DECam),
% Magellan:Baade (IMACS),
Rubin:Simonyi (LSSTCam)
% VLT:VST (OmegaCAM)
% CXO
}

%% Similar to \facility{}, there is the optional \software command to allow authors a place to specify which programs were used during the creation of  the manuscript. Authors should list each code and include either a citation or url to the code inside ()s when available.

\software{
        astropy \citep{astropy2013,astropy2018,astropy2022}, 
        {\tt astrometry.net} \citep{langAstrometryNetBlindAstrometric2010}, % already in acks 1/31/2023 COC
        {\tt astroquery} \citep{astroqueryginsburg}, 
        {\tt ChatGPT} (coding) \citep{OpenAI2023ChatGPT}, % per ChatGPT
        {\tt find\_orb} \citep{FindOrb},
        {\tt Grammarly}, %
        {\tt GRSS} \citep{makadia2025gauss}, 
        {\tt JPL Horizons} \citep{giorginiJPLsOnLineSolar1996}, % already in acks, text 1/31/2023 COC
        {\tt Layup}\footnote{\url{https://github.com/Smithsonian/layup}}, 
        {\tt Matplotlib} \citep{hunterMatplotlib2DGraphics2007},
        {\tt numba} \citep{lam2015numba}, 
        {\tt NumPy} \citep{harrisArrayProgrammingNumPy2020},
        {\tt OverLeaf}, 
        {\tt pandas} \citep{rebackPandasdevPandasPandas2022}, % limiting to 1 citation, removing older mckinneyDataStructuresStatistical2010 1/31/2023 COC
        {\tt PyTorch} \citep{paszkePyTorchImperativeStyle2019}, % 1/17/2024 COC
        {\tt rubin\_scheduler} \citep{2019AJ....157..151N, peter_yoachim_2025_15742506}, 
        {\tt rubin\_sim} \citep{2014SPIE.9149E..0BJ, peter_yoachim_2025_15368965}, 
        {\tt SAOImageDS9} \citep{joyeNewFeaturesSAOImage2006},
        {\tt SCARLETT} \citep{melchior2018_scarlet},
        {\tt SciPy} \citep{virtanenSciPy10Fundamental2020}
        {\tt Siril}\footnote{\url{https://siril.org}}, % actually not used in this work (but will be for the full paper) so removing 1/31/2023 COC
        {\tt SkyBot} \citep{berthierSkyBoTNewVO2006}, % cited in acks already
        {\tt sbpy} \citep{mommert2019_sbpy},
        {\tt Sorcha} \citep{Merritt2025,Holman2025},
        {\tt uncertainties} (version 3.0.2), a {\tt python} package for calculations with uncertainties by E.~O.\ Lebigot\footnote{\url{http://pythonhosted.org/uncertainties/}},
}

%%%%%%%%%%%%%%%%%%%%%%%%%%%%%%%%%%%%%%%%%%%%%%%%%%%%%%%%%
%\clearpage
\appendix

% Images are now in the main text but 
% \section{Additional Images}\label{appendix:images}
\setcounter{figure}{0}
\renewcommand{\thefigure}{A\arabic{figure}}
\setcounter{table}{0}
\renewcommand{\thetable}{A\arabic{table}}

This appendix shows additional images of \objname{} from \rubinobs{}, in Figures \ref{fig:gallery2},  \ref{fig:gallery3}, \ref{fig:gallery4}, and \ref{fig:gallery5}; the first (precovery) gallery is found in Figure \ref{fig:gallery}. All images were produced by initially following the methods outlined in \cite{chandlerActiveAsteroidsCitizen2024}. We also reprojected images such that north is up and east left (Rubin images are acquired at varying rotation angles) and mosaicked the images for cases where more than one chip was in the FOV. In all cases, the on-sky positions were provided by the JPL Horizons ephemeris service.

% do not change figsize!
\renewcommand{\figsize}{0.18}
\begin{figure*}
    \centering
        % \input{gallery2_5x5}
        %% Generated on 2025-10-06T22:17:33.260211+00:00 UTC
\begin{tabular}{ccccc}
\labelpicB{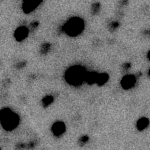}{}{01:53:12}{\figsize}{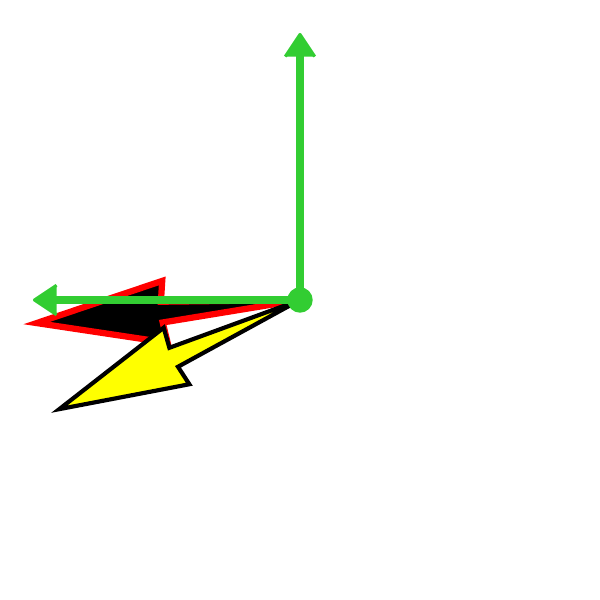} & \labelpicB{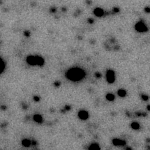}{}{03:03:56}{\figsize}{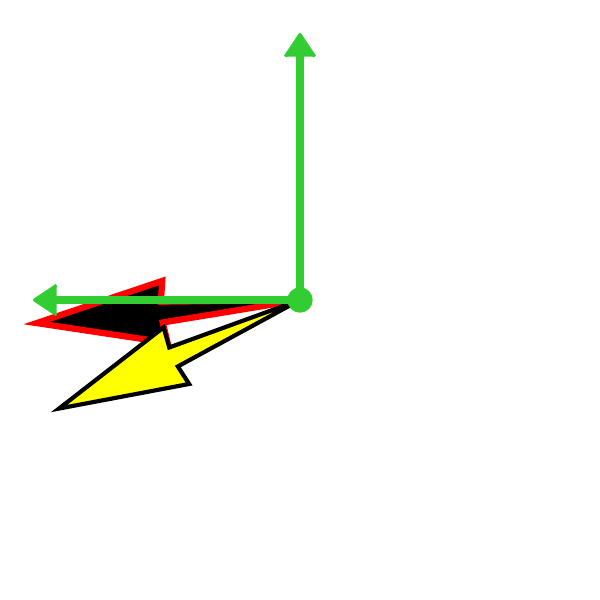} & \labelpicB{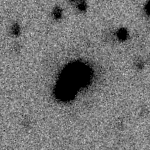}{}{06:25:00}{\figsize}{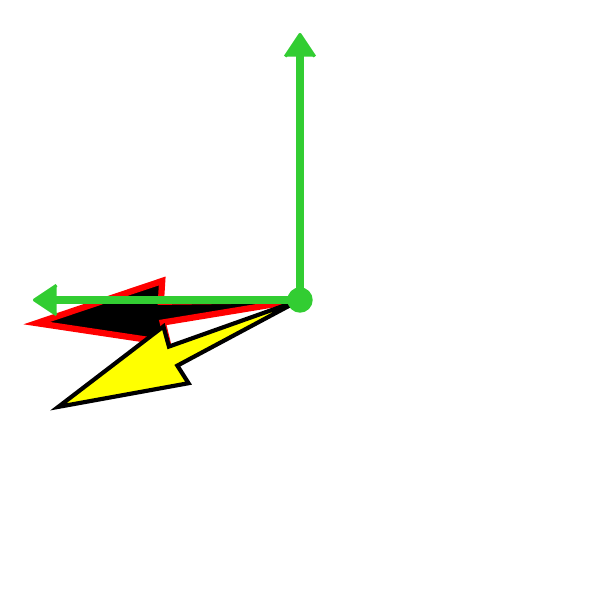} & \labelpicB{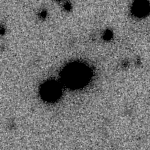}{}{06:27:17}{\figsize}{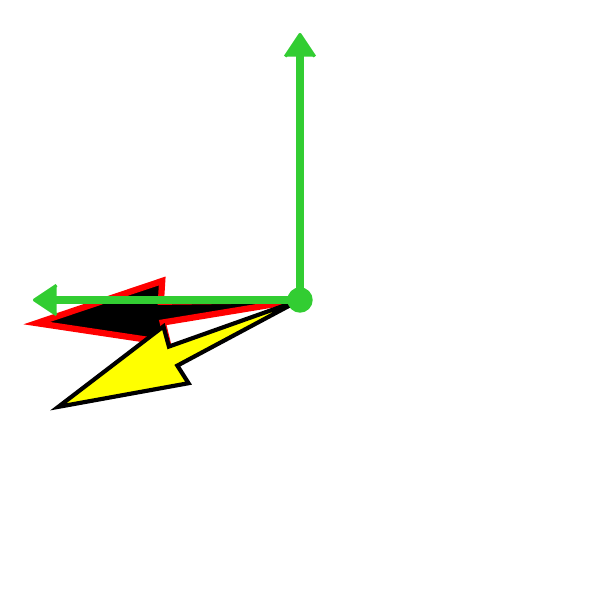} & \labelpicB{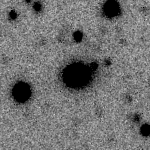}{}{06:32:03}{\figsize}{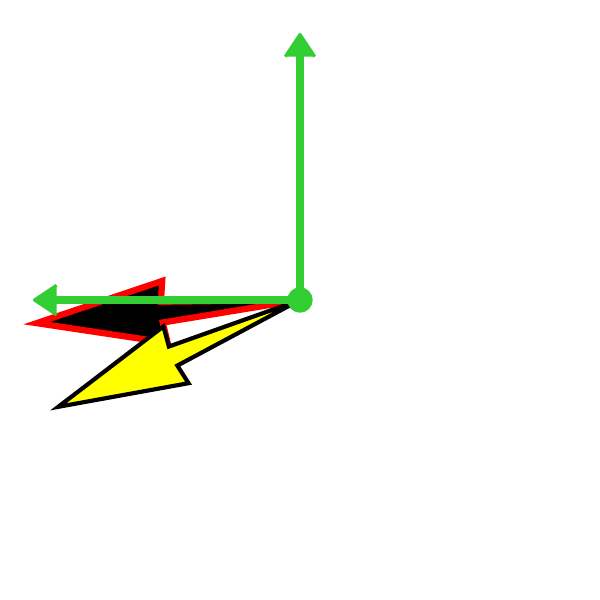} \\
\labelpicB{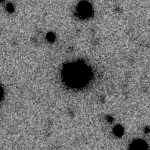}{}{06:36:22}{\figsize}{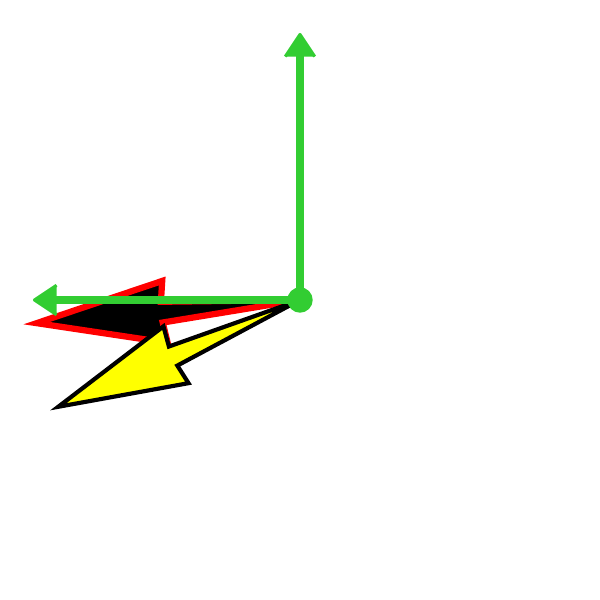} & \labelpicB{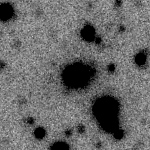}{}{06:48:53}{\figsize}{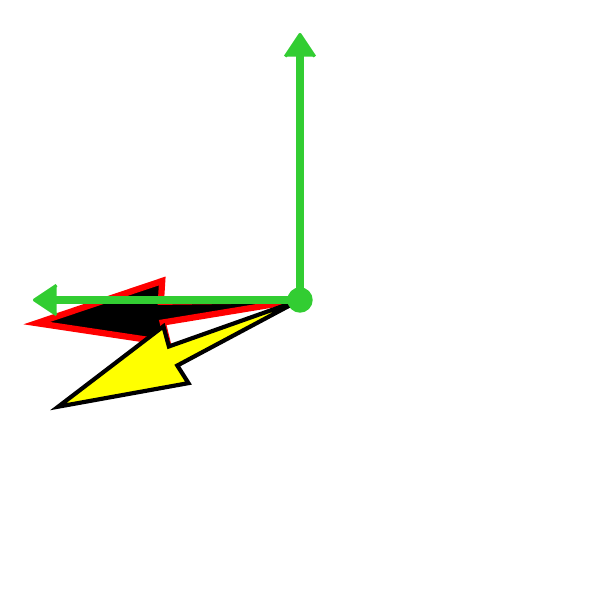} & \labelpicB{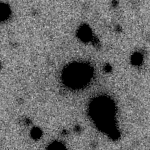}{}{06:49:27}{\figsize}{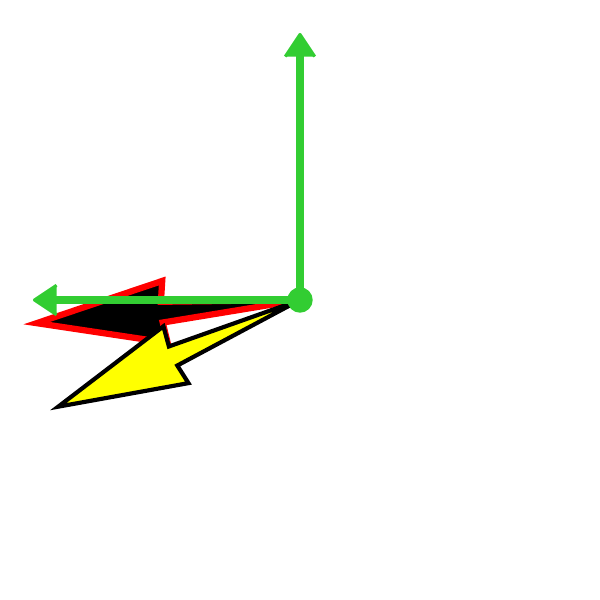} & \labelpicB{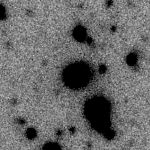}{}{06:50:17}{\figsize}{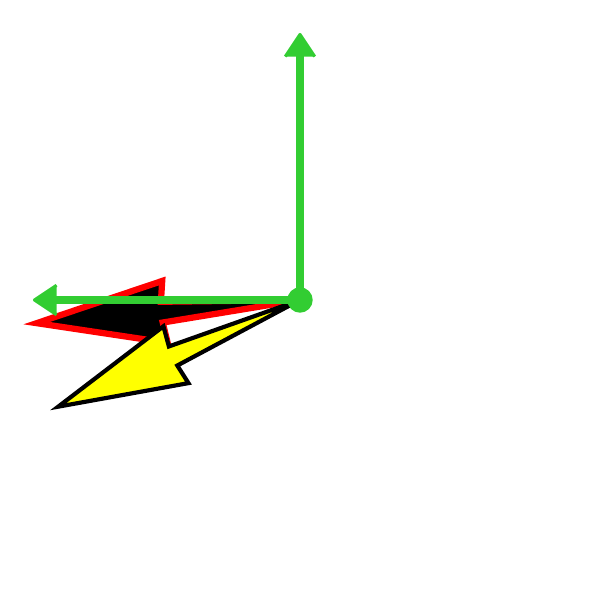} & \labelpicB{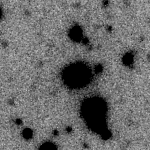}{}{06:50:52}{\figsize}{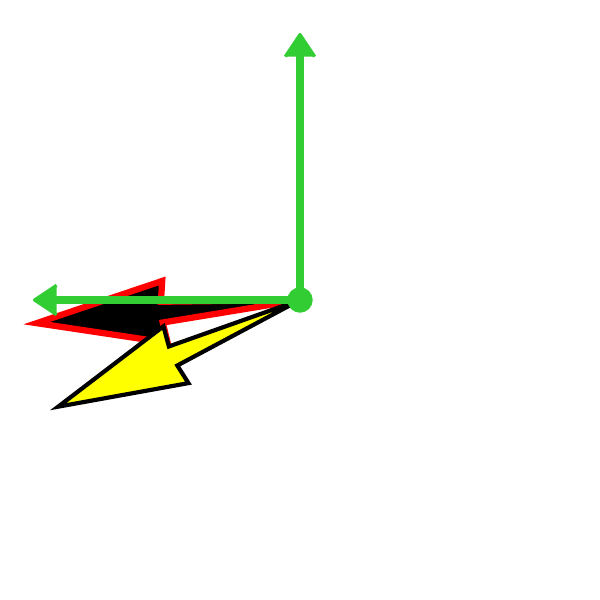} \\
\labelpicB{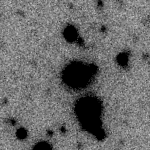}{}{06:51:47}{\figsize}{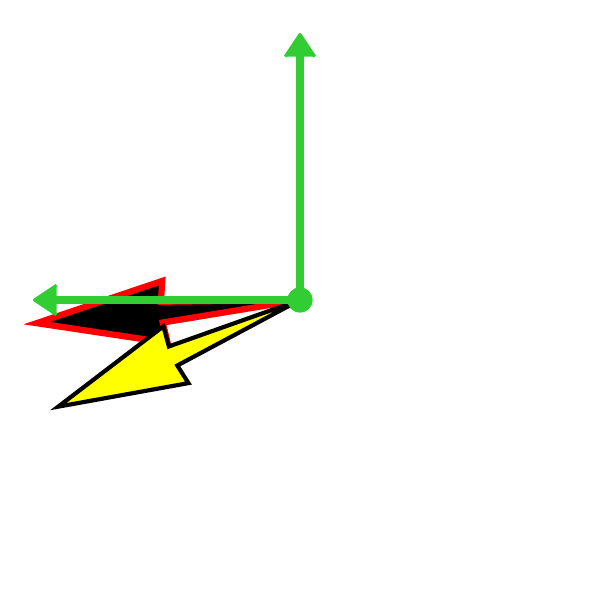} & \labelpicB{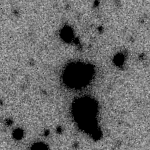}{}{06:52:21}{\figsize}{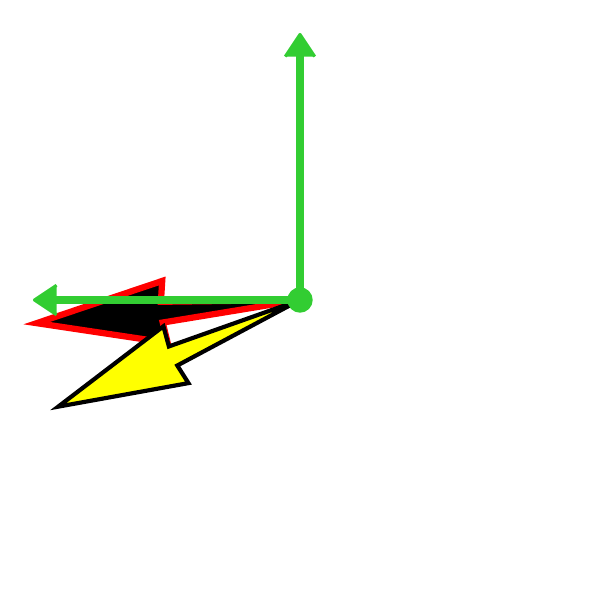} & \labelpicB{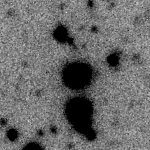}{}{06:53:16}{\figsize}{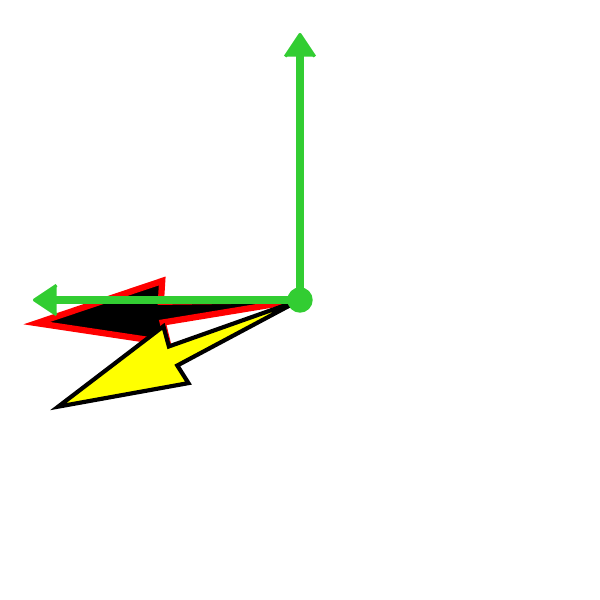} & \labelpicB{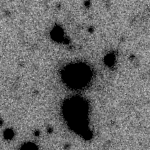}{}{06:53:51}{\figsize}{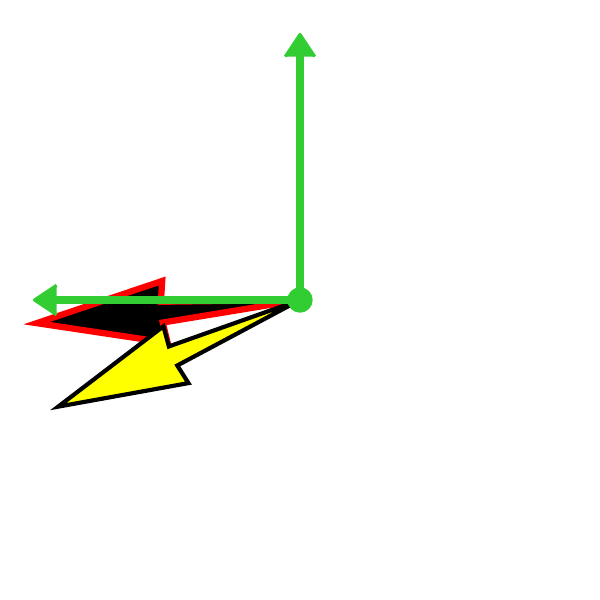} & \labelpicB{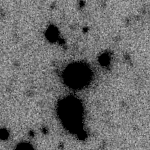}{}{06:54:40}{\figsize}{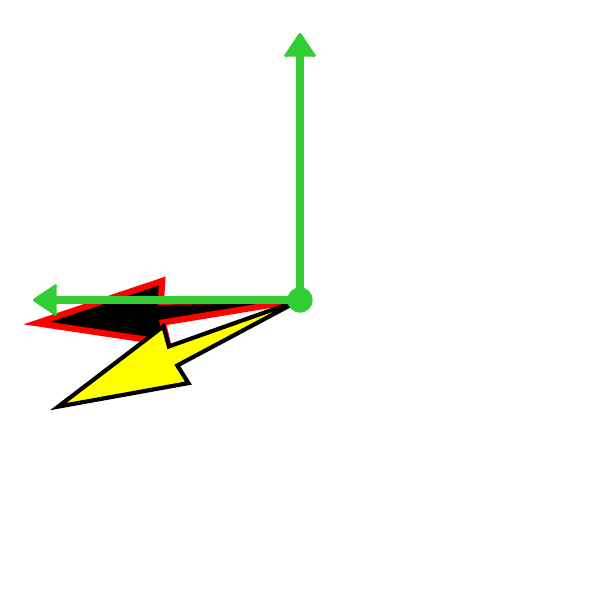} \\
\labelpicB{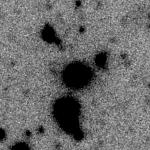}{}{06:55:15}{\figsize}{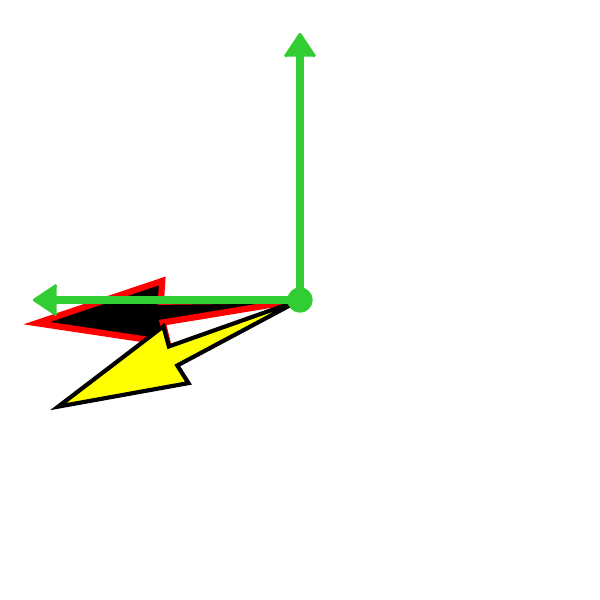} & \labelpicB{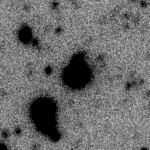}{}{06:58:56}{\figsize}{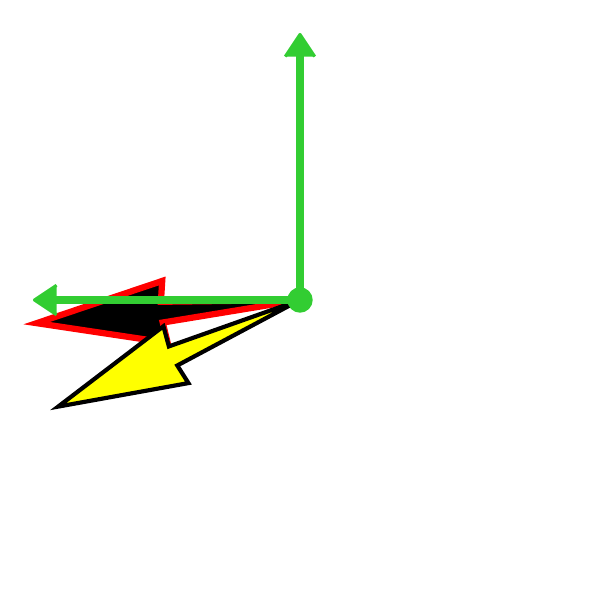} & \labelpicB{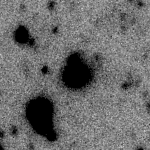}{}{06:59:30}{\figsize}{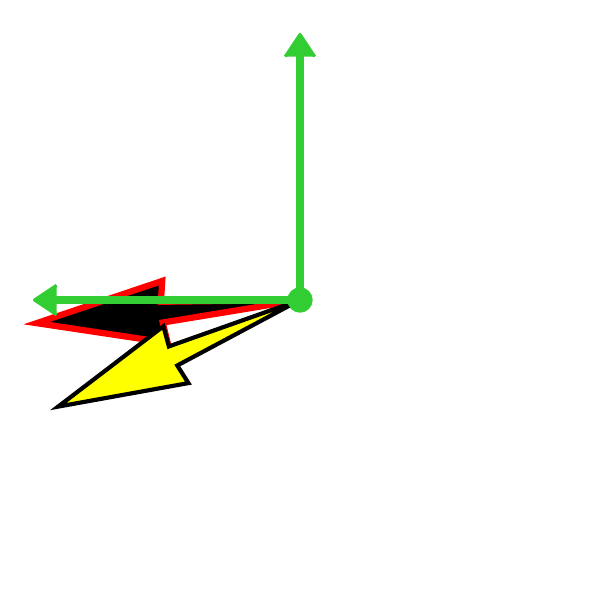} & \labelpicB{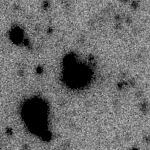}{}{07:00:20}{\figsize}{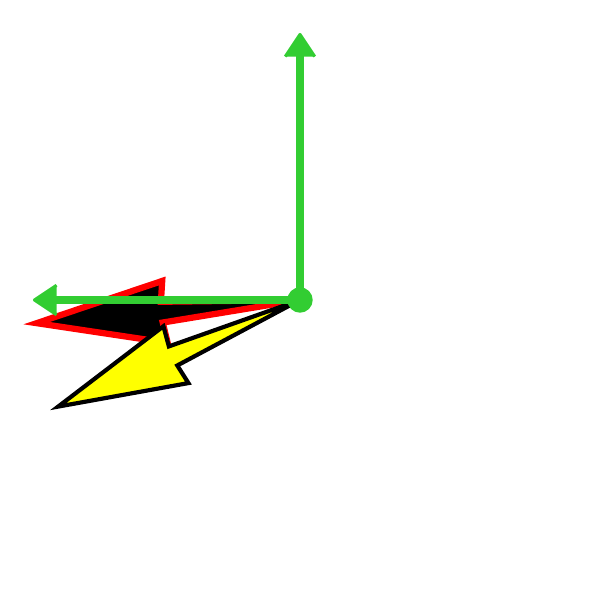} & \labelpicB{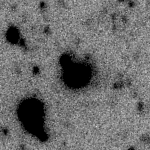}{}{07:00:54}{\figsize}{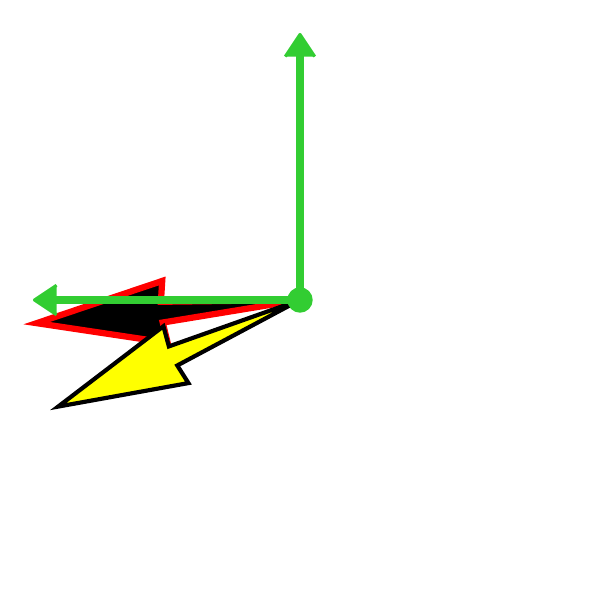} \\
\labelpicB{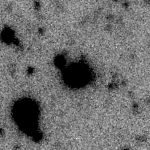}{}{07:01:46}{\figsize}{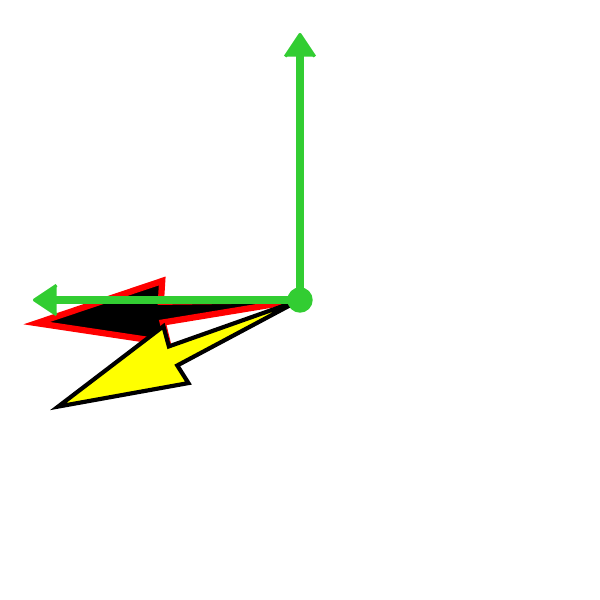} & \labelpicB{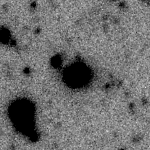}{}{07:02:20}{\figsize}{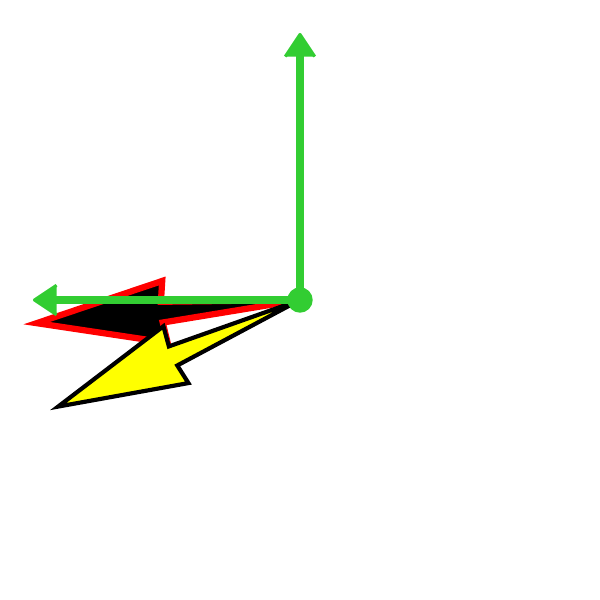} & \labelpicB{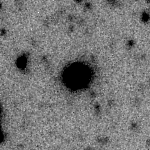}{}{07:07:48}{\figsize}{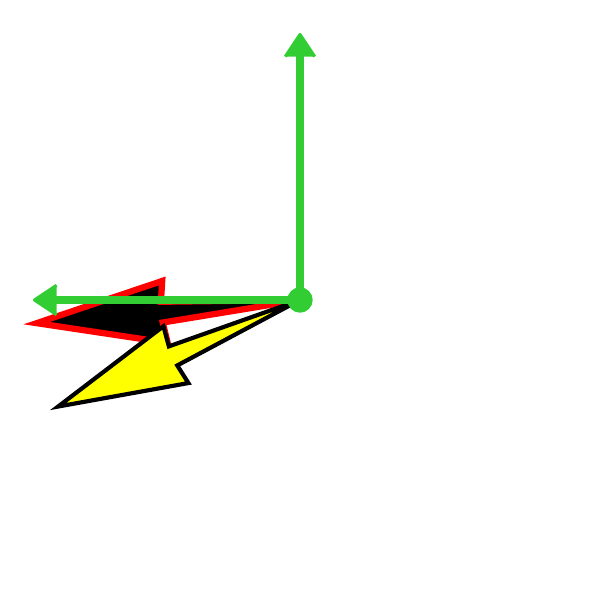} & \labelpicB{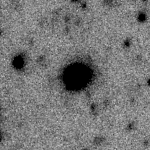}{}{07:08:22}{\figsize}{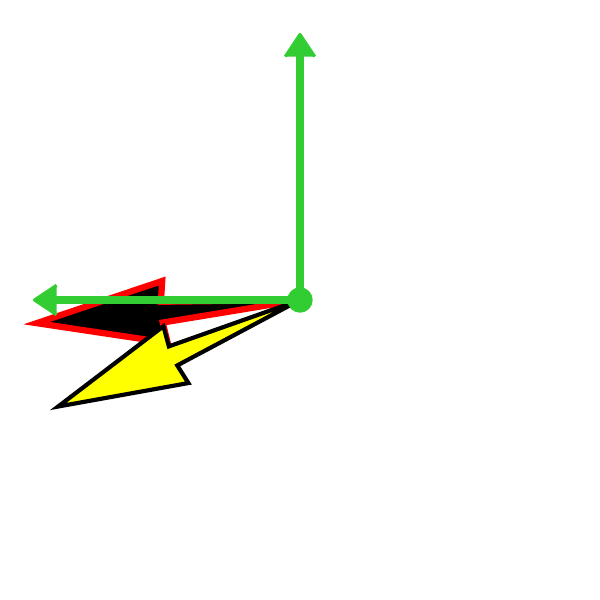} & \labelpicB{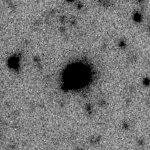}{}{07:09:07}{\figsize}{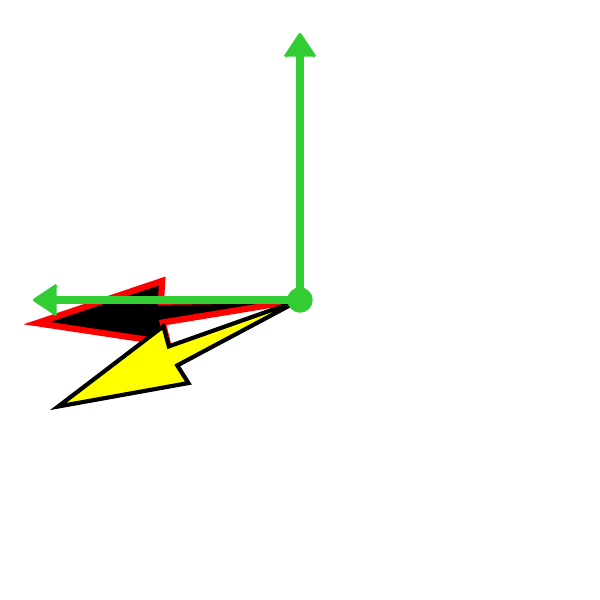} \\
\labelpicB{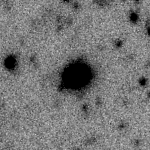}{}{07:09:41}{\figsize}{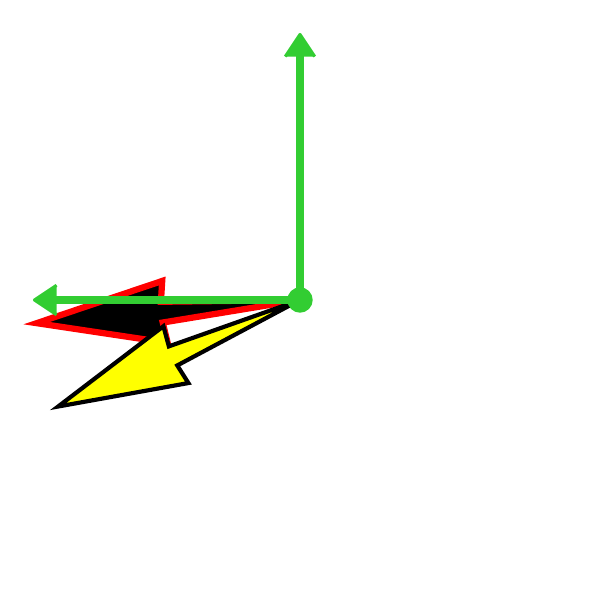} & \labelpicB{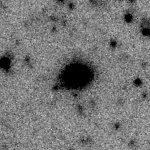}{}{07:10:31}{\figsize}{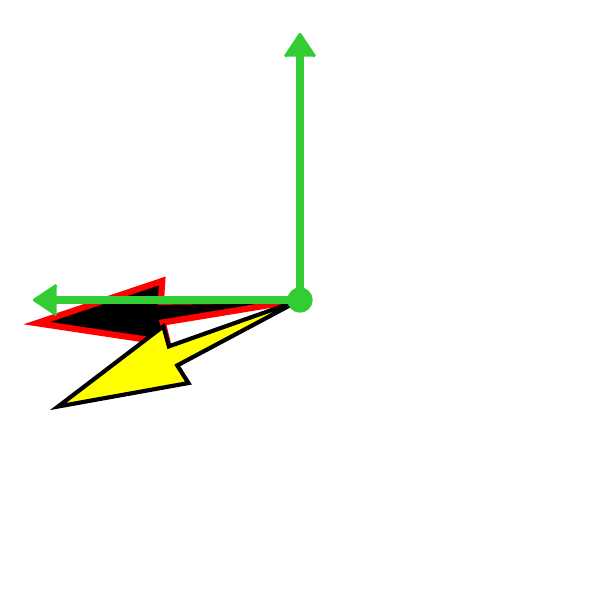} & \labelpicB{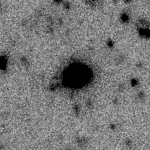}{}{07:11:05}{\figsize}{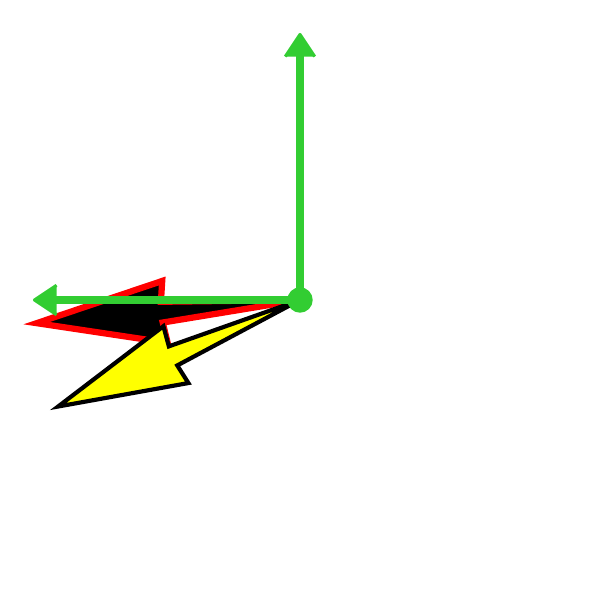} & \labelpicB{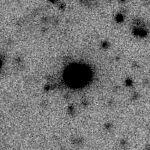}{}{07:11:55}{\figsize}{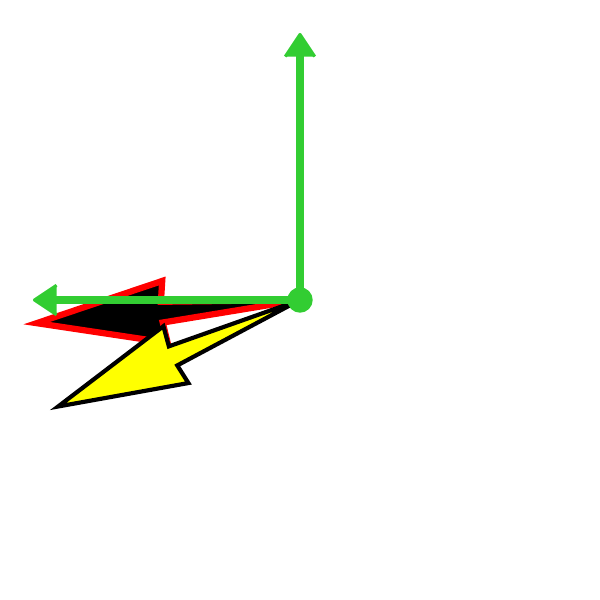} & \labelpicB{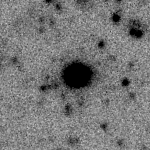}{}{07:12:30}{\figsize}{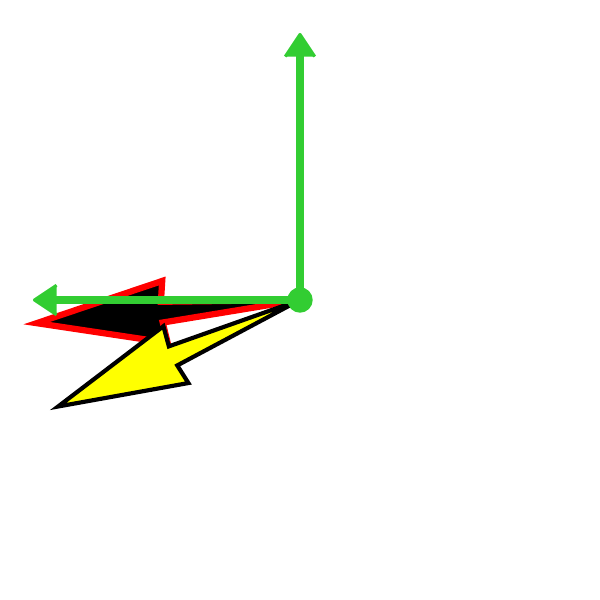} \\
\labelpicB{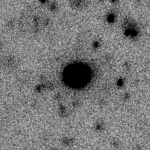}{}{07:13:19}{\figsize}{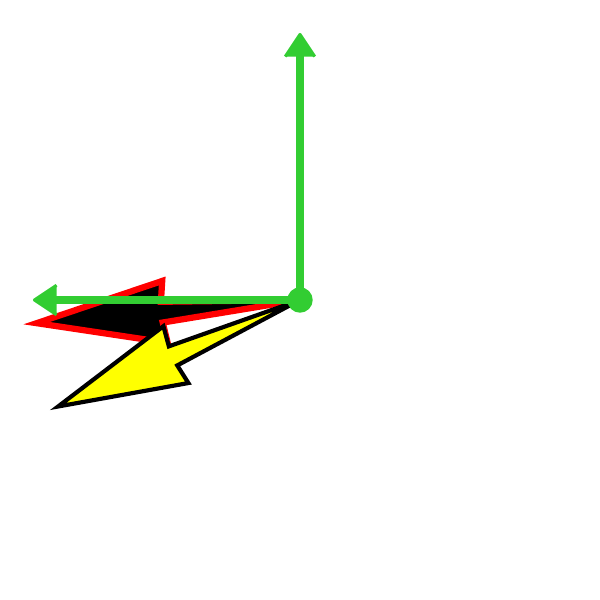} & \labelpicB{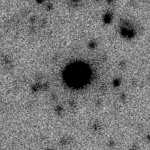}{}{07:13:53}{\figsize}{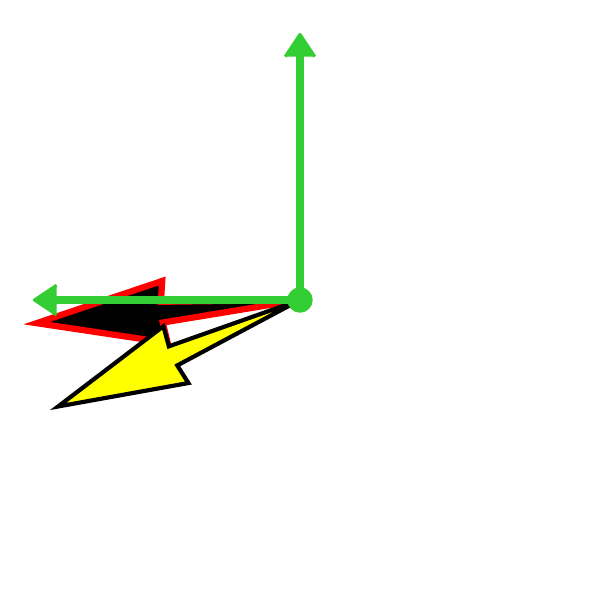} \\
\end{tabular}

%% Label-to-date mapping:
%% \textbf{()} 2025-07-03 01:53:12
%% \textbf{()} 2025-07-03 03:03:56
%% \textbf{()} 2025-07-03 06:25:00
%% \textbf{()} 2025-07-03 06:27:17
%% \textbf{()} 2025-07-03 06:32:03
%% \textbf{()} 2025-07-03 06:36:22
%% \textbf{()} 2025-07-03 06:48:53
%% \textbf{()} 2025-07-03 06:49:27
%% \textbf{()} 2025-07-03 06:50:17
%% \textbf{()} 2025-07-03 06:50:52
%% \textbf{()} 2025-07-03 06:51:47
%% \textbf{()} 2025-07-03 06:52:21
%% \textbf{()} 2025-07-03 06:53:16
%% \textbf{()} 2025-07-03 06:53:51
%% \textbf{()} 2025-07-03 06:54:40
%% \textbf{()} 2025-07-03 06:55:15
%% \textbf{()} 2025-07-03 06:58:56
%% \textbf{()} 2025-07-03 06:59:30
%% \textbf{()} 2025-07-03 07:00:20
%% \textbf{()} 2025-07-03 07:00:54
%% \textbf{()} 2025-07-03 07:01:46
%% \textbf{()} 2025-07-03 07:02:20
%% \textbf{()} 2025-07-03 07:07:48
%% \textbf{()} 2025-07-03 07:08:22
%% \textbf{()} 2025-07-03 07:09:07
%% \textbf{()} 2025-07-03 07:09:41
%% \textbf{()} 2025-07-03 07:10:31
%% \textbf{()} 2025-07-03 07:11:05
%% \textbf{()} 2025-07-03 07:11:55
%% \textbf{()} 2025-07-03 07:12:30
%% \textbf{()} 2025-07-03 07:13:19
%% \textbf{()} 2025-07-03 07:13:53

    \caption{\objname{} (centered and indicated by magenta ticks) imaged on UT 2025 July 3 by the \rubinobs{} as a part of commissioning. All images are 30~s $r$-band exposures with eorth is up and east left (green arrows), and the anti-solar $-\odot$ (yellow arrow) and anti-motion $-v$ (red-outlined black arrow) vectors shown inset at the top-left. The \ac{TAI} exposure midpoint times are given.
%     \textbf{(a)} 2025-07-03 06:27:17
% \textbf{(b)} 2025-07-03 06:48:53. 
% \textbf{(c)} 2025-07-03 06:50:52. 
% \textbf{(d)} 2025-07-03 06:51:47. 
% \textbf{(e)} 2025-07-03 06:52:21. 
% \textbf{(f)} 2025-07-03 06:53:16. 
% \textbf{(g)} 2025-07-03 06:53:51. 
% \textbf{(h)} 2025-07-03 06:54:40. 
% \textbf{(i)} 2025-07-03 06:55:15. 
% \textbf{(j)} 2025-07-03 07:00:20. 
% \textbf{(k)} 2025-07-03 07:00:54. 
% \textbf{(l)} 2025-07-03 07:01:46. 
% \textbf{(m)} 2025-07-03 07:02:20. 
% \textbf{(n)} 2025-07-03 07:07:48. 
% \textbf{(o)} 2025-07-03 07:08:22. 
% \textbf{(p)} 2025-07-03 07:09:07. 
% \textbf{(q)} 2025-07-03 07:10:31. 
% \textbf{(r)} 2025-07-03 07:11:05. 
% \textbf{(s)} 2025-07-03 07:11:55. 
% \textbf{(t)} 2025-07-03 07:12:30. 
% \textbf{(u)} 2025-07-03 07:13:19. 
% \textbf{(v)} 2025-07-03 07:13:53. 
    }
    \label{fig:gallery2}
\end{figure*}

\renewcommand{\figsize}{0.18}
\begin{figure*}
    \centering
    % \input{gallery3_4X}
    % \input{gallery3/gallery.tex}
    %% Generated on 2025-10-06T22:49:00.504423+00:00 UTC
\begin{tabular}{ccccc}
\labelpicBb{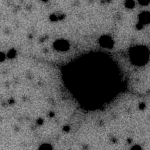}{}{2025-07-04}{\figsize}{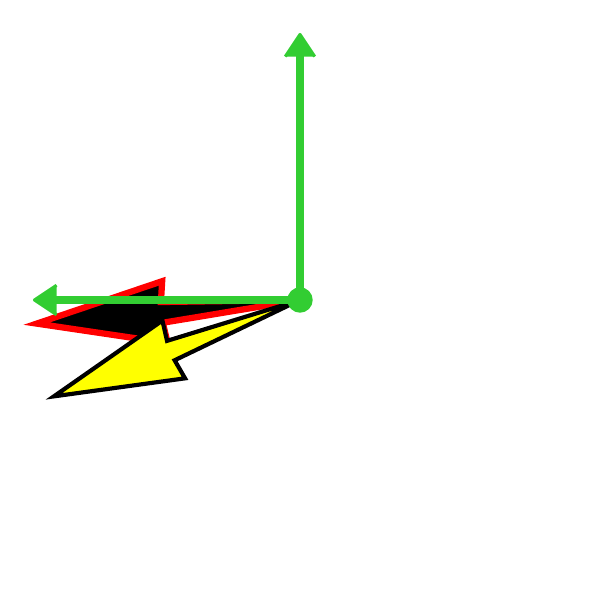} & \labelpicBb{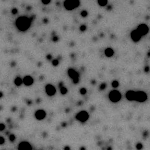}{}{2025-07-05}{\figsize}{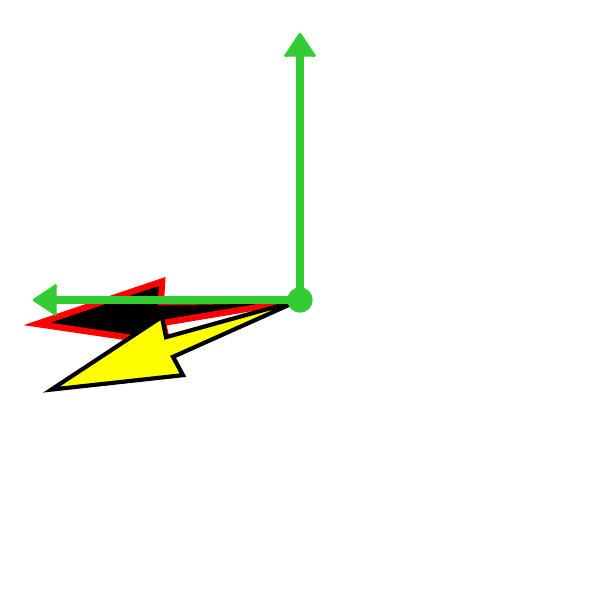} & \labelpicBb{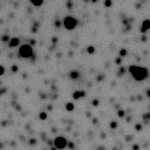}{}{2025-07-05}{\figsize}{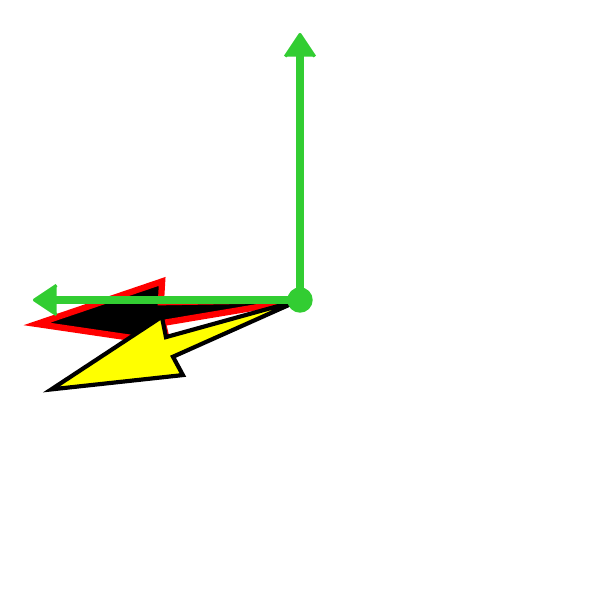} & 
% \labelpicBb{3I_2025-07-06T23-34-15_LSSTCam_2025070600033_41_warp_cropped_cropped.png}{}{2025-07-06}{\figsize}{3I_2025-07-06T23-34-15_LSSTCam_2025070600033_41_arrows.pdf} & 
\labelpicBb{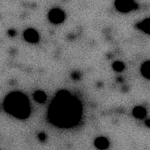}{}{2025-07-06}{\figsize}{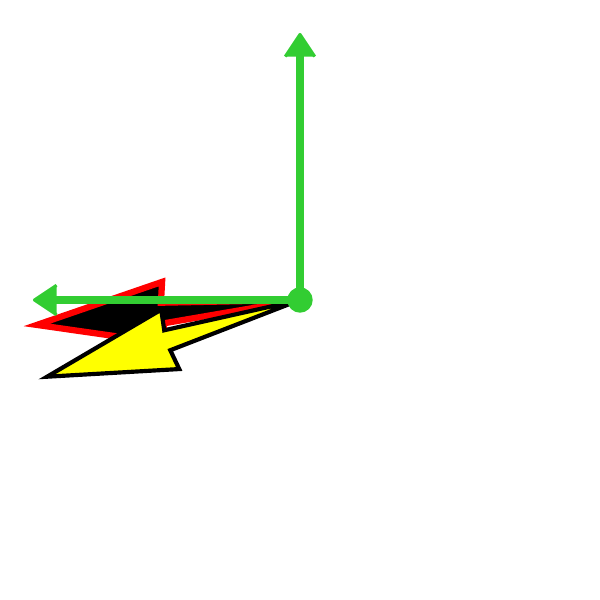} &
% \labelpicBb{3I_2025-07-07T00-12-48_LSSTCam_2025070600092_41_warp_cropped_cropped.png}{}{2025-07-07}{\figsize}{3I_2025-07-07T00-12-48_LSSTCam_2025070600092_41_arrows.pdf} \\
 \labelpicBb{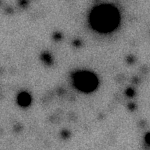}{}{2025-07-07}{\figsize}{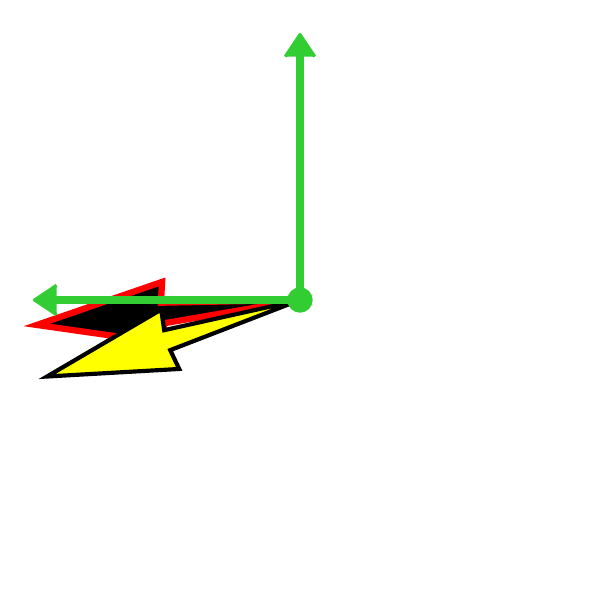} \\ 
 \labelpicBb{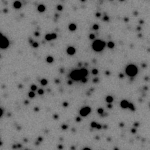}{}{2025-07-12}{\figsize}{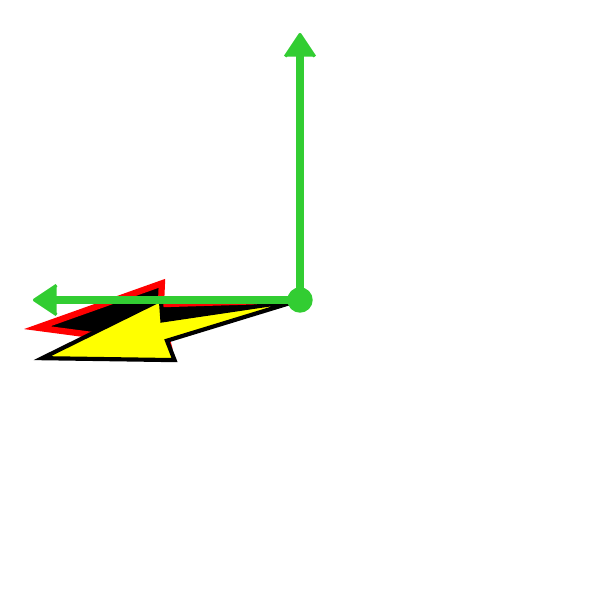} & \labelpicBb{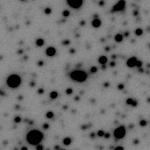}{}{2025-07-12}{\figsize}{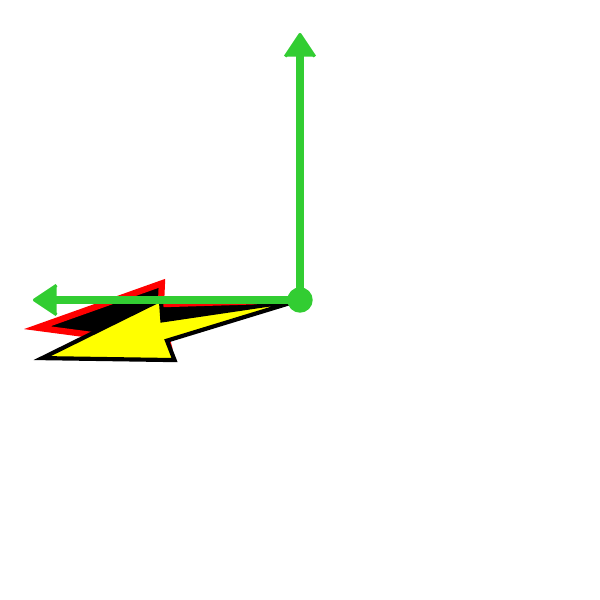} & \labelpicBb{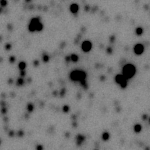}{}{2025-07-12}{\figsize}{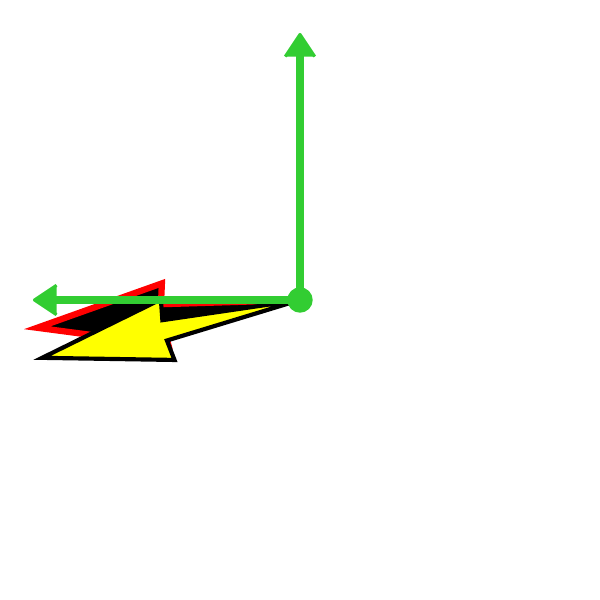} & 
\labelpicBb{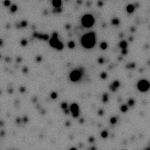}{}{2025-07-12}{\figsize}{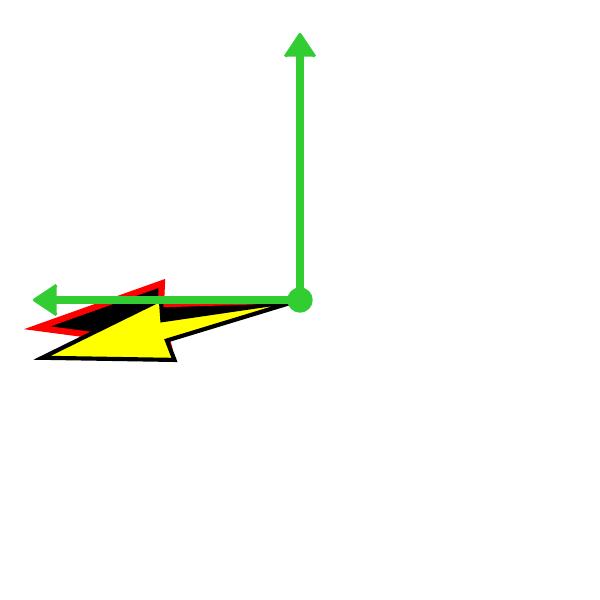} \\
\end{tabular}

%% Label-to-date mapping:
%% \textbf{()} 2025-07-04 06:36:15
%% \textbf{()} 2025-07-05 01:23:40
%% \textbf{()} 2025-07-05 01:56:38
%% \textbf{()} 2025-07-06 23:34:15
%% \textbf{()} 2025-07-06 23:53:08
%% \textbf{()} 2025-07-07 00:12:48
%% \textbf{()} 2025-07-07 00:44:03
%% \textbf{()} 2025-07-12 01:12:32
%% \textbf{()} 2025-07-12 01:46:44
%% \textbf{()} 2025-07-12 04:39:06
%% \textbf{()} 2025-07-12 05:13:58

    \caption{\objname{} (center, indicated by magenta ticks) observations after UT 2025 July 3 and before the \acl{ToO} of UT 2025 July 13. North is up, and east left (green arrows), with the anti-solar (yellow arrow) and anti-motion (black arrow with red outline) shown.}
    \label{fig:gallery3}
\end{figure*}

\renewcommand{\figsize}{0.18}
\begin{figure*}
    \centering
    %% Generated on 2025-10-06T22:41:02.471516+00:00 UTC
\begin{tabular}{ccccc}
\labelpicB{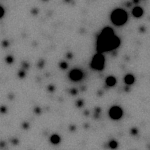}{}{00:35:43}{\figsize}{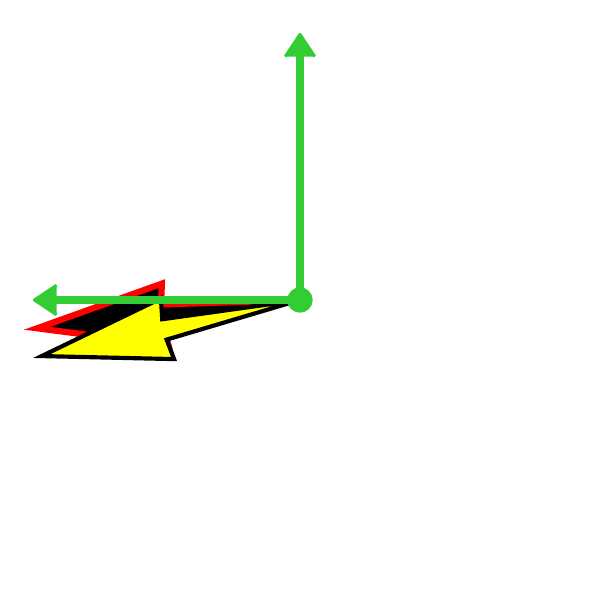} & \labelpicB{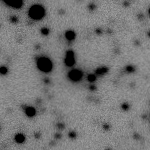}{}{03:13:21}{\figsize}{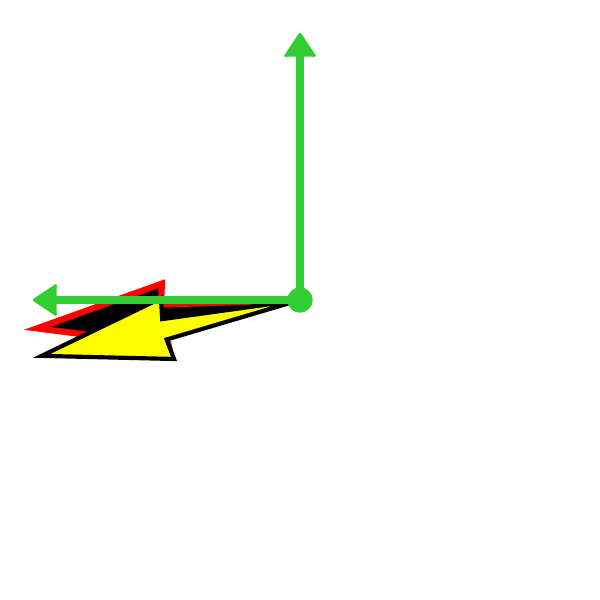} & \labelpicB{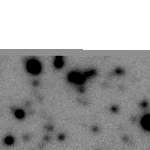}{}{03:14:43}{\figsize}{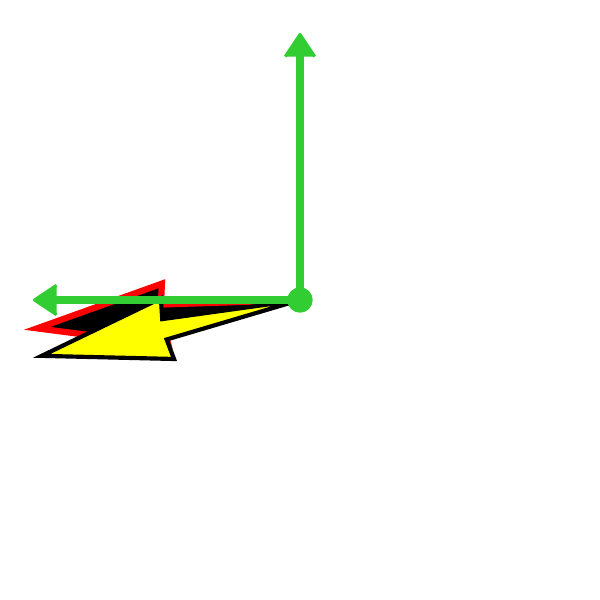} & \labelpicB{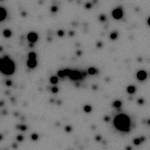}{}{03:18:35}{\figsize}{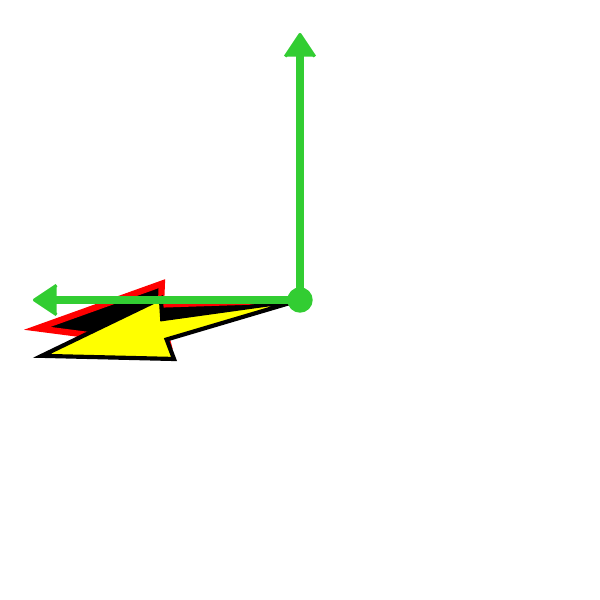} & \labelpicB{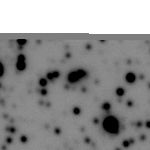}{}{03:19:57}{\figsize}{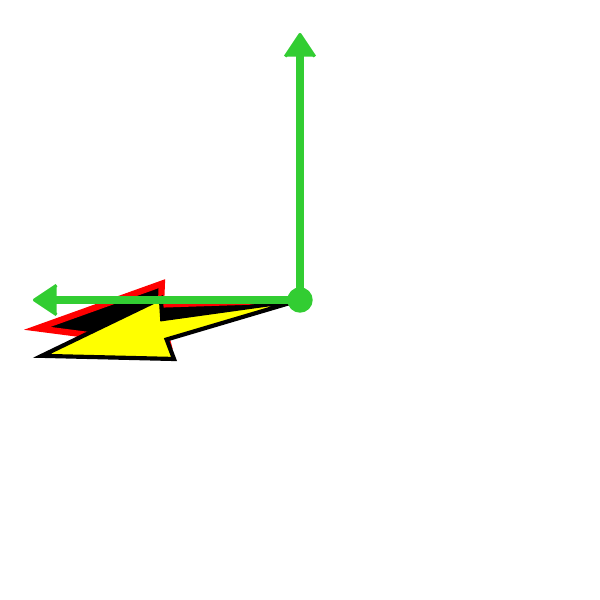} \\
\labelpicB{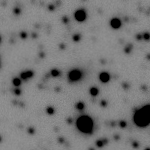}{}{03:23:44}{\figsize}{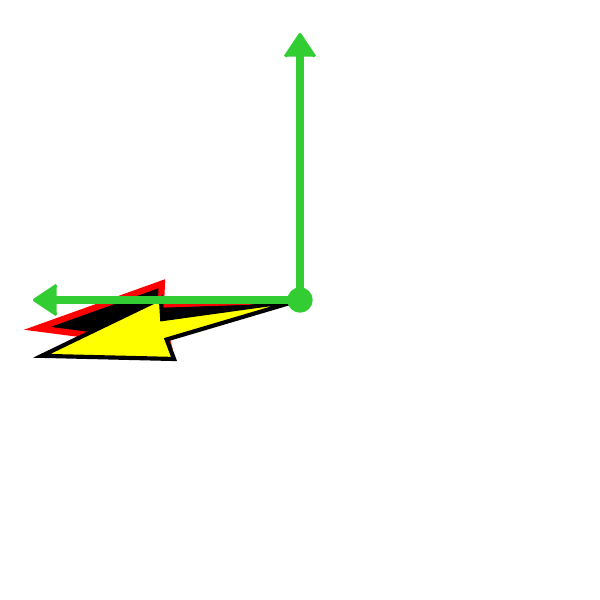} & \labelpicB{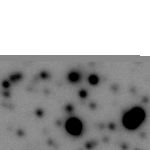}{}{03:25:05}{\figsize}{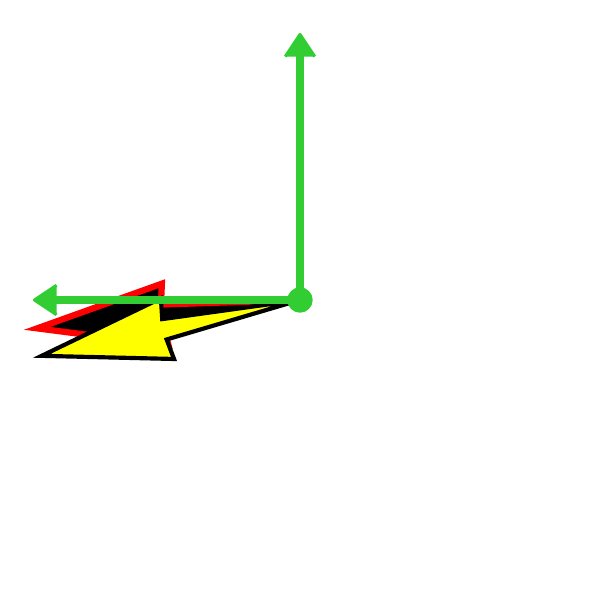} & \labelpicB{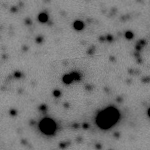}{}{03:28:52}{\figsize}{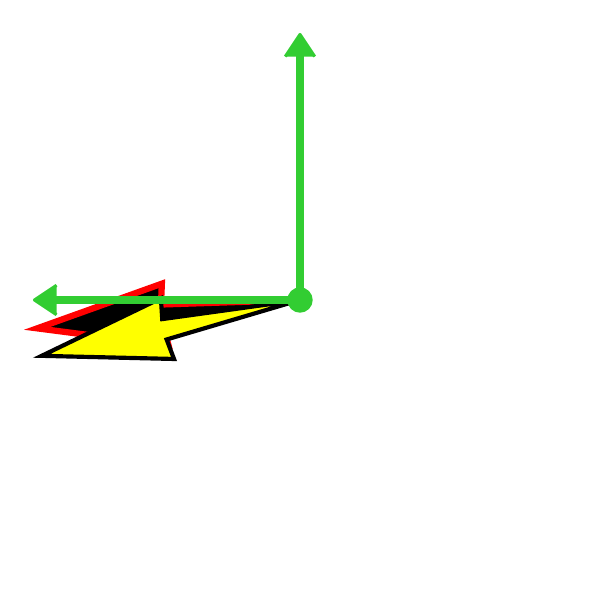} & \labelpicB{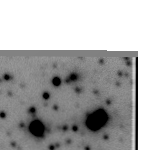}{}{03:30:13}{\figsize}{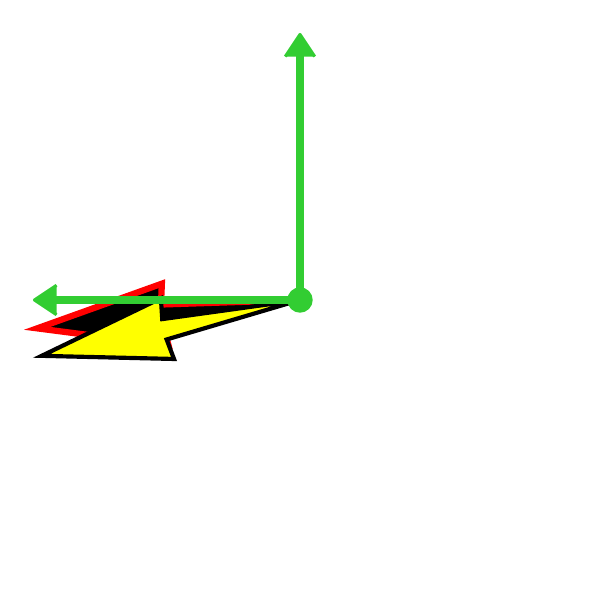} & \labelpicB{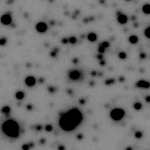}{}{03:33:59}{\figsize}{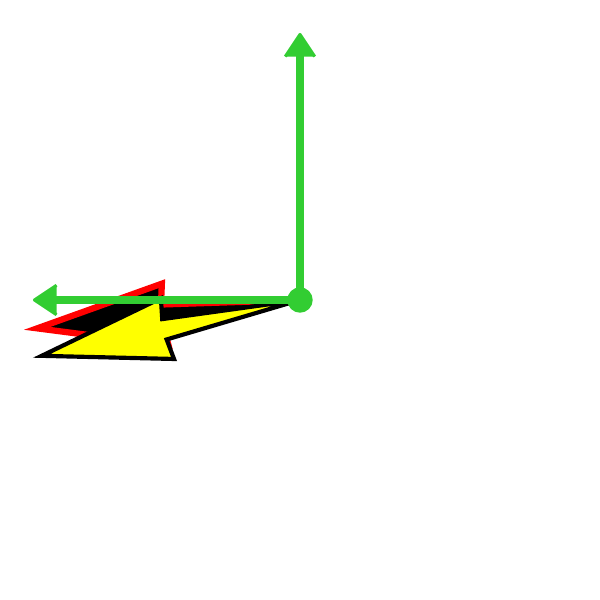} \\
\labelpicB{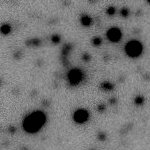}{}{03:38:59}{\figsize}{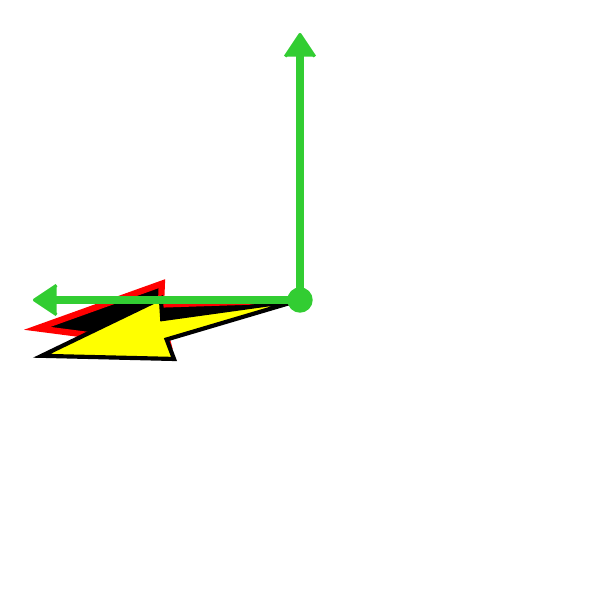} & \labelpicB{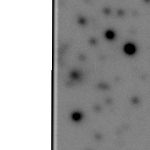}{}{03:39:38}{\figsize}{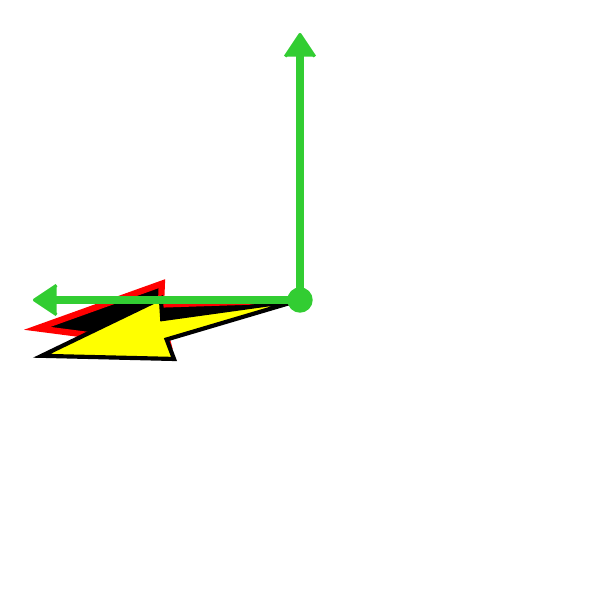} & \labelpicB{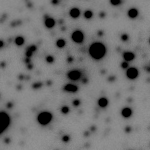}{}{03:44:01}{\figsize}{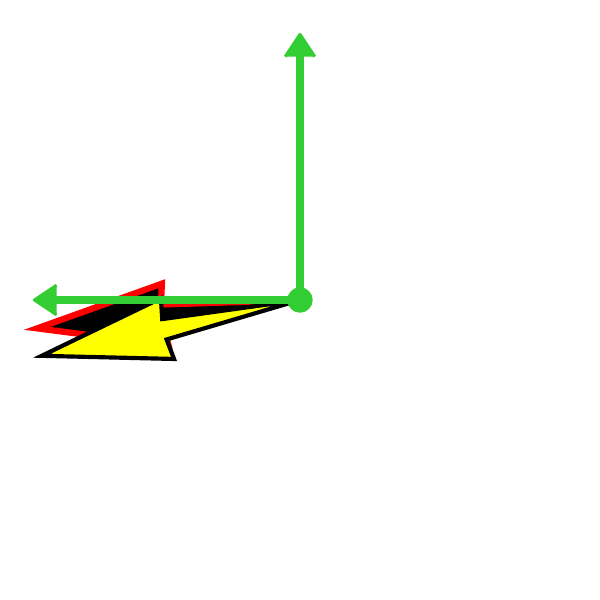} & \labelpicB{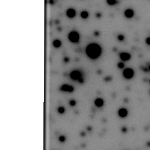}{}{03:44:39}{\figsize}{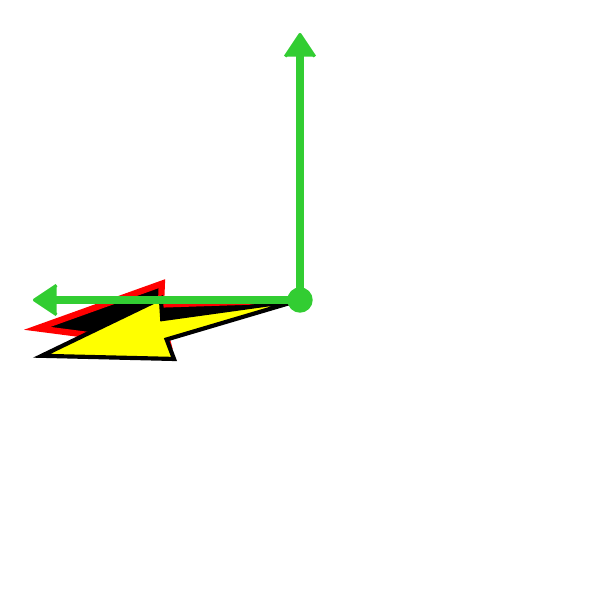} & \labelpicB{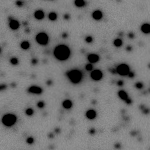}{}{03:48:58}{\figsize}{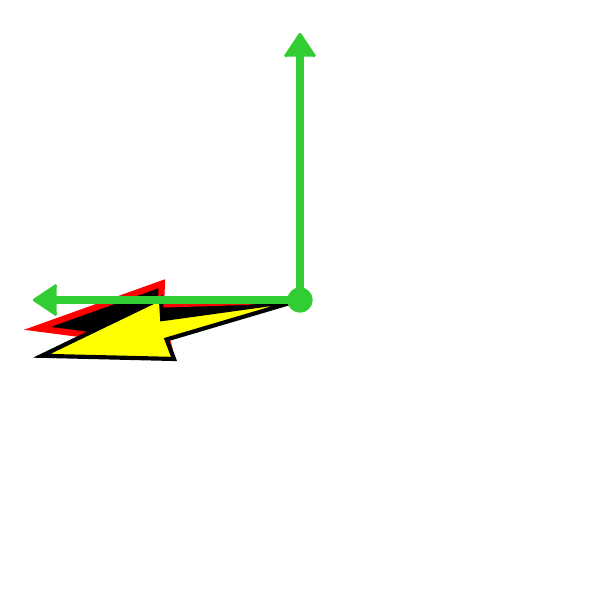} \\
\labelpicB{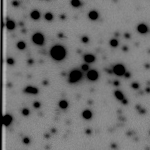}{}{03:49:36}{\figsize}{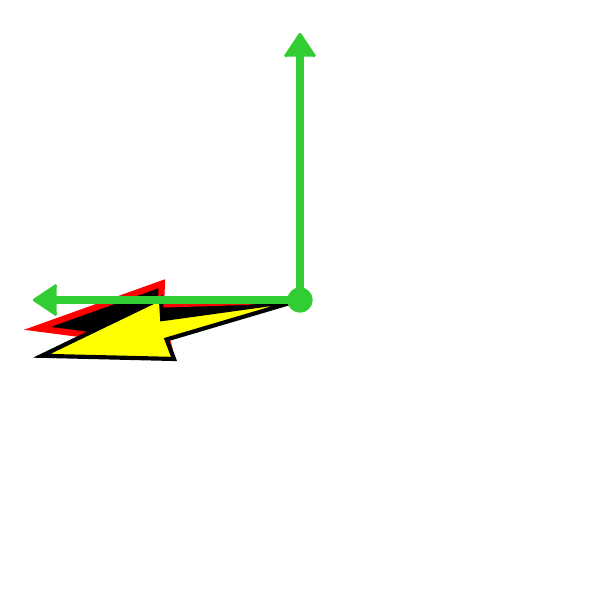} & \labelpicB{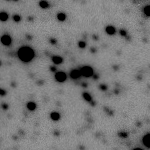}{}{03:53:56}{\figsize}{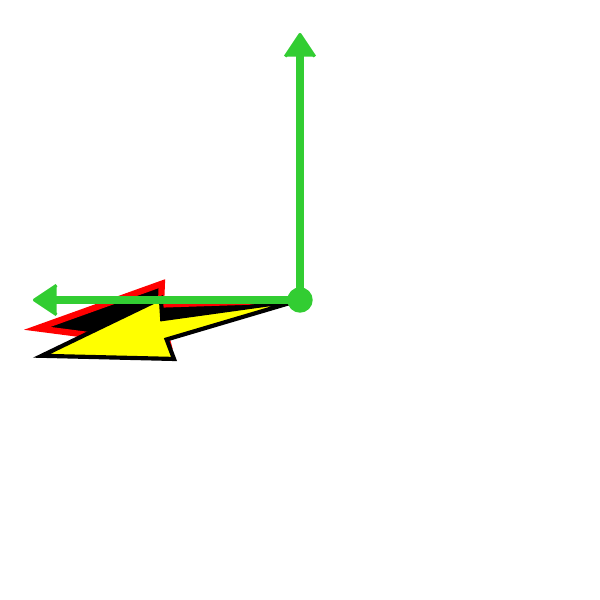} & \labelpicB{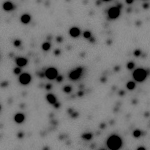}{}{03:58:53}{\figsize}{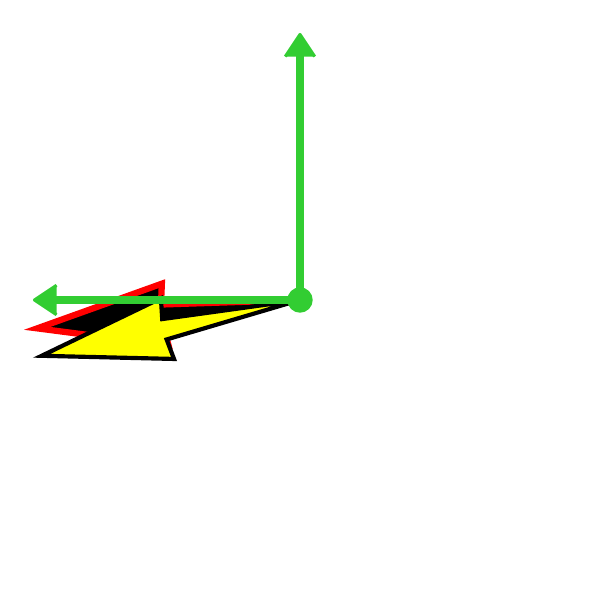} & \labelpicB{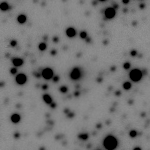}{}{03:59:31}{\figsize}{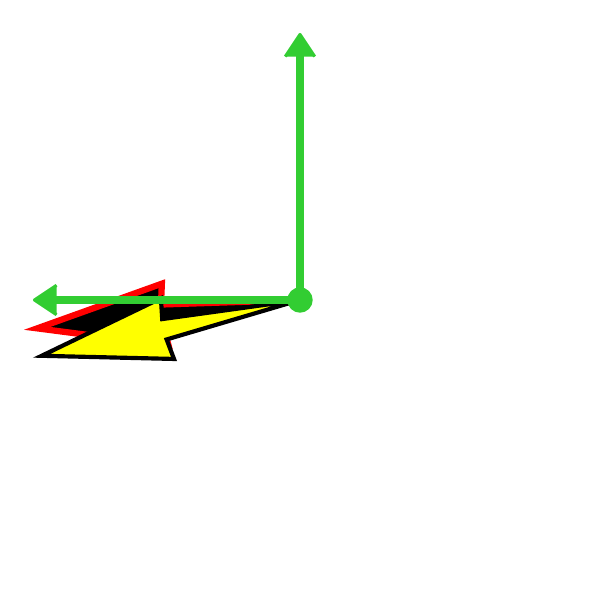} & \labelpicB{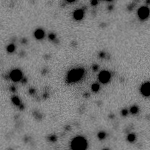}{}{04:03:46}{\figsize}{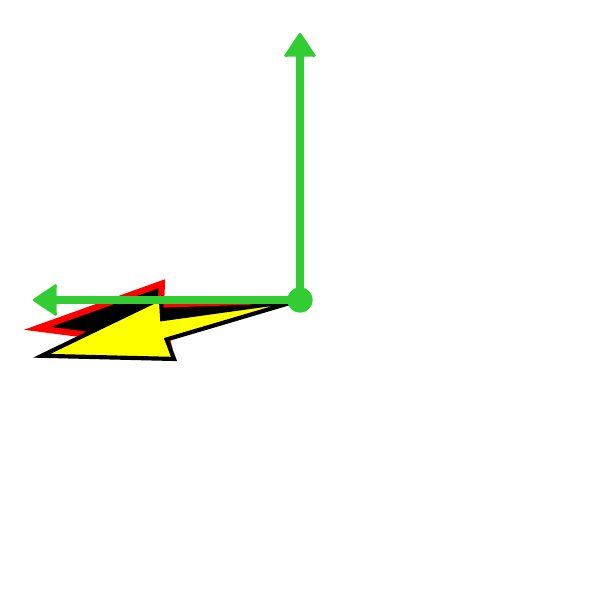} \\
\labelpicB{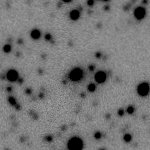}{}{04:04:24}{\figsize}{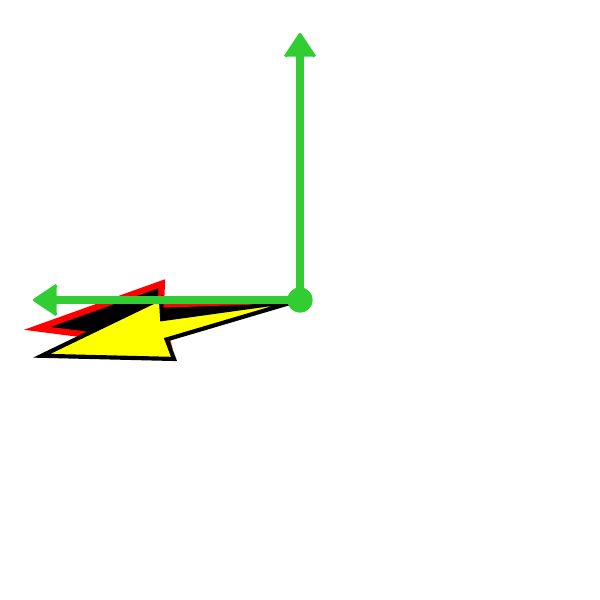} & \labelpicB{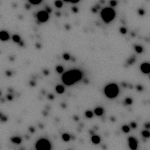}{}{04:08:42}{\figsize}{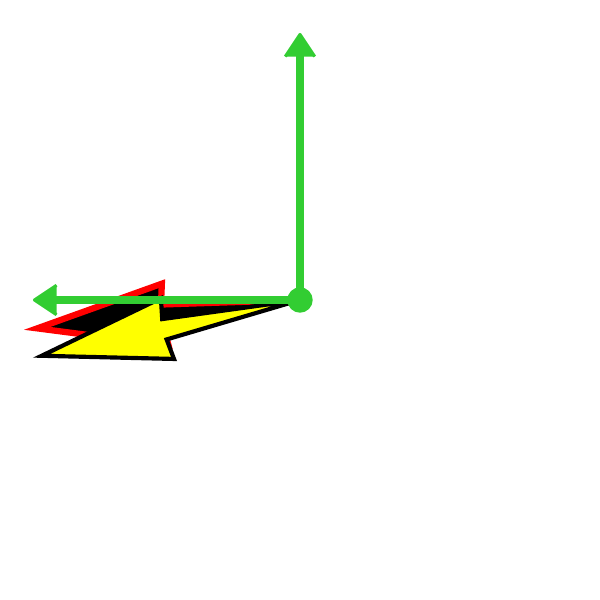} & \labelpicB{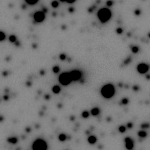}{}{04:09:20}{\figsize}{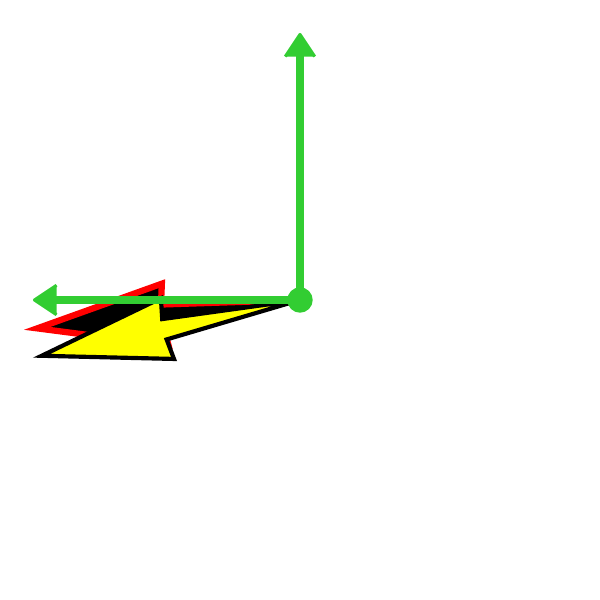} & \labelpicB{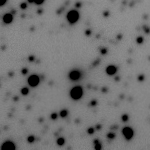}{}{04:13:34}{\figsize}{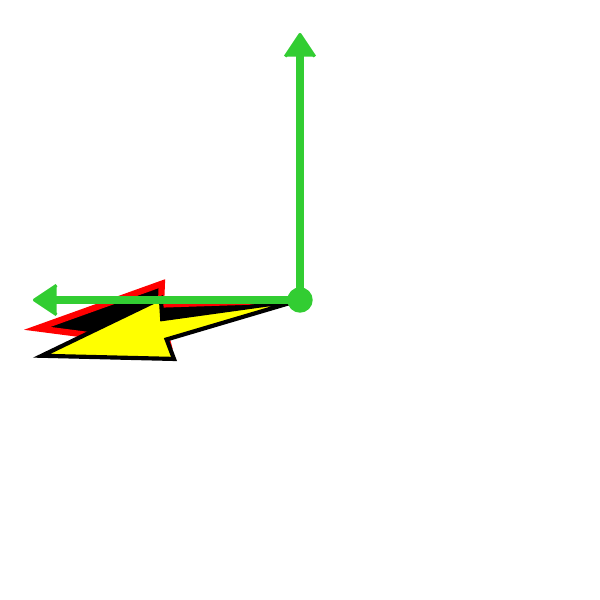} & \labelpicB{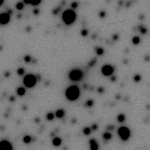}{}{04:14:11}{\figsize}{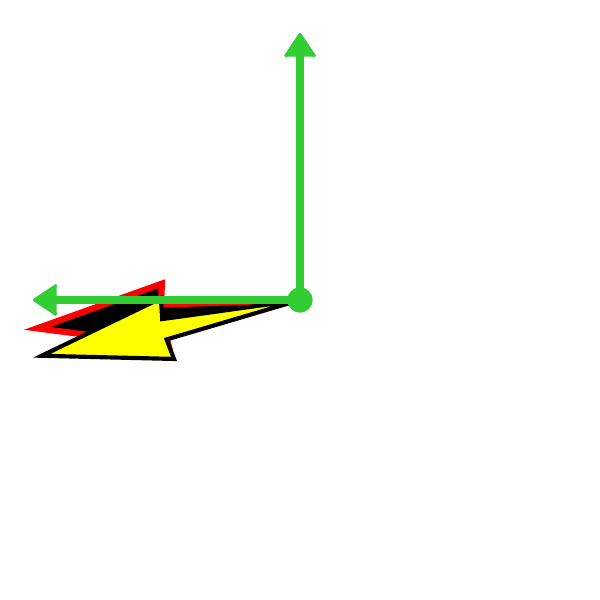} \\
\labelpicB{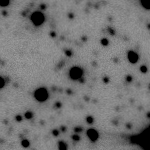}{}{04:18:26}{\figsize}{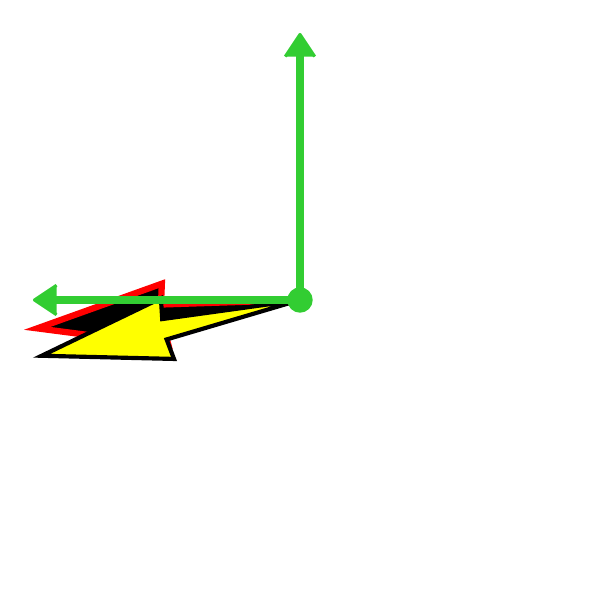} & \labelpicB{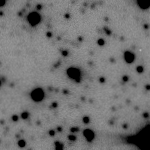}{}{04:19:04}{\figsize}{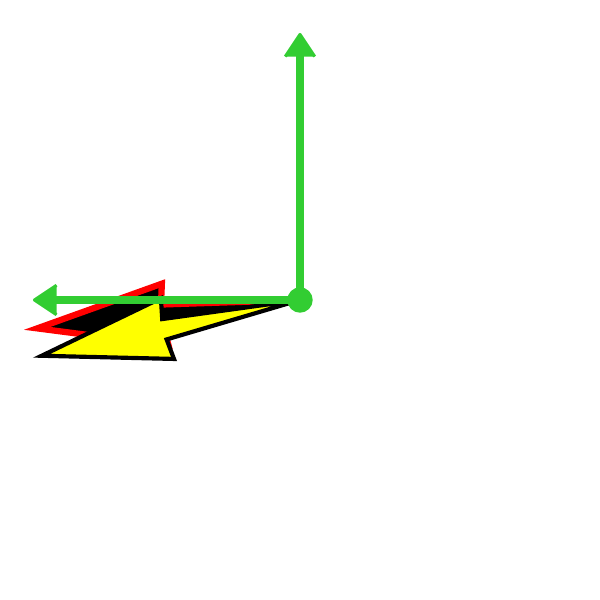} & \labelpicB{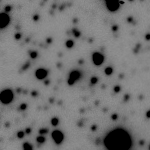}{}{04:23:19}{\figsize}{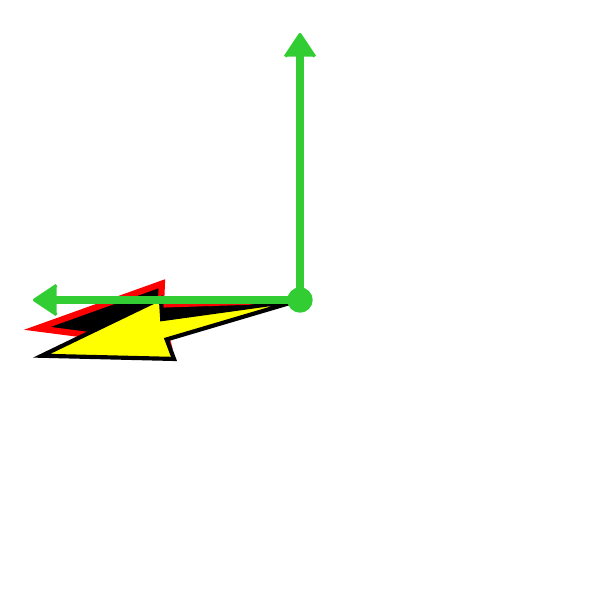} & \labelpicB{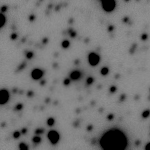}{}{04:23:57}{\figsize}{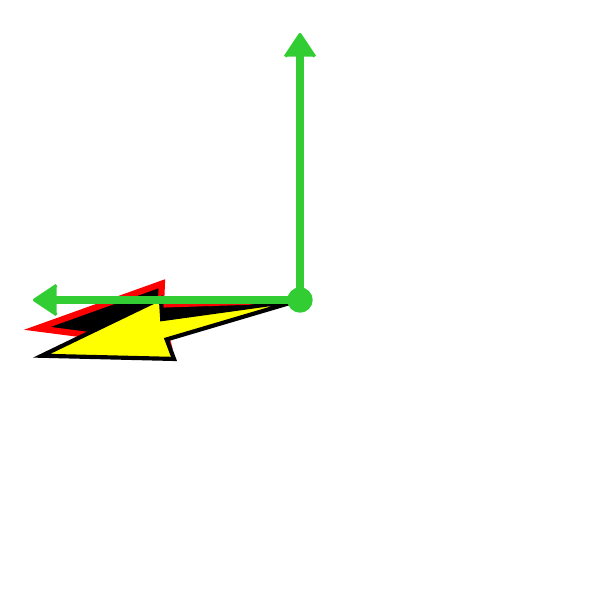} & \labelpicB{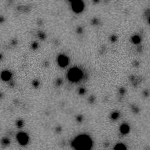}{}{04:28:07}{\figsize}{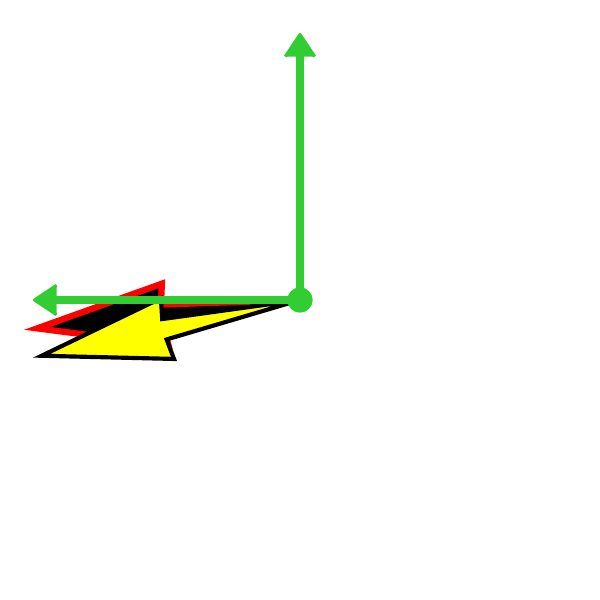} \\
\labelpicB{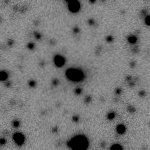}{}{04:28:45}{\figsize}{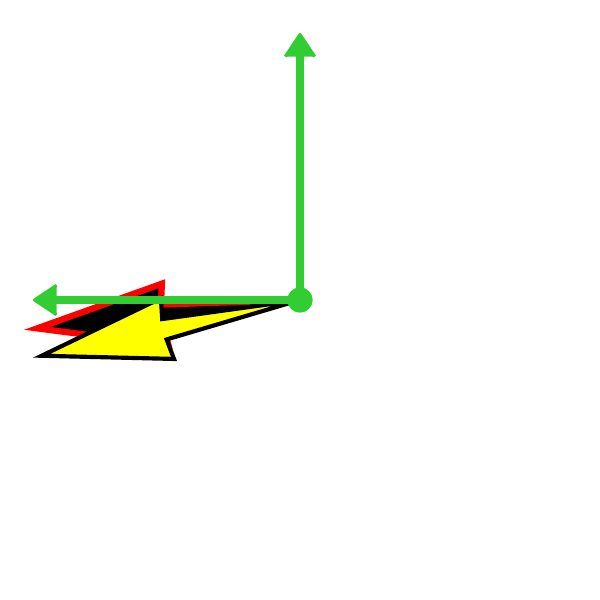} & \labelpicB{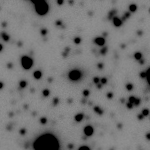}{}{04:33:06}{\figsize}{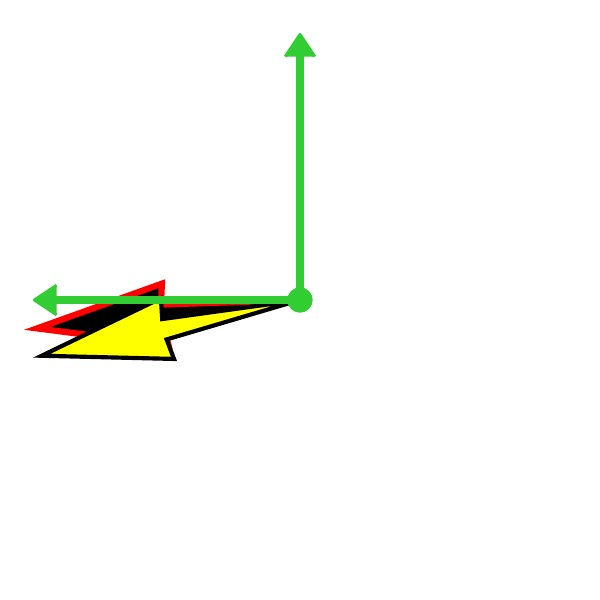} & \labelpicB{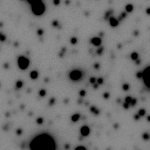}{}{04:33:44}{\figsize}{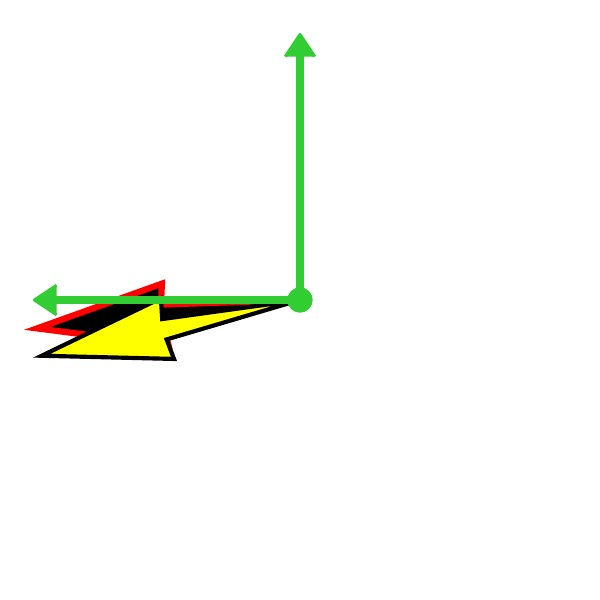} & \labelpicB{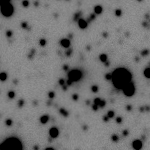}{}{04:37:54}{\figsize}{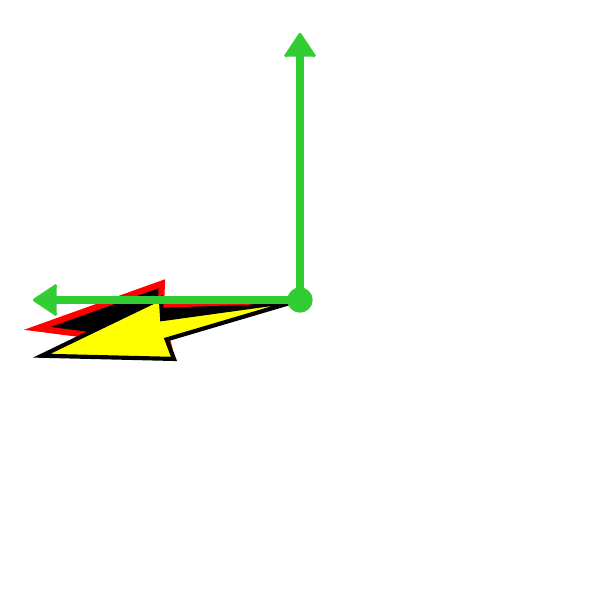} & \labelpicB{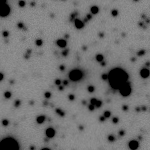}{}{04:38:32}{\figsize}{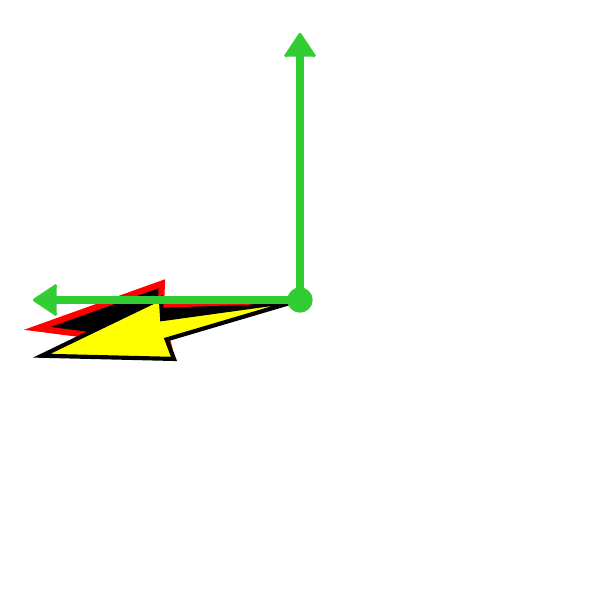} \\
\end{tabular}

%% Label-to-date mapping:
%% \textbf{()} 2025-07-13 00:35:43
%% \textbf{()} 2025-07-13 03:13:21
%% \textbf{()} 2025-07-13 03:14:43
%% \textbf{()} 2025-07-13 03:18:35
%% \textbf{()} 2025-07-13 03:19:57
%% \textbf{()} 2025-07-13 03:23:44
%% \textbf{()} 2025-07-13 03:25:05
%% \textbf{()} 2025-07-13 03:28:52
%% \textbf{()} 2025-07-13 03:30:13
%% \textbf{()} 2025-07-13 03:33:59
%% \textbf{()} 2025-07-13 03:38:59
%% \textbf{()} 2025-07-13 03:39:38
%% \textbf{()} 2025-07-13 03:44:01
%% \textbf{()} 2025-07-13 03:44:39
%% \textbf{()} 2025-07-13 03:48:58
%% \textbf{()} 2025-07-13 03:49:36
%% \textbf{()} 2025-07-13 03:53:56
%% \textbf{()} 2025-07-13 03:58:53
%% \textbf{()} 2025-07-13 03:59:31
%% \textbf{()} 2025-07-13 04:03:46
%% \textbf{()} 2025-07-13 04:04:24
%% \textbf{()} 2025-07-13 04:08:42
%% \textbf{()} 2025-07-13 04:09:20
%% \textbf{()} 2025-07-13 04:13:34
%% \textbf{()} 2025-07-13 04:14:11
%% \textbf{()} 2025-07-13 04:18:26
%% \textbf{()} 2025-07-13 04:19:04
%% \textbf{()} 2025-07-13 04:23:19
%% \textbf{()} 2025-07-13 04:23:57
%% \textbf{()} 2025-07-13 04:28:07
%% \textbf{()} 2025-07-13 04:28:45
%% \textbf{()} 2025-07-13 04:33:06
%% \textbf{()} 2025-07-13 04:33:44
%% \textbf{()} 2025-07-13 04:37:54
%% \textbf{()} 2025-07-13 04:38:32

    \caption{\objname{} (center, indicated by magenta ticks) \acl{ToO} observations from UT 2025 July 13. North is up, and east left (green arrows), with the anti-solar (yellow arrow) and anti-motion (black arrow with red outline) shown.}
    \label{fig:gallery4}
\end{figure*}

\renewcommand{\figsize}{0.18}
\begin{figure*}
    \centering
    %% Generated on 2025-10-06T22:57:46.023637+00:00 UTC
\begin{tabular}{ccccc}
\labelpicBb{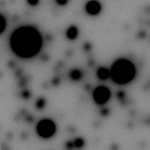}{a}{2025-07-14}{\figsize}{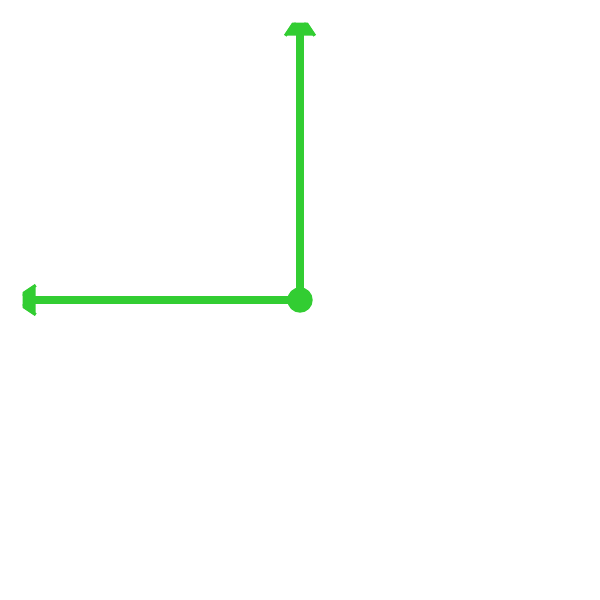} & \labelpicBb{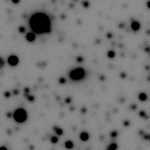}{b}{2025-07-14}{\figsize}{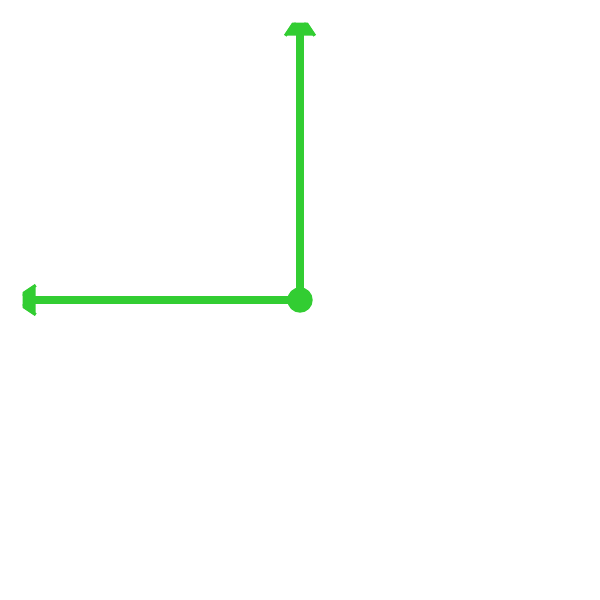} & \labelpicBb{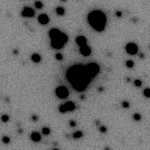}{c}{2025-07-19}{\figsize}{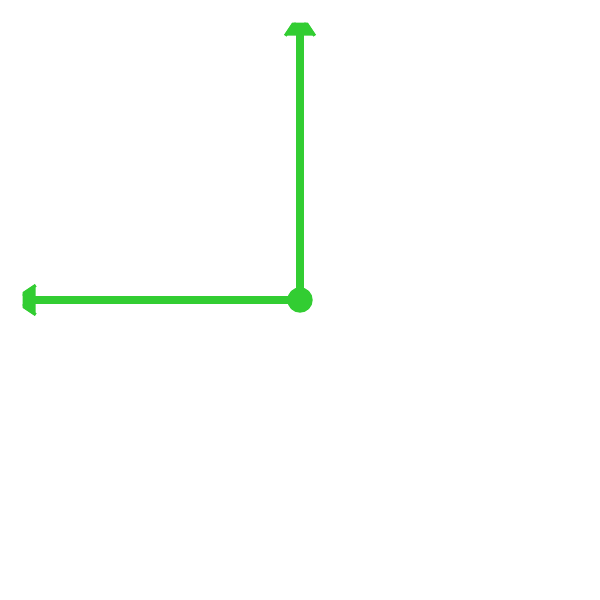} & \labelpicBb{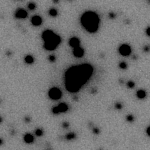}{d}{2025-07-19}{\figsize}{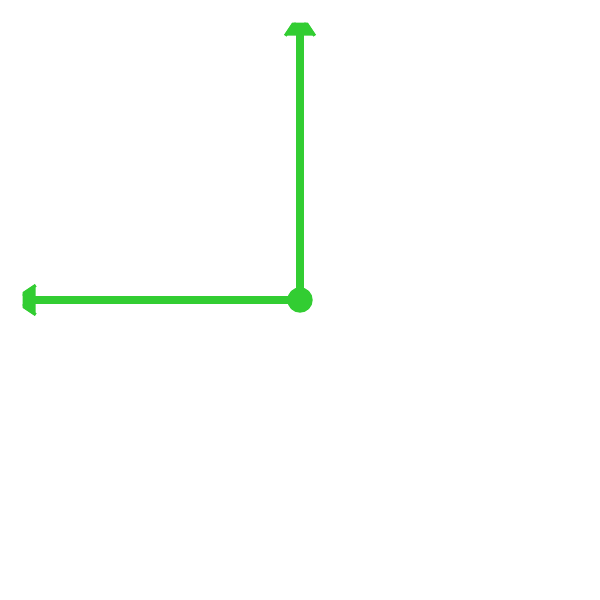} & \labelpicBb{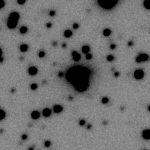}{e}{2025-07-19}{\figsize}{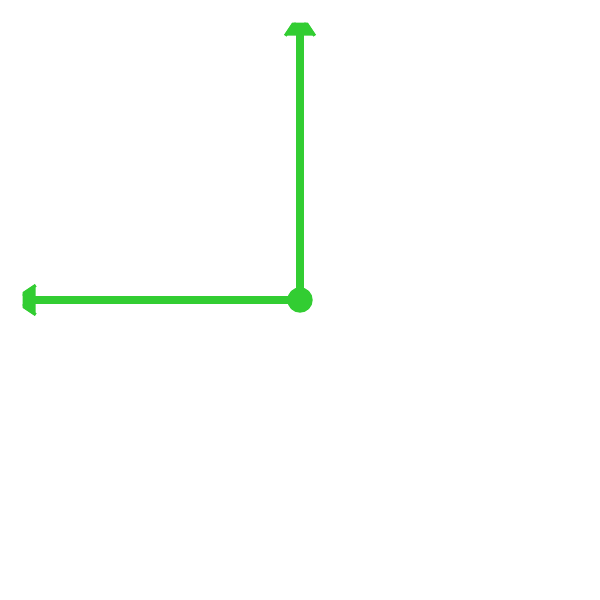} \\
\labelpicBb{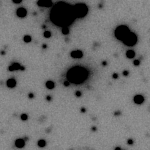}{f}{2025-07-19}{\figsize}{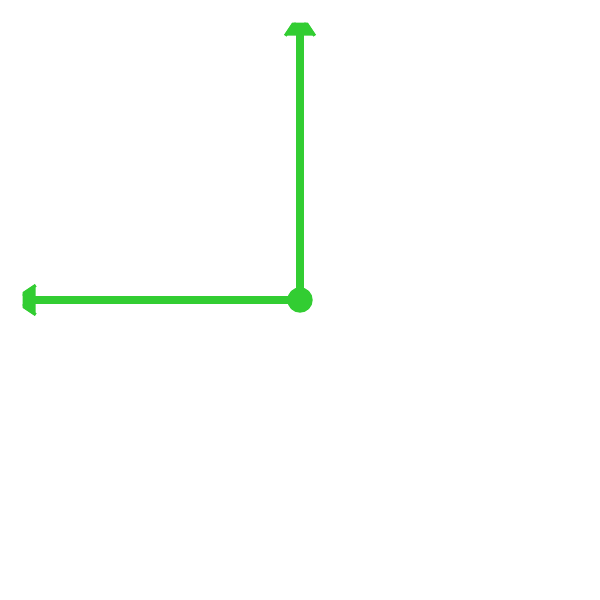} & \labelpicBb{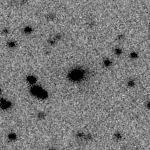}{g}{2025-07-20}{\figsize}{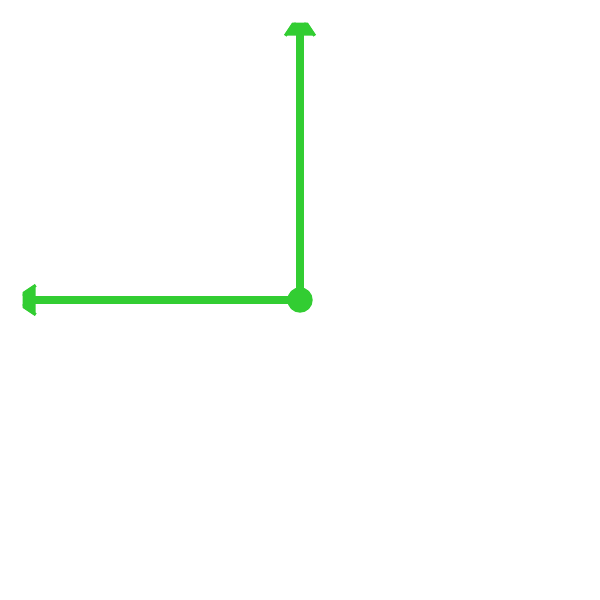} & \labelpicBb{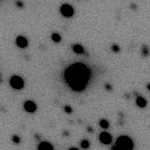}{h}{2025-07-20}{\figsize}{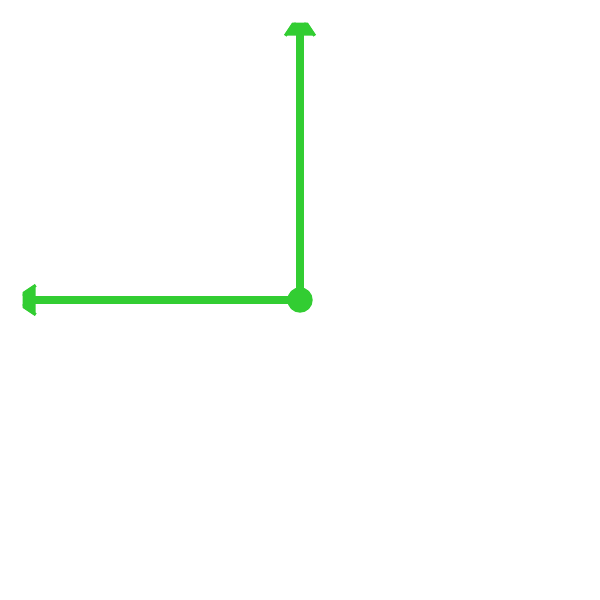} & \labelpicBb{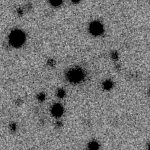}{i}{2025-07-20}{\figsize}{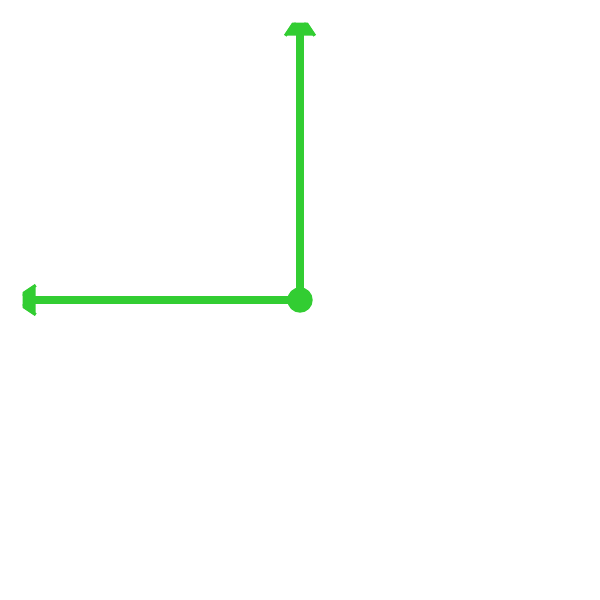} \\
\end{tabular}

%% Label-to-date mapping:
%% \textbf{(a)} 2025-07-14 04:22:11
%% \textbf{(b)} 2025-07-14 04:56:49
%% \textbf{(c)} 2025-07-19 00:57:02
%% \textbf{(d)} 2025-07-19 00:57:45
%% \textbf{(e)} 2025-07-19 23:25:43
%% \textbf{(f)} 2025-07-19 23:42:18
%% \textbf{(g)} 2025-07-20 00:05:16
%% \textbf{(h)} 2025-07-20 03:33:14
%% \textbf{(i)} 2025-07-20 04:15:08

    \caption{\objname{} (center, indicated by magenta ticks) serendipitous observations from UT 2025 July 14 onward. North is up, east is left (green arrows), with the anti-solar (yellow arrow) and anti-motion (black arrow with red outline) shown.}
    \label{fig:gallery5}
\end{figure*}

% \clearpage
\bibliography{3I}

@ARTICLE{deLaFuenteMarcos2025,
       author = {{de la Fuente Marcos}, R. and {Alarcon}, M.~R. and {Licandro}, J. and {Serra-Ricart}, M. and {de Le{\'o}n}, J. and {de la Fuente Marcos}, C. and {Lombardi}, G. and {Tejero}, A. and {Cabrera-Lavers}, A. and {Guerra Arencibia}, S. and {Ruiz Cejudo}, I.},
        title = "{Assessing interstellar comet 3I/ATLAS with the 10.4 m Gran Telescopio Canarias and the Two-meter Twin Telescope}",
      journal = {\aap},
         year = 2025,
        month = aug,
       volume = {700},
          eid = {L9},
        pages = {L9},
          doi = {10.1051/0004-6361/202556439},
archivePrefix = {arXiv},
       eprint = {2507.12922},
 primaryClass = {astro-ph.EP},
       adsurl = {https://ui.adsabs.harvard.edu/abs/2025A&A...700L...9D}
}

@INPROCEEDINGS{2024SPIE13094E..09S,
       author = {{Stalder}, Brian and {Munoz}, Freddy and {Aguilar}, Christian and {Araya}, Claudio and {Aubel}, Karla and {Barr}, Jeffrey and {Borstad}, Anthony and {Claver}, Charles and {Clements}, Andy W. and {Constanzo}, Julio and {Corvetto}, Giovanni and {Coughlin}, Eric and {Daruich}, Felipe and {Dennihy}, Erik and {Drass}, Holger and {Esteves}, Johnny and {F{\'a}brega}, Juan and {Fanning}, Kevin and {Ferguson}, Peter and {Fernandez}, Manuelangel Garcia and {Fernandez}, Miguel and {Fernandez Lobon}, Pelayo and {Fisher-Levine}, Merlin and {Gamundi}, Samuel Bellver and {Garcia}, Julen and {Gonzalez}, Ivan and {Harris}, Ronald and {Herrera}, Hernan and {Hoblitt}, Joshua and {Izpizua}, Alberto and {Jimenez Mejias}, David and {Johnson}, Brian and {Kang}, Yijung and {Kelkar}, Kshitija and {Lage}, Craig and {Lopez Toro}, Juan and {Maulen}, Guido and {Mills}, David and {Mills}, Neill and {Neill}, Doug and {Nunez}, Oscar and {Ordenes}, Ian and {Orellana}, Juan and {Park}, HyeYun and {Quint}, Bruno and {Reil}, Kevin and {Reinking}, Heinrich and {Reuter}, Michael and {Ribeiro}, Tiago and {Rojas}, Rodrigo and {Romero}, Sandra and {Romero Casas}, Francisco Manuel and {Sanmartim}, David and {Saunders}, Clare and {Schoening}, Bill and {Sebag}, Jacques and {Serrano}, Eduardo and {Shestakov}, Adrian and {Shugart}, Alysha and {Silva}, Christian and {Siruno}, Kevin and {Sotuela}, Ioana and {Tache}, Anthony and {Tapia}, Diego and {Thomas}, Sandrine and {Tighe}, Roberto and {Tsai}, Te-Wei and {Urbach}, Elana and {Vergara}, Luis and {Walter}, Christopher},
        title = "{Rubin Observatory Simonyi Survey Telescope integrated mount performance}",
    booktitle = {Ground-based and Airborne Telescopes X},
         year = 2024,
       editor = {{Marshall}, Heather K. and {Spyromilio}, Jason and {Usuda}, Tomonori},
       series = {Society of Photo-Optical Instrumentation Engineers (SPIE) Conference Series},
       volume = {13094},
        month = aug,
          eid = {1309409},
        pages = {1309409},
          doi = {10.1117/12.3019266},
       adsurl = {https://ui.adsabs.harvard.edu/abs/2024SPIE13094E..09S}
}

@software{FindOrb,
       author = {{Gray}, Bill},
        title = "{Find\_Orb: Orbit determination from observations}",
 howpublished = {Astrophysics Source Code Library, record ascl:2202.016},
         year = 2022,
        month = feb,
          eid = {ascl:2202.016},
       adsurl = {https://ui.adsabs.harvard.edu/abs/2022ascl.soft02016G}
}

@INPROCEEDINGS{Roodman2024,
       author = {{Roodman}, A. and {Rasmussen}, A. and {Bradshaw}, A. and {Charles}, E. and {Chiang}, J. and {Digel}, S.~W. and {Dubois}, R. and {Johnson}, A.~S. and {Kahn}, S. and {Liang}, S. and {Marshall}, S. and {Neal}, H. and {Plazas}, A.~A. and {Reil}, K. and {Rykoff}, E. and {Schindler}, R. and {Schutt}, T. and {Utsumi}, Y. and {Bogart}, T. and {Bond}, T. and {Bowdish}, B. and {Cisneros}, S. and {Eisner}, A. and {Freytag}, M. and {Hascall}, D. and {Lange}, T. and {Lazarte}, J.~C. and {Lopez}, M. and {Mendez}, C. and {Newbry}, S. and {Nordby}, M. and {Onoprienko}, D. and {Osier}, S. and {Pollek}, H. and {Qiu}, B. and {Saxton}, O. and {Tether}, S. and {Thayer}, G. and {Turri}, M. and {Banovetz}, J. and {O'Connor}, P. and {Riot}, V. and {Wolfe}, J. and {Lage}, C. and {Polin}, D. and {Snyder}, A. and {Tyson}, A. and {Nichols}, R. and {Ritz}, S. and {Shestakov}, A. and {Wood}, D. and {Broughton}, A. and {Park}, H. and {Esteves}, J. and {Barrau}, A. and {Bregeon}, J. and {Combet}, C. and {Dargaud}, G. and {Lagorio}, E. and {Migliore}, M. and {Vezzu}, F. and {Antilogus}, P. and {Astier}, P. and {Daubard}, G. and {Juramy}, C. and {Laporte}, D. and {Guillemin}, T. and {Aubourg}, E. and {Boucaud}, A. and {Parisel}, C. and {Virieux}, F. and {Breugnon}, P. and {Karst}, P. and {Marini}, A. and {Fisher-Levine}, M. and {Waters}, C.},
        title = "{LSST camera verification testing and characterization}",
    booktitle = {Ground-based and Airborne Instrumentation for Astronomy X},
         year = 2024,
       editor = {{Bryant}, Julia J. and {Motohara}, Kentaro and {Vernet}, Jo{\"e}l. R.~D.},
       series = {Society of Photo-Optical Instrumentation Engineers (SPIE) Conference Series},
       volume = {13096},
        month = jul,
          eid = {130961S},
        pages = {130961S},
          doi = {10.1117/12.3019698},
       adsurl = {https://ui.adsabs.harvard.edu/abs/2024SPIE13096E..1SR}
}

@INPROCEEDINGS{Juric2017,
       author = {{Juri{\'c}}, M. and {Kantor}, J. and {Lim}, K. -T. and {Lupton}, R.~H. and {Dubois-Felsmann}, G. and {Jenness}, T. and {Axelrod}, T.~S. and {Aleksi{\'c}}, J. and {Allsman}, R.~A. and {AlSayyad}, Y. and {Alt}, J. and {Armstrong}, R. and {Basney}, J. and {Becker}, A.~C. and {Becla}, J. and {Biswas}, R. and {Bosch}, J. and {Boutigny}, D. and {Kind}, M.~C. and {Ciardi}, D.~R. and {Connolly}, A.~J. and {Daniel}, S.~F. and {Daues}, G.~E. and {Economou}, F. and {Chiang}, H. -F. and {Fausti}, A. and {Fisher-Levine}, M. and {Freemon}, D.~M. and {Gris}, P. and {Hernandez}, F. and {Hoblitt}, J. and {Ivezi{\'c}}, Z. and {Jammes}, F. and {Jevremovi{\'c}}, D. and {Jones}, R.~L. and {Kalmbach}, J.~B. and {Kasliwal}, V.~P. and {Krughoff}, K.~S. and {Lurie}, J. and {Lust}, N.~B. and {MacArthur}, L.~A. and {Melchior}, P. and {Moeyens}, J. and {Nidever}, D.~L. and {Owen}, R. and {Parejko}, J.~K. and {Peterson}, J.~M. and {Petravick}, D. and {Pietrowicz}, S.~R. and {Price}, P.~A. and {Reiss}, D.~J. and {Shaw}, R.~A. and {Sick}, J. and {Slater}, C.~T. and {Strauss}, M.~A. and {Sullivan}, I.~S. and {Swinbank}, J.~D. and {Van Dyk}, S. and {Vuj{\v{c}}i{\'c}}, V. and {Withers}, A. and {Yoachim}, P.},
        title = "{The LSST Data Management System}",
    booktitle = {Astronomical Data Analysis Software and Systems XXV},
         year = 2017,
       editor = {{Lorente}, N.~P.~F. and {Shortridge}, K. and {Wayth}, R.},
       series = {Astronomical Society of the Pacific Conference Series},
       volume = {512},
        month = dec,
        pages = {279},
          doi = {10.48550/arXiv.1512.07914},
archivePrefix = {arXiv},
       eprint = {1512.07914},
 primaryClass = {astro-ph.IM},
       adsurl = {https://ui.adsabs.harvard.edu/abs/2017ASPC..512..279J}
}

@misc{LSSTCam,
    author = "{SLAC National Accelerator Laboratory} and {NSF-DOE Vera C. Rubin Observatory}",
    doi = "10.71929/rubin/2571927",
    url = "https://www.osti.gov//servlets/purl/2571927",
    language = "en",
    title = "{The LSST Camera (LSSTCam)}",
    publisher = "SLAC National Accelerator Laboratory (SLAC), Menlo Park, CA (United States)",
    year = "2025"
}

@TechReport{PSTN-019,
    author = "{Rubin Observatory Science Pipelines Developers}",
    title = "{The LSST Science Pipelines Software: Optical Survey Pipeline Reduction and Analysis Environment}",
    institution = "{NSF-DOE Vera C. Rubin Observatory}",
    year = "2025",
    month = "July",
    handle = "PSTN-019",
    type = "{Project Science Technical Note}",
    number = "PSTN-019",
    doi = "10.71929/rubin/2570545",
    url = "https://pstn-019.lsst.io/"
}

@misc{LSSTComCam,
    author = "{SLAC National Accelerator Laboratory} and {NSF-DOE Vera C. Rubin Observatory}",
    doi = "10.71929/rubin/2561361",
    url = "https://www.osti.gov//servlets/purl/2561361",
    language = "en",
    title = "{The LSST Commissioning Camera (LSSTComCam)}",
    publisher = "SLAC National Accelerator Laboratory (SLAC), Menlo Park, CA (United States)",
    year = "2024"
}

@article{makadia2025gauss,
  title={Gauss-radau small-body simulator (GRSS): an open-source library for planetary defense},
  author={Makadia, Rahil and Farnocchia, Davide and Chesley, Steven R and Eggl, Siegfried},
  journal={The Planetary Science Journal},
  volume={6},
  number={4},
  pages={85},
  year={2025},
  publisher={IOP Publishing}
}

@article{eggl2020star,
  title={Star catalog position and proper motion corrections in asteroid astrometry II: The Gaia era},
  author={Eggl, Siegfried and Farnocchia, Davide and Chamberlin, Alan B and Chesley, Steven R},
  journal={Icarus},
  volume={339},
  pages={113596},
  year={2020},
  publisher={Elsevier}
}

@article{hsiehMainbeltCometsPanSTARRS12015,
  title = {The Main-Belt Comets: {{The Pan-STARRS1}} Perspective},
  author = {Hsieh, Henry H. and Denneau, Larry and Wainscoat, Richard J. and Sch{\"o}rghofer, Norbert and Bolin, Bryce and Fitzsimmons, Alan and Jedicke, Robert and Kleyna, Jan and Micheli, Marco and Vere{\v s}, Peter and Kaiser, Nicholas and Chambers, Kenneth C. and Burgett, William S. and Flewelling, Heather and Hodapp, Klaus W. and Magnier, Eugene A. and Morgan, Jeffrey S. and Price, Paul A. and Tonry, John L. and Waters, Christopher},
  year = {2015},
  month = mar,
  journal = {Icarus},
  volume = {248},
  pages = {289--312},
  doi = {10.1016/j.icarus.2014.10.031},
  langid = {english}
}

@article{chandlerCometaryActivityDiscovered2020b,
  title = {Cometary {{Activity Discovered}} on a {{Distant Centaur}}: {{A Nonaqueous Sublimation Mechanism}}},
  shorttitle = {Cometary {{Activity Discovered}} on a {{Distant Centaur}}},
  author = {Chandler, Colin Orion and Kueny, Jay K. and Trujillo, Chadwick A. and Trilling, David E. and Oldroyd, William J.},
  year = {2020},
  month = apr,
  journal = {The Astrophysical Journal Letters},
  volume = {892},
  pages = {L38},
  issn = {0004-637X},
  doi = {10/gg36xz},
  urldate = {2020-07-07}
}

@article{verevs2017statistical,
  title={Statistical analysis of astrometric errors for the most productive asteroid surveys},
  author={Vere{\v{s}}, Peter and Farnocchia, Davide and Chesley, Steven R and Chamberlin, Alan B},
  journal={Icarus},
  volume={296},
  pages={139--149},
  year={2017},
  publisher={Elsevier}
}

@ARTICLE{ahearn1984_bowell,
   author = {{A'Hearn}, M.~F. and {Schleicher}, D.~G. and {Millis}, R.~L. and 
	{Feldman}, P.~D. and {Thompson}, D.~T.},
    title = "{Comet Bowell 1980b}",
  journal = {\aj},
     year = 1984,
    month = apr,
   volume = 89,
    pages = {579-591},
      doi = {10.1086/113552},
   adsurl = {http://adsabs.harvard.edu/abs/1984AJ.....89..579A}
}

@ARTICLE{alvarezcandal2025_3I,
       author = {{Alvarez-Candal}, A. and {Rizos}, J.~L. and {Lara}, L.~M. and {Santos-Sanz}, P. and {Gutierrez}, P.~J. and {Ortiz}, J.~L. and {Morales}, N. and {de Le{\'o}n}, J.},
        title = "{X-SHOOTER Spectrum of Comet C/2025 N1: Insights into a Distant Interstellar Visitor}",
      journal = {arXiv e-prints},
         year = 2025,
        month = jul,
          eid = {arXiv:2507.07312},
        pages = {arXiv:2507.07312},
          doi = {10.48550/arXiv.2507.07312},
archivePrefix = {arXiv},
       eprint = {2507.07312},
 primaryClass = {astro-ph.EP},
       adsurl = {https://ui.adsabs.harvard.edu/abs/2025arXiv250707312A}
}

@inproceedings{berthierSkyBoTNewVO2006,
  title = {{{SkyBoT}}, a New {{VO}} Service to Identify {{Solar System}} Objects},
  booktitle = {Astronomical {{Data Analysis Software}} and {{Systems XV ASP Conference Series}}},
  author = {Berthier, J. and Vachier, F. and Thuillot, W. and Fernique, P. and Ochsenbein, F. and Genova, F. and Lainey, V. and Arlot, J. E. and Gabriel, C. and Arviset, C. and Ponz, D. and Solano, E.},
  year = {2006},
  month = jul,
  volume = {351},
  pages = {367},
  publisher = {Astronomical Society of the Pacific},
  address = {Orem, UT}
}

@ARTICLE{bolin2020_borisov,
       author = {{Bolin}, Bryce T. and {Lisse}, Carey M. and {Kasliwal}, Mansi M. and {Quimby}, Robert and {Tan}, Hanjie and {Copperwheat}, Chris M. and {Lin}, Zhong-Yi and {Morbidelli}, Alessandro and {Abe}, Lyu and {Bendjoya}, Philippe and {Burdge}, Kevin B. and {Coughlin}, Michael and {Fremling}, Christoffer and {Itoh}, Ryosuke and {Koss}, Michael and {Masci}, Frank J. and {Maeno}, Syota and {Mamajek}, Eric E. and {Marocco}, Federico and {Murata}, Katsuhiro and {Rivet}, Jean-Pierre and {Sitko}, Michael L. and {Stern}, Daniel and {Vernet}, David and {Walters}, Richard and {Yan}, Lin and {Andreoni}, Igor and {Bhalerao}, Varun and {Bodewits}, Dennis and {De}, Kishalay and {Deshmukh}, Kunal P. and {Bellm}, Eric C. and {Blagorodnova}, Nadejda and {Buzasi}, Derek and {Cenko}, S. Bradley and {Chang}, Chan-Kao and {Chojnowski}, Drew and {Dekany}, Richard and {Duev}, Dmitry A. and {Graham}, Matthew and {Juri{\'c}}, Mario and {Kulkarni}, Shrinivas R. and {Kupfer}, Thomas and {Mahabal}, Ashish and {Neill}, James D. and {Ngeow}, Chow-Choong and {Penprase}, Bryan and {Riddle}, Reed and {Rodriguez}, Hector and {Smith}, Roger M. and {Rosnet}, Philippe and {Sollerman}, Jesper and {Soumagnac}, Maayane T.},
        title = "{Characterization of the Nucleus, Morphology, and Activity of Interstellar Comet 2I/Borisov by Optical and Near-infrared GROWTH, Apache Point, IRTF, ZTF, and Keck Observations}",
      journal = {\aj},
         year = 2020,
        month = jul,
       volume = {160},
       number = {1},
          eid = {26},
        pages = {26},
          doi = {10.3847/1538-3881/ab9305},
archivePrefix = {arXiv},
       eprint = {1910.14004},
 primaryClass = {astro-ph.EP},
       adsurl = {https://ui.adsabs.harvard.edu/abs/2020AJ....160...26B}
}

@ARTICLE{bottke2023_kbevolution,
       author = {{Bottke}, William F. and {Vokrouhlick{\'y}}, David and {Marshall}, Raphael and {Nesvorn{\'y}}, David and {Morbidelli}, Alessandro and {Deienno}, Rogerio and {Marchi}, Simone and {Dones}, Luke and {Levison}, Harold F.},
        title = "{The Collisional Evolution of the Primordial Kuiper Belt, Its Destabilized Population, and the Trojan Asteroids}",
      journal = {\psj},
         year = 2023,
        month = sep,
       volume = {4},
       number = {9},
          eid = {168},
        pages = {168},
          doi = {10.3847/PSJ/ace7cd},
archivePrefix = {arXiv},
       eprint = {2307.07089},
 primaryClass = {astro-ph.EP},
       adsurl = {https://ui.adsabs.harvard.edu/abs/2023PSJ.....4..168B}
}

@ARTICLE{buie2018_lucytrojanLCs,
       author = {{Buie}, Marc W. and {Zangari}, Amanda M. and {Marchi}, Simone and {Levison}, Harold F. and {Mottola}, Stefano},
        title = "{Light Curves of Lucy Targets: Leucus and Polymele}",
      journal = {\aj},
         year = 2018,
        month = jun,
       volume = {155},
       number = {6},
          eid = {245},
        pages = {245},
          doi = {10.3847/1538-3881/aabd81},
       adsurl = {https://ui.adsabs.harvard.edu/abs/2018AJ....155..245B}
}

@techreport{deen2025cbet5578,
  title = {Central Bureau Electronic Telegram No.5578: {{Comet C}}/2025 {{N1}} ({{ATLAS}}) = {{3I}}/{{ATLAS}}},
  author = {{S. Deen} and {A. Hale} and {H. Sato} and {F. Romanov} and {R. Weryk} and {R. Wainscoat} and {J. Silva} and {N. Manset} and {S. Nakano} and {Daniel W. E. Green}},
  year = {2025},
  month = jul,
  number = {CBET 5578},
  address = {Hoffman Lab 209, Harvard University, Cambridge, MA 02138, U.S.A.},
  institution = {Central Bureau for Astronomical Telegrams}
}

@article{giorginiJPLsOnLineSolar1996,
  title = {{{JPL}}'s {{On-Line Solar System Data Service}}},
  author = {Giorgini, J. D. and Yeomans, D. K. and Chamberlin, A. B. and Chodas, P. W. and Jacobson, R. A. and Keesey, M. S. and Lieske, J. H. and Ostro, S. J. and Standish, E. M. and Wimberly, R. N.},
  year = {1996},
  month = sep,
  journal = {American Astronomical Society},
  volume = {28},
  pages = {25.04-},
  url = {https://hdl.handle.net/2014/27350},
  langid = {english}
}

@article{harrisArrayProgrammingNumPy2020,
  title = {Array Programming with {{NumPy}}},
  author = {Harris, Charles R. and Millman, K. Jarrod and {van der Walt}, St{\'e}fan J. and Gommers, Ralf and Virtanen, Pauli and Cournapeau, David and Wieser, Eric and Taylor, Julian and Berg, Sebastian and Smith, Nathaniel J. and Kern, Robert and Picus, Matti and Hoyer, Stephan and {van Kerkwijk}, Marten H. and Brett, Matthew and Haldane, Allan and {del R{\'i}o}, Jaime Fern{\'a}ndez and Wiebe, Mark and Peterson, Pearu and {G{\'e}rard-Marchant}, Pierre and Sheppard, Kevin and Reddy, Tyler and Weckesser, Warren and Abbasi, Hameer and Gohlke, Christoph and Oliphant, Travis E.},
  year = {2020},
  month = sep,
  journal = {Nature},
  volume = {585},
  number = {7825},
  pages = {357--362},
  publisher = {Nature Publishing Group},
  issn = {1476-4687},
  doi = {10.1038/s41586-020-2649-2},
  urldate = {2022-06-11},
  copyright = {2020 The Author(s)},
  langid = {english}
}

@ARTICLE{hsieh2011_176p,
   author = {{Hsieh}, H.~H. and {Ishiguro}, M. and {Lacerda}, P. and {Jewitt}, D.},
    title = "{Physical Properties of Main-belt Comet 176P/LINEAR}",
  journal = {\aj},
archivePrefix = "arXiv",
   eprint = {1105.0944},
 primaryClass = "astro-ph.EP",
     year = 2011,
    month = jul,
   volume = 142,
      eid = {29},
    pages = {29},
      doi = {10.1088/0004-6256/142/1/29},
   adsurl = {http://adsabs.harvard.edu/abs/2011AJ....142...29H}
}

@article{hunterMatplotlib2DGraphics2007,
  title = {Matplotlib: {{A 2D Graphics Environment}}},
  shorttitle = {Matplotlib},
  author = {Hunter, John D.},
  year = {2007},
  month = may,
  journal = {Computing in Science \& Engineering},
  volume = {9},
  number = {3},
  pages = {90--95},
  issn = {1558-366X},
  doi = {10.1109/MCSE.2007.55}
}

@ARTICLE{jewitt2020_borisovnucleus,
       author = {{Jewitt}, David and {Hui}, Man-To and {Kim}, Yoonyoung and {Mutchler}, Max and {Weaver}, Harold and {Agarwal}, Jessica},
        title = "{The Nucleus of Interstellar Comet 2I/Borisov}",
      journal = {\apjl},
         year = 2020,
        month = jan,
       volume = {888},
       number = {2},
          eid = {L23},
        pages = {L23},
          doi = {10.3847/2041-8213/ab621b},
archivePrefix = {arXiv},
       eprint = {1912.05422},
 primaryClass = {astro-ph.EP},
       adsurl = {https://ui.adsabs.harvard.edu/abs/2020ApJ...888L..23J}
}

@inproceedings{joyeNewFeaturesSAOImage2006,
  title = {New {{Features}} of {{SAOImage DS9}}},
  booktitle = {Astronomical {{Data Analysis Software}} and {{Systems XV ASP Conference Series}}},
  author = {Joye, W. A.},
  year = {2006},
  month = jul,
  volume = {351},
  pages = {574-}
}

@misc{kurucz1993_modelatmospheres1,
  author    = {Kurucz, R.~L.},
  title     = {{VizieR Online Data Catalog: Model Atmospheres (Kurucz, 1979)}},
  year      = {1993},
  month     = {oct},
  version   = {VI/39},
  publisher = {VizieR Online Data Catalog},
  url       = {https://ui.adsabs.harvard.edu/abs/1993yCat.6039....0K},
  note      = {Originally published in 1979 as ApJS, 40, 1; provided by the SAO/NASA Astrophysics Data System}
}

@article{langAstrometryNetBlindAstrometric2010,
  title = {Astrometry.{{Net}}: {{Blind Astrometric Calibration}} of {{Arbitrary Astronomical Images}}},
  author = {Lang, D. and Hogg, D. W. and Mierle, K. and Blanton, M. and Roweis, S.},
  year = {2010},
  month = may,
  journal = {Astronomical Journal},
  volume = {139},
  number = {5},
  pages = {1782--1800},
  issn = {0004-6256},
  doi = {10.1088/0004-6256/139/5/1782},
  langid = {english}
}

@ARTICLE{melchior2018_scarlet,
       author = {{Melchior}, P. and {Moolekamp}, F. and {Jerdee}, M. and {Armstrong}, R. and {Sun}, A. -L. and {Bosch}, J. and {Lupton}, R.},
        title = "{SCARLET: Source separation in multi-band images by Constrained Matrix Factorization}",
      journal = {Astronomy and Computing},
         year = 2018,
        month = jul,
       volume = {24},
          eid = {129},
        pages = {129},
          doi = {10.1016/j.ascom.2018.07.001},
archivePrefix = {arXiv},
       eprint = {1802.10157},
 primaryClass = {astro-ph.IM},
       adsurl = {https://ui.adsabs.harvard.edu/abs/2018A&C....24..129M}
}

@ARTICLE{ochsenbein2000_vizier,
       author = {{Ochsenbein}, F. and {Bauer}, P. and {Marcout}, J.},
        title = "{The VizieR database of astronomical catalogues}",
      journal = {\aaps},
         year = 2000,
        month = apr,
       volume = {143},
        pages = {23-32},
          doi = {10.1051/aas:2000169},
archivePrefix = {arXiv},
       eprint = {astro-ph/0002122},
 primaryClass = {astro-ph},
       adsurl = {https://ui.adsabs.harvard.edu/abs/2000A&AS..143...23O}
}

@misc{OpenAI2023ChatGPT,
  title = {{{ChatGPT}} ({{Mar}} 14, 2023 Version)},
  author = {{OpenAI}},
  year = {2023}
}

@inproceedings{paszkePyTorchImperativeStyle2019,
  title = {{{PyTorch}}: {{An Imperative Style}}, {{High-Performance Deep Learning Library}}},
  shorttitle = {{{PyTorch}}},
  booktitle = {Advances in {{Neural Information Processing Systems}}},
  author = {Paszke, Adam and Gross, Sam and Massa, Francisco and Lerer, Adam and Bradbury, James and Chanan, Gregory and Killeen, Trevor and Lin, Zeming and Gimelshein, Natalia and Antiga, Luca and Desmaison, Alban and Kopf, Andreas and Yang, Edward and DeVito, Zachary and Raison, Martin and Tejani, Alykhan and Chilamkurthy, Sasank and Steiner, Benoit and Fang, Lu and Bai, Junjie and Chintala, Soumith},
  year = {2019},
  volume = {32},
  publisher = {Curran Associates, Inc.},
  urldate = {2024-01-18}
}

@misc{rebackPandasdevPandasPandas2022,
  title = {Pandas-Dev/Pandas: {{Pandas}} 1.4.2},
  shorttitle = {Pandas-Dev/Pandas},
  author = {Reback, Jeff and {jbrockmendel} and McKinney, Wes and den Bossche, Joris Van and Augspurger, Tom and Roeschke, Matthew and Hawkins, Simon and Cloud, Phillip and {gfyoung} and Sinhrks and Hoefler, Patrick and Klein, Adam and Petersen, Terji and Tratner, Jeff and She, Chang and Ayd, William and Naveh, Shahar and Darbyshire, J. H. M. and Garcia, Marc and Shadrach, Richard and Schendel, Jeremy and Hayden, Andy and Saxton, Daniel and Gorelli, Marco Edward and Li, Fangchen and Zeitlin, Matthew and Jancauskas, Vytautas and McMaster, Ali and W{\"o}rtwein, Torsten and Battiston, Pietro},
  year = {2022},
  month = apr,
  doi = {10.5281/zenodo.6408044},
  urldate = {2022-06-11},
  howpublished = {Zenodo}
}

@ARTICLE{astropy2013,
       author = {{Astropy Collaboration} and {Robitaille}, Thomas P. and {Tollerud}, Erik J. and {Greenfield}, Perry and {Droettboom}, Michael and {Bray}, Erik and {Aldcroft}, Tom and {Davis}, Matt and {Ginsburg}, Adam and {Price-Whelan}, Adrian M. and {Kerzendorf}, Wolfgang E. and {Conley}, Alexander and {Crighton}, Neil and {Barbary}, Kyle and {Muna}, Demitri and {Ferguson}, Henry and {Grollier}, Fr{\'e}d{\'e}ric and {Parikh}, Madhura M. and {Nair}, Prasanth H. and {Unther}, Hans M. and {Deil}, Christoph and {Woillez}, Julien and {Conseil}, Simon and {Kramer}, Roban and {Turner}, James E.~H. and {Singer}, Leo and {Fox}, Ryan and {Weaver}, Benjamin A. and {Zabalza}, Victor and {Edwards}, Zachary I. and {Azalee Bostroem}, K. and {Burke}, D.~J. and {Casey}, Andrew R. and {Crawford}, Steven M. and {Dencheva}, Nadia and {Ely}, Justin and {Jenness}, Tim and {Labrie}, Kathleen and {Lim}, Pey Lian and {Pierfederici}, Francesco and {Pontzen}, Andrew and {Ptak}, Andy and {Refsdal}, Brian and {Servillat}, Mathieu and {Streicher}, Ole},
        title = "{Astropy: A community Python package for astronomy}",
      journal = {\aap},
         year = 2013,
        month = oct,
       volume = {558},
          eid = {A33},
        pages = {A33},
          doi = {10.1051/0004-6361/201322068},
archivePrefix = {arXiv},
       eprint = {1307.6212},
 primaryClass = {astro-ph.IM},
       adsurl = {https://ui.adsabs.harvard.edu/abs/2013A&A...558A..33A}
}

@ARTICLE{astropy2018,
       author = {{Astropy Collaboration} and {Price-Whelan}, A.~M. and {Sip{\H{o}}cz}, B.~M. and {G{\"u}nther}, H.~M. and {Lim}, P.~L. and {Crawford}, S.~M. and {Conseil}, S. and {Shupe}, D.~L. and {Craig}, M.~W. and {Dencheva}, N. and {Ginsburg}, A. and {VanderPlas}, J.~T. and {Bradley}, L.~D. and {P{\'e}rez-Su{\'a}rez}, D. and {de Val-Borro}, M. and {Aldcroft}, T.~L. and {Cruz}, K.~L. and {Robitaille}, T.~P. and {Tollerud}, E.~J. and {Ardelean}, C. and {Babej}, T. and {Bach}, Y.~P. and {Bachetti}, M. and {Bakanov}, A.~V. and {Bamford}, S.~P. and {Barentsen}, G. and {Barmby}, P. and {Baumbach}, A. and {Berry}, K.~L. and {Biscani}, F. and {Boquien}, M. and {Bostroem}, K.~A. and {Bouma}, L.~G. and {Brammer}, G.~B. and {Bray}, E.~M. and {Breytenbach}, H. and {Buddelmeijer}, H. and {Burke}, D.~J. and {Calderone}, G. and {Cano Rodr{\'\i}guez}, J.~L. and {Cara}, M. and {Cardoso}, J.~V.~M. and {Cheedella}, S. and {Copin}, Y. and {Corrales}, L. and {Crichton}, D. and {D'Avella}, D. and {Deil}, C. and {Depagne}, {\'E}. and {Dietrich}, J.~P. and {Donath}, A. and {Droettboom}, M. and {Earl}, N. and {Erben}, T. and {Fabbro}, S. and {Ferreira}, L.~A. and {Finethy}, T. and {Fox}, R.~T. and {Garrison}, L.~H. and {Gibbons}, S.~L.~J. and {Goldstein}, D.~A. and {Gommers}, R. and {Greco}, J.~P. and {Greenfield}, P. and {Groener}, A.~M. and {Grollier}, F. and {Hagen}, A. and {Hirst}, P. and {Homeier}, D. and {Horton}, A.~J. and {Hosseinzadeh}, G. and {Hu}, L. and {Hunkeler}, J.~S. and {Ivezi{\'c}}, {\v{Z}}. and {Jain}, A. and {Jenness}, T. and {Kanarek}, G. and {Kendrew}, S. and {Kern}, N.~S. and {Kerzendorf}, W.~E. and {Khvalko}, A. and {King}, J. and {Kirkby}, D. and {Kulkarni}, A.~M. and {Kumar}, A. and {Lee}, A. and {Lenz}, D. and {Littlefair}, S.~P. and {Ma}, Z. and {Macleod}, D.~M. and {Mastropietro}, M. and {McCully}, C. and {Montagnac}, S. and {Morris}, B.~M. and {Mueller}, M. and {Mumford}, S.~J. and {Muna}, D. and {Murphy}, N.~A. and {Nelson}, S. and {Nguyen}, G.~H. and {Ninan}, J.~P. and {N{\"o}the}, M. and {Ogaz}, S. and {Oh}, S. and {Parejko}, J.~K. and {Parley}, N. and {Pascual}, S. and {Patil}, R. and {Patil}, A.~A. and {Plunkett}, A.~L. and {Prochaska}, J.~X. and {Rastogi}, T. and {Reddy Janga}, V. and {Sabater}, J. and {Sakurikar}, P. and {Seifert}, M. and {Sherbert}, L.~E. and {Sherwood-Taylor}, H. and {Shih}, A.~Y. and {Sick}, J. and {Silbiger}, M.~T. and {Singanamalla}, S. and {Singer}, L.~P. and {Sladen}, P.~H. and {Sooley}, K.~A. and {Sornarajah}, S. and {Streicher}, O. and {Teuben}, P. and {Thomas}, S.~W. and {Tremblay}, G.~R. and {Turner}, J.~E.~H. and {Terr{\'o}n}, V. and {van Kerkwijk}, M.~H. and {de la Vega}, A. and {Watkins}, L.~L. and {Weaver}, B.~A. and {Whitmore}, J.~B. and {Woillez}, J. and {Zabalza}, V. and {Astropy Contributors}},
        title = "{The Astropy Project: Building an Open-science Project and Status of the v2.0 Core Package}",
      journal = {\aj},
         year = 2018,
        month = sep,
       volume = {156},
       number = {3},
          eid = {123},
        pages = {123},
          doi = {10.3847/1538-3881/aabc4f},
archivePrefix = {arXiv},
       eprint = {1801.02634},
 primaryClass = {astro-ph.IM},
       adsurl = {https://ui.adsabs.harvard.edu/abs/2018AJ....156..123A}
}

@ARTICLE{astropy2022,
       author = {{Astropy Collaboration} and {Price-Whelan}, Adrian M. and {Lim}, Pey Lian and {Earl}, Nicholas and {Starkman}, Nathaniel and {Bradley}, Larry and {Shupe}, David L. and {Patil}, Aarya A. and {Corrales}, Lia and {Brasseur}, C.~E. and {N{\"o}the}, Maximilian and {Donath}, Axel and {Tollerud}, Erik and {Morris}, Brett M. and {Ginsburg}, Adam and {Vaher}, Eero and {Weaver}, Benjamin A. and {Tocknell}, James and {Jamieson}, William and {van Kerkwijk}, Marten H. and {Robitaille}, Thomas P. and {Merry}, Bruce and {Bachetti}, Matteo and {G{\"u}nther}, H. Moritz and {Aldcroft}, Thomas L. and {Alvarado-Montes}, Jaime A. and {Archibald}, Anne M. and {B{\'o}di}, Attila and {Bapat}, Shreyas and {Barentsen}, Geert and {Baz{\'a}n}, Juanjo and {Biswas}, Manish and {Boquien}, M{\'e}d{\'e}ric and {Burke}, D.~J. and {Cara}, Daria and {Cara}, Mihai and {Conroy}, Kyle E. and {Conseil}, Simon and {Craig}, Matthew W. and {Cross}, Robert M. and {Cruz}, Kelle L. and {D'Eugenio}, Francesco and {Dencheva}, Nadia and {Devillepoix}, Hadrien A.~R. and {Dietrich}, J{\"o}rg P. and {Eigenbrot}, Arthur Davis and {Erben}, Thomas and {Ferreira}, Leonardo and {Foreman-Mackey}, Daniel and {Fox}, Ryan and {Freij}, Nabil and {Garg}, Suyog and {Geda}, Robel and {Glattly}, Lauren and {Gondhalekar}, Yash and {Gordon}, Karl D. and {Grant}, David and {Greenfield}, Perry and {Groener}, Austen M. and {Guest}, Steve and {Gurovich}, Sebastian and {Handberg}, Rasmus and {Hart}, Akeem and {Hatfield-Dodds}, Zac and {Homeier}, Derek and {Hosseinzadeh}, Griffin and {Jenness}, Tim and {Jones}, Craig K. and {Joseph}, Prajwel and {Kalmbach}, J. Bryce and {Karamehmetoglu}, Emir and {Ka{\l}uszy{\'n}ski}, Miko{\l}aj and {Kelley}, Michael S.~P. and {Kern}, Nicholas and {Kerzendorf}, Wolfgang E. and {Koch}, Eric W. and {Kulumani}, Shankar and {Lee}, Antony and {Ly}, Chun and {Ma}, Zhiyuan and {MacBride}, Conor and {Maljaars}, Jakob M. and {Muna}, Demitri and {Murphy}, N.~A. and {Norman}, Henrik and {O'Steen}, Richard and {Oman}, Kyle A. and {Pacifici}, Camilla and {Pascual}, Sergio and {Pascual-Granado}, J. and {Patil}, Rohit R. and {Perren}, Gabriel I. and {Pickering}, Timothy E. and {Rastogi}, Tanuj and {Roulston}, Benjamin R. and {Ryan}, Daniel F. and {Rykoff}, Eli S. and {Sabater}, Jose and {Sakurikar}, Parikshit and {Salgado}, Jes{\'u}s and {Sanghi}, Aniket and {Saunders}, Nicholas and {Savchenko}, Volodymyr and {Schwardt}, Ludwig and {Seifert-Eckert}, Michael and {Shih}, Albert Y. and {Jain}, Anany Shrey and {Shukla}, Gyanendra and {Sick}, Jonathan and {Simpson}, Chris and {Singanamalla}, Sudheesh and {Singer}, Leo P. and {Singhal}, Jaladh and {Sinha}, Manodeep and {Sip{\H{o}}cz}, Brigitta M. and {Spitler}, Lee R. and {Stansby}, David and {Streicher}, Ole and {{\v{S}}umak}, Jani and {Swinbank}, John D. and {Taranu}, Dan S. and {Tewary}, Nikita and {Tremblay}, Grant R. and {de Val-Borro}, Miguel and {Van Kooten}, Samuel J. and {Vasovi{\'c}}, Zlatan and {Verma}, Shresth and {de Miranda Cardoso}, Jos{\'e} Vin{\'\i}cius and {Williams}, Peter K.~G. and {Wilson}, Tom J. and {Winkel}, Benjamin and {Wood-Vasey}, W.~M. and {Xue}, Rui and {Yoachim}, Peter and {Zhang}, Chen and {Zonca}, Andrea and {Astropy Project Contributors}},
        title = "{The Astropy Project: Sustaining and Growing a Community-oriented Open-source Project and the Latest Major Release (v5.0) of the Core Package}",
      journal = {\apj},
         year = 2022,
        month = aug,
       volume = {935},
       number = {2},
          eid = {167},
        pages = {167},
          doi = {10.3847/1538-4357/ac7c74},
archivePrefix = {arXiv},
       eprint = {2206.14220},
 primaryClass = {astro-ph.IM},
       adsurl = {https://ui.adsabs.harvard.edu/abs/2022ApJ...935..167A}
}

@article{virtanenSciPy10Fundamental2020,
  title = {{{SciPy}} 1.0: Fundamental Algorithms for Scientific Computing in {{Python}}},
  shorttitle = {{{SciPy}} 1.0},
  author = {Virtanen, Pauli and Gommers, Ralf and Oliphant, Travis E. and Haberland, Matt and Reddy, Tyler and Cournapeau, David and Burovski, Evgeni and Peterson, Pearu and Weckesser, Warren and Bright, Jonathan and {van der Walt}, St{\'e}fan J. and Brett, Matthew and Wilson, Joshua and Millman, K. Jarrod and Mayorov, Nikolay and Nelson, Andrew R. J. and Jones, Eric and Kern, Robert and Larson, Eric and Carey, C. J. and Polat, {\.I}lhan and Feng, Yu and Moore, Eric W. and VanderPlas, Jake and Laxalde, Denis and Perktold, Josef and Cimrman, Robert and Henriksen, Ian and Quintero, E. A. and Harris, Charles R. and Archibald, Anne M. and Ribeiro, Ant{\^o}nio H. and Pedregosa, Fabian and {van Mulbregt}, Paul},
  year = {2020},
  month = mar,
  journal = {Nature Methods},
  volume = {17},
  number = {3},
  pages = {261--272},
  publisher = {Nature Publishing Group},
  issn = {1548-7105},
  doi = {10.1038/s41592-019-0686-2},
  urldate = {2022-06-11},
  copyright = {2020 The Author(s)},
  langid = {english}
}

@INCOLLECTION{fitzsimmonsreviewcomets3,
       author = {{Fitzsimmons}, Alan and {Meech}, Karen and {Matr{\`a}}, Luca and {Pfalzner}, Susanne},
        title = "{Interstellar Objects and Exocomets}",
    booktitle = {Comets III},
         year = 2024,
       editor = {{Meech}, Karen. J. and {Combi}, Michael. R. and {Bockel{\'e}e-Morvan}, Dominique and {Raymodn}, Sean. N. and {Zolensky}, Michael. E.},
        pages = {731-766},
          doi = {10.2458/azu_uapress_9780816553631-ch022},
       adsurl = {https://ui.adsabs.harvard.edu/abs/2024come.book..731F}
}

@ARTICLE{seligmanmororeview,
       author = {{Seligman}, Darryl Z. and {Moro-Mart{\'\i}n}, Amaya},
        title = "{Interstellar objects}",
      journal = {Contemporary Physics},
         year = 2022,
        month = jul,
       volume = {63},
       number = {3},
        pages = {200-232},
          doi = {10.1080/00107514.2023.2203976},
archivePrefix = {arXiv},
       eprint = {2304.00568},
 primaryClass = {astro-ph.EP},
       adsurl = {https://ui.adsabs.harvard.edu/abs/2022ConPh..63..200S}
}

@ARTICLE{jewitteseligman,
       author = {{Jewitt}, David and {Seligman}, Darryl Z.},
        title = "{The Interstellar Interlopers}",
      journal = {\araa},
         year = 2023,
        month = aug,
       volume = {61},
        pages = {197-236},
          doi = {10.1146/annurev-astro-071221-054221},
archivePrefix = {arXiv},
       eprint = {2209.08182},
 primaryClass = {astro-ph.EP},
       adsurl = {https://ui.adsabs.harvard.edu/abs/2023ARA&A..61..197J}
}

@ARTICLE{mororeview,
       author = {{Moro-Mart{\'\i}n}, Amaya},
        title = "{Interstellar planetesimals}",
      journal = {arXiv e-prints},
         year = 2022,
        month = may,
          eid = {arXiv:2205.04277},
        pages = {arXiv:2205.04277},
          doi = {10.48550/arXiv.2205.04277},
archivePrefix = {arXiv},
       eprint = {2205.04277},
 primaryClass = {astro-ph.EP},
       adsurl = {https://ui.adsabs.harvard.edu/abs/2022arXiv220504277M}
}

@ARTICLE{Borisov2019,
  author = {{Borisov}, G. and {Durig}, D.~T. and {Sato}, H. and {et~al.}},
  title = "{Comet C/2019 Q4 (Borisov)}",
  journal = {Central Bureau Electronic Telegrams},
  year = 2019,
  number = 4666,
  pages = 1
}

@ARTICLE{champagne2025_3I,
       author = {{Champagne}, Chansey and {McClure}, Lucas and {Emery}, Joshua and {Bauer}, James M. and {Kareta}, Theodore and {Sharkey}, Benjamin N. L. and {Connelly}, Michael and {Rayner}, J. and {Thomas}, C. and {Reddy}, Vishnu},
        title = "{Initial Results from NASA/IRTF Observations of Interstellar Comet 3I/ATLAS}",
      journal = {The Astronomer's Telegram},
         year = "2025",
        month = "Jul",
       volume = {17283},
        pages = {1},
}

@ARTICLE{cooklsstiso,
       author = {{Cook}, Nathaniel V. and {Ragozzine}, Darin and {Granvik}, Mikael and {Stephens}, Denise C.},
        title = "{Realistic Detectability of Close Interstellar Comets}",
      journal = {\apj},
         year = 2016,
        month = jul,
       volume = {825},
       number = {1},
          eid = {51},
        pages = {51},
          doi = {10.3847/0004-637X/825/1/51},
archivePrefix = {arXiv},
       eprint = {1607.08162},
 primaryClass = {astro-ph.EP},
       adsurl = {https://ui.adsabs.harvard.edu/abs/2016ApJ...825...51C}
}

@ARTICLE{Denneau20253I,
       author = {{Denneau}, L. and {Siverd}, R. and {Tonry}, J. and {Weiland}, H. and {Erasmus}, N. and {Fitzsimmons}, A. and {Robinson}, J.},
        title = "{3I/ATLAS = C/2025 N1 (ATLAS)}",
      journal = {MPEC},
         year = 2025,
        month = jul,
       number = {2025-N12}
}

@ARTICLE{engelhardtlsstiso,
       author = {{Engelhardt}, Toni and {Jedicke}, Robert and {Vere{\v{s}}}, Peter and {Fitzsimmons}, Alan and {Denneau}, Larry and {Beshore}, Ed and {Meinke}, Bonnie},
        title = "{An Observational Upper Limit on the Interstellar Number Density of Asteroids and Comets}",
      journal = {\aj},
         year = 2017,
        month = mar,
       volume = {153},
       number = {3},
          eid = {133},
        pages = {133},
          doi = {10.3847/1538-3881/aa5c8a},
archivePrefix = {arXiv},
       eprint = {1702.02237},
 primaryClass = {astro-ph.EP},
       adsurl = {https://ui.adsabs.harvard.edu/abs/2017AJ....153..133E}
}

@ARTICLE{ivezic2019,
       author = {{Ivezi{\'c}}, {\v{Z}}eljko and {Kahn}, Steven M. and {Tyson}, J. Anthony and {Abel}, Bob and {Acosta}, Emily and {Allsman}, Robyn and {Alonso}, David and {AlSayyad}, Yusra and {Anderson}, Scott F. and {Andrew}, John and {Angel}, James Roger P. and {Angeli}, George Z. and {Ansari}, Reza and {Antilogus}, Pierre and {Araujo}, Constanza and {Armstrong}, Robert and {Arndt}, Kirk T. and {Astier}, Pierre and {Aubourg}, {\'E}ric and {Auza}, Nicole and {Axelrod}, Tim S. and {Bard}, Deborah J. and {Barr}, Jeff D. and {Barrau}, Aurelian and {Bartlett}, James G. and {Bauer}, Amanda E. and {Bauman}, Brian J. and {Baumont}, Sylvain and {Bechtol}, Ellen and {Bechtol}, Keith and {Becker}, Andrew C. and {Becla}, Jacek and {Beldica}, Cristina and {Bellavia}, Steve and {Bianco}, Federica B. and {Biswas}, Rahul and {Blanc}, Guillaume and {Blazek}, Jonathan and {Blandford}, Roger D. and {Bloom}, Josh S. and {Bogart}, Joanne and {Bond}, Tim W. and {Booth}, Michael T. and {Borgland}, Anders W. and {Borne}, Kirk and {Bosch}, James F. and {Boutigny}, Dominique and {Brackett}, Craig A. and {Bradshaw}, Andrew and {Brandt}, William Nielsen and {Brown}, Michael E. and {Bullock}, James S. and {Burchat}, Patricia and {Burke}, David L. and {Cagnoli}, Gianpietro and {Calabrese}, Daniel and {Callahan}, Shawn and {Callen}, Alice L. and {Carlin}, Jeffrey L. and {Carlson}, Erin L. and {Chandrasekharan}, Srinivasan and {Charles-Emerson}, Glenaver and {Chesley}, Steve and {Cheu}, Elliott C. and {Chiang}, Hsin-Fang and {Chiang}, James and {Chirino}, Carol and {Chow}, Derek and {Ciardi}, David R. and {Claver}, Charles F. and {Cohen-Tanugi}, Johann and {Cockrum}, Joseph J. and {Coles}, Rebecca and {Connolly}, Andrew J. and {Cook}, Kem H. and {Cooray}, Asantha and {Covey}, Kevin R. and {Cribbs}, Chris and {Cui}, Wei and {Cutri}, Roc and {Daly}, Philip N. and {Daniel}, Scott F. and {Daruich}, Felipe and {Daubard}, Guillaume and {Daues}, Greg and {Dawson}, William and {Delgado}, Francisco and {Dellapenna}, Alfred and {de Peyster}, Robert and {de Val-Borro}, Miguel and {Digel}, Seth W. and {Doherty}, Peter and {Dubois}, Richard and {Dubois-Felsmann}, Gregory P. and {Durech}, Josef and {Economou}, Frossie and {Eifler}, Tim and {Eracleous}, Michael and {Emmons}, Benjamin L. and {Fausti Neto}, Angelo and {Ferguson}, Henry and {Figueroa}, Enrique and {Fisher-Levine}, Merlin and {Focke}, Warren and {Foss}, Michael D. and {Frank}, James and {Freemon}, Michael D. and {Gangler}, Emmanuel and {Gawiser}, Eric and {Geary}, John C. and {Gee}, Perry and {Geha}, Marla and {Gessner}, Charles J.~B. and {Gibson}, Robert R. and {Gilmore}, D. Kirk and {Glanzman}, Thomas and {Glick}, William and {Goldina}, Tatiana and {Goldstein}, Daniel A. and {Goodenow}, Iain and {Graham}, Melissa L. and {Gressler}, William J. and {Gris}, Philippe and {Guy}, Leanne P. and {Guyonnet}, Augustin and {Haller}, Gunther and {Harris}, Ron and {Hascall}, Patrick A. and {Haupt}, Justine and {Hernandez}, Fabio and {Herrmann}, Sven and {Hileman}, Edward and {Hoblitt}, Joshua and {Hodgson}, John A. and {Hogan}, Craig and {Howard}, James D. and {Huang}, Dajun and {Huffer}, Michael E. and {Ingraham}, Patrick and {Innes}, Walter R. and {Jacoby}, Suzanne H. and {Jain}, Bhuvnesh and {Jammes}, Fabrice and {Jee}, M. James and {Jenness}, Tim and {Jernigan}, Garrett and {Jevremovi{\'c}}, Darko and {Johns}, Kenneth and {Johnson}, Anthony S. and {Johnson}, Margaret W.~G. and {Jones}, R. Lynne and {Juramy-Gilles}, Claire and {Juri{\'c}}, Mario and {Kalirai}, Jason S. and {Kallivayalil}, Nitya J. and {Kalmbach}, Bryce and {Kantor}, Jeffrey P. and {Karst}, Pierre and {Kasliwal}, Mansi M. and {Kelly}, Heather and {Kessler}, Richard and {Kinnison}, Veronica and {Kirkby}, David and {Knox}, Lloyd and {Kotov}, Ivan V. and {Krabbendam}, Victor L. and {Krughoff}, K. Simon and {Kub{\'a}nek}, Petr and {Kuczewski}, John and {Kulkarni}, Shri and {Ku}, John and {Kurita}, Nadine R. and {Lage}, Craig S. and {Lambert}, Ron and {Lange}, Travis and {Langton}, J. Brian and {Le Guillou}, Laurent and {Levine}, Deborah and {Liang}, Ming and {Lim}, Kian-Tat and {Lintott}, Chris J. and {Long}, Kevin E. and {Lopez}, Margaux and {Lotz}, Paul J. and {Lupton}, Robert H. and {Lust}, Nate B. and {MacArthur}, Lauren A. and {Mahabal}, Ashish and {Mandelbaum}, Rachel and {Markiewicz}, Thomas W. and {Marsh}, Darren S. and {Marshall}, Philip J. and {Marshall}, Stuart and {May}, Morgan and {McKercher}, Robert and {McQueen}, Michelle and {Meyers}, Joshua and {Migliore}, Myriam and {Miller}, Michelle and {Mills}, David J.},
        title = "{LSST: From Science Drivers to Reference Design and Anticipated Data Products}",
      journal = {\apj},
         year = 2019,
        month = mar,
       volume = {873},
       number = {2},
          eid = {111},
        pages = {111},
          doi = {10.3847/1538-4357/ab042c},
archivePrefix = {arXiv},
       eprint = {0805.2366},
 primaryClass = {astro-ph},
       adsurl = {https://ui.adsabs.harvard.edu/abs/2019ApJ...873..111I}
}

@ARTICLE{ATLASmainpaperperchance,
       author = {{Tonry}, J.~L. and {Denneau}, L. and {Heinze}, A.~N. and {Stalder}, B. and {Smith}, K.~W. and {Smartt}, S.~J. and {Stubbs}, C.~W. and {Weiland}, H.~J. and {Rest}, A.},
        title = "{ATLAS: A High-cadence All-sky Survey System}",
      journal = {\pasp},
         year = 2018,
        month = jun,
       volume = {130},
       number = {988},
        pages = {064505},
          doi = {10.1088/1538-3873/aabadf},
archivePrefix = {arXiv},
       eprint = {1802.00879},
 primaryClass = {astro-ph.IM},
       adsurl = {https://ui.adsabs.harvard.edu/abs/2018PASP..130f4505T}
}

@MISC{Bianco2022,
  author = {{Bianco}, F.~B. and {Jones}, L. and {Ivezi{\'c}}, \v{Z}. and {Ritz}, S. and {Rubin Project Science Team}},
  title = "{Updated estimates of the Rubin system throughput and expected LSST image depth}",
  howpublished = {Rubin Observatory Project Document PSTN-054},
  year = 2022,
  note = {Published 21 June 2022},
  url = {https://pstn-054.lsst.io}
}

@ARTICLE{meech2017_oumuamua,
       author = {{Meech}, Karen J. and {Weryk}, Robert and {Micheli}, Marco and {Kleyna}, Jan T. and {Hainaut}, Olivier R. and {Jedicke}, Robert and {Wainscoat}, Richard J. and {Chambers}, Kenneth C. and {Keane}, Jacqueline V. and {Petric}, Andreea and {Denneau}, Larry and {Magnier}, Eugene and {Berger}, Travis and {Huber}, Mark E. and {Flewelling}, Heather and {Waters}, Chris and {Schunova-Lilly}, Eva and {Chastel}, Serge},
        title = "{A brief visit from a red and extremely elongated interstellar asteroid}",
      journal = {\nat},
         year = 2017,
        month = dec,
       volume = {552},
       number = {7685},
        pages = {378-381},
          doi = {10.1038/nature25020},
       adsurl = {https://ui.adsabs.harvard.edu/abs/2017Natur.552..378M}
}

@ARTICLE{ye2017_oumuamua,
       author = {{Ye}, Quan-Zhi and {Zhang}, Qicheng and {Kelley}, Michael S.~P. and {Brown}, Peter G.},
        title = "{1I/2017 U1 ({\textquoteleft}Oumuamua) is Hot: Imaging, Spectroscopy, and Search of Meteor Activity}",
      journal = {\apjl},
         year = 2017,
        month = dec,
       volume = {851},
       number = {1},
          eid = {L5},
        pages = {L5},
          doi = {10.3847/2041-8213/aa9a34},
archivePrefix = {arXiv},
       eprint = {1711.02320},
 primaryClass = {astro-ph.EP},
       adsurl = {https://ui.adsabs.harvard.edu/abs/2017ApJ...851L...5Y}
}

@ARTICLE{vavilov2019_oumuamua,
       author = {{Vavilov}, Dmitrii E. and {Medvedev}, Yurii D.},
        title = "{Dust bombardment can explain the extremely elongated shape of 1I/'Oumuamua and the lack of interstellar objects}",
      journal = {\mnras},
         year = 2019,
        month = mar,
       volume = {484},
       number = {1},
        pages = {L75-L78},
          doi = {10.1093/mnrasl/sly244},
archivePrefix = {arXiv},
       eprint = {1812.11334},
 primaryClass = {astro-ph.EP},
       adsurl = {https://ui.adsabs.harvard.edu/abs/2019MNRAS.484L..75V}
}

@ARTICLE{jewitt2017_oumuamua,
       author = {{Jewitt}, David and {Luu}, Jane and {Rajagopal}, Jayadev and {Kotulla}, Ralf and {Ridgway}, Susan and {Liu}, Wilson and {Augusteijn}, Thomas},
        title = "{Interstellar Interloper 1I/2017 U1: Observations from the NOT and WIYN Telescopes}",
      journal = {\apjl},
         year = 2017,
        month = dec,
       volume = {850},
       number = {2},
          eid = {L36},
        pages = {L36},
          doi = {10.3847/2041-8213/aa9b2f},
archivePrefix = {arXiv},
       eprint = {1711.05687},
 primaryClass = {astro-ph.EP},
       adsurl = {https://ui.adsabs.harvard.edu/abs/2017ApJ...850L..36J}
}

@ARTICLE{trilling2018_oumuamua,
       author = {{Trilling}, David E. and {Mommert}, Michael and {Hora}, Joseph L. and {Farnocchia}, Davide and {Chodas}, Paul and {Giorgini}, Jon and {Smith}, Howard A. and {Carey}, Sean and {Lisse}, Carey M. and {Werner}, Michael and {McNeill}, Andrew and {Chesley}, Steven R. and {Emery}, Joshua P. and {Fazio}, Giovanni and {Fernandez}, Yanga R. and {Harris}, Alan and {Marengo}, Massimo and {Mueller}, Michael and {Roegge}, Alissa and {Smith}, Nathan and {Weaver}, H.~A. and {Meech}, Karen and {Micheli}, Marco},
        title = "{Spitzer Observations of Interstellar Object 1I/{\textquoteleft}Oumuamua}",
      journal = {\aj},
         year = 2018,
        month = dec,
       volume = {156},
       number = {6},
          eid = {261},
        pages = {261},
          doi = {10.3847/1538-3881/aae88f},
archivePrefix = {arXiv},
       eprint = {1811.08072},
 primaryClass = {astro-ph.EP},
       adsurl = {https://ui.adsabs.harvard.edu/abs/2018AJ....156..261T}
}

@ARTICLE{micheli2018_oumuamua,
       author = {{Micheli}, Marco and {Farnocchia}, Davide and {Meech}, Karen J. and {Buie}, Marc W. and {Hainaut}, Olivier R. and {Prialnik}, Dina and {Sch{\"o}rghofer}, Norbert and {Weaver}, Harold A. and {Chodas}, Paul W. and {Kleyna}, Jan T. and {Weryk}, Robert and {Wainscoat}, Richard J. and {Ebeling}, Harald and {Keane}, Jacqueline V. and {Chambers}, Kenneth C. and {Koschny}, Detlef and {Petropoulos}, Anastassios E.},
        title = "{Non-gravitational acceleration in the trajectory of 1I/2017 U1 ('Oumuamua)}",
      journal = {\nat},
         year = 2018,
        month = jun,
       volume = {559},
        pages = {223-226},
          doi = {10.1038/s41586-018-0254-4},
       adsurl = {https://ui.adsabs.harvard.edu/abs/2018Natur.559..223M}
}

@ARTICLE{oumuamuaissi2019_oumuamua,
       author = {{'Oumuamua ISSI Team} and {Bannister}, Michele T. and {Bhandare}, Asmita and {Dybczy{\'n}ski}, Piotr A. and {Fitzsimmons}, Alan and {Guilbert-Lepoutre}, Aur{\'e}lie and {Jedicke}, Robert and {Knight}, Matthew M. and {Meech}, Karen J. and {McNeill}, Andrew and {Pfalzner}, Susanne and {Raymond}, Sean N. and {Snodgrass}, Collin and {Trilling}, David E. and {Ye}, Quanzhi},
        title = "{The natural history of `Oumuamua}",
      journal = {Nature Astronomy},
         year = 2019,
        month = jul,
       volume = {3},
        pages = {594-602},
          doi = {10.1038/s41550-019-0816-x},
archivePrefix = {arXiv},
       eprint = {1907.01910},
 primaryClass = {astro-ph.EP},
       adsurl = {https://ui.adsabs.harvard.edu/abs/2019NatAs...3..594O}
}

@ARTICLE{drahus2017_oumuamua,
       author = {{Drahus}, Michal and {Guzik}, Piotr and {Waniak}, Waclaw and {Handzlik}, Barbara and {Kurowski}, Sebastian and {Xu}, Siyi},
        title = "{Tumbling motion of 1I/'Oumuamua reveals body's violent past}",
      journal = {arXiv e-prints},
         year = 2017,
        month = dec,
          eid = {arXiv:1712.00437},
        pages = {arXiv:1712.00437},
          doi = {10.48550/arXiv.1712.00437},
archivePrefix = {arXiv},
       eprint = {1712.00437},
 primaryClass = {astro-ph.EP},
       adsurl = {https://ui.adsabs.harvard.edu/abs/2017arXiv171200437D}
}

@ARTICLE{knight2017_oumuamua,
       author = {{Knight}, Matthew M. and {Protopapa}, Silvia and {Kelley}, Michael S.~P. and {Farnham}, Tony L. and {Bauer}, James M. and {Bodewits}, Dennis and {Feaga}, Lori M. and {Sunshine}, Jessica M.},
        title = "{On the Rotation Period and Shape of the Hyperbolic Asteroid 1I/{\textquoteleft}Oumuamua (2017 U1) from Its Lightcurve}",
      journal = {\apjl},
         year = 2017,
        month = dec,
       volume = {851},
       number = {2},
          eid = {L31},
        pages = {L31},
          doi = {10.3847/2041-8213/aa9d81},
archivePrefix = {arXiv},
       eprint = {1711.01402},
 primaryClass = {astro-ph.EP},
       adsurl = {https://ui.adsabs.harvard.edu/abs/2017ApJ...851L..31K}
}

@ARTICLE{bolin2018_1I,
       author = {{Bolin}, Bryce T. and {Weaver}, Harold A. and {Fernandez}, Yanga R. and {Lisse}, Carey M. and {Huppenkothen}, Daniela and {Jones}, R. Lynne and {Juri{\'c}}, Mario and {Moeyens}, Joachim and {Schambeau}, Charles A. and {Slater}, Colin. T. and {Ivezi{\'c}}, {\v{Z}}eljko and {Connolly}, Andrew J.},
        title = "{APO Time-resolved Color Photometry of Highly Elongated Interstellar Object 1I/{\textquoteleft}Oumuamua}",
      journal = {\apjl},
         year = 2018,
        month = jan,
       volume = {852},
       number = {1},
          eid = {L2},
        pages = {L2},
          doi = {10.3847/2041-8213/aaa0c9},
archivePrefix = {arXiv},
       eprint = {1711.04927},
 primaryClass = {astro-ph.EP},
       adsurl = {https://ui.adsabs.harvard.edu/abs/2018ApJ...852L...2B}
}

@ARTICLE{farnham2021_C2014UN271,
       author = {{Farnham}, Tony L. and {Kelley}, Michael S.~P. and {Bauer}, James M.},
        title = "{Early Activity in Comet C/2014 UN271 Bernardinelli-Bernstein as Observed by TESS}",
      journal = {\psj},
         year = 2021,
        month = dec,
       volume = {2},
       number = {6},
          eid = {236},
        pages = {236},
          doi = {10.3847/PSJ/ac323d},
       adsurl = {https://ui.adsabs.harvard.edu/abs/2021PSJ.....2..236F}
}

@ARTICLE{fraser2018_oumuamua,
       author = {{Fraser}, Wesley C. and {Pravec}, Petr and {Fitzsimmons}, Alan and {Lacerda}, Pedro and {Bannister}, Michele T. and {Snodgrass}, Colin and {Smoli{\'c}}, Igor},
        title = "{The tumbling rotational state of 1I/`Oumuamua}",
      journal = {Nature Astronomy},
         year = 2018,
        month = feb,
       volume = {2},
        pages = {383-386},
          doi = {10.1038/s41550-018-0398-z},
archivePrefix = {arXiv},
       eprint = {1711.11530},
 primaryClass = {astro-ph.EP},
       adsurl = {https://ui.adsabs.harvard.edu/abs/2018NatAs...2..383F}
}

@ARTICLE{hardorp1980_sun3,
   author = {{Hardorp}, J.},
    title = "{The sun among the stars. III - Energy distributions of 16 northern G-type stars and the solar flux calibration}",
  journal = {\aap},
     year = 1980,
    month = nov,
   volume = 91,
    pages = {221-232},
   adsurl = {http://adsabs.harvard.edu/abs/1980A%26A....91..221H}
}

@ARTICLE{hsieh2025_358p,
       author = {{Hsieh}, Henry H. and {Noonan}, John W. and {Kelley}, Michael S.~P. and {Bodewits}, Dennis and {Pittichov{\'a}}, Jana and {Thirouin}, Audrey and {Micheli}, Marco and {Knight}, Matthew M. and {Bannister}, Michele T. and {Chandler}, Colin O. and {Holt}, Carrie E. and {Hopkins}, Matthew J. and {Kim}, Yaeji and {Moskovitz}, Nicholas A. and {Oldroyd}, William J. and {Patterson}, Jack and {Sheppard}, Scott S. and {Tan}, Nicole and {Trujillo}, Chadwick A. and {Ye}, Quanzhi},
        title = "{The Volatile Composition and Activity Evolution of Main-belt Comet 358P/PANSTARRS}",
      journal = {\psj},
         year = 2025,
        month = jan,
       volume = {6},
       number = {1},
          eid = {3},
        pages = {3},
          doi = {10.3847/PSJ/ad9199},
archivePrefix = {arXiv},
       eprint = {2411.07435},
 primaryClass = {astro-ph.EP},
       adsurl = {https://ui.adsabs.harvard.edu/abs/2025PSJ.....6....3H}
}

@ARTICLE{jordi2006_filtertransformations,
       author = {{Jordi}, K. and {Grebel}, E.~K. and {Ammon}, K.},
        title = "{Empirical color transformations between SDSS photometry and other photometric systems}",
      journal = {\aap},
         year = 2006,
        month = dec,
       volume = {460},
       number = {1},
        pages = {339-347},
          doi = {10.1051/0004-6361:20066082},
archivePrefix = {arXiv},
       eprint = {astro-ph/0609121},
 primaryClass = {astro-ph},
       adsurl = {https://ui.adsabs.harvard.edu/abs/2006A&A...460..339J}
}

@ARTICLE{mcneill2018_oumuamua,
       author = {{McNeill}, Andrew and {Trilling}, David E. and {Mommert}, Michael},
        title = "{Constraints on the Density and Internal Strength of 1I/{\textquoteright}Oumuamua}",
      journal = {\apjl},
         year = 2018,
        month = apr,
       volume = {857},
       number = {1},
          eid = {L1},
        pages = {L1},
          doi = {10.3847/2041-8213/aab9ab},
archivePrefix = {arXiv},
       eprint = {1803.09864},
 primaryClass = {astro-ph.EP},
       adsurl = {https://ui.adsabs.harvard.edu/abs/2018ApJ...857L...1M}
}

@ARTICLE{belton2018_oumuamua,
       author = {{Belton}, Michael J.~S. and {Hainaut}, Olivier R. and {Meech}, Karen J. and {Mueller}, Beatrice E.~A. and {Kleyna}, Jan T. and {Weaver}, Harold A. and {Buie}, Marc W. and {Drahus}, Micha{\l} and {Guzik}, Piotr and {Wainscoat}, Richard J. and {Waniak}, Wac{\l}aw and {Handzlik}, Barbara and {Kurowski}, Sebastian and {Xu}, Siyi and {Sheppard}, Scott S. and {Micheli}, Marco and {Ebeling}, Harald and {Keane}, Jacqueline V.},
        title = "{The Excited Spin State of 1I/2017 U1 {\textquoteleft}Oumuamua}",
      journal = {\apjl},
         year = 2018,
        month = apr,
       volume = {856},
       number = {2},
          eid = {L21},
        pages = {L21},
          doi = {10.3847/2041-8213/aab370},
archivePrefix = {arXiv},
       eprint = {1804.03471},
 primaryClass = {astro-ph.EP},
       adsurl = {https://ui.adsabs.harvard.edu/abs/2018ApJ...856L..21B}
}

@INCOLLECTION{knight2024_cometnuclei,
       author = {{Knight}, Matthew M. and {Kokotanekova}, Rosita and {Samarasinha}, Nalin H.},
        title = "{Physical and Surface Properties of Comet Nuclei from Remote Observations}",
    booktitle = {Comets III},
         year = 2024,
       editor = {{Meech}, Karen. J. and {Combi}, Michael. R. and {Bockel{\'e}e-Morvan}, Dominique and {Raymodn}, Sean. N. and {Zolensky}, Michael. E.},
        pages = {361-404},
        publisher = {University of Arizona Press},
       adsurl = {https://ui.adsabs.harvard.edu/abs/2024come.book..361K}
}

@ARTICLE{maschenko2019_oumuamua,
       author = {{Mashchenko}, Sergey},
        title = "{Modelling the light curve of `Oumuamua: evidence for torque and disc-like shape}",
      journal = {\mnras},
         year = 2019,
        month = nov,
       volume = {489},
       number = {3},
        pages = {3003-3021},
          doi = {10.1093/mnras/stz2380},
archivePrefix = {arXiv},
       eprint = {1906.03696},
 primaryClass = {astro-ph.EP},
       adsurl = {https://ui.adsabs.harvard.edu/abs/2019MNRAS.489.3003M}
}

@ARTICLE{jewitt2019_borisov,
       author = {{Jewitt}, David and {Luu}, Jane},
        title = "{Initial Characterization of Interstellar Comet 2I/2019 Q4 (Borisov)}",
      journal = {\apjl},
         year = 2019,
        month = dec,
       volume = {886},
       number = {2},
          eid = {L29},
        pages = {L29},
          doi = {10.3847/2041-8213/ab530b},
archivePrefix = {arXiv},
       eprint = {1910.02547},
 primaryClass = {astro-ph.EP},
       adsurl = {https://ui.adsabs.harvard.edu/abs/2019ApJ...886L..29J}
}

@article{fitzsimmons2019_borisov,
   title={Detection of CN Gas in Interstellar Object 2I/Borisov},
   volume={885},
   ISSN={2041-8213},
   url={http://dx.doi.org/10.3847/2041-8213/ab49fc},
   DOI={10.3847/2041-8213/ab49fc},
   number={1},
   journal={The Astrophysical Journal Letters},
   publisher={American Astronomical Society},
   author={Fitzsimmons, Alan and Hainaut, Olivier and Meech, Karen J. and Jehin, Emmanuel and Moulane, Youssef and Opitom, Cyrielle and Yang, Bin and Keane, Jacqueline V. and Kleyna, Jan T. and Micheli, Marco and Snodgrass, Colin},
   year={2019},
   month=oct, pages={L9} }

@ARTICLE{ye2020_borisov,
       author = {{Ye}, Quanzhi and {Kelley}, Michael S.~P. and {Bolin}, Bryce T. and {Bodewits}, Dennis and {Farnocchia}, Davide and {Masci}, Frank J. and {Meech}, Karen J. and {Micheli}, Marco and {Weryk}, Robert and {Bellm}, Eric C. and {Christensen}, Eric and {Dekany}, Richard and {Delacroix}, Alexandre and {Graham}, Matthew J. and {Kulkarni}, Shrinivas R. and {Laher}, Russ R. and {Rusholme}, Ben and {Smith}, Roger M.},
        title = "{Pre-discovery Activity of New Interstellar Comet 2I/Borisov beyond 5 au}",
      journal = {\aj},
         year = 2020,
        month = feb,
       volume = {159},
       number = {2},
          eid = {77},
        pages = {77},
          doi = {10.3847/1538-3881/ab659b},
archivePrefix = {arXiv},
       eprint = {1911.05902},
 primaryClass = {astro-ph.EP},
       adsurl = {https://ui.adsabs.harvard.edu/abs/2020AJ....159...77Y}
}

@ARTICLE{mckay,
       author = {{McKay}, Adam J. and {Cochran}, Anita L. and {Dello Russo}, Neil and {DiSanti}, Michael A.},
        title = "{Detection of a Water Tracer in Interstellar Comet 2I/Borisov}",
      journal = {\apjl},
         year = 2020,
        month = jan,
       volume = {889},
       number = {1},
          eid = {L10},
        pages = {L10},
          doi = {10.3847/2041-8213/ab64ed},
archivePrefix = {arXiv},
       eprint = {1910.12785},
 primaryClass = {astro-ph.EP},
       adsurl = {https://ui.adsabs.harvard.edu/abs/2020ApJ...889L..10M}
}

@ARTICLE{guzik2020_borisov,
       author = {{Guzik}, Piotr and {Drahus}, Micha{\l} and {Rusek}, Krzysztof and {Waniak}, Wac{\l}aw and {Cannizzaro}, Giacomo and {Pastor-Marazuela}, In{\'e}s},
        title = "{Initial characterization of interstellar comet 2I/Borisov}",
      journal = {Nature Astronomy},
         year = 2020,
        month = jan,
       volume = {4},
        pages = {53-57},
          doi = {10.1038/s41550-019-0931-8},
archivePrefix = {arXiv},
       eprint = {1909.05851},
 primaryClass = {astro-ph.EP},
       adsurl = {https://ui.adsabs.harvard.edu/abs/2020NatAs...4...53G}
}

@ARTICLE{hui,
       author = {{Hui}, Man-To and {Ye}, Quan-Zhi and {F{\"o}hring}, Dora and {Hung}, Denise and {Tholen}, David J.},
        title = "{Physical Characterization of Interstellar Comet 2I/2019 Q4 (Borisov)}",
      journal = {\aj},
         year = 2020,
        month = aug,
       volume = {160},
       number = {2},
          eid = {92},
        pages = {92},
          doi = {10.3847/1538-3881/ab9df8},
       adsurl = {https://ui.adsabs.harvard.edu/abs/2020AJ....160...92H}
}

@ARTICLE{kimanisotropy,
       author = {{Kim}, Yoonyoung and {Jewitt}, David and {Mutchler}, Max and {Agarwal}, Jessica and {Hui}, Man-To and {Weaver}, Harold},
        title = "{Coma Anisotropy and the Rotation Pole of Interstellar Comet 2I/Borisov}",
      journal = {\apjl},
         year = 2020,
        month = jun,
       volume = {895},
       number = {2},
          eid = {L34},
        pages = {L34},
          doi = {10.3847/2041-8213/ab9228},
archivePrefix = {arXiv},
       eprint = {2005.02468},
 primaryClass = {astro-ph.EP},
       adsurl = {https://ui.adsabs.harvard.edu/abs/2020ApJ...895L..34K}
}

@ARTICLE{cremonese,
       author = {{Cremonese}, G. and {Fulle}, M. and {Cambianica}, P. and {Munaretto}, G. and {Capria}, M.~T. and {La Forgia}, F. and {Lazzarin}, M. and {Migliorini}, A. and {Boschin}, W. and {Milani}, G. and {Aletti}, A. and {Arlic}, G. and {Bacci}, P. and {Bacci}, R. and {Bryssinck}, E. and {Carosati}, D. and {Castellano}, D. and {Buzzi}, L. and {Di Rubbo}, S. and {Facchini}, M. and {Guido}, E. and {Kugel}, F. and {Ligustri}, R. and {Maestripieri}, M. and {Mantero}, A. and {Nicolas}, J. and {Ochner}, P. and {Perrella}, C. and {Trabatti}, R. and {Valvasori}, A.},
        title = "{Dust Environment Model of the Interstellar Comet 2I/Borisov}",
      journal = {\apjl},
         year = 2020,
        month = apr,
       volume = {893},
       number = {1},
          eid = {L12},
        pages = {L12},
          doi = {10.3847/2041-8213/ab8455},
       adsurl = {https://ui.adsabs.harvard.edu/abs/2020ApJ...893L..12C}
}

@article{Yang_2021,
   title={Compact pebbles and the evolution of volatiles in the interstellar comet 2I/Borisov},
   volume={5},
   ISSN={2397-3366},
   url={http://dx.doi.org/10.1038/s41550-021-01336-w},
   DOI={10.1038/s41550-021-01336-w},
   number={6},
   journal={Nature Astronomy},
   publisher={Springer Science and Business Media LLC},
   author={Yang, Bin and Li, Aigen and Cordiner, Martin A. and Chang, Chin-Shin and Hainaut, Olivier R. and Williams, Jonathan P. and Meech, Karen J. and Keane, Jacqueline V. and Villard, Eric},
   year={2021},
   month=mar, pages={586–593} }

@MISC{Graham2022,
  author = {{Graham}, M.},
  title = "{The Rubin Data Products, Abridged (2022)}",
  howpublished = {Zenodo, Version 2},
  year = 2022,
  doi = {10.5281/zenodo.7011229},
  url = {https://doi.org/10.5281/zenodo.7011229},
  note = {Presentation, published 19 August 2022}
}

@MISC{Juric2023,
  author = {{Juri{\'c}}, M. and {Axelrod}, T. and {Becker}, A.~C. and {Becla}, J. and {Bellm}, E. and {Bosch}, J.~F. and {Ciardi}, D. and {Connolly}, A.~J. and {Dubois-Felsmann}, G.~P. and {Economou}, F. and {Freemon}, M. and {Gelman}, M. and {Gill}, R. and {Graham}, M. and {Guy}, L.~P. and {Ivezić}, Ž. and {Jenness}, T. and {Kantor}, J. and {Krughoff}, K.~S. and {Lim}, K.-T. and {Lupton}, R.~H. and {Mueller}, F. and {Nidever}, D. and {O'Mullane}, W. and {Patterson}, M. and {Petravick}, D. and {Shaw}, D. and {Slater}, C. and {Strauss}, M. and {Swinbank}, J. and {Tyson}, J.~A. and {Wood-Vasey}, M. and {Wu}, X.},
  title = "{Data Products Definition Document (LSE-163)}",
  howpublished = {LSST Project Document LSE-163, v3.9},
  year = 2023,
  note = {Published 10 July 2023},
  url = {https://ls.st/LSE-163}
}

@INPROCEEDINGS{maybeSCOUT,
       author = {{Farnocchia}, Davide and {Chesley}, Steven R. and {Chamberlin}, Alan B.},
        title = "{Scout: orbit analysis and hazard assessment for NEOCP objects}",
    booktitle = {AAS/Division for Planetary Sciences Meeting Abstracts \#48},
         year = 2016,
       series = {AAS/Division for Planetary Sciences Meeting Abstracts},
       volume = {48},
        month = oct,
          eid = {305.03},
        pages = {305.03},
       adsurl = {https://ui.adsabs.harvard.edu/abs/2016DPS....4830503F}
}

@ARTICLE{sorcha,
       author = {{Holman}, Matthew J. and {Bernardinelli}, Pedro H. and {Schwamb}, Megan E. and {Juri{\'c}}, Mario and {Oldag}, Drew and {West}, Maxine and {Napier}, Kevin J. and {Merritt}, Stephanie R. and {Fedorets}, Grigori and {Cornwall}, Samuel and {Kurlander}, Jacob A. and {Eggl}, Siegfried and {Kubica}, Jeremy and {Kiker}, Kathleen and {Murtagh}, Joseph and {Naidu}, Shantanu P. and {Chandler}, Colin Orion},
        title = "{Sorcha: Optimized Solar System Ephemeris Generation}",
      journal = {arXiv e-prints},
         year = 2025,
        month = jun,
          eid = {arXiv:2506.02140},
        pages = {arXiv:2506.02140},
          doi = {10.48550/arXiv.2506.02140},
archivePrefix = {arXiv},
       eprint = {2506.02140},
 primaryClass = {astro-ph.EP},
       adsurl = {https://ui.adsabs.harvard.edu/abs/2025arXiv250602140H}
}

@ARTICLE{dorsey2025_isomodel,
       author = {{Dorsey}, Rosemary C. and {Hopkins}, Matthew J. and {Bannister}, Michele T. and {Lawler}, Samantha M. and {Lintott}, Chris and {Parker}, Alex H. and {Forbes}, John C.},
        title = "{The visibility of the {\={O}}tautahi-Oxford interstellar object population model in LSST}",
      journal = {arXiv e-prints},
         year = 2025,
        month = feb,
          eid = {arXiv:2502.16741},
        pages = {arXiv:2502.16741},
          doi = {10.48550/arXiv.2502.16741},
archivePrefix = {arXiv},
       eprint = {2502.16741},
 primaryClass = {astro-ph.EP},
       adsurl = {https://ui.adsabs.harvard.edu/abs/2025arXiv250216741D}
}

@ARTICLE{bodewits2020_borisov,
       author = {{Bodewits}, D. and {Noonan}, J.~W. and {Feldman}, P.~D. and {Bannister}, M.~T. and {Farnocchia}, D. and {Harris}, W.~M. and {Li}, J. -Y. and {Mandt}, K.~E. and {Parker}, J. Wm. and {Xing}, Z. -X.},
        title = "{The carbon monoxide-rich interstellar comet 2I/Borisov}",
      journal = {Nature Astronomy},
         year = 2020,
        month = apr,
       volume = {4},
        pages = {867-871},
          doi = {10.1038/s41550-020-1095-2},
archivePrefix = {arXiv},
       eprint = {2004.08972},
 primaryClass = {astro-ph.EP},
       adsurl = {https://ui.adsabs.harvard.edu/abs/2020NatAs...4..867B}
}

@ARTICLE{cordiner,
       author = {{Cordiner}, M.~A. and {Milam}, S.~N. and {Biver}, N. and {Bockel{\'e}e-Morvan}, D. and {Roth}, N.~X. and {Bergin}, E.~A. and {Jehin}, E. and {Remijan}, A.~J. and {Charnley}, S.~B. and {Mumma}, M.~J. and {Boissier}, J. and {Crovisier}, J. and {Paganini}, L. and {Kuan}, Y. -J. and {Lis}, D.~C.},
        title = "{Unusually high CO abundance of the first active interstellar comet}",
      journal = {Nature Astronomy},
         year = 2020,
        month = apr,
       volume = {4},
        pages = {861-866},
          doi = {10.1038/s41550-020-1087-2},
archivePrefix = {arXiv},
       eprint = {2004.09586},
 primaryClass = {astro-ph.EP},
       adsurl = {https://ui.adsabs.harvard.edu/abs/2020NatAs...4..861C}
}

@ARTICLE{seligman2025_3I,
       author = {{Seligman}, Darryl Z. and {Micheli}, Marco and {Farnocchia}, Davide and {Denneau}, Larry and {Noonan}, John W. and {Hsieh}, Henry H. and {Santana-Ros}, Toni and {Tonry}, John and {Auchettl}, Katie and {Conversi}, Luca and {Devog{\`e}le}, Maxime and {Faggioli}, Laura and {Feinstein}, Adina D. and {Fenucci}, Marco and {Ferrais}, Marin and {Frincke}, Tessa and {Gillon}, Michael and {Hainaut}, Olivier R. and {Hart}, Kyle and {Hoffman}, Andrew and {Holt}, Carrie E. and {Hoogendam}, Willem B. and {Huber}, Mark E. and {Jehin}, Emmanuel and {Kareta}, Theodore and {Keane}, Jacqueline V. and {Kelley}, Michael S.~P. and {Lister}, Tim and {Mandt}, Kathleen and {Manfroid}, Jean and {Mar{\v{c}}eta}, Du{\v{s}}an and {Meech}, Karen J. and {Amine Miftah}, Mohamed and {Morgan}, Marvin and {Oca{\~n}a}, Francisco and {Pe{\~n}a-Asensio}, Eloy and {Shappee}, Benjamin J. and {Siverd}, Robert J. and {Taylor}, Aster G. and {Tucker}, Michael A. and {Wainscoat}, Richard and {Weryk}, Robert and {Wray}, James J. and {Yaginuma}, Atsuhiro and {Yang}, Bin and {Ye}, Quanzhi and {Zhang}, Qicheng},
        title = "{Discovery and Preliminary Characterization of a Third Interstellar Object: 3I/ATLAS}",
      journal = {\apjl},
         year = 2025,
        month = aug,
       volume = {989},
       number = {2},
          eid = {L36},
        pages = {L36},
          doi = {10.3847/2041-8213/adf49a},
archivePrefix = {arXiv},
       eprint = {2507.02757},
 primaryClass = {astro-ph.EP},
       adsurl = {https://ui.adsabs.harvard.edu/abs/2025ApJ...989L..36S}
}

@ARTICLE{opitom_3I,
       author = {{Opitom}, Cyrielle and {Snodgrass}, Colin and {Jehin}, Emmanuel and {Bannister}, Michele T. and {Bufanda}, Erica and {Deam}, Sophie E. and {Dorsey}, Rosemary and {Ferrais}, Marin and {Hmiddouch}, Said and {Knight}, Matthew M. and {Kokotanekova}, Rosita and {Leicester}, Brayden and {Marsset}, Micha{\"e}l and {Murphy}, Brian and {Okoth}, Vincent and {Ridden-Harper}, Ryan and {Vander Donckt}, Mathieu and {Ferellec}, L{\'e}a and {Hutsemekers}, Damien and {Lippi}, Manuela and {Manfroid}, Jean and {Benkhaldoun}, Zouhair},
        title = "{Snapshot of a new interstellar comet: 3I/ATLAS has a red and featureless spectrum}",
      journal = {arXiv e-prints},
         year = 2025,
        month = jul,
          eid = {arXiv:2507.05226},
        pages = {arXiv:2507.05226},
          doi = {10.48550/arXiv.2507.05226},
archivePrefix = {arXiv},
       eprint = {2507.05226},
 primaryClass = {astro-ph.EP},
       adsurl = {https://ui.adsabs.harvard.edu/abs/2025arXiv250705226O}
}

@ARTICLE{bolin2025_3I,
       author = {{Bolin}, Bryce T. and {Belyakov}, Matthew and {Fremling}, Christoffer and {Graham}, Matthew J. and {Abdelaziz}, Ahmed M. and {Elhosseiny}, Eslam and {Gray}, Candace L. and {Ingebretsen}, Carl and {Jewett}, Gracyn and {Lisse}, Carey M. and {Karpov}, Sergey and {Kilic}, Mukremin and {Ma{\v{s}}ek}, Martin and {Molham}, Mona and {Roderick}, Diana and {Takey}, Ali and {Abron}, Laura-May and {Coughlin}, Michael W. and {Hsieh}, Cheng-Han and {Noll}, Keith S. and {Wong}, Ian},
        title = "{Interstellar comet 3I/ATLAS: discovery and physical description}",
      journal = {\mnras},
         year = 2025,
        month = sep,
       volume = {542},
       number = {1},
        pages = {L139-L143},
          doi = {10.1093/mnrasl/slaf078},
archivePrefix = {arXiv},
       eprint = {2507.05252},
 primaryClass = {astro-ph.EP},
       adsurl = {https://ui.adsabs.harvard.edu/abs/2025MNRAS.542L.139B}
}

@misc{hopkins2025differentstar3iatlascontext,
      title={From a Different Star: 3I/ATLAS in the context of the \={O}tautahi-Oxford interstellar object population model}, 
      author={Matthew J. Hopkins and Rosemary C. Dorsey and John C. Forbes and Michele T. Bannister and Chris J. Lintott and Brayden Leicester},
      year={2025},
      eprint={2507.05318},
      archivePrefix={arXiv},
      primaryClass={astro-ph.EP},
      url={https://arxiv.org/abs/2507.05318}, 
}

@ARTICLE{hopkins1,
       author = {{Hopkins}, Matthew J. and {Bannister}, Michele T. and {Lintott}, Chris},
        title = "{Predicting Interstellar Object Chemodynamics with Gaia}",
      journal = {\aj},
         year = 2025,
        month = feb,
       volume = {169},
       number = {2},
          eid = {78},
        pages = {78},
          doi = {10.3847/1538-3881/ad9eb3},
archivePrefix = {arXiv},
       eprint = {2402.04904},
 primaryClass = {astro-ph.EP},
       adsurl = {https://ui.adsabs.harvard.edu/abs/2025AJ....169...78H}
}

@ARTICLE{kokotanekova2017_jfcrotation,
   author = {{Kokotanekova}, R. and {Snodgrass}, C. and {Lacerda}, P. and 
	{Green}, S.~F. and {Lowry}, S.~C. and {Fern{\'a}ndez}, Y.~R. and 
	{Tubiana}, C. and {Fitzsimmons}, A. and {Hsieh}, H.~H.},
    title = "{Rotation of cometary nuclei: new light curves and an update of the ensemble properties of Jupiter-family comets}",
  journal = {\mnras},
archivePrefix = "arXiv",
   eprint = {1707.02133},
 primaryClass = "astro-ph.EP",
     year = 2017,
    month = nov,
   volume = 471,
    pages = {2974-3007},
      doi = {10.1093/mnras/stx1716},
   adsurl = {http://adsabs.harvard.edu/abs/2017MNRAS.471.2974K}
}

@ARTICLE{marcetaseligman,
       author = {{Mar{\v{c}}eta}, Du{\v{s}}an and {Seligman}, Darryl Z.},
        title = "{Synthetic Detections of Interstellar Objects with the Rubin Observatory Legacy Survey of Space and Time}",
      journal = {\psj},
         year = 2023,
        month = dec,
       volume = {4},
       number = {12},
          eid = {230},
        pages = {230},
          doi = {10.3847/PSJ/ad08c1},
archivePrefix = {arXiv},
       eprint = {2310.17575},
 primaryClass = {astro-ph.EP},
       adsurl = {https://ui.adsabs.harvard.edu/abs/2023PSJ.....4..230M}
}

@misc{loeb2025commentdiscoverypreliminarycharacterization,
      title={Comment on "Discovery and Preliminary Characterization of a Third Interstellar Object: 3I/ATLAS" [arXiv:2507.02757]}, 
      author={Abraham Loeb},
      year={2025},
      eprint={2507.05881},
      archivePrefix={arXiv},
      primaryClass={astro-ph.EP},
      url={https://arxiv.org/abs/2507.05881}, 
}

@ARTICLE{marcus2007_cometphasefunction,
       author = {{Marcus}, Joseph N.},
        title = "{Forward-Scattering Enhancement of Comet Brightness. I. Background and Model}",
      journal = {International Comet Quarterly},
         year = 2007,
        month = apr,
       volume = {29},
        pages = {39-66},
       adsurl = {https://ui.adsabs.harvard.edu/abs/2007ICQ....29...39M}
}

@ARTICLE{luu1992_neoprofiles,
       author = {{Luu}, Jane X. and {Jewitt}, David C.},
        title = "{High resolution surface brightness profiles of near-earth asteroids}",
      journal = {\icarus},
         year = 1992,
        month = jun,
       volume = {97},
       number = {2},
        pages = {276-287},
          doi = {10.1016/0019-1035(92)90134-S},
       adsurl = {https://ui.adsabs.harvard.edu/abs/1992Icar...97..276L}
}

@ARTICLE{hsieh2005_phaethon,
   author = {{Hsieh}, H.~H. and {Jewitt}, D.},
    title = "{Search for Activity in 3200 Phaethon}",
  journal = {\apj},
     year = 2005,
    month = may,
   volume = 624,
    pages = {1093-1096},
      doi = {10.1086/429250},
   adsurl = {http://adsabs.harvard.edu/abs/2005ApJ...624.1093H}
}

@ARTICLE{astroqueryginsburg,
       author = {{Ginsburg}, Adam and {Sip{\H{o}}cz}, Brigitta M. and {Brasseur}, C.~E. and {Cowperthwaite}, Philip S. and {Craig}, Matthew W. and {Deil}, Christoph and {Guillochon}, James and {Guzman}, Giannina and {Liedtke}, Simon and {Lian Lim}, Pey and {Lockhart}, Kelly E. and {Mommert}, Michael and {Morris}, Brett M. and {Norman}, Henrik and {Parikh}, Madhura and {Persson}, Magnus V. and {Robitaille}, Thomas P. and {Segovia}, Juan-Carlos and {Singer}, Leo P. and {Tollerud}, Erik J. and {de Val-Borro}, Miguel and {Valtchanov}, Ivan and {Woillez}, Julien and {Astroquery Collaboration} and {a subset of astropy Collaboration}},
        title = "{astroquery: An Astronomical Web-querying Package in Python}",
      journal = {\aj},
         year = 2019,
        month = mar,
       volume = {157},
       number = {3},
          eid = {98},
        pages = {98},
          doi = {10.3847/1538-3881/aafc33},
archivePrefix = {arXiv},
       eprint = {1901.04520},
 primaryClass = {astro-ph.IM},
       adsurl = {https://ui.adsabs.harvard.edu/abs/2019AJ....157...98G}
}

@ARTICLE{finson1968_cometdustmodeling1,
   author = {{Finson}, M.~J. and {Probstein}, R.~F.},
    title = "{A theory of dust comets. I. Model and equations}",
  journal = {\apj},
     year = 1968,
    month = oct,
   volume = 154,
    pages = {327-352},
      doi = {10.1086/149761},
   adsurl = {http://adsabs.harvard.edu/abs/1968ApJ...154..327F}
}

@ARTICLE{finson1968_cometdustmodeling2,
   author = {{Finson}, M.~L. and {Probstein}, R.~F.},
    title = "{A theory of dust comets. II. Results for Comet Arend-Roland}",
  journal = {\apj},
     year = 1968,
    month = oct,
   volume = 154,
    pages = {353-380},
      doi = {10.1086/149762},
   adsurl = {http://adsabs.harvard.edu/abs/1968ApJ...154..353F}
}

@ARTICLE{mommert2019_sbpy,
       author = {{Mommert}, Michael and {Kelley}, Michael and {de Val-Borro}, Miguel and {Li}, Jian-Yang and {Guzman}, Giannina and {Sip{\H{o}}cz}, Brigitta and {{\v{D}}urech}, Josef and {Granvik}, Mikael and {Grundy}, Will and {Moskovitz}, Nick and {Penttil{\"a}}, Antti and {Samarasinha}, Nalin},
        title = "{sbpy: A Python module for small-body planetary astronomy}",
      journal = {The Journal of Open Source Software},
         year = 2019,
        month = aug,
       volume = {4},
       number = {38},
          eid = {1426},
        pages = {1426},
          doi = {10.21105/joss.01426},
       adsurl = {https://ui.adsabs.harvard.edu/abs/2019JOSS....4.1426M}
}

@MISC{Jewitt2025,
  author = {{Jewitt}, D. and {Luu}, J.},
  title = "{Interstellar Interloper C/2025 N1 is Active}",
  howpublished = {The Astronomer's Telegram, No. 17263},
  year = 2025,
  note = {Published 3 Jul 2025, 13:40 UT},
  url = {https://www.astronomerstelegram.org/?read=17263}
}

@MISC{Alarcon2025,
  author = {{Alarcon}, M.~R. and {Serra-Ricart}, M. and {Licandro}, J. and {Guerra Arencibia}, S. and {Ruiz Cejudo}, I. and {Trujillo}, I.},
  title = "{Deep g'-band Imaging of Interstellar Comet 3I/ATLAS from the Two-meter Twin Telescope (TTT)}",
  howpublished = {The Astronomer's Telegram, No. 17264},
  year = 2025,
  note = {Published 3 Jul 2025, 18:33 UT},
  url = {https://www.astronomerstelegram.org/?read=17264}
}

@ARTICLE{Merritt2025,
       author = {{Merritt}, Stephanie R. and {Fedorets}, Grigori and {Schwamb}, Megan E. and {Cornwall}, Samuel and {Bernardinelli}, Pedro H. and {Juric}, Mario and {Holman}, Matthew J. and {Kurlander}, Jacob A. and {Eggl}, Siegfried and {Oldag}, Drew and {West}, Maxine and {Kubica}, Jeremy and {Murtagh}, Joseph and {Jones}, R. Lynne and {Yoachim}, Peter and {Lyttle}, Ryan R. and {Kelley}, Michael S.~P. and {Moeyens}, Joachim and {Kiker}, Kathleen and {Naidu}, Shantanu P. and {Snodgrass}, Colin and {Matthews}, Shannon M. and {Chandler}, Colin Orion},
        title = "{Sorcha: A Solar System Survey Simulator for the Legacy Survey of Space and Time}",
      journal = {arXiv e-prints},
         year = 2025,
        month = jun,
          eid = {arXiv:2506.02804},
        pages = {arXiv:2506.02804},
          doi = {10.48550/arXiv.2506.02804},
archivePrefix = {arXiv},
       eprint = {2506.02804},
 primaryClass = {astro-ph.EP},
       adsurl = {https://ui.adsabs.harvard.edu/abs/2025arXiv250602804M}
}

@ARTICLE{petit2023_hotkbos,
       author = {{Petit}, Jean-Marc and {Gladman}, Brett and {Kavelaars}, J.~J. and {Bannister}, Michele T. and {Alexandersen}, Mike and {Volk}, Kathryn and {Chen}, Ying-Tung},
        title = "{The Hot Main Kuiper Belt Size Distribution from OSSOS}",
      journal = {\apjl},
         year = 2023,
        month = apr,
       volume = {947},
       number = {1},
          eid = {L4},
        pages = {L4},
          doi = {10.3847/2041-8213/acc525},
archivePrefix = {arXiv},
       eprint = {2309.01530},
 primaryClass = {astro-ph.EP},
       adsurl = {https://ui.adsabs.harvard.edu/abs/2023ApJ...947L...4P}
}

@ARTICLE{Holman2025,
       author = {{Holman}, Matthew J. and {Bernardinelli}, Pedro H. and {Schwamb}, Megan E. and {Juri{\'c}}, Mario and {Oldag}, Drew and {West}, Maxine and {Napier}, Kevin J. and {Merritt}, Stephanie R. and {Fedorets}, Grigori and {Cornwall}, Samuel and {Kurlander}, Jacob A. and {Eggl}, Siegfried and {Kubica}, Jeremy and {Kiker}, Kathleen and {Murtagh}, Joseph and {Naidu}, Shantanu P. and {Chandler}, Colin Orion},
        title = "{Sorcha: Optimized Solar System Ephemeris Generation}",
      journal = {arXiv e-prints},
         year = 2025,
        month = jun,
          eid = {arXiv:2506.02140},
        pages = {arXiv:2506.02140},
          doi = {10.48550/arXiv.2506.02140},
archivePrefix = {arXiv},
       eprint = {2506.02140},
 primaryClass = {astro-ph.EP},
       adsurl = {https://ui.adsabs.harvard.edu/abs/2025arXiv250602140H}
}

@article{Bertin1996,
  title = {{{SExtractor}}: {{Software}} for Source Extraction},
  author = {Bertin, E. and Arnouts, S.},
  year = {1996},
  month = jun,
  journal = {Astronomy and Astrophysics Supplement Series},
  volume = {117},
  number = {2},
  pages = {393--404},
  issn = {0365-0138},
  doi = {10.1051/aas:1996164},
  urldate = {2020-02-04}
}

@ARTICLE{Bannister:2017,
       author = {{Bannister}, Michele T. and {Schwamb}, Megan E. and {Fraser}, Wesley C.
        and {Marsset}, Michael and {Fitzsimmons}, Alan and {Benecchi},
        Susan D. and {Lacerda}, Pedro and {Pike}, Rosemary E. and
        {Kavelaars}, J.~J. and {Smith}, Adam B. and {Stewart}, Sunny O.
        and {Wang}, Shiang-Yu and {Lehner}, Matthew J.},
        title = "{Col-OSSOS: Colors of the Interstellar Planetesimal
        1I/{\textquoteleft}Oumuamua}",
      journal = {\apj},
         year = 2017,
        month = Dec,
       volume = {851},
          eid = {L38},
        pages = {L38},
          doi = {10.3847/2041-8213/aaa07c},
 primaryClass = {astro-ph.EP},
       adsurl = {https://ui.adsabs.harvard.edu/#abs/2017ApJ...851L..38B}
}

@ARTICLE{Deam:2025,
       author = {{Deam}, Sophie E. and {Bannister}, Michele T. and {Opitom}, Cyrielle and {Knight}, Matthew M. and {Ridden-Harper}, Ryan and {Seligman}, Darryl Z. and {Fitzsimmons}, Alan and {Guilbert-Lepoutre}, Aur{\'e}lie and {Jehin}, Emmanuel and {Jorda}, Laurent and {Marsset}, Michael and {Moulane}, Youssef and {Rousselot}, Philippe and {Vernazza}, Pierre and {Yang}, Bin},
        title = "{A portrait throughout perihelion of the NH$_2$-rich interstellar comet 2I/Borisov}",
      journal = {arXiv e-prints},
         year = 2025,
        month = jul,
          eid = {arXiv:2507.05051},
        pages = {arXiv:2507.05051},
          doi = {10.48550/arXiv.2507.05051},
archivePrefix = {arXiv},
       eprint = {2507.05051},
 primaryClass = {astro-ph.EP},
       adsurl = {https://ui.adsabs.harvard.edu/abs/2025arXiv250705051D}
}

@article{chandlerActiveAsteroidsCitizen2024,
  title = {The {{Active Asteroids Citizen Science Program}}: {{Overview}} and {{First Results}}},
  shorttitle = {The {{Active Asteroids Citizen Science Program}}},
  author = {Chandler, Colin Orion and Trujillo, Chadwick A. and Oldroyd, William J. and Kueny, Jay K. and Burris, William A. and Hsieh, Henry H. and DeSpain, Jarod A. and Sedaghat, Nima and Sheppard, Scott S. and Farrell, Kennedy A. and Trilling, David E. and Gustafsson, Annika and Magbanua, Mark Jesus Mendoza and Mazzucato, Michele T. and Bosch, Milton K. D. and {Shaw-Diaz}, Tiffany and Gonano, Virgilio and Lamperti, Al and {da Silva Campos}, Jos{\'e} A. and Goodwin, Brian L. and Terentev, Ivan A. and Dukes, Charles J. A. and Deen, Sam},
  year = {2024},
  month = apr,
  journal = {The Astronomical Journal},
  volume = {167},
  pages = {156},
  issn = {0004-6256},
  doi = {10.3847/1538-3881/ad1de2},
  urldate = {2024-03-21}
}

@ARTICLE{zogy,
       author = {{Zackay}, Barak and {Ofek}, Eran O. and {Gal-Yam}, Avishay},
        title = "{Proper Image Subtraction{\textemdash}Optimal Transient Detection, Photometry, and Hypothesis Testing}",
      journal = {\apj},
         year = 2016,
        month = oct,
       volume = {830},
       number = {1},
          eid = {27},
        pages = {27},
          doi = {10.3847/0004-637X/830/1/27},
archivePrefix = {arXiv},
       eprint = {1601.02655},
 primaryClass = {astro-ph.IM},
       adsurl = {https://ui.adsabs.harvard.edu/abs/2016ApJ...830...27Z}
}

@ARTICLE{schleicher1998_halley,
       author = {{Schleicher}, David G. and {Millis}, Robert L. and {Birch}, Peter V.},
        title = "{Narrowband Photometry of Comet P/Halley: Variation with Heliocentric Distance, Season, and Solar Phase Angle}",
      journal = {\icarus},
         year = "1998",
        month = "Apr",
       volume = {132},
       number = {2},
        pages = {397-417},
          doi = {10.1006/icar.1997.5902},
       adsurl = {https://ui.adsabs.harvard.edu/abs/1998Icar..132..397S}
}

@ARTICLE{schleicher2011_sw3,
       author = {{Schleicher}, David G. and {Bair}, Allison N.},
        title = "{The Composition of the Interior of Comet 73P/Schwassmann-Wachmann 3: Results from Narrowband Photometry of Multiple Components}",
      journal = {\aj},
         year = "2011",
        month = "Jun",
       volume = {141},
       number = {6},
          eid = {177},
        pages = {177},
          doi = {10.1088/0004-6256/141/6/177},
       adsurl = {https://ui.adsabs.harvard.edu/abs/2011AJ....141..177S}
}

@ARTICLE{sedaghat,
       author = {{Sedaghat}, Nima and {Mahabal}, Ashish},
        title = "{Effective image differencing with convolutional neural networks for real-time transient hunting}",
      journal = {\mnras},
         year = 2018,
        month = jun,
       volume = {476},
       number = {4},
        pages = {5365-5376},
          doi = {10.1093/mnras/sty613},
archivePrefix = {arXiv},
       eprint = {1710.01422},
 primaryClass = {astro-ph.IM},
       adsurl = {https://ui.adsabs.harvard.edu/abs/2018MNRAS.476.5365S}
}

@ARTICLE{alardlupton,
       author = {{Alard}, C. and {Lupton}, Robert H.},
        title = "{A Method for Optimal Image Subtraction}",
      journal = {\apj},
         year = 1998,
        month = aug,
       volume = {503},
       number = {1},
        pages = {325-331},
          doi = {10.1086/305984},
archivePrefix = {arXiv},
       eprint = {astro-ph/9712287},
 primaryClass = {astro-ph},
       adsurl = {https://ui.adsabs.harvard.edu/abs/1998ApJ...503..325A}
}

@ARTICLE{bernardinelli2023photometry,
       author = {{Bernardinelli}, Pedro H. and {Bernstein}, Gary M. and {Jindal}, Nicholas and {Abbott}, T.~M.~C. and {Aguena}, M. and {Alves}, O. and {Andrade-Oliveira}, F. and {Annis}, J. and {Bacon}, D. and {Bertin}, E. and {Brooks}, D. and {Burke}, D.~L. and {Carnero Rosell}, A. and {Carrasco Kind}, M. and {Carretero}, J. and {da Costa}, L.~N. and {Pereira}, M.~E.~S. and {Davis}, T.~M. and {Desai}, S. and {Diehl}, H.~T. and {Doel}, P. and {Everett}, S. and {Ferrero}, I. and {Friedel}, D. and {Frieman}, J. and {Garc{\'\i}a-Bellido}, J. and {Giannini}, G. and {Gruen}, D. and {Herner}, K. and {Hinton}, S.~R. and {Hollowood}, D.~L. and {Honscheid}, K. and {James}, D.~J. and {Kuehn}, K. and {Marshall}, J.~L. and {Mena-Fern{\'a}ndez}, J. and {Menanteau}, F. and {Miquel}, R. and {Ogando}, R.~L.~C. and {Palmese}, A. and {Pieres}, A. and {Plazas Malag{\'o}n}, A.~A. and {Raveri}, M. and {Sanchez}, E. and {Sevilla-Noarbe}, I. and {Smith}, M. and {Suchyta}, E. and {Swanson}, M.~E.~C. and {Tarle}, G. and {To}, C. and {Walker}, A.~R. and {Wiseman}, P. and {Zhang}, Y. and {DES Collaboration}},
        title = "{Photometry of Outer Solar System Objects from the Dark Energy Survey. I. Photometric Methods, Light-curve Distributions, and Trans-Neptunian Binaries}",
      journal = {\apjs},
         year = 2023,
        month = nov,
       volume = {269},
       number = {1},
          eid = {18},
        pages = {18},
          doi = {10.3847/1538-4365/acf6bf},
archivePrefix = {arXiv},
       eprint = {2304.03017},
 primaryClass = {astro-ph.EP},
       adsurl = {https://ui.adsabs.harvard.edu/abs/2023ApJS..269...18B}
}

@ARTICLE{Williams2017,
  author = {{Williams}, G.~V. and {Sato}, H. and {Sarneczky}, K. and {et~al.}},
  title = "{Minor Planets 2017 SN 33 and 2017 U1}",
  journal = {Central Bureau Electronic Telegrams},
  year = 2017,
  number = 4450,
  pages = 1
}

@misc{RTN-095,
    author = "{Vera C. Rubin Observatory}",
    title = "{The Vera C. Rubin Observatory Data Preview 1}",
    institution = "{NSF-DOE Vera C. Rubin Observatory}",
    year = "2025",
    month = "July",
    handle = "RTN-095",
    type = "{Technical Note}",
    number = "RTN-095",
    doi = "10.71929/rubin/2570536",
    url = "https://rtn-095.lsst.io/"
}

@misc{taylor2025kinematicage3iatlasimplications,
      title={The Kinematic Age of 3I/ATLAS and its Implications for Early Planet Formation}, 
      author={Aster G. Taylor and Darryl Z. Seligman},
      year={2025},
      eprint={2507.08111},
      archivePrefix={arXiv},
      primaryClass={astro-ph.EP},
      url={https://arxiv.org/abs/2507.08111}, 
}

@TechReport{SITCOMTN-005,
    author = "Claver, Chuck and Bauer, Amanda and Bechtol, Keith and Bellm, Eric and Blum, Robert and Bosch, Jim and Clements, Andy and Connolly, Andrew and Guy, Leanne and Ivezi\'{c}, {\v Z}eljko and Lupton, Robert and Ritz, Steve and O'Mullane, William and Thomas, Sandrine and Drass, Holger and Jones, Lynne",
    title = "{Construction Completeness and Operations Readiness Criteria}",
    institution = "{NSF-DOE Vera C. Rubin Observatory}",
    year = "2025",
    month = "June",
    handle = "SITCOMTN-005",
    type = "{Commissioning Technical Note}",
    number = "SITCOMTN-005",
    url = "https://sitcomtn-005.lsst.io/"
}

@INPROCEEDINGS{lam2015numba,
       author = {{Lam}, Siu Kwan and {Pitrou}, Antoine and {Seibert}, Stanley},
        title = "{Numba: A LLVM-based Python JIT Compiler}",
    booktitle = {Proc. Second Workshop on the LLVM Compiler Infrastructure in HPC},
         year = 2015,
        month = nov,
        pages = {1-6},
          doi = {10.1145/2833157.2833162},
       adsurl = {https://ui.adsabs.harvard.edu/abs/2015llvm.confE...1L}
}

@article{Bosch_HSC,
       author = {{Bosch}, James and {Armstrong}, Robert and {Bickerton}, Steven and
         {Furusawa}, Hisanori and {Ikeda}, Hiroyuki and {Koike}, Michitaro and
         {Lupton}, Robert and {Mineo}, Sogo and {Price}, Paul and
         {Takata}, Tadafumi and {Tanaka}, Masayuki and {Yasuda}, Naoki and
         {AlSayyad}, Yusra and {Becker}, Andrew C. and {Coulton}, William and
         {Coupon}, Jean and {Garmilla}, Jose and {Huang}, Song and
         {Krughoff}, K. Simon and {Lang}, Dustin and {Leauthaud}, Alexie and
         {Lim}, Kian-Tat and {Lust}, Nate B. and {MacArthur}, Lauren A. and {Mand
        elbaum}, Rachel and {Miyatake}, Hironao and {Miyazaki}, Satoshi and
         {Murata}, Ryoma and {More}, Surhud and {Okura}, Yuki and
         {Owen}, Russell and {Swinbank}, John D. and {Strauss}, Michael A. and
         {Yamada}, Yoshihiko and {Yamanoi}, Hitomi},
        title = "{The Hyper Suprime-Cam software pipeline}",
      journal = {\pasj},
         year = "2018",
        month = Jan,
       volume = {70},
          eid = {S5},
        pages = {S5},
          doi = {10.1093/pasj/psx080},
archivePrefix = {arXiv},
       eprint = {1705.06766},
 primaryClass = {astro-ph.IM},
       adsurl = {https://ui.adsabs.harvard.edu/abs/2018PASJ...70S...5B}
}

@ARTICLE{Rodionov1998_B2,
       author = {{Rodionov}, A.~V. and {Jorda}, L. and {Jones}, G.~H. and {Crifo}, J.~F. and {Colas}, F. and {Lecacheux}, J.},
        title = "{Comet Hyakutake Gas Arcs: First Observational Evidence of Standing Shock Waves in a Cometary Coma}",
      journal = {\icarus},
         year = 1998,
        month = dec,
       volume = {136},
       number = {2},
        pages = {232-267},
          doi = {10.1006/icar.1998.6010},
       adsurl = {https://ui.adsabs.harvard.edu/abs/1998Icar..136..232R}
}

@ARTICLE{shi2018_67Pconcavities,
       author = {{Shi}, X. and {Hu}, X. and {Mottola}, S. and {Sierks}, H. and {Keller}, H.~U. and {Rose}, M. and {G{\"u}ttler}, C. and {Fulle}, M. and {Fornasier}, S. and {Agarwal}, J. and {Pajola}, M. and {Tubiana}, C. and {Bodewits}, D. and {Barbieri}, C. and {Lamy}, P.~L. and {Rodrigo}, R. and {Koschny}, D. and {Barucci}, M.~A. and {Bertaux}, J. -L. and {Bertini}, I. and {Boudreault}, S. and {Cremonese}, G. and {Da Deppo}, V. and {Davidsson}, B. and {Debei}, S. and {De Cecco}, M. and {Deller}, J. and {Groussin}, O. and {Guti{\'e}rrez}, P.~J. and {Hviid}, S.~F. and {Ip}, W. -H. and {Jorda}, L. and {Knollenberg}, J. and {Kovacs}, G. and {Kramm}, J. -R. and {K{\"u}hrt}, E. and {K{\"u}ppers}, M. and {Lara}, L.~M. and {Lazzarin}, M. and {Lopez-Moreno}, J.~J. and {Marzari}, F. and {Naletto}, G. and {Oklay}, N. and {Toth}, I. and {Vincent}, J. -B.},
        title = "{Coma morphology of comet 67P controlled by insolation over irregular nucleus}",
      journal = {Nature Astronomy},
         year = 2018,
        month = may,
       volume = {2},
        pages = {562-567},
          doi = {10.1038/s41550-018-0481-5},
       adsurl = {https://ui.adsabs.harvard.edu/abs/2018NatAs...2..562S}
}

@ARTICLE{leon2020,
       author = {{de Le{\'o}n}, J. and {Licandro}, J. and {de la Fuente Marcos}, C. and {de la Fuente Marcos}, R. and {Lara}, L.~M. and {Moreno}, F. and {Pinilla-Alonso}, N. and {Serra-Ricart}, M. and {De Pr{\'a}}, M. and {Tozzi}, G.~P. and {Souza-Feliciano}, A.~C. and {Popescu}, M. and {Scarpa}, R. and {Font Serra}, J. and {Geier}, S. and {Lorenzi}, V. and {Harutyunyan}, A. and {Cabrera-Lavers}, A.},
        title = "{Visible and near-infrared observations of interstellar comet 2I/Borisov with the 10.4-m GTC and the 3.6-m TNG telescopes}",
      journal = {\mnras},
         year = 2020,
        month = jun,
       volume = {495},
       number = {2},
        pages = {2053-2062},
          doi = {10.1093/mnras/staa1190},
archivePrefix = {arXiv},
       eprint = {2005.00786},
 primaryClass = {astro-ph.EP},
       adsurl = {https://ui.adsabs.harvard.edu/abs/2020MNRAS.495.2053D}
}

@ARTICLE{leon2019,
       author = {{de Le{\'o}n}, Julia and {Licandro}, Javier and {Serra-Ricart}, Miquel and {Cabrera-Lavers}, Antonio and {Font Serra}, Joan and {Scarpa}, Riccardo and {de la Fuente Marcos}, Carlos and {de la Fuente Marcos}, Ra{\'u}l},
        title = "{Interstellar Visitors: A Physical Characterization of Comet C/2019 Q4 (Borisov) with OSIRIS at the 10.4{\,}m GTC}",
      journal = {Research Notes of the American Astronomical Society},
         year = 2019,
        month = sep,
       volume = {3},
       number = {9},
          eid = {131},
        pages = {131},
          doi = {10.3847/2515-5172/ab449c},
       adsurl = {https://ui.adsabs.harvard.edu/abs/2019RNAAS...3..131D}
}

@ARTICLE{prodan2024,
       author = {{Prodan}, George P. and {Popescu}, Marcel and {Licandro}, Javier and {Akhlaghi}, Mohammad and {de Le{\'o}n}, Julia and {Tatsumi}, Eri and {Pastrav}, Bogdan Adrian and {Hibbert}, Jacob M. and {V{\v{a}}duvescu}, Ovidiu and {Simion}, Nicolae Gabriel and {Pall{\'e}}, Enric and {Narita}, Norio and {Fukui}, Akihiko and {Murgas}, Felipe},
        title = "{Pre-perihelion monitoring of interstellar comet 2I/Borisov}",
      journal = {\mnras},
         year = 2024,
        month = apr,
       volume = {529},
       number = {4},
        pages = {3521-3535},
          doi = {10.1093/mnras/stae539},
archivePrefix = {arXiv},
       eprint = {2402.12428},
 primaryClass = {astro-ph.EP},
       adsurl = {https://ui.adsabs.harvard.edu/abs/2024MNRAS.529.3521P}
}

@ARTICLE{alvarezcandal2006,
       author = {{Alvarez-Candal}, A. and {Licandro}, J.},
        title = "{The size distribution of asteroids in cometary orbits and related populations}",
      journal = {\aap},
         year = 2006,
        month = nov,
       volume = {458},
       number = {3},
        pages = {1007-1011},
          doi = {10.1051/0004-6361:20064971},
       adsurl = {https://ui.adsabs.harvard.edu/abs/2006A&A...458.1007A}
}

@INPROCEEDINGS{2014SPIE.9149E..0BJ,
   author = {{Jones}, R.~L. and {Yoachim}, P. and {Chandrasekharan}, S. and {Connolly}, A.~J. and {Cook}, K.~H. and {Ivezic}, {\v Z}. and  {Krughoff}, K.~S. and {Petry}, C. and {Ridgway}, S.~T.},
    title = "{The LSST metrics analysis framework (MAF)}",
    series = {Society of Photo-Optical Instrumentation Engineers (SPIE) Conference Series},
    booktitle = {Observatory Operations: Strategies, Processes, and Systems V},
    editor = {Alison B. Peck and Chris R. Benn and Robert L. Seaman},
    year = 2014,
    volume = 9149,
    eid = {91490B},
    pages = {0},
    doi = {10.1117/12.2056835},
    adsurl = {http://adsabs.harvard.edu/abs/2014SPIE.9149E..0BJ}
}

@software{peter_yoachim_2025_15368965,
  author       = {Peter Yoachim and
                  Lynne Jones and
                  Eric H. Neilsen, Jr. and
                  Tiago and
                  John Parejko and
                  Eric Bellm and
                  Rachel Street and
                  Jeff Carlin and
                  Humna and
                  Matthew R. Becker and
                  pgris and
                  erykoff and
                  Loredana Prisinzano and
                  Erik Dennihy and
                  Giovanni A. Gollotti and
                  Jonathan Sick and
                  lmptc and
                  LI and
                  Natasha Abrams and
                  Roberto J. Assef and
                  Leanne Guy and
                  Ross and
                  Katja Bricman and
                  Johan Bregeon and
                  Kian-Tat Lim and
                  Michael Kelley and
                  Igor Andreoni},
  title        = {lsst/rubin\_sim: v2.2.4},
  month        = may,
  year         = 2025,
  publisher    = {Zenodo},
  version      = {v2.2.4},
  doi          = {10.5281/zenodo.15368965},
  url          = {https://doi.org/10.5281/zenodo.15368965},
  swhid        = {swh:1:dir:5af5919756f84d2bf9470ccb114582e3aa92bfe3
                   ;origin=https://doi.org/10.5281/zenodo.7087822;vis
                   it=swh:1:snp:2e8977767b6c389a3a28789861b70c066f73b
                   734;anchor=swh:1:rel:d906fa8a449db09c551889053d5f4
                   9bba84e1646;path=lsst-rubin\_sim-6eda8b6
                  },
}

@ARTICLE{2019AJ....157..151N,
       author = {{Naghib}, Elahesadat and {Yoachim}, Peter and {Vanderbei}, Robert J. and {Connolly}, Andrew J. and {Jones}, R. Lynne},
        title = "{A Framework for Telescope Schedulers: With Applications to the Large Synoptic Survey Telescope}",
      journal = {\aj},
         year = 2019,
        month = apr,
       volume = {157},
       number = {4},
          eid = {151},
        pages = {151},
          doi = {10.3847/1538-3881/aafece},
archivePrefix = {arXiv},
       eprint = {1810.04815},
 primaryClass = {astro-ph.IM},
       adsurl = {https://ui.adsabs.harvard.edu/abs/2019AJ....157..151N}
}

@software{peter_yoachim_2025_15742506,
  author       = {Peter Yoachim and
                  Lynne Jones and
                  Eric H. Neilsen, Jr. and
                  Sean MacBride and
                  Keith Bechtol and
                  Matthew R. Becker and
                  Ross and
                  Humna},
  title        = {lsst/rubin\_scheduler: v3.11.0},
  month        = jun,
  year         = 2025,
  publisher    = {Zenodo},
  version      = {v3.11.0},
  doi          = {10.5281/zenodo.15742506},
  url          = {https://doi.org/10.5281/zenodo.15742506},
  swhid        = {swh:1:dir:cdec70472ddec70b436c7a6e1674ead0fdb3de91
                   ;origin=https://doi.org/10.5281/zenodo.10076770;vi
                   sit=swh:1:snp:4c836ce7f29e643d8b0bb8217c59c3f50014
                   b4d1;anchor=swh:1:rel:9fecbc40e0cf32508707112ea5ec
                   1c0a22303deb;path=lsst-rubin\_scheduler-2059e7d
                  },
}

@misc{jones_2025_15128504,
  author       = {Jones, Lynne and
                  Bianco, Federica Bettina and
                  Yoachim, Peter and
                  Neilsen, Eric},
  title        = {https://survey-strategy.lsst.io},
  month        = apr,
  year         = 2025,
  publisher    = {Zenodo},
  version      = {v0.1.0},
  doi          = {10.5281/zenodo.15128504},
  url          = {https://doi.org/10.5281/zenodo.15128504},
}

@software{peter_yoachim_2025_14920193,
  author       = {Peter Yoachim},
  title        = {lsst-sims/sims\_featureScheduler\_runs4.3: Initial
                   Release
                  },
  month        = feb,
  year         = 2025,
  publisher    = {Zenodo},
  version      = {1.0},
  doi          = {10.5281/zenodo.14920193},
  url          = {https://doi.org/10.5281/zenodo.14920193},
  swhid        = {swh:1:dir:45ef6ed2572dcba7d2f9205ddec606c3ce93fdb1
                   ;origin=https://doi.org/10.5281/zenodo.14920192;vi
                   sit=swh:1:snp:8de4e7ee2729aaccb1ffdb76a4f36c276494
                   39e0;anchor=swh:1:rel:5ffbbd269d0a184fc7612fd4640c
                   273ada8476d1;path=lsst-sims-
                   sims\_featureScheduler\_runs4.3-c4b09b9
                  },
}

@TechReport{DMTN-227,
    author = "Lim, Kian-Tat",
    title = "{The Consolidated Database of Image Metadata}",
    institution = "{NSF-DOE Vera C. Rubin Observatory}",
    year = "2025",
    month = "April",
    handle = "DMTN-227",
    type = "{Data Management Technical Note}",
    number = "DMTN-227",
    url = "https://dmtn-227.lsst.io/"
}

@ARTICLE{ahearn1995,
       author = {{A'Hearn}, Michael F. and {Millis}, Robert C. and {Schleicher}, David O. and {Osip}, David J. and {Birch}, Peter V.},
        title = "{The ensemble properties of comets: Results from narrowband photometry of 85 comets, 1976-1992.}",
      journal = {\icarus},
         year = 1995,
        month = dec,
       volume = {118},
       number = {2},
        pages = {223-270},
          doi = {10.1006/icar.1995.1190},
       adsurl = {https://ui.adsabs.harvard.edu/abs/1995Icar..118..223A}
}

@misc{hoblittITTN009SummitTime2022,
  title = {{{ITTN-009}}: {{Summit Time Synchronization}}},
  author = {Hoblitt, Joshua and Thebo, Adrien},
  year = 2022,
  month = nov,
  urldate = {2025-11-15},
  howpublished = {https://ittn-009.lsst.io/}
}

@INPROCEEDINGS{2004DPS....36.3416T,
       author = {{Tholen}, D.~J. and {Chesley}, S.~R.},
        title = "{Groundbased Ephemeris Development Effort for StarDust Target Wild 2}",
    booktitle = {AAS/Division for Planetary Sciences Meeting Abstracts \#36},
         year = 2004,
       series = {AAS/Division for Planetary Sciences Meeting Abstracts},
       volume = {36},
        month = nov,
          eid = {34.16},
        pages = {34.16},
       adsurl = {https://ui.adsabs.harvard.edu/abs/2004DPS....36.3416T}
}

@ARTICLE{2021Icar..35814276F,
       author = {{Farnocchia}, Davide and {Bellerose}, Julie and {Bhaskaran}, Shyam and {Micheli}, Marco and {Weryk}, Robert},
        title = "{High-fidelity comet 67P ephemeris and predictions based on Rosetta data}",
      journal = {\icarus},
         year = 2021,
        month = apr,
       volume = {358},
          eid = {114276},
        pages = {114276},
          doi = {10.1016/j.icarus.2020.114276},
       adsurl = {https://ui.adsabs.harvard.edu/abs/2021Icar..35814276F}
}

@ARTICLE{2022PSJ.....3..156F,
       author = {{Farnocchia}, Davide and {Reddy}, Vishnu and {Bauer}, James M. and {Warner}, Elizabeth M. and {Micheli}, Marco and {Payne}, Matthew J. and {Farnham}, Tony and {Kelley}, Michael S. and {Balam}, David D. and {Barkov}, Anatoly P. and {Berte{\c{s}}teanu}, Daniel and {Birlan}, Mirel and {Bolin}, Bryce T. and {Brucker}, Melissa J. and {Buzzi}, Luca and {Chambers}, Kenneth C. and {Demetz}, Lukas and {Djupvik}, Anlaug A. and {Elenin}, Leonid and {Fini}, Paolo and {Flynn}, Randy and {Galli}, Gianni and {Gao}, Xing and {G{\c{e}}dek}, Marcin and {Granvik}, Mikael and {Hasubick}, Werner and {Ivanov}, Alexander L. and {Ivanov}, Viktor A. and {Ivanova}, Natalya V. and {Jaques}, Crist{\'o}v{\~a}o and {Kasikov}, Anni and {Kim}, Myung-Jin and {Lane}, David and {Lee}, Hee-Jae and {Li}, Bin and {Li}, Fan and {Lister}, Tim and {Lysenko}, Vadim E. and {Magnier}, Eugene A. and {Mahomed}, Nawaz and {McCormick}, Jennie and {Moon}, Darrel and {Nastasi}, Alessandro and {Nedelcu}, Dan A. and {Neue}, Guenther and {Petrescu}, Elisabeta and {Popescu}, Marcel and {Prosperi}, Enrico and {Reszelewski}, Rafa{\l} and {Roh}, Dong-Goo and {Romanov}, Filipp D. and {Santana-Ros}, Toni and {Schmalz}, Anastasia and {Schmalz}, Sergei and {Scotti}, James V. and {Seaman}, Robert and {Sioulas}, Nick and {Sonka}, Adrian B. and {Tholen}, David J. and {Trelia}, Madalina M. and {Wainscoat}, Richard and {Wang}, Xin and {Wells}, Guy and {Weryk}, Robert and {Yakovenko}, Nikolai A. and {Ye}, Quanzhi and {Yim}, Hong-Suh and {Zhai}, Chengxing and {Zhang}, Chen and {Zhao}, Haibin and {Zhu}, Tinglei and {{\.Z}o{\l}nowski}, Micha{\l}},
        title = "{International Asteroid Warning Network Timing Campaign: 2019 XS}",
      journal = {\psj},
         year = 2022,
        month = jul,
       volume = {3},
       number = {7},
          eid = {156},
        pages = {156},
          doi = {10.3847/PSJ/ac7224},
       adsurl = {https://ui.adsabs.harvard.edu/abs/2022PSJ.....3..156F}
}

@ARTICLE{quaia,
       author = {{Storey-Fisher}, Kate and {Hogg}, David W. and {Rix}, Hans-Walter and {Eilers}, Anna-Christina and {Fabbian}, Giulio and {Blanton}, Michael R. and {Alonso}, David},
        title = "{Quaia, the Gaia-unWISE Quasar Catalog: An All-sky Spectroscopic Quasar Sample}",
      journal = {\apj},
         year = 2024,
        month = mar,
       volume = {964},
       number = {1},
          eid = {69},
        pages = {69},
          doi = {10.3847/1538-4357/ad1328},
archivePrefix = {arXiv},
       eprint = {2306.17749},
 primaryClass = {astro-ph.GA},
       adsurl = {https://ui.adsabs.harvard.edu/abs/2024ApJ...964...69S}
}

@TechReport{RTN-011,
    author = "Guy, Leanne P. and Bechtol, Keith and Bellm, Eric and Blum, Bob and Dubois-Felsmann, Gregory P. and Graham, Melissa L.",
    title = "{Rubin Observatory Plans for an Early Science Program}",
    institution = "{NSF-DOE Vera C. Rubin Observatory}",
    year = "2025",
    month = "May",
    handle = "RTN-011",
    type = "{Rubin Technical Note}",
    number = "RTN-011",
    url = "https://rtn-011.lsst.io/"
}

@ARTICLE{jewitt2025_hst3I,
       author = {{Jewitt}, David and {Hui}, Man-To and {Mutchler}, Max and {Kim}, Yoonyoung and {Agarwal}, Jessica},
        title = "{Hubble Space Telescope Observations of the Interstellar Interloper 3I/ATLAS}",
      journal = {\apjl},
         year = 2025,
        month = sep,
       volume = {990},
       number = {1},
          eid = {L2},
        pages = {L2},
          doi = {10.3847/2041-8213/adf8d8},
archivePrefix = {arXiv},
       eprint = {2508.02934},
 primaryClass = {astro-ph.EP},
       adsurl = {https://ui.adsabs.harvard.edu/abs/2025ApJ...990L...2J}
}

@ARTICLE{manToHui2026,
       author = {{Hui}, Man-To and {Jewitt}, David and {Mutchler}, Max J. and {Agarwal}, Jessica and {Kim}, Yoonyoung},
        title = "{Nucleus and Postperihelion Activity of Interstellar Object 3I/ATLAS Observed by the Hubble Space Telescope}",
      journal = {\apjl},
         year = 2026,
        month = mar,
       volume = {999},
       number = {2},
          eid = {L37},
        pages = {L37},
          doi = {10.3847/2041-8213/ae471c},
archivePrefix = {arXiv},
       eprint = {2601.21569},
 primaryClass = {astro-ph.EP},
       adsurl = {https://ui.adsabs.harvard.edu/abs/2026ApJ...999L..37H}
}
\bibliographystyle{aasjournal}

%% This command is needed to show the entire author+affiliation list when the collaboration and author truncation commands are used.  It has to go at the end of the manuscript.
%\allauthors

%% Include this line if you are using the \added, \replaced, \deleted commands to see a summary list of all changes at the end of the article.
%\listofchanges

% \section*{Acronyms}
% \label{sec:acronyms}
% \begin{acronym}

\begin{acronym}[JSONP]\itemsep0pt % 9/11/2023 COC -- single space
\acro{2MASS}{Two Micron All Sky Survey}
\acro{AA}{active asteroid}
\acro{ACO}{asteroid on a cometary orbit}
\acro{AI}{artificial intelligence}
\acro{API}{Application Programming Interface}
\acro{APT}{Aperture Photometry Tool}
\acro{ARC}{Astrophysical Research Consortium}
\acro{ARCTIC}{Astrophysical Research Consortium Telescope Imaging Camera}
\acro{AOS}{Active Optics System}
\acro{APO}{Apache Point Observatory}
\acro{ARO}{Atmospheric Research Observatory}
\acro{AstOrb}{Asteroid Orbital Elements Database}
\acro{ASU}{Arizona Statue University}
\acro{ATLAS}{Asteroid Terrestrial-impact Last Alert System}
\acro{AURA}{Association of Universities for Research in Astronomy}
\acro{BASS}{Beijing-Arizona Sky Survey}
\acro{BLT}{Barry Lutz Telescope}
\acro{CADC}{Canadian Astronomy Data Centre}
\acro{CASU}{Cambridge Astronomy Survey Unit}
\acro{CATCH}{Comet Asteroid Telescopic Catalog Hub}
\acro{CBAT}{Central Bureau for Astronomical Telegrams}
\acro{CBET}{Central Bureau for Electronic Telegrams}
\acro{CCD}{charge-coupled device}
\acro{CEA}{Commissariat a l'Energes Atomique}
\acro{CFHT}{Canada-France-Hawaii Telescope}
\acro{CFITSIO}{C Flexible Image Transport System Input Output}
\acro{CNEOS}{Center for Near Earth Object Studies} % 9/11/2023 COC
\acro{CNRS}{Centre National de la Recherche Scientifique}
\acro{CPU}{Central Processing Unit}
\acro{CTIO}{Cerro Tololo Inter-American Observatory}
\acro{DAPNIA}{Département d'Astrophysique, de physique des Particules, de physique Nucléaire et de l'Instrumentation Associée}
\acro{DART}{Double Asteroid Redirection Test}
\acro{DDF}{Deep Drilling Field}
\acro{DDT}{Director's Discretionary Time}
\acro{DECaLS}{Dark Energy Camera Legacy Survey}
\acro{DECam}{Dark Energy Camera}
\acro{DES}{Dark Energy Survey}
\acro{DESI}{Dark Energy Spectroscopic Instrument}
\acro{DCR}{differential chromatic refraction}
\acro{DCT}{Discovery Channel Telescope}
\acro{DOE}{Department of Energy}
\acro{DR}{Data Release}
\acro{DS9}{Deep Space Nine}
\acro{ESA}{European Space Agency}
\acro{ESO}{European Southern Observatory}
\acro{ETC}{exposure time calculator}
\acro{ETH}{Eidgenössische Technische Hochschule}
\acro{FAQ}{frequently asked questions}
\acro{FITS}{Flexible Image Transport System}
\acro{FOV}{field of view}
\acro{FTN}{Faulkes Telescope North}
\acro{GEODSS}{Ground-Based Electro-Optical Deep Space Surveillance}
\acro{GIF}{Graphic Interchange Format}
\acro{GMOS}{Gemini Multi-Object Spectrograph}
\acro{GPU}{Graphics Processing Unit}
\acro{GRFP}{Graduate Research Fellowship Program}
\acro{GRSS}{Gauss-Radau Small-body Simulator }
\acro{HARVEST}{Hunting for Activity in Repositories with Vetting-Enhanced Search Techniques}
\acro{HSC}{Hyper Suprime-Cam}
\acro{HST}{Hubble Space Telescope}
\acro{IAU}{International Astronomical Union}
\acro{IMACS}{Inamori-Magellan Areal Camera and Spectrograph}
\acro{IMB}{inner Main-belt}
\acro{IMCCE}{Institut de Mécanique Céleste et de Calcul des Éphémérides}
\acro{INAF}{Istituto Nazionale di Astrofisica}
\acro{INT}{Isaac Newton Telescopes}
\acro{IP}{Internet Protocol}
\acro{IRSA}{Infrared Science Archive}
\acro{IRTF}{Infrared Telescope Facility}
\acro{ISR}{Instrumental Signature Removal}
\acro{ISO}{Interstellar Object}
\acro{ISO2}{International Organization for Standardization} % 8/29/2025 COC
\acro{ITC}{integration time calculator}
\acro{JAXA}{Japan Aerospace Exploration Agency}
\acro{JD}{Julian Date}
\acro{JFC}{Jupiter-family comet}
\acro{JPL}{Jet Propulsion Laboratory}
\acro{KBO}{Kuiper Belt object}
\acro{KOA}{Keck Observatory Archive}
\acro{KPNO}{Kitt Peak National Observatory}
\acro{LBC}{Large Binocular Camera}
%\acro{LCO}{Las Campanas Observatory}
\acro{LCO}{Las Cumbres Observatory}
\acro{LBCB}{Large Binocular Camera Blue}
\acro{LBCR}{Large Binocular Camera Red}
\acro{LBT}{Large Binocular Telescope}
\acro{LDT}{Lowell Discovery Telescope}
\acro{LINCC}{LSST Interdisciplinary Network for Collaboration and Computing}
\acro{LINEAR}{Lincoln Near-Earth Asteroid Research}
\acro{LMI}{Large Monolithic Imager}
\acro{LONEOS}{Lowell Observatory Near-Earth-Object Search}
\acro{LSST}{Legacy Survey of Space and Time}
\acro{MBC}{Main-belt comet}
\acro{MGIO}{Mount Graham International Observatory}
\acro{MITHNEOS}{MIT-Hawaii Near-Earth Object Spectroscopic Survey}
\acro{ML}{machine learning}
\acro{MMB}{middle Main-belt}
\acro{MMR}{mean-motion resonance}
\acro{MOST}{Moving Object Search Tool}
\acro{MPEC}{Minor Planet Electronic Circular} % 9/10/2024 COC
\acro{MzLS}{Mayall z-band Legacy Survey}
\acro{MPC}{Minor Planet Center}
\acro{MUSE}{Multi Unit Spectroscopic Explorer}
\acro{MuSCAT3}{Multicolor Simultaneous Camera for studying Atmospheres of Transiting exoplanets}
\acro{NAU}{Northern Arizona University}
\acro{NEA}{near-Earth asteroid}
\acro{NEAT}{Near-Earth Asteroid Tracking}
\acro{NEATM}{Near Earth Asteroid Thermal Model}
\acro{NEO}{near-Earth object}
\acro{NEOCP}{Near-Earth Object Confirmation Page}
\acro{NEOWISE}{Near-Earth Object Wide-field Infrared Survey Explorer}
\acro{NGPS}{Next Generation Palomar Spectrograph}
\acro{NIHTS}{Near-Infrared High-Throughput Spectrograph}
\acro{NOAO}{National Optical Astronomy Observatory}
\acro{NOIRLab}{National Optical and Infrared Laboratory}
\acro{NRC}{National Research Council}
\acro{OMB}{outer Main-belt}
\acro{OSIRIS-REx}{Origins, Spectral Interpretation, Resource Identification, Security, Regolith Explorer}
\acro{NSF}{National Science Foundation}
\acro{PA}{position angle}
\acro{PANSTARRS}{Panoramic Survey Telescope and Rapid Response System}
\acro{Pan-STARRS1}{Panoramic Survey Telescope and Rapid Response System}
\acro{PANSTARRS1}{Panoramic Survey Telescope and Rapid Response System}
\acro{PI}{Principal Investigator}
\acro{PNG}{Portable Network Graphics}
\acro{PSI}{Planetary Science Institute}
\acro{PSF}{point spread function}
\acro{PTF}{Palomar Transient Factory}
\acro{QH}{Quasi-Hilda}
\acro{QHA}{Quasi-Hilda Asteroid}
\acro{QHC}{Quasi-Hilda Comet}
\acro{QHO}{Quasi-Hilda Object}
\acro{RA}{Right Ascension}
\acro{REU}{Research Experiences for Undergraduates}
\acro{RMS}{root-mean-square}
\acro{RNAAS}{Research Notes of the American Astronomical Society}
\acro{SAFARI}{Searching Asteroids For Activity Revealing Indicators}
\acro{SDSS}{Sloan Digital Sky Survey}
\acro{SMOKA}{Subaru Mitaka Okayama Kiso Archive}
\acro{SAO}{Smithsonian Astrophysical Observatory}
\acro{SBDB}{Small Body Database}
\acro{SDSS DR-9}{Sloan Digital Sky Survey Data Release Nine}
\acro{SFD}{Size-Frequency Distribution}
\acro{SLAC}{Stanford Linear Accelerator Center}
\acro{SOAR}{Southern Astrophysical Research Telescope}
\acro{SNR}{signal to noise ratio}
\acro{SSOIS}{Solar System Object Information Search}
\acro{SSP}{Solar System Processing}
\acro{SQL}{Structured Query Language}
\acro{SUP}{Suprime Cam}
\acro{SV}{Science Validation}
\acro{SwRI}{Southwestern Research Institute}
\acro{TAI}{Temps Atomique International}
\acro{TAP}{Telescope Access Program}
\acro{TBTs}{Test Bed Telescopes}
\acro{TNO}{Trans-Neptunian object}
\acro{ToO}{Target of Opportunity}
\acro{TRAPPIST}{Transiting Planets and Planetesimals Small Telescope}
\acro{UA}{University of Arizona}
\acro{UT}{Universal Time}
\acro{UCSC}{University of California Santa Cruz}
\acro{UCSF}{University of California San Francisco}
\acro{VATT}{Vatican Advanced Technology Telescope}
\acro{VIA}{Virtual Institute of Astrophysics}
\acro{VIRCam}{VISTA InfraRed Camera}
\acro{VISTA}{Visible and Infrared Survey Telescope for Astronomy}
\acro{VLT}{Very Large Telescope}
\acro{VST}{Very Large Telescope (VLT) Survey Telescope}
\acro{WEB}{Water Enriched Block}
\acro{WFC}{Wide Field Camera}
\acro{WIRCam}{Wide-field Infrared Camera}
\acro{WISE}{Wide-field Infrared Survey Explorer}
\acro{WCS}{World Coordinate System}
\acro{YORP}{Yarkovsky--O'Keefe--Radzievskii--Paddack}
\acro{ZTF}{Zwicky Transient Facility}
\end{acronym}
\end{document}